\documentclass[12pt, a4paper]{article}

\usepackage[utf8]{inputenc}
\usepackage[T1]{fontenc}
\usepackage{lmodern}
\usepackage{setspace}
\usepackage{titlesec}
\usepackage{titling}
\usepackage{graphicx}
\usepackage{subcaption}
\usepackage{amssymb}
\usepackage{amsfonts}
\usepackage{amsmath}
\usepackage{amsthm}
\usepackage{mathtools}
\usepackage[colorlinks = false, hidelinks]{hyperref}
\usepackage[english]{babel}
\usepackage{setspace}
\usepackage{multirow}
\usepackage{array}
\usepackage{float}
\usepackage{bbm}
\usepackage{microtype, fancyhdr}
\usepackage{amsfonts}
\usepackage[style = authoryear, backend = biber, maxcitenames = 2]{biblatex}
\usepackage{cleveref}
\usepackage{tabularx}
\usepackage{ragged2e}
\usepackage{xcolor}
\usepackage{tikz}
\usepackage{threeparttable}
\usetikzlibrary{arrows.meta, positioning, shapes.geometric, calc}
\usepackage[letterpaper,top=2cm,bottom=2cm,left=3cm,right=3cm,marginparwidth=1.75cm]{geometry}

\definecolor{UBCred}{RGB}{170, 36, 63}
\definecolor{UBCred2}{RGB}{228, 120, 120}
\definecolor{darkgreen}{rgb}{0,0.5,0}

\usepackage{tabularray}
\usepackage{codehigh}
\usepackage[normalem]{ulem}
\UseTblrLibrary{booktabs}
\UseTblrLibrary{siunitx}

\NewTableCommand{\tinytableDefineColor}[3]{\definecolor{#1}{#2}{#3}}

\newcommand{\E}{\mathbb{E}} 

\title{\textbf{When the Fed Speaks: Dynamics and Forecasts of the Volatility Surface}}
\author{Łukasz Adamski\textsuperscript{1} \and Robert Ślepaczuk\textsuperscript{2}}
\date{\today}

\begin{document}

\maketitle
\thispagestyle{empty}

\begin{abstract}

Our primary goal is to forecast and empirically examine the evolution of the implied volatility (IV) surface, 
with particular focus on the dates of scheduled meetings of the Federal Open Market Committee (FOMC). 

Firstly, we check if IV increases before the announcement and if thes effect is stronger for short-dated, out-the-money (OTM) options in high volatility regimes. 
In the second part, we turn the focus to verifying if the ML framework can beat the benchmark random walk in forecasting this effect. 
A feature related to dates of scheduled FOMC meetings augments the model, which allows us to discover if it can learn the effect of elevated pre-announcement uncertainty.

Our contribution relies mainly on the quantitative prediction of the pre-announcement effect and the inclusion of exogenous information inside the ML framework used for the IV surface forecasting. 
It is also on of the first attempts to apply ML models directly on the IV surface without relying on dimensionality reduction.
To achieve this, we employ a convolutional two-dimensional LSTM model, which is capable of learning spatio-temporal signals in the surface. 

Our analysis reveals that the edge of the ML framework can be limited due to the noisy characteristics of the IV surface. 
Nevertheless, our study reinforces the perspective that ML models can effectively forecast the IV surface also during abnormal days. 

\end{abstract}

\vfill

\setlength{\abovedisplayskip}{0pt}
\setlength{\abovedisplayshortskip}{0pt}

\noindent\rule{\textwidth}{0.4pt}
\begin{justify}
    \footnotesize
        \textsuperscript{1}\, PhD candidate University of Warsaw, Faculty of Economic Sciences, Department of Quantitative Finance and Machine Learning, Quantitative Finance Research Group (QFRG), Warsaw, Poland. \\
        \noindent 
        \textsuperscript{2}\,Professor at University of Warsaw, Faculty of Economic Sciences, Department of Quantitative Finance and Machine Learning, Quantitative Finance Research Group (QFRG), Warsaw, Poland.
\end{justify}

\newpage


\section{Introduction}\label{sec:int}

At the core of financial decision-making lies a fundamental challenge. Investors must price assets in an economy they can observe only imperfectly.
Frequent announcements and economic news are closely tracked by investors around the world since they may shed some light into this otherwise hidden world, thus 
influencing the behaviour of the market. Indeed, it has been observed that such announcements usually have a two-stage influence on financial assets.
First of all, the uncertainty inherent to an announcement and a lack of knowledge regarding a final data which will be revealed supposedly induces a premium as investors bear more undiversifiable risk.
Secondly, the exact signal revealed may cause a jump in prices depending on the number of variables like sentiment and the actual perceived meaning of an announcement.

In this paper, we investigate the first stage of the announcement effect, namely the pre-announcement premium. 
Particular emphasis is placed on monetary policy decisions announced by the Federal Open Market Committee (FOMC) and their impact on the U.S. options market, 
specifically on options written on the S\&P 500 index and the induced implied volatility (IV) surface. 
This choice of instrument is especially relevant, as uncertainty and changes in market expectations are often reflected first in derivatives markets, 
which are inherently forward-looking. Furthermore, the focus on FOMC announcements is motivated by prior empirical evidence documenting the particularly strong 
magnitude of the pre-FOMC effect in equity markets, as shown, for example, in \parencite{LuccaMoench}.

The analysis is divided into two parts. The first one is rather empirical and allows to statistically describe and understand the pre-announcement effect. 
We state a number of hypotheses related to this aspect:

    \begin{enumerate}
        \item The implied volatility does not change before nor during the FOMC conference.
        \item The pre-announcement effect is the same for the entire IV surface.
    \end{enumerate}

The second part constitutes a forecasting exercise where we attempt to quantitatively capture and predict the pre-announcement effect using Machine Learning (ML) framework.
Therefore, the hypotheses for this aspect are mainly connected to the performance of ML models, namely we hypothesise that:

    \begin{enumerate}
        \item ML framework provides the same prediction power both globally and during policy announcement days as the benchmark random walk.
        \item A model fit is the same across the entire IV surface.
        \item Adding the date of FOMC meetings as a feature does not change overall model performance nor during policy annoucement days.
    \end{enumerate}

These hypotheses are designed to address several key research questions that guide the analysis undertaken in this study. 
Specifically, we examine:

    \begin{enumerate}
        \item How the implied volatility surface evolves in the periods preceding and surrounding U.S. monetary policy announcements in particular does it increases?
        \item Is the pre-announcement effect stronger for short-dated options?
    \end{enumerate}

Furthermore, we assess the extent to which ML models are able to capture abnormal dynamics in the implied volatility surface during policy announcement periods, especially:

    \begin{enumerate}
        \item Does ML framework beat the benchmark random walk in prediction power both globally and during policy announcement days?
        \item Does model fit more closely short-dated, ATM part of the IV surface?
        \item Can incorporating information about announcement dates meaningfully improve the predictive performance of these models?
    \end{enumerate}

As such, the main contribution of our work lies in the quantitative prediction of the pre-announcement effect of elevated implied volatility in derivatives markets.
We also study the whole IV surface jointly, in contrast to previous works which focused solely on at-the-money (ATM) IV or VIX.
In addition, not only do we employ more advanced, ML models to forecast the evolution of IV surface,
but also include purely exogenous information inside these models, while previous studies focused rather on the past information encoded in prices of options.
Furthermore, we conduct a much more comprehensive and explainable diagnostic analysis of the fitted models including statistical significance of achieved forecasting accuracy with respect to benchmarks. 
Finally, from the methodological standpoint this is a first attempt to fit ML models directly on the IV surface interpolated and extrapolated to a fixed grid
without resorting to dimensionality reduction. 

The work is stuctured as follows. The following \Cref{sec:lit} and \Cref{sec:data} discuss the existing literature and the data used. 
These sections are followed by the description of the applied methodology in \Cref{sec:met}, divided into the issue of dealing with the IV on a sparse grid,
the forecasting task and the implemented ML architecture. In \Cref{sec:res} the results of the empirical analysis and the forecasting study are presented and discussed.
\Cref{sec:conc} concludes.

\subsection{Literature Review}\label{sec:lit}

\subsubsection{Pre-announcement and announcement effects in financial markets}

The literature related to the whole paper can be divided into several parts, the first one pertaining to the impact of various announcements on asset prices. 
The majority of the literature focuses on the equity spot announcement premium, 
one of the first entries being \parencite{Beaver}, followed by e.g. \parencite{Cohen} or \parencite{LamontFrazzini} who both focus on companies' earnings announcements 
and find positive premia for the uncertainty related to the announcement. 
When it comes to macroeconomic announcements specifically \parencite{LuccaMoench} document a positive equity premium only for monetary policy decisions of FOMC. 
However, the premium is not always present as observed by \parencite{Kurov} which may be explained by the amount of the overall uncertainty in the economy.
Moreover, other studies show that other announcements can also account for as much as 11\% excess equity returns per annum as shown by \parencite{SavorWilson}. 
Other markets also feature similar characteristics as evidenced by \parencite{FlemingRemolona} for U.S. treasuries and by \parencite{Andersen} for FX.

This strand of literature is complemented by a large number of studies which analyse the response of asset prices after scheduled announcements. 
It is particularly striking that the announcement effect (both pre- and post-) is generally observed in different markets.
For instance \parencite{JiangLo}, as well as \parencite{Dungey} identify price jumps in U.S. Treasury bond prices during macroeconomic news announcement for the enire term structure, 
while \parencite{LahayeLaurent} focus on bond, stock index futures and exchange rates finding similar jump occurences. 
\parencite{Lee} confirms that these discountinuities are mostly observed for non-farm US payroll reports and FOMC announcements. 
In \parencite{McQueenRoley} it was shown that the announcement effect is particularly high during recessions.
These jumps in prices are naturally connected to spikes in realised volatility documented in \parencite{JonesLamont} for Treasury bond volatility, 
\parencite{Plihal} for FX and \parencite{ChanGrey} for equity indices. 

Apart from the spot market and volatility, there is also a vast literature studying derivatives market. 
One of the first evidence that implied volatility in contrast to realised volatility increases before earnings announcements was gathered in \parencite{PattelWolfson1979} and \parencite{PattelWolfson1981}. 
New studies replicate this result, for example \parencite{Dubinsky} who focus on earnings announcements and find that they cause 6\% at-the-money implied volatility spike. 
In turn, \parencite{EderingtonLee} study interest rate derivatives and find that implied volatility drops sharply after the announcement. 
There is also some evidence that for example elections feature similar behaviour of financial assets as shown in \parencite{Kelly}.

Consequently, it is evident that economic announcements, especially interest rate policy decisions immensely influence the market dynamics. 
The question remains to what extent this behaviour is predictable.

\subsubsection{Implied volatility surface construction}

On the other note, our paper focuses on the IV and its dynamics, therefore building on the previous methodologies used to construct the entire IV surface based
on the observation of prices of options. They can be subdivided into four distinct parts:

\begin{itemize}
    \item Principal Component Analysis (PCA) or other methods of dimensionality reduction
    \item Parametric models
    \item Non-paramatric methods (like splines or kernel regression)
    \item Machine learning
\end{itemize}

The classical representantive of the first type of methodology is \parencite{Cont} who apply PCA to study the dynamics of IV discovering three first eigenmodes explain most of the variation.
Parametric models are vastly used both in academia and industry, the first example being \parencite{Dumas} who proposed a polynomial function to fit the entire IV surface at once
known as Ad-Hoc-Black-Scholes model. More advanced methods were developed in \parencite{Gatheral} where authors propose a no-arbitrage parametrisation of the IV surface.
A smooth IV surface can also be constructed non-parametrically using splines like in \parencite{Orosi} or \parencite{Fengler}. Last but not least, ML methods
can be taken advantage of in various ways to achieve the task. 
Indeed, \parencite{Horvath} train a deep learning model to approximate the pricing function of a complex rough volatility model, while
\parencite{Almeida} applies a neural network model to correct the error of a simple polynomial model used to approiximate IV surface. 
On the other hand, \parencite{Bergeron} employ a variational autoencoder to extract latent features from IV surface which is a method similar in spirit to dimensionality reduction.

Selecting one of these methodologies is a key step in our approach, as it requires the IV surface to lie on a fixed, pre-defined set of points.
These transformed observations will become a direct input to ML models, in contrast to other studies which either employed dimensionality reduction and 
trained models on extracted latent features or constructed IV surfaces parametrically and trained the models on time series of fitted parameters.
Therefore, we use three distinct methods, to first of all, check which most accurately represents the original surface and performs the best in terms of forecasting, 
as well as to ensure that choice of one particular methodology does not biases the final results.

\subsubsection{Financial time-series forecasting}

Finally, as a large contribution of our paper stems from the quantitative prediction of IV surface especially during economic announcements, it can placed
within the literature which focuses on IV forecasting. Since IV is not a simple one-dimensional time series, there is very little econometric studies which attempt to predict it.
Nevertheless, \parencite{FenglerHardle} created a factor model acting as an input for a vector autoregressive process. VAR was also employed in \parencite{Goncalves}.

However, especially lately ML models have become a standard tool in financial forecasting, especially in crypto-currency markets. For instance, \parencite{GhadariLSTMcrypto}
and \parencite{ParkLSTMcrypto} design algorithmic trading systems based on LSTM neural networks which will also be used in our study. 
ML models have also been applied to study and predict realised volatility in \parencite{FiszederVolatility} and \parencite{SongVolatility} who apply 
deep neural networks, recurrent neural networks, as well as, more traditional ML models like random forest. 

This effectiveness also spills over to forecasting IV surface as ML models allow to capture non-linear dependencies in an efficient way. 
They can also deal with large dimensionality of IV surface. The studies use a plethora of different models and architectures.
\parencite{Colangelo} apply regression trees, while \parencite{Chen} deep learning with attention mechanism. 
As a complement, \parencite{Zhang} employ autoencoder architecture to construct IV surface and later predict latent factors using LSTM model.
On the other hand, \parencite{ChenGrith} inteprolate IV surface using splines and predict it using neural tangent kernel, 
while \parencite{Bloch} create IV surface using SVI model and forecast it using convolutional 2D LSTM model. 
The general conclusion of most of these studies is that ML framework can vastly improve the accuracy of predictions and outperform benchmarks.

What differentiates the forecasting part of our study from the aforementioned ones is the augmentation of the model with exogeneous information.
Since we deal with pre-scheduled economic announcements and hypothesise that they influence IV even before the actual release of information, we can easily
include this information as an additional feature of the model. We hypothesise not only will ML models learn this abnormal behaviour but also that it will
meaningfully improve the forecasting accuracy. 
Furthermore, we forecast directly the entire IV surface on a grid in contrast to latent factors or parameters of interpolating models,
which allows for a higher interpretability and simpler architecture.

\subsection{Data}\label{sec:data}

The data used for this analysis was obtained from Chicago Board Options Exchange (CBOE) market data service \parencite{CBOE} (see \cref{sec:data_code_availability} for more details). 
It covers price quotes of options on the S\&P 500 index in one minute intervals during the daily trading session between 9:30am to 4:00pm ET enriched by the price of the underlying instrument.
For statistical analysis the data in five minute intervals was used, while for forecasting in daily intervals with snapshots at 2pm ET. 
To filter incorrect data points a set of no-arbitrage and sanity check conditions have been applied to options data, namely:

    \begin{itemize}
        \item Bid price lower or equal than ask price
        \item Bid price higher than 0
        \item Moneyness higher than 0
        \item Implied volatility higher than 0
        \item Convexity and monotonicity in strike
    \end{itemize}

Furthermore, to focus on liquid data points we consider only options with open interest higher than $0$.
Since options which mature on Wednesdays (the days of FOMC coferences) have been introduced only in 2016 our dataset starts in January that year and covers years until 2025 included. 
In case of forecasting we divide the dataset into three intervals - training data between 2016 and 2023 included, validation data is the year 2024 and testing data is the year 2025.
The risk-free interest rate is proxied by readily available yield of U.S. Treasury Securities at 5-Year Constant Maturity extracted from \parencite{fred}.
Although this introduces a maturity mismatch, its effect on log-moneyness is limited to predominantly short-dated options considered and 
is unlikely to materially affect the surface-forecasting results.

This particular hour has been chosen since FOMC has around 8 scheduled meetings per year, which last two days from Tuesday to Wednesday. 
They always finish with the press conference led by FED chair at 2pm EST. A calendar of these meetings was constructed based on the Fed website \parencite{FED}.

\section{Methodology}\label{sec:met}

In this section we describe the methodology we apply throughout the paper, especially related to the construction of IV surface and the ML architecture.


\subsection{Implied volatility}\label{sec:IV}

For the sake of completeness, we will briefly describe what is meant by implied volatility of an option.

Suppose the strike price is denoted by $K$ and maturity by $T$. 
The baseline model used to value options has been introduced in \parencite{BlackScholes}, which yields the price $C_\text{BS}(K,T,\sigma)$ in terms of volatility parameter $\sigma$.
Implied volatility is a mapping from maturities and strike prices to $\mathbb{R}_+$ such that the Black-Scholes formula produces a price consistent with the market price $C^*(K, T)$:

\begin{equation*}
    \sigma^{IV}: (K, T) \to \sigma^{IV}(K, T)  \quad \text{s.t.} \quad C_{\text{BS}}(K,T,\sigma^{\text{IV}}(K,T)) = C^*(K, T)
\end{equation*}

Therefore, we can view implied volatility as a bivariate surface which will be a key object in this study.

\subsection{Interpolation/extrapolation of IV surface}\label{sec:iv_int_ext}

As it has been stated before, working with implied volatility data is challenging. 
The prices of options and the underlying instruments adjust almost continuously and the grid of moneyness and maturities evolves over time which results in a sparse grid.
Therefore, implied volatility surface needs to be somehow constructed so as to fit the raw data as close as possible at each point in time. 
As discussed before there are several ways to address this issue. In this analysis we consider three methods, namely:
    
    \begin{itemize}
        \item Linear interpolation representing a non-parametric approach
        \item Stochastic Volatility Inspired (SVI) model introduced in \parencite{Gatheral} representing a parametric approach
        \item Ad-hoc-Black-Scholes representing a global, simple parametric approach
    \end{itemize}

It has to be noted we do not impose any particular no-arbitrage conditions since our task is not to obtain an arbitrage-free surface per se, but to match the original observations
and check the ability of ML models to learn the patterns hidden in the surface.

Furthermore, as mentioned previously we have selected these methods to ensure that a specific interpolation or extrapolation does not biases the results. 
We also do not choose one method with the best fit to the data 
because it could very well happen that slightly less accurate representation (albeit smoother) is more suitable for forecasting and will achieve a better performance.

To ensure consistency and give a little rigour to our setup 
denote the set of maturities by $\mathbb{T}_t = \{T_1,T_2,\dots, T_J\}$ and the set of strikes available per fixed maturity $T_{t,j}$ 
by $\mathbb{K}_t^j = \{K_1,K_2,\dots, K_I\}$ both observed at time $t$. These sets describe the real, raw data points.
    
However, we need a fixed grid of points which will not change over time.
Denote the grid of fixed strikes and maturities as $\mathbb{K}^* \times \mathbb{T}^*$. 
We focus on the most liquid part of the surface. As such $\mathbb{K}^* = \{K^*_1,\dots,0,\dots,K^*_N\}$ lies in the range of $[-0.2,0.2]$ with a denser grid around $0$ 
such that it has a size of $41$ with point $0$ concatenated 
and $\mathbb{T}^* = \{T^*_1,\dots,T^*_M\} = \{0,1,\dots,20\}$. Therefore, we use $M=21$ maturities and $N=41$ strikes at each point in time.
Also define $k = \ln\left(\frac{S_te^{rT}}{K}\right)$ as log-forward-moneyness, $T = T-t$ as remaining maturity and $w(K,T) = (\sigma^{\text{IV}}_t(K,T))^2 T$ as total variance.

\subsubsection{Linear interpolation}\label{sec:lin_int}

We start with linear interpolation, which is the most rudimentary method of representing the surface. 
Although it should fit the data almost perfectly by definition, it is very rigid and susceptible to noisy points.

For each $t$ we begin with interpolating the surface per fixed maturity $T_{t,j}$.  
Therefore the fitted total variance at $(K^*_n,T_{t,j})$ with $K_{t,i-1} \leq K^*_n \leq K_{t,i}$ at time $t$ is equal to:
    
\begin{equation*}
    w_t(K^*_n,T_{t,j}) = \frac{w_t(K_{i-1},T_{t,j}) \left(K_{t,i} - K_{t,i-1}\right) + w_t(K_{i},T_{t,j})\left(K^*_n - K_{t,i-1}\right)}{K_{t,i} - K_{t, i-1}}
\end{equation*}

Then we proceed with interpolating the IV surface in maturity per already fitted strikes. 
Therefore for a fixed $K^*_n \in \mathbb{K}^*$ the fitted IV at $(K^*_n,T^*_m)$ with $T_{t,j-1} \leq T^*_m \leq T_{t,j}$ is equal to:

\begin{equation*}
    w_t(K^*_n,T^*_m) = \frac{w_t(K^*_n,T_{t,j-1}) \left(T_{t,j} - T_{t,j-1}\right) + w_t(K^*_n,T_{t,j})\left(T^*_m - T_{t,j-1}\right)}{T_{t,j} - T_{t, j-1}}
\end{equation*}

Then we extract the IV back as $\sigma^{\text{IV}}_t(K^*_n,T^*_m) = \sqrt{\frac{1}{T}w_t(K^*_n,T^*_m)}$.

If the strikes at any given time were available in the wider range than $[-0.2,0.2]$ 
this procedure would produce interpolated IV surface $\sigma^{\text{IV}}_t(K^*_n,T^*_m)$ for all pairs $(K^*_n,T^*_m) \in \mathbb{K}^* \times \mathbb{T}^*$.

However, at some dates the data especially in the in-the-money wing is scarce. 
Because it has been proven in \parencite{LeeMoment} that wings of implied volatility surface are always bounded by a linear function 
we extrapolate wings linearly using a slope estimate using a finite difference between edge points spaced out by 2 to ensure a higher robustness:

\begin{equation*}
    \Delta w_t(K^j_{I},T_{t,j}) = \frac{w_t(K^j_{I},T_{t,j}) - w_t(K^j_{I-2},T_{t,j})}{K^j_{I} - K^j_{I-2}}
\end{equation*}
\begin{equation*}
    \Delta w_t(K^j_1,T_{t,j}) = \frac{w_t(K^j_3,T_{t,j}) - w_t(K^j_1,T_{t,j})}{K^j_1 - K^j_3}
\end{equation*}

\begin{figure}[H]
    \centering
    \includegraphics[width = \linewidth]{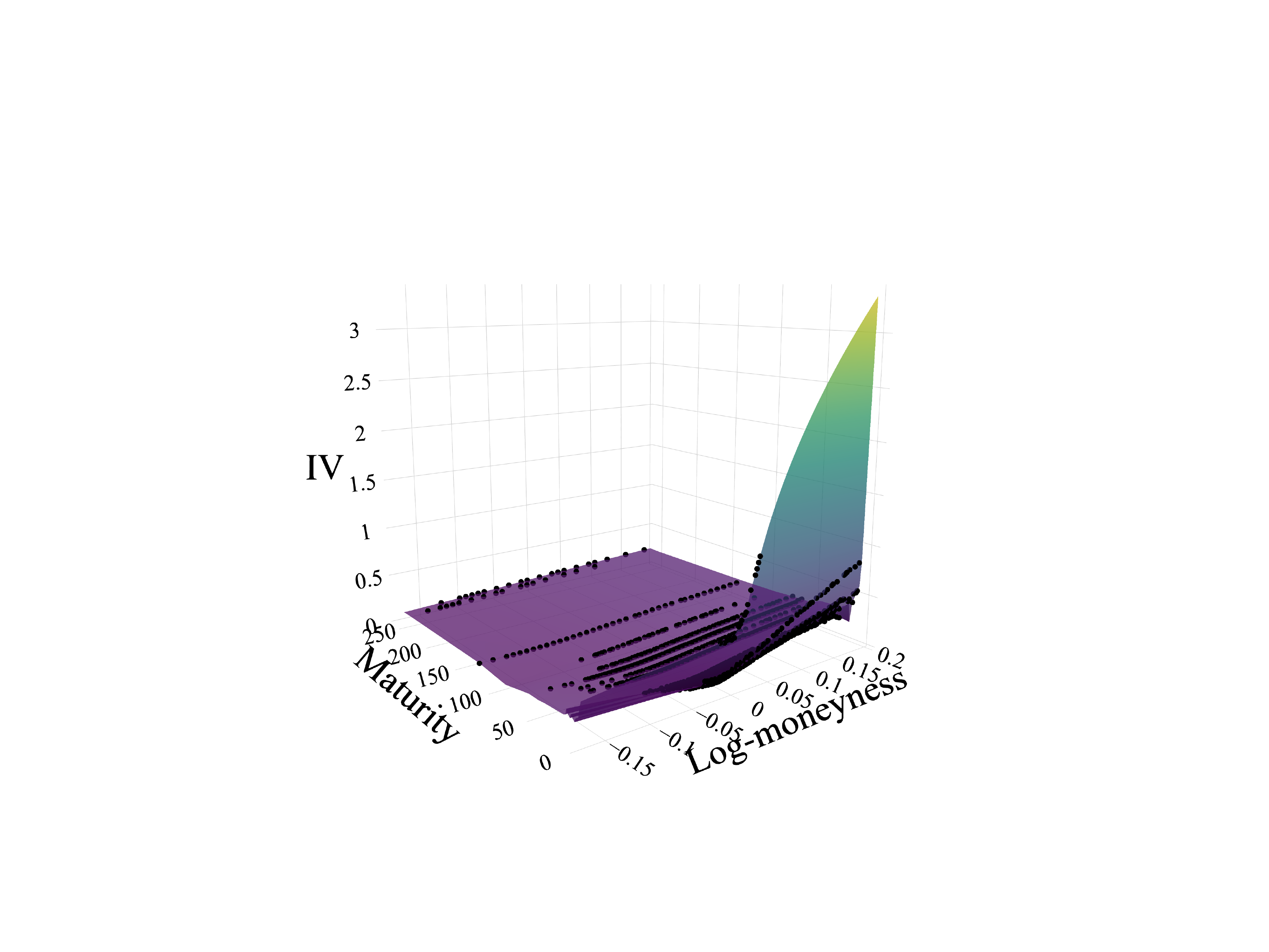}
    \caption{The implied volatility surface fitted using linear interpolation and linear extrapolation of wings - 
    call options on S\&P 500 index as observed at 2pm on 16 May 2024.}
\end{figure}

\subsubsection{Ad-hoc Black Scholes (AHBS)}\label{sec:AHBS}

For the next two methods we turn to parametric approaches. The first one is also rather simple and  aimed at representing the IV surface smoothly and globally at the cost of a worse fit.
It does not require fitting any function per slices of maturities. The whole surface is fitted at once.

The method introduced in \parencite{Dumas} relies on fitting the polynomial function in strike and moneyness. The surface is described by:

    \begin{equation*}
        \sigma^{\text{IV}}(k, T) = \alpha_0 + \alpha_1 k + \alpha_2 k^2 + \alpha_3 T + \alpha_4 T ^2 + \alpha_5 k T + \varepsilon
    \end{equation*}

This function can be fitted using least squares at each time $t$ using the whole grid $\bigcup_{j=1}^{J}{K^j_t \times \{T_{t,j}\}}$ as input points. 
Then the IV can be evaluated at all points $K^* \times T^*$.

\begin{figure}[H]
    \centering
    \includegraphics[width = \linewidth]{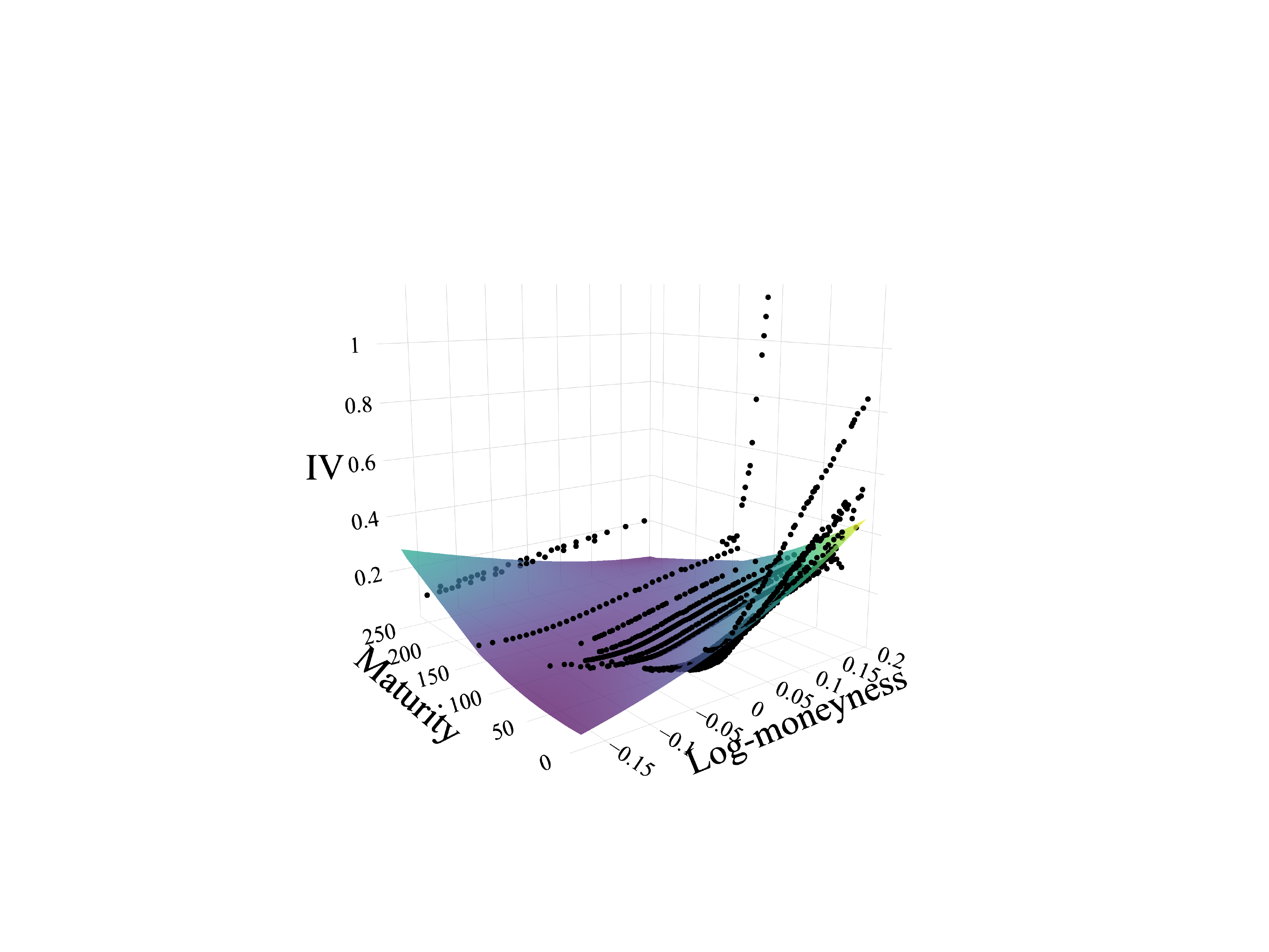}
    \caption{The implied volatility surface with fitted AHBS model - call options on S\&P 500 index as observed at 2pm on 16 May 2024.}
\end{figure}

\subsubsection{Stochastic Volatility Inspired (SVI)}\label{sec:SVI}

The final method considered is also the most complex. 
It is quite possibly the most flexible method out of the three while it leverages a sufficient number of parameters so as to provide a reasonable fit to the observed data.
This approach introduced in \parencite{Gatheral} also can be extended so as to exclude static arbitrage.

The surface is described by a function:

\begin{equation*}
    w(k, T) = a + b \left(\rho(k-m)+\sqrt{(k-m)^2+\sigma^2}\right)
\end{equation*}

where $a,m \in \mathbb{R}$, $b \geq 0$, $|\rho| \leq 1$ and $\sigma > 0$.

This function can be fitted using any optimising scheme at each time $t$ per each maturity slice $T_{t,j}$. 
Then the surface is interpolated linearly in maturity per fitted strike slices as described in \Cref{sec:lin_int}.
Finally, the IV can be evaluated at all points $K^* \times T^*$ as $\sigma^{\text{IV}}(k^*, T^*) = \sqrt{\frac{w(k^*, T^*)}{T^*}}$ with one caveat - for $0$ DTE options
the formula would imply division by $0$. Therefore, we treat such options as having $0.5$ DTE 
which is reasonable given we use 2 pm snapshot and helps to stabilise the fitting procedure. 

\begin{figure}[H]
    \centering
    \includegraphics[width = 0.75\linewidth]{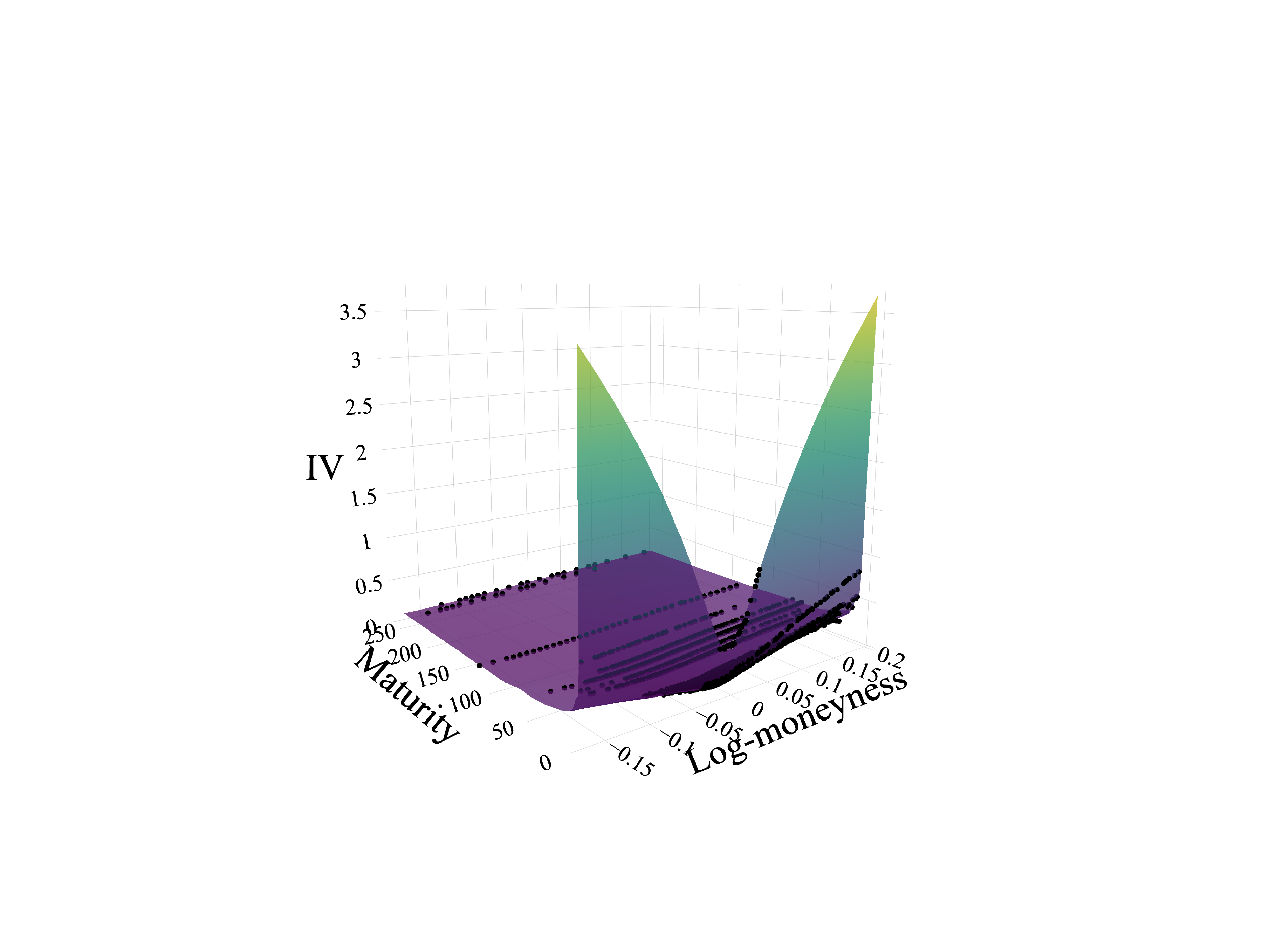}
    \caption{The implied volatility surface with fitted SVI model - call options on S\&P 500 index as observed at 2pm on 16 May 2024.}
\end{figure}


\subsection{Forecasting task}\label{sec:for_task}

Having overcome the challenge of dealing with the sparse grid we can turn our attention toward forecasting the fixed grid IV surface $\sigma^{\text{IV}}_t(K^*,T^*)$ 
using its past observations up to time $t-l$ in the first stage.
In the second stage also exogeneous dummy variable $X_{t}$ indicating the day of the FOMC conference with $1$ is introduced.
To capture the post-announcement reaction of the market the day after the conference the dummy variable is set to $-1$.
    
Therefore, we attempt to approximate $\sigma^{\text{IV}}_t(K^*,T^*)$ using a (nonlinear) function $f$:

\begin{equation*}
    \sigma^{\text{IV}}_t(K^*,T^*) = f\left(\sigma^{\text{IV}}_{t-1}(K^*,T^*), \dots,\sigma^{\text{IV}}_{t-l}(K^*,T^*), X_t\right) + \varepsilon_t
\end{equation*}

The goal of the ML models will be to fit $f$ resulting in trained network $\hat{f}$.

Moreover, to check the long-term performance of the model we forecast the IV for horizons $h=\{1,2,5,10\}$. 
The forecasts are generated by passing the forecasted values as additional inputs which is an example of an autoregressive method of generating predictions.
The choice of this method over a direct generation of the sequence of forecasts up to $h=10$ comes down to its simplicity and the lack of need to change the architecture. 
Therefore, $h$-step forecast is generated as:  

\begin{equation*}
    \hat{\sigma}^{\text{IV}}_{t+h}(K^*,T^*) = \hat{f}\left(\hat{\sigma}^{\text{IV}}_{t+h-1}(K^*,T^*), \dots, \hat{\sigma}^{\text{IV}}_{t+h-l}(K^*,T^*), X_{t+h}\right)
\end{equation*}

Note that since $X$ is known ahead of time since the FOMC conferences are pre-scheduled, the values of the dummy variable do not have to be forecasted, 
which constitutes a huge advantage of using this information inside any modelling framework, 
provided that it carries a meaningful information and significantly influences the IV.

Fitted ML models are benchmarked to a commonly used random walk (RW) prediction, 
i.e. the prediction of IV at any $(K^*, T^*)$ is a value at this point from the previous day:

\begin{equation*}
    \hat{\sigma}^{\text{IV}}_t(K^*,T^*) = \sigma^{\text{IV}}_{t-1}(K^*,T^*)
\end{equation*}

For majority of tasks this benchmark is quite easy to beat by ML methods, 
however in case of financial time series, which are inherently noisy, susceptible to shocks and 
if markets are efficient even harder to predict, this benchmark is hypothesised to be a strong competitor.

As far as the evaluation of forecasts is concerned, we will employ two metrics - 
root mean squared error (RMSE) and mean percentage error (MPE) to account for bias and sign error. 
RMSE for a $1$-step prediction is defined as follows: 

\begin{equation*}
    \text{RMSE} = \sqrt{\frac{1}{NM}\sum_{i=1}^{N}{\sum_{j=1}^{M}{\left(\sigma^{\text{IV}}_t(K^*_i,T^*_j) - \hat{\sigma}^{\text{IV}}_t(K^*_i,T^*_j)\right)^2}}}
\end{equation*}

MPE for a $1$-step prediction is defined as follows: 

\begin{equation*}
    \text{MPE} = \frac{1}{NM}\sum_{i=1}^{N}{\sum_{j=1}^{M}{\frac{\sigma^{\text{IV}}_t(K^*_i,T^*_j) - \hat{\sigma}^{\text{IV}}_t(K^*_i,T^*_j)}{\sigma^{\text{IV}}_t(K^*_i,T^*_j)}}}
\end{equation*}

Evaluation metrics for longer time horizons are defined analogously.

\subsection{ML architecture}\label{sec:ML_arch}

\subsubsection{Long Short-Term Memory (LSTM)}\label{sec:LSTM}

Regular neural networks (NN) have become incredibly useful for performing various tasks, however, their ability to handle time-series is very limited.
As an answer to this issue, recurrent neural networks (RNNs) have been introduced. 
They differ from regular NNs in their ability to feed back the output of a neuron at one time step as an input to the network at the next time step.
Therefore, they can handle sequential data, where the order of observations matter. 

Unfortunately, the implementation of these neteworks encountered the issue of vanishing gradient, which has been solved in \parencite{Hochreiter} 
through Long Short-Term Memory (LSTM) network. It is designed to capture long-term temporal dependencies in the data. A key parameter of the model is
lookback $l$ which determines the number of iterations per time-step, therefore, how far in the past the model reaches. 
At each time $t$ each cell of the network is updated according to these equations in terms of input $x_t$:

\begin{itemize}
    \item Memory cell $c_t = f_t \odot c_{t-1} + i_t \odot \tilde{c}_t$ 
            where $\tilde{c}_t = \tanh(W_c x_t + U_c h_{t-1} + b_c)$
    \item Hidden state $h_t = o_t \odot \tanh(c_t)$
    \item Three gates:
        \begin{itemize}
            \item Input gate $i_t = \sigma(W_i x_t + U_i h_{t-1} + b_i)$
            \item Forget gate $f_t = \sigma(W_f x_t + U_f h_{t-1} + b_f)$
            \item Output gate $o_t = \sigma(W_ox_t + U_oh_{t-1} + b_o)$
        \end{itemize}
\end{itemize}

where $\odot$ denotes Hadamard product and $W$, $U$ are matrices with learnable parameters. 

\subsubsection{Convolutional LSTM}\label{sec:conv_LSTM}

Although LSTMs have become a standard implementation for dealing with sequential data, including time-series, 
the standard version ignores spatial dependencies in the data 
while implied volatility surface is inherently 2-dimensional and can be represented by a small number of latent factors as shown in \parencite{Cont}. 
This issue can be solved through convolution, often used for image processing,
which was first introduced to LSTM in \parencite{Shi}. Since then, such networks have been applied to a number of tasks,
including forecasting the charter times in the shipping industry in \parencite{MoShipping}, cyclones and weather patterns in \parencite{YangCyclones} and 
COVID-19 infections in \parencite{SumitroCOVID} to name a few. In this kind of architecture all matrix multiplications are replaced by a convolution operation $*$:

\begin{itemize}
    \item $c_t = f_t \odot c_{t-1} + i_t \odot \tilde{c}_t$ 
            where $\tilde{c}_t = \tanh(W_c * x_t + U_c * h_{t-1} + b_c)$
    \item Hidden state $h_t = o_t \odot \tanh(c_t)$
    \item $i_t = \sigma(W_i * x_t + U_i * h_{t-1} + b_i)$
    \item Forget gate $f_t = \sigma(W_f * x_t + U_f * h_{t-1} + b_f)$
    \item Output gate $o_t = \sigma(W_o * x_t + U_o * h_{t-1} + b_o)$
\end{itemize}

In our case the inputs to 2D ConvLSTM are 3D tensors of the shape $x_t \in \mathbb{R}^{l \times N \times M}_+$, i.e. lookback and number of fixed grid strikes and maturities,
hence the convolutional 2D LSTM implementation.

A big upside of convolutions is their ability to act locally across $(K,T)$ grid allowing the hidden state to capture spatial dependencies. 
Since implied volatility does not change uniformly, this type of architecure should learn the correlations between different parts of the surface. 
Naturally LSTM architecture should handle temporal dependencies. 
As a result, IV surface is essentially treated by us as a moving picture which should account for a better generalisation and smoother, more consistent forecasts.

\subsection{Model specification}\label{sec:model_spec}

Similarly to all neural networks, convolutional neural networks require the specification of several hyperparameters. 
By trial and error approach the final values chosen are:

\begin{table}[H]
    \centering
    \caption{Hyperparameters of the implemented 2D convolutional LSTM.}
    \label{tab:params}
    \begin{tabular}{l c}
    \toprule
    \textbf{Parameter} &  \textbf{Value} \\
    \midrule
    Lookback $l$ & $5$  \\
    Learning rate $\eta$ & $0.00005$  \\
    Kernel dimension (the size of convolutional matrix) $k$ & $3$  \\
    Filter size $F$ (dimension of the hidden state space) & $32$  \\
    Batch size $b$ & $32$   \\
    \bottomrule
    \end{tabular}
\end{table}

The optimiser is Adam and the objective function is Mean Squared Error (MSE) defined similarly to RMSE discussed in \Cref{sec:for_task}.
In order to help model to generalise well, we apply regularisation in the form of $0.1$ dropout rate for regular connections and the same rate for recurrent connections. 
The specfic architecture of our network follows the simplified encoding-forecasting structure based on \parencite{Shi} shown in the diagram below:

\vspace{5mm}

\begin{center}
\begin{tikzpicture}[
    arr/.style={-{Stealth[length=5pt,width=4pt]}, thick, black!70},
    lbl/.style={font=\small\itshape},
    dlbl/.style={font=\footnotesize, black!55},
    scale = 0.6
  ]

\def\zx{0.50}
\def\zy{0.28}
\def\PW{3.4}
\def\PD{2.2}

\newcommand{\drawplane}[7]{%
  \coordinate (#1/FL) at ({(#2)     + (#4)*\zx}, {(#3)     + (#4)*\zy});
  \coordinate (#1/FR) at ({(#2)+\PW + (#4)*\zx}, {(#3)     + (#4)*\zy});
  \coordinate (#1/BR) at ({(#2)+\PW + (#4+\PD)*\zx}, {(#3) + (#4+\PD)*\zy});
  \coordinate (#1/BL) at ({(#2)     + (#4+\PD)*\zx}, {(#3) + (#4+\PD)*\zy});
  \filldraw[fill=#5, fill opacity=#7, draw=#6, line width=0.55pt]
    (#1/FL) -- (#1/FR) -- (#1/BR) -- (#1/BL) -- cycle;
}

\drawplane{inp}{0}{0}{0}{white}{black!55}{1.0}
\node[lbl, below=3pt] at ($(inp/FL)!0.5!(inp/FR)$) {\textit{Input}};

\drawplane{c1b}{0}{2.2}{0}{white}{black!28}{0.50}
\drawplane{c1} {0}{2.6}{0}{white}{black!60}{0.92}
\node[lbl,  right=6pt] at ($(c1/FR)+(1,0)$)  {$\mathit{ConvLSTM2D}_1$};
\node[dlbl, right=4pt] at ($(c1/FR)+(1,-0.50)$) {32 filters,\ $3\!\times\!3$ kernel,\ tanh};

\drawplane{c2b}{0}{4.6}{0}{white}{black!28}{0.50}
\drawplane{c2} {0}{5.0}{0}{white}{black!60}{0.92}
\node[lbl,  right=6pt] at ($(c2/FR)+(1,0)$)  {$\mathit{ConvLSTM2D}_2$};
\node[dlbl, right=4pt] at ($(c2/FR)+(1,-0.50)$) {32 filters,\ $3\!\times\!3$ kernel,\ tanh};

\drawplane{cv}{0}{7.1}{0}{white}{black!55}{0.88}
\node[lbl,  right=6pt] at ($(cv/FR)+(1,0)$)  {$\mathit{Conv2D}$};
\node[dlbl, right=4pt] at ($(cv/FR)+(1,-0.50)$) {1 filter,\ $1\!\times\!1$ kernel,\ linear};

\drawplane{out}{0}{8.8}{0}{black!8}{black!65}{1.0}
\node[lbl,  right=4pt] at ($(out/FR)+(1,0)$)  {\textit{Prediction}};

\draw[arr] ($(inp/FL)!0.5!(inp/FR)$) -- ($(c1b/FL)!0.5!(c1b/FR)$);
\draw[arr] ($(c1/FL)!0.5!(c1/FR)$)  -- ($(c2b/FL)!0.5!(c2b/FR)$);
\draw[arr] ($(c2/FL)!0.5!(c2/FR)$)  -- ($(cv/FL)!0.5!(cv/FR)$);
\draw[arr] ($(cv/FL)!0.5!(cv/FR)$)  -- ($(out/FL)!0.5!(out/FR)$);

\end{tikzpicture}
\end{center}

\vspace{5mm}

This results in $112,033$ learnable parameters.

This type of architecture can only handle the IV surface as an input. 
To include exogeneous variables, like the announcement dummy denoting dates of FOMC conferences, it has to be modified.
We could include the announcement dummy as an additional channel paired with each observation per all strikes $K^*_n$ and maturities $T^*_m$. 
However, a more efficient approach is to include another branch of the network which will adjust every point on the surface via a dense layer if there is an announcement, 
meaning a dummy variable $X_t = 1$. Since we expect a sharp drop in IV the day after the announcement 
we also denote this day via -1 making the dummy variable, in fact, a categorical one. 
The first branch then handles only the endogeneous information embedded in IV surface. 
The two branches are then merged together through addition. This results in the type of architecture displayed in the diagram below:

\vspace{5mm}

\begin{tikzpicture}[
    >=Stealth,
    thick,
    neuron/.style={
        circle,
        draw=black!80,
        minimum size=7mm,
        line width=0.6pt
    },
    arr/.style={
        -{Stealth[length=5pt,width=4pt]},
        line width=0.8pt
    },
    lbl/.style={
        font=\small\itshape,
        align=center
    },
    dlbl/.style={
        font=\scriptsize,
        black!55,
        align=left
    },
]


\def\zx{0.50}
\def\zy{0.28}
\def\PW{3.4}
\def\PD{2.2}

\newcommand{\drawplane}[7]{%
  \coordinate (#1/FL) at ({(#2)     + (#4)*\zx}, {(#3)     + (#4)*\zy});
  \coordinate (#1/FR) at ({(#2)+\PW + (#4)*\zx}, {(#3)     + (#4)*\zy});
  \coordinate (#1/BR) at ({(#2)+\PW + (#4+\PD)*\zx}, {(#3) + (#4+\PD)*\zy});
  \coordinate (#1/BL) at ({(#2)     + (#4+\PD)*\zx}, {(#3) + (#4+\PD)*\zy});
  \filldraw[fill=#5, fill opacity=#7, draw=#6, line width=0.55pt]
    (#1/FL) -- (#1/FR) -- (#1/BR) -- (#1/BL) -- cycle;
}


\drawplane{inp}{0}{0}{0}{white}{black!55}{1.0}
\node[lbl, below=3pt] at ($(inp/FL)!0.5!(inp/FR)$) {\textit{Surface input}};

\drawplane{c1b}{0}{2.2}{0}{white}{black!28}{0.50}
\drawplane{c1} {0}{2.6}{0}{white}{black!60}{0.92}
\node[lbl,  right=6pt]  at ($(c1/FR)+(1,0)$)      {$\mathit{ConvLSTM2D}_1$};
\node[dlbl, right=4pt]  at ($(c1/FR)+(1,-0.50)$)  {32 filters,\ $3\!\times\!3$ kernel,\ tanh};

\drawplane{c2b}{0}{4.6}{0}{white}{black!28}{0.50}
\drawplane{c2} {0}{5.0}{0}{white}{black!60}{0.92}
\node[lbl,  right=6pt]  at ($(c2/FR)+(1,0)$)      {$\mathit{ConvLSTM2D}_2$};
\node[dlbl, right=4pt]  at ($(c2/FR)+(1,-0.50)$)  {32 filters,\ $3\!\times\!3$ kernel,\ tanh};

\drawplane{cv}{0}{7.1}{0}{white}{black!55}{0.88}
\node[lbl,  right=6pt]  at ($(cv/FR)+(1,0)$)      {$\mathit{Conv2D}$};
\node[dlbl, right=4pt]  at ($(cv/FR)+(1,-0.50)$)  {1 filter,\ $1\!\times\!1$ kernel,\ linear};


\node[neuron] (macro) at (11.0, 0) {$X_t$};
\node[lbl, below=10pt] at (macro) {\textit{Macro input}};

\node[neuron] (l1) at (11.0, 2.0) {};
\node[neuron] (l2) at (11.0, 3.2) {};
\node[font=\large] at (11.0, 4.0) {$\vdots$};
\node[neuron] (l3) at (11.0, 4.8) {};
\node[neuron] (l4) at (11.0, 6.0) {};

\node[lbl,  left=8pt] at (10.5, 4.1)  {$\mathit{Dense}_{1}$};
\node[dlbl, left=6pt] at (10.5, 3.65) {64 units,\ ReLU};

\node[neuron] (r1) at (13.8, 2.0) {};
\node[neuron] (r2) at (13.8, 3.2) {};
\node[font=\large] at (13.8, 4.0) {$\vdots$};
\node[neuron] (r3) at (13.8, 4.8) {};
\node[neuron] (r4) at (13.8, 6.0) {};

\node[lbl,  right=8pt] at (14.3, 4.1)  {$\mathit{Dense}_{2}$};
\node[dlbl, right=6pt] at (14.3, 3.65) {$N\times M$ units,\ linear};


\draw[arr] (macro) -- (l1);

\foreach \i in {1,2,3,4}{
    \foreach \j in {1,2,3,4}{
        \draw[black!35, line width=0.4pt] (l\i) -- (r\j);
    }
}


\node[
    circle,
    draw=black,
    minimum size=9mm,
    line width=0.8pt,
    font=\large
] (add) at (7.0, 10.5) {$+$};


\drawplane{out}{5.30}{12.2}{0}{black!8}{black!65}{1.0}
\node[lbl, right=6pt] at ($(out/FR)+(1,0)$) {\textit{Prediction}};


\draw[arr]
    ($(inp/FL)!0.5!(inp/FR)$) -- ($(c1b/FL)!0.5!(c1b/FR)$);

\draw[arr]
    ($(c1/FL)!0.5!(c1/FR)$) -- ($(c2b/FL)!0.5!(c2b/FR)$);

\draw[arr]
    ($(c2/FL)!0.5!(c2/FR)$) -- ($(cv/FL)!0.5!(cv/FR)$);

\draw[arr]
    ($(cv/BL)!0.5!(cv/BR)$) -- (add.210);


\draw[arr] (r4.north) -- (add.330);


\draw[arr]
    (add.north) -- ($(out/FL)!0.5!(out/FR)$);

\end{tikzpicture}

\vspace{5mm}

We use $64$ neurons in the first dense layer. This results in $168,126$ learnable parameters. The function $f$ for this kind of architecture can be decomposed roughly as:

\begin{equation*}
    f\left(\sigma^{\text{IV}}_{t-1}(K^*,T^*), \dots,\sigma^{\text{IV}}_{t-l}(K^*,T^*), X_t\right) = 
    g\left(\sigma^{\text{IV}}_{t-1}(K^*,T^*), \dots,\sigma^{\text{IV}}_{t-l}(K^*,T^*)\right) + h\left(X_t\right)
\end{equation*}
\section{Results}\label{sec:res}

In this section we turn to the results of our study. We start with the empirical analysis necessary to get the intuition regarding the reaction of the options market
to the event we consider - the FOMC conference. It is important to note that throughout the empirical analysis we use the original data on the original grid.

\subsection{Empirical analysis}\label{sec:res_emp}

    Before we conduct more formal statistical tests to check the first three hypotheses we turn to visual analysis. 
    Throughout this subsection in order for the analysis to be tractable we focus mainly on at-the-money (ATM) IV, getting rid of the moneyness dimension, 
    defined as the IV with log-moneyness $k$ closest to $0$ (i.e. moneyness closest to $1$) per fixed maturity $T_j$:

    \begin{equation*}
        \sigma^{\text{ATM IV}}_t(T_j) = \sigma^{\text{IV}}_t(K^{\text{ATM}}_{t,j},T_j) \quad \text{where} \quad K^{\text{ATM}}_{t,j} = \text{arg}\min_{K \in \mathbb{K}_t^j} |k|
    \end{equation*}

    We also stick to call options, since put options feature a very similar reaction. All plots are gathered in \Cref{sec:app_figures}.
    In subsequent paragraphs we also compare the ATM IV to realised volatility 
    which is defined as rolling annualised standard deviation of returns on the underlying (S\&P 500) with the window of size one hour.

\begin{figure}[H]
        \centering
        \includegraphics[width=0.8\linewidth]{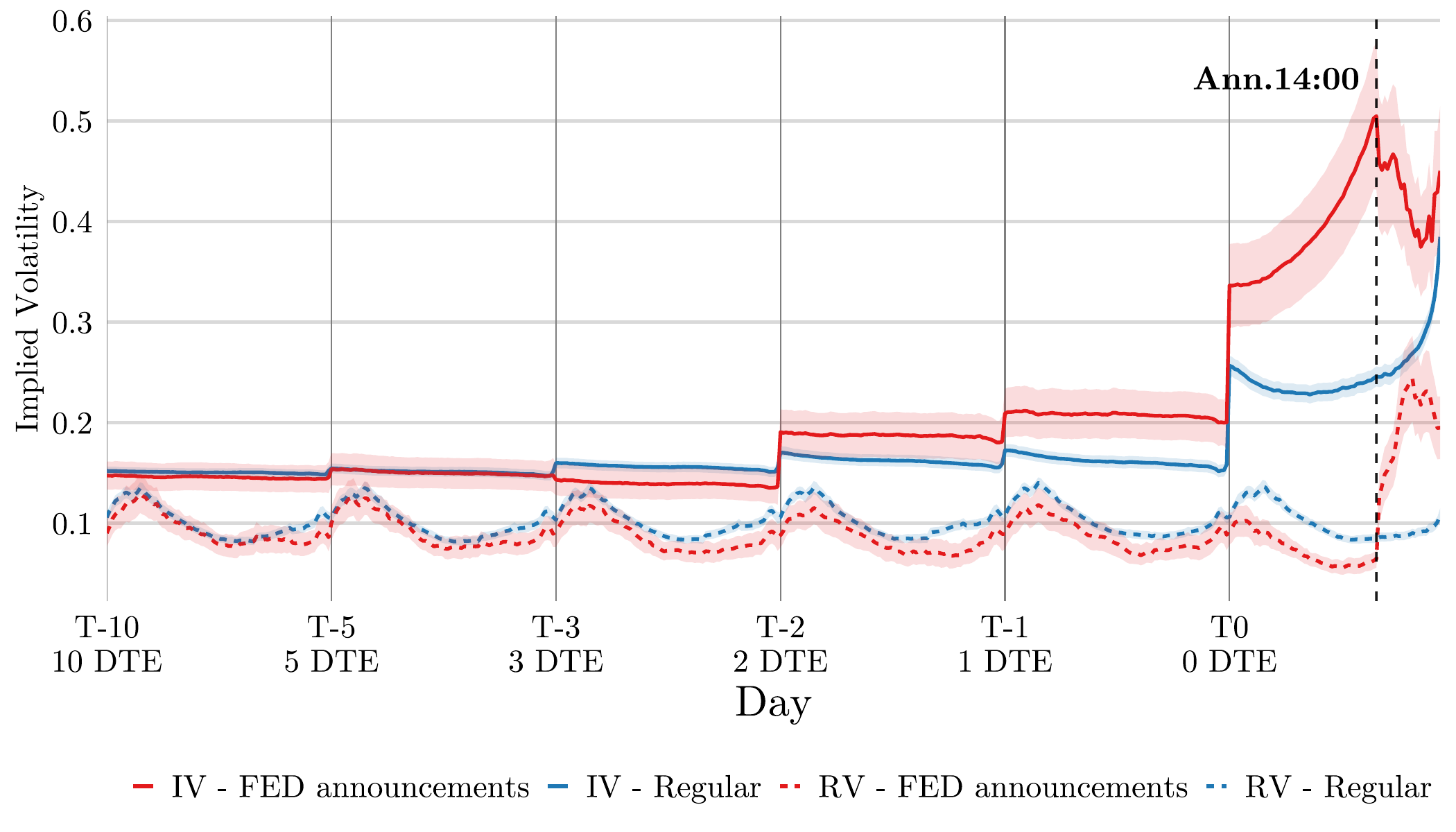}
        \caption{Average ATM Implied and Realised Volatility for call options expiring on the day of announcement denoted by $T_0$. 
        The ribbon is the standard 95\% confidence interval.}\label{fig:Call0DTE_+0_full}
    \end{figure}

    Firstly, we display plots of the average $\sigma^{\text{ATM IV}}_t(T_j)$ for $T_j = \{0,2,10\}$, as well as the realised volatility
    in the window around the FOMC conference.
    As it can be seen in \Cref{fig:Call0DTE_+0_full} for very short dated options on average the implied volatility starts to rise in comparison to regular days
    approximately two days before the announcement. The highest increase in IV even up to 50\% can be observed on the day of the announcement up to 2pm EST when the conference starts.
    Even before the conference ends the implied volatility starts a steady decline due to possibly the resolution of uncertainty 
    which stands in contrast to the usual exploding IV of options nearing maturity.
    What is also striking is the fact that the realised volatility only starts rising after the start of the FOMC conference. 
    
    This plot highlights the two-fold reaction of the market to the news. Even before the actual information is revealed the uncertainty accumulates which
    cannot be diversified away in this case. 
    Standard asset pricing theory in the spirit of \parencite{Sharpe} asserts that instruments bearing higher systematic risk are associated with higher premiums. 
    Quite possibly, as derivatives are forward-looking instruments and investors anticipate higher volatility and
    the higher probability of extreme events, options become more attractive and hence more expensive which raises the ATM IV.
    In contrast, the spot market only reacts to the news when they happen which raises the realised volatility later.  

    As discussed in \parencite{LuccaMoench} standard asset pricing theory is hard to reconcile with the fact that the flow of information only spikes at the announcement,
    while the premiums in the underlying are earned before the announcement is made. This short analysis suggests, that trading flow is indeed high even before the FOMC conference, 
    just not in the spot market. In our view, it is the fact the conference will be held and not the signal received what increases uncertainty, systematic risk and 
    therefore, induces higher premia, which is also reflected in option prices.

\begin{figure}[H]
        \centering
        \includegraphics[width=0.8\linewidth]{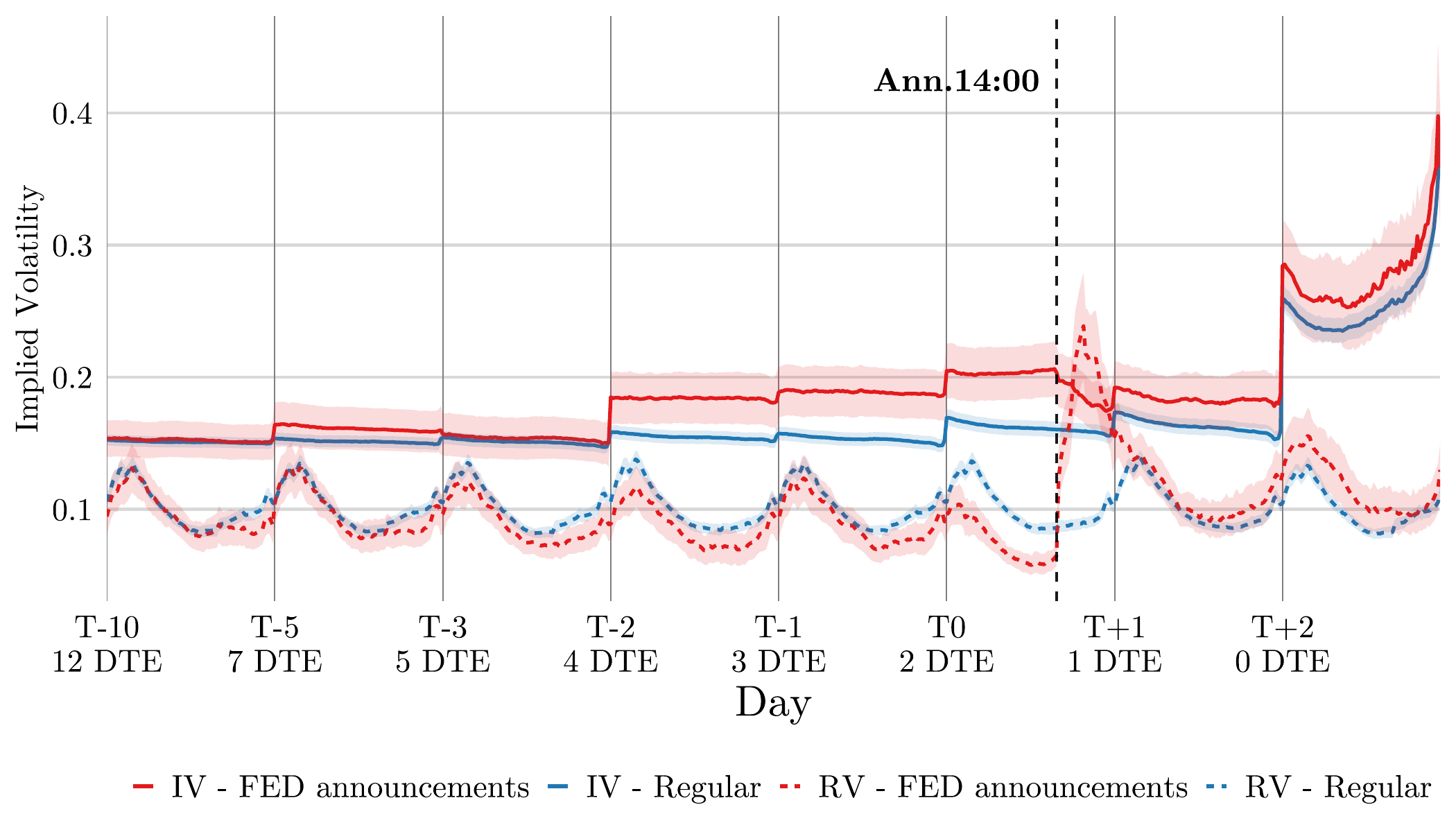}
        \caption{Average ATM Implied and Realised Volatility for call options expiring 2 trading days after the announcement denoted by $T_0$.
        The ribbon is the standard 95\% confidence interval.}\label{fig:Call0DTE_+2_full}
    \end{figure}

    Very similar effect, albeit much less pronounced, can be observed for options with a slightly longer maturity, as shown in \Cref{fig:Call0DTE_+2_full}.
    It suggests that the information content of the FOMC conference is deemed by investors as rather having a short-time impact on the equity market at least on average.
    On the one hand, one could expect that monetary policy decisions have profound long-term implications on the economy and consequently on companies and their stock.
    This in turn should make the long-term options also more expensive and implied volatility higher.
    On the other, we attribute the decreasing strength of the effect with maturity to crowding out - 
    in the short-run the market seems to react sharply to the news and not to the policy.
    However, over time the news itself has a much smaller impact, while the consequences of the policy are uncertain and take time to take effect.
    Furthermore, there are plenty of other news and events that affect equity prices and the longer the horizon, the more such events can occur.
    Finally, options with maturities longer than a month cover more than one FOMC meeting over their lifespan. 
    Therefore, the significance of the closest one is greatly diminished. 

\begin{figure}[H]
        \centering
        \includegraphics[width=0.8\linewidth]{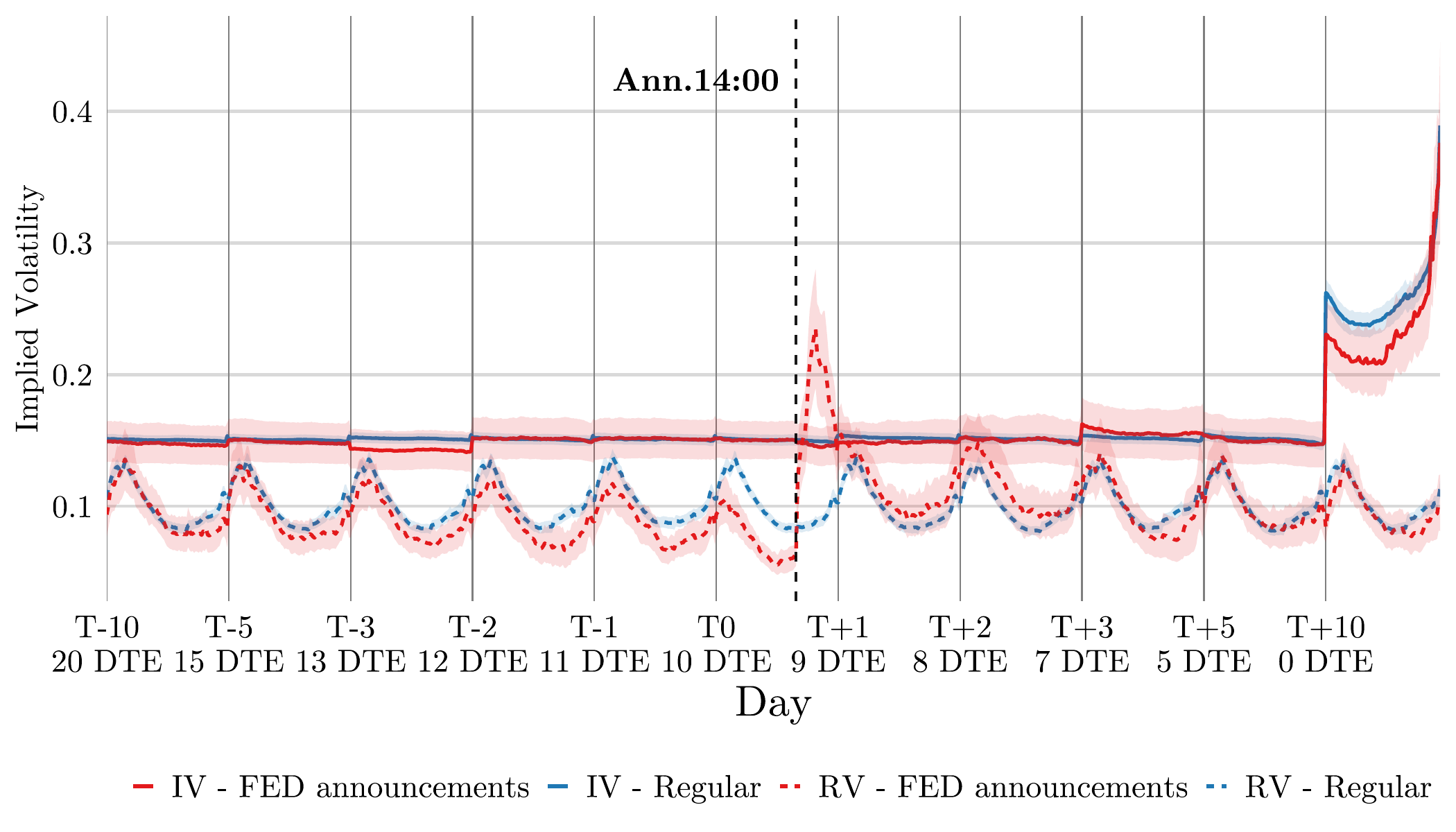}
        \caption{Average ATM Implied and Realised Volatility for call options expiring 10 trading days after the announcement denoted by $T_0$.
        The ribbon is the standard 95\% confidence interval.}\label{fig:Call0DTE_+10_full}
    \end{figure}

    For the sake of completeness we also show the plot for options expiring 10 days after the announcement is made. 
    As depicted in \Cref{fig:Call0DTE_+10_full} there is little to no effect of elevated ATM IV around and before the conference. 
    ATM IV is almost equal to its average over the whole dataset. 
    All these results suggest a very sharply declining term structure of the pre-announcement effect.

    To give this analysis a little more rigour and verify our hypotheses, 
    we test more formally for the change in IV during the days of FOMC conferences using Mann-Whitney U-test which only requires the independence of both samples.
    The null and alternative of the test suited for our analysis are:

    \begin{equation*}
        \begin{aligned}
            \begin{cases}
                H_0:\text{The distribution of IV at the date of the announcement is the same} \\
                \text{as the distribution of IV during all other periods.} \\
                H_1:\text{The distribution of IV at the date of the announcement is shifted to the right} \\
                \text{from the distribution of IV during all other periods.}
            \end{cases}
        \end{aligned}
    \end{equation*}

    The following \Cref{tab:MW_test_0DTE} shows the p-values of the test across log-moneyness $k$ for $0$ DTE call options, since the graphical analysis suggested
    that the impact of interest rate policy announcement is the highest for short-maturity instruments. 
    We apply the U-test across 3 days preceding the FOMC conference in 2 hour intervals, as well as on the day of the conference itself.  
    As it can be observed at 5\% significance level we obtain the evidence to reject null hypotheses mostly one day before the conference. 
    The closer to the announcement, the more significant the results. P-values also create a pyramid pattern meaning that the ATM IV starts to shift the earliest.
    These results suggest that the distribution of IV at the date of the announcement shifts to the right in comparison to the distribution of IV during all other periods 
    and this shift is the strongest for ATM IV.
    As such, these tests would allow us to reject our first hypotheses, i.e. the implied volatility does not change before nor during the FOMC conference.
    At the same time, this test points towards our projections set forth in the first research question, which stated that the IV increases prior to the conference.
    In contrast these results also bring some evidence against the second hypothesis, 
    while suggesting the effect is stronger for ATM options. Simultaneously, it stands in contrast to our expectations regarding the concentration of the effect in the wings.
    However, this investigation covers the part of IV surface only very close to ATM point due to poorer data quality and liquidity in the wings.

    \begin{table}[H]
\centering
\caption{The Mann-Whitney U test around FED interest rates announcements across log-moneyness $k$ for call options expiring on the day of the announcement.\\
        \textit{$H_0$: The distribution of IV at the date of the announcement is the same as the distribution of IV during all other periods. \\
        $H_1$: The distribution of IV at the date of the announcement is shifted to the right from the distribution of IV during all other periods.}}\label{tab:MW_test_0DTE}
\centering
\fontsize{8}{12}\selectfont
\begin{tabular}[t]{>{}l>{}lccllccllc}
\toprule
\textbf{DTE} & \textbf{Time} & $k$ = -0.02 & $k$ = -0.015 & $k$ = -0.01 & $k$ = -0.005 & $k$ = 0 & $k$ = 0.005 & $k$ = 0.01 & $k$ = 0.015 & $k$ = 0.02\\
\cmidrule{1-11}
\midrule
 & \textbf{10:00} & 0.397 & 0.441 & 0.454 & 0.496 & 0.931 & 0.457 & 0.417 & 0.410 & 0.372\\

 & \textbf{12:00} & 0.460 & 0.467 & 0.477 & 0.480 & 0.924 & 0.447 & 0.445 & 0.445 & 0.437\\

 & \textbf{14:00} & 0.401 & 0.372 & 0.402 & 0.398 & 0.904 & 0.363 & 0.326 & 0.348 & 0.318\\

\multirow{-4}{*}{\raggedright\arraybackslash \textbf{3}} & \textbf{16:00} & 0.489 & 0.465 & 0.496 & 0.497 & 0.930 & 0.481 & 0.506 & 0.538 & 0.610\\
\cmidrule{1-11}
 & \textbf{10:00} & 0.358 & 0.288 & 0.276 & 0.298 & \textbf{\textcolor{UBCred}{0.004**}} & 0.231 & 0.236 & 0.234 & 0.217\\

 & \textbf{12:00} & 0.324 & 0.228 & 0.183 & 0.169 & \textbf{\textcolor{UBCred}{0.001***}} & 0.120 & 0.104 & 0.094 & 0.108\\

 & \textbf{14:00} & 0.415 & 0.265 & 0.195 & 0.173 & \textbf{\textcolor{UBCred}{0.001***}} & 0.119 & 0.120 & 0.132 & 0.186\\

\multirow{-4}{*}{\raggedright\arraybackslash \textbf{2}} & \textbf{16:00} & 0.184 & 0.108 & 0.092 & 0.095 & \textbf{\textcolor{UBCred}{0.000***}} & 0.089 & 0.068 & 0.104 & 0.112\\
\cmidrule{1-11}
 & \textbf{10:00} & 0.080 & \textbf{\textcolor{UBCred}{0.029*}} & \textbf{\textcolor{UBCred}{0.018*}} & \textbf{\textcolor{UBCred}{0.012*}} & \textbf{\textcolor{UBCred}{0.000***}} & \textbf{\textcolor{UBCred}{0.009**}} & \textbf{\textcolor{UBCred}{0.013*}} & \textbf{\textcolor{UBCred}{0.009**}} & \textbf{\textcolor{UBCred}{0.028*}}\\

 & \textbf{12:00} & \textbf{\textcolor{UBCred}{0.049*}} & \textbf{\textcolor{UBCred}{0.013*}} & \textbf{\textcolor{UBCred}{0.004**}} & \textbf{\textcolor{UBCred}{0.003**}} & \textbf{\textcolor{UBCred}{0.000***}} & \textbf{\textcolor{UBCred}{0.002**}} & \textbf{\textcolor{UBCred}{0.002**}} & \textbf{\textcolor{UBCred}{0.006**}} & \textbf{\textcolor{UBCred}{0.018*}}\\

 & \textbf{14:00} & \textbf{\textcolor{UBCred}{0.031*}} & \textbf{\textcolor{UBCred}{0.008**}} & \textbf{\textcolor{UBCred}{0.003**}} & \textbf{\textcolor{UBCred}{0.002**}} & \textbf{\textcolor{UBCred}{0.000***}} & \textbf{\textcolor{UBCred}{0.003**}} & \textbf{\textcolor{UBCred}{0.003**}} & \textbf{\textcolor{UBCred}{0.011*}} & 0.059\\

\multirow{-4}{*}{\raggedright\arraybackslash \textbf{1}} & \textbf{16:00} & \textbf{\textcolor{UBCred}{0.011*}} & \textbf{\textcolor{UBCred}{0.002**}} & \textbf{\textcolor{UBCred}{0.000***}} & \textbf{\textcolor{UBCred}{0.000***}} & \textbf{\textcolor{UBCred}{0.000***}} & \textbf{\textcolor{UBCred}{0.000***}} & \textbf{\textcolor{UBCred}{0.002**}} & \textbf{\textcolor{UBCred}{0.018*}} & 0.078\\
\cmidrule{1-11}
 & \textbf{10:00} & \textbf{\textcolor{UBCred}{0.000***}} & \textbf{\textcolor{UBCred}{0.000***}} & \textbf{\textcolor{UBCred}{0.000***}} & \textbf{\textcolor{UBCred}{0.000***}} & \textbf{\textcolor{UBCred}{0.000***}} & \textbf{\textcolor{UBCred}{0.000***}} & \textbf{\textcolor{UBCred}{0.000***}} & \textbf{\textcolor{UBCred}{0.000***}} & \textbf{\textcolor{UBCred}{0.000***}}\\

 & \textbf{12:00} & \textbf{\textcolor{UBCred}{0.000***}} & \textbf{\textcolor{UBCred}{0.000***}} & \textbf{\textcolor{UBCred}{0.000***}} & \textbf{\textcolor{UBCred}{0.000***}} & \textbf{\textcolor{UBCred}{0.000***}} & \textbf{\textcolor{UBCred}{0.000***}} & \textbf{\textcolor{UBCred}{0.000***}} & \textbf{\textcolor{UBCred}{0.000***}} & \textbf{\textcolor{UBCred}{0.000***}}\\

\multirow{-3}{*}{\raggedright\arraybackslash \textbf{0}} & \textbf{14:00} & \textbf{\textcolor{UBCred}{0.000***}} & \textbf{\textcolor{UBCred}{0.000***}} & \textbf{\textcolor{UBCred}{0.000***}} & \textbf{\textcolor{UBCred}{0.000***}} & \textbf{\textcolor{UBCred}{0.000***}} & \textbf{\textcolor{UBCred}{0.000***}} & \textbf{\textcolor{UBCred}{0.000***}} & \textbf{\textcolor{UBCred}{0.000***}} & \textbf{\textcolor{UBCred}{0.000***}}\\
\bottomrule

\end{tabular}
\begin{tablenotes}
    \footnotesize
    \item *, **, and *** denote significance at 5\%, 1\%, and 0.1\% levels, respectively.
    \item Colored cells are significant at at least 5\% significance level.
    \item The FOMC conference starts at 2pm at 0 DTE.
\end{tablenotes}
\end{table}

    A similar analysis for call options expiring 2 days after the announcement whose results are gathered in \Cref{tab:MW_test_2DTE} also 
    confirms our conclusions stemming from graphical analysis. The tests for put options and for longer-maturity options are collected in \Cref{sec:app_tests}
    and ommitted here due to the fact they do not bring any more substantial evidence.
    The pattern is similar to 0 DTE options, yet the p-values are much higher and significant results at 5\% level start to appear much later and closer to the conference. 
    This suggests that the shift of IV begins earlier for short-maturity options, 
    which allows us to reject the second hypothesis stating that pre-announcement effect is the same across the whole surface. As our third research question suggested
    it is actually stronger for short-dated options. 

    \begin{table}[H]
\centering
\caption{The Mann-Whitney U test around FED interest rates announcements across log-moneyness $k$ for call options expiring 2 days after the announcement.
        \textit{$H_0$: The distribution of IV at the date of the announcement is the same as the distribution of IV during all other periods. \\
        $H_1$: The distribution of IV at the date of the announcement is shifted to the right from the distribution of IV during all other periods.}}\label{tab:MW_test_2DTE}
\centering
\fontsize{8}{12}\selectfont
\begin{tabular}[t]{>{}l>{}lccllccllc}
\toprule
\textbf{DTE} & \textbf{Time} & $k$ = -0.02 & $k$ = -0.015 & $k$ = -0.01 & $k$ = -0.005 & $k$ = 0 & $k$ = 0.005 & $k$ = 0.01 & $k$ = 0.015 & $k$ = 0.02\\
\cmidrule{1-11}
\midrule
 & \textbf{10:00} & 0.471 & 0.473 & 0.459 & 0.474 & 0.328 & 0.464 & 0.466 & 0.422 & 0.367\\

 & \textbf{12:00} & 0.471 & 0.464 & 0.463 & 0.470 & 0.303 & 0.422 & 0.416 & 0.430 & 0.420\\

 & \textbf{14:00} & 0.408 & 0.415 & 0.413 & 0.408 & 0.247 & 0.375 & 0.375 & 0.346 & 0.292\\

\multirow{-4}{*}{\raggedright\arraybackslash \textbf{5}} & \textbf{16:00} & 0.490 & 0.511 & 0.508 & 0.500 & 0.343 & 0.511 & 0.476 & 0.513 & 0.505\\
\cmidrule{1-11}
 & \textbf{10:00} & 0.359 & 0.366 & 0.365 & 0.379 & \textbf{\textcolor{UBCred}{0.000***}} & 0.286 & 0.281 & 0.230 & 0.169\\

 & \textbf{12:00} & 0.269 & 0.265 & 0.242 & 0.226 & \textbf{\textcolor{UBCred}{0.000***}} & 0.190 & 0.163 & 0.141 & 0.107\\

 & \textbf{14:00} & 0.274 & 0.282 & 0.276 & 0.263 & \textbf{\textcolor{UBCred}{0.000***}} & 0.203 & 0.161 & 0.158 & 0.154\\

\multirow{-4}{*}{\raggedright\arraybackslash \textbf{4}} & \textbf{16:00} & 0.156 & 0.186 & 0.180 & 0.154 & \textbf{\textcolor{UBCred}{0.000***}} & 0.130 & 0.118 & 0.069 & 0.069\\
\cmidrule{1-11}
 & \textbf{10:00} & 0.078 & 0.102 & 0.113 & 0.105 & \textbf{\textcolor{UBCred}{0.000***}} & 0.088 & 0.083 & \textbf{\textcolor{UBCred}{0.047*}} & \textbf{\textcolor{UBCred}{0.040*}}\\

 & \textbf{12:00} & 0.055 & \textbf{\textcolor{UBCred}{0.047*}} & 0.050 & 0.052 & \textbf{\textcolor{UBCred}{0.000***}} & \textbf{\textcolor{UBCred}{0.046*}} & \textbf{\textcolor{UBCred}{0.045*}} & \textbf{\textcolor{UBCred}{0.040*}} & \textbf{\textcolor{UBCred}{0.041*}}\\

 & \textbf{14:00} & \textbf{\textcolor{UBCred}{0.037*}} & \textbf{\textcolor{UBCred}{0.044*}} & \textbf{\textcolor{UBCred}{0.050*}} & 0.054 & \textbf{\textcolor{UBCred}{0.000***}} & 0.058 & 0.059 & 0.062 & 0.085\\

\multirow{-4}{*}{\raggedright\arraybackslash \textbf{3}} & \textbf{16:00} & \textbf{\textcolor{UBCred}{0.012*}} & \textbf{\textcolor{UBCred}{0.018*}} & \textbf{\textcolor{UBCred}{0.026*}} & \textbf{\textcolor{UBCred}{0.038*}} & \textbf{\textcolor{UBCred}{0.000***}} & \textbf{\textcolor{UBCred}{0.033*}} & \textbf{\textcolor{UBCred}{0.024*}} & \textbf{\textcolor{UBCred}{0.023*}} & \textbf{\textcolor{UBCred}{0.042*}}\\
\cmidrule{1-11}
 & \textbf{10:00} & \textbf{\textcolor{UBCred}{0.005**}} & \textbf{\textcolor{UBCred}{0.004**}} & \textbf{\textcolor{UBCred}{0.006**}} & \textbf{\textcolor{UBCred}{0.010**}} & \textbf{\textcolor{UBCred}{0.000***}} & \textbf{\textcolor{UBCred}{0.008**}} & \textbf{\textcolor{UBCred}{0.007**}} & \textbf{\textcolor{UBCred}{0.004**}} & \textbf{\textcolor{UBCred}{0.002**}}\\

 & \textbf{12:00} & \textbf{\textcolor{UBCred}{0.001**}} & \textbf{\textcolor{UBCred}{0.001***}} & \textbf{\textcolor{UBCred}{0.001**}} & \textbf{\textcolor{UBCred}{0.002**}} & \textbf{\textcolor{UBCred}{0.000***}} & \textbf{\textcolor{UBCred}{0.002**}} & \textbf{\textcolor{UBCred}{0.001**}} & \textbf{\textcolor{UBCred}{0.001***}} & \textbf{\textcolor{UBCred}{0.001***}}\\

 & \textbf{14:00} & \textbf{\textcolor{UBCred}{0.000***}} & \textbf{\textcolor{UBCred}{0.000***}} & \textbf{\textcolor{UBCred}{0.000***}} & \textbf{\textcolor{UBCred}{0.001**}} & \textbf{\textcolor{UBCred}{0.000***}} & \textbf{\textcolor{UBCred}{0.001***}} & \textbf{\textcolor{UBCred}{0.001***}} & \textbf{\textcolor{UBCred}{0.000***}} & \textbf{\textcolor{UBCred}{0.000***}}\\

\multirow{-4}{*}{\raggedright\arraybackslash \textbf{2}} & \textbf{16:00} & 0.095 & 0.079 & 0.107 & 0.146 & \textbf{\textcolor{UBCred}{0.000***}} & 0.170 & 0.131 & 0.092 & 0.115\\
\bottomrule
\end{tabular}
\begin{tablenotes}
    \small
    \item *, **, and *** denote significance at 5\%, 1\%, and 0.1\% levels, respectively.
    \item Colored cells are significant at at least 5\% significance level.
    \item The FOMC conference starts at 2pm at 2 DTE.
\end{tablenotes}
\end{table}

\subsection{Forecasting IV surface using ML}\label{sec:res_for}

\subsubsection{Forecasting accuracy}

    In the second part of our study we review the performance of ML models in IV surface forecasting. 
    
    First of all, we check the fit of IV surface construction methods explained in \Cref{sec:iv_int_ext}, 
    which will allow us to decompose the performance of ML models in terms of the error stemming from the imperfect interpolation/extrapolation and the forecasting error.
    Not surprisingly, the closest fit both in terms of RMSE and MPE according to \Cref{tab:iv_model_errors} can be attributed to linear interpolation by definition, 
    even though SVI fares almost the same. The fit of AHBS parametrisation is the worst.
    
    \begin{table}[H] \centering
\caption{Overall errors for IV parametric interpolating models}
\label{tab:iv_model_errors}
\begin{tabular}{lcccc}
\toprule
 & \multicolumn{2}{c}{RMSE} & \multicolumn{2}{c}{MPE} \\
Type & Call & Put & Call & Put \\
model &  &  &  &  \\
\midrule
\textbf{AHBS} & 0.21 & 0.14 & 0.01 & -0.11 \\
\textbf{SVI} & 0.02 & 0.01 & -0.01 & -0.01 \\
\textbf{linear} & 0 (by construction) & 0 (by construction) & 0 (by construction) & 0 (by construction)\\
\bottomrule
\end{tabular}
\end{table}

    Having constructed the IV surface on the fixed grid, we can finally apply ML models on this data.
    In the first stage, we start with the baseline model which leverages solely the information from past IV surface observations. 
    To ensure robustness against stochasticity of the applied neural network we report the mean metric across $6$ randomly selected seeds with the corresponding standard deviation.
    
    The error metrics RMSE and MPE presented in \Cref{sec:for_task} are displayed in \Cref{tab:rmse_iv_surface_convLSTM} and \Cref{tab:MPE_iv_surface_convLSTM}, respectively.
    When it comes to RMSE both for call options and put options for shorter prediction horizons the error is the smallest for convolutional LSTM model which 
    learns on the surface constructed using SVI parametrisation. 
    The gain in performance when compared to benchmark random walk is tangible and in horizon $1$ amounts to around 8\% for call options and 9\% for put options. 
    Moreover, the drop in performance in longer time horizons is rather small even though so is the case for random walk, 
    hence the lack of the gain in comparison to benchmark in case of longer horizons for put options. These results are virtually the same across each seed.
    What's worth noting is that the general magnitude of error of the ML model is closely tied to the interpolating error of the underlying method used to construct IV surface.
    Nevertheless, these results show applying ML framework to forecast IV surface results in generally a better or similar performance than the benchmark in terms of RMSE 
    regardless of the method of interpolation/extrapolation. Finally, we note that a perfect fit of linear interpolation in fact impacts the forecasting a little negatively,
    possibly due to overfitting to noisy points and sometimes implausible extrapolation of wings which results in a poor generalisation. 
 
    \begin{table}[H]
\caption{Comparison of out-of-sample mean RMSE for ConvLSTM and benchmarks. Standard deviation is displayed inside the brackets.}
\label{tab:rmse_iv_surface_convLSTM}
\fontsize{10}{14}\selectfont
\begin{tabular}{llllllllll}
\toprule
 & Type & \multicolumn{4}{c}{Call} & \multicolumn{4}{c}{Put} \\
 & Horizon & 1 & 2 & 5 & 10 & 1 & 2 & 5 & 10 \\
Model type & Model &  &  &  &  &  &  &  &  \\
\midrule
\multirow[t]{3}{*}{\textbf{ConvLSTM}} & \textbf{AHBS} & \begin{tabular}{@{}c@{}}0.208\\(0.001)\end{tabular} & \begin{tabular}{@{}c@{}}0.210\\(0.001)\end{tabular} & \begin{tabular}{@{}c@{}}0.212\\(0.002)\end{tabular} & \begin{tabular}{@{}c@{}}0.216\\(0.003)\end{tabular} & \begin{tabular}{@{}c@{}}0.155\\(0.001)\end{tabular} & \begin{tabular}{@{}c@{}}0.156\\(0.001)\end{tabular} & \begin{tabular}{@{}c@{}}0.161\\(0.003)\end{tabular} & \begin{tabular}{@{}c@{}}0.167\\(0.006)\end{tabular} \\
\textbf{} & \textbf{SVI} & \begin{tabular}{@{}c@{}}\textbf{0.085}\\(0.003)\end{tabular} & \begin{tabular}{@{}c@{}}\textbf{0.087}\\(0.002)\end{tabular} & \begin{tabular}{@{}c@{}}\textbf{0.102}\\(0.001)\end{tabular} & \begin{tabular}{@{}c@{}}0.117\\(0.004)\end{tabular} & \begin{tabular}{@{}c@{}}\textbf{0.077}\\(0.003)\end{tabular} & \begin{tabular}{@{}c@{}}\textbf{0.081}\\(0.003)\end{tabular} & \begin{tabular}{@{}c@{}}0.097\\(0.005)\end{tabular} & \begin{tabular}{@{}c@{}}0.114\\(0.009)\end{tabular} \\
\textbf{} & \textbf{linear} & \begin{tabular}{@{}c@{}}0.116\\(0.004)\end{tabular} & \begin{tabular}{@{}c@{}}0.118\\(0.004)\end{tabular} & \begin{tabular}{@{}c@{}}0.133\\(0.005)\end{tabular} & \begin{tabular}{@{}c@{}}0.149\\(0.009)\end{tabular} & \begin{tabular}{@{}c@{}}0.087\\(0.002)\end{tabular} & \begin{tabular}{@{}c@{}}0.091\\(0.002)\end{tabular} & \begin{tabular}{@{}c@{}}0.109\\(0.003)\end{tabular} & \begin{tabular}{@{}c@{}}0.127\\(0.006)\end{tabular} \\
\cline{1-10}
\multirow[t]{3}{*}{\textbf{Random Walk}} & \textbf{AHBS} & 0.208 & 0.211 & 0.215 & 0.219 & 0.155 & 0.156 & 0.162 & 0.167 \\
\textbf{} & \textbf{SVI} & 0.092 & 0.105 & 0.107 & \textbf{0.116} & 0.084 & 0.088 & \textbf{0.093} & \textbf{0.107} \\
\textbf{} & \textbf{linear} & 0.124 & 0.134 & 0.137 & 0.146 & 0.102 & 0.112 & 0.111 & 0.126 \\
\cline{1-10}
\bottomrule
\end{tabular}
\end{table}

    When it comes to MPE, SVI again achieves the best performance in most of the cases. 
    However, in case of this metric one can conclude that most models achieve a somewhat similar performance to its benchmark counterparts, 
    especially taking into account large standard deviation of this metric across seeds.

    It is also important to highlight that random walk systematically underestimates the IV surface, 
    which can be probably explained by the inability to capture immense unexpected shocks. 
    On the other hand, convolutional LSTM tends to overestimate the IV surface on average except for AHBS construction method as it tries to overcorrect.
    This pattern can also be due to the memory property of LSTM - past shocks are remembered by the network and elevate the IV in the future.
    We will try to explore these explanations in subsequent paragraphs by deepening the diagnostics of the model.

    \begin{table}[H] \centering
\caption{Comparison of out-of-sample mean MPE for ConvLSTM and benchmarks. Standard deviation is displayed inside the brackets.}
\label{tab:MPE_iv_surface_convLSTM}
\fontsize{10}{14}\selectfont
\begin{tabular}{llllllllll}
\toprule
 & Type & \multicolumn{4}{c}{Call} & \multicolumn{4}{c}{Put} \\
 & Horizon & 1 & 2 & 5 & 10 & 1 & 2 & 5 & 10 \\
Model type & Model &  &  &  &  &  &  &  &  \\
\midrule
\multirow[t]{3}{*}{\textbf{ConvLSTM}} & \textbf{AHBS} & \begin{tabular}{@{}c@{}}\textcolor{red}{-0.068}\\(0.057)\end{tabular} & \begin{tabular}{@{}c@{}}\textcolor{red}{-0.072}\\(0.076)\end{tabular} & \begin{tabular}{@{}c@{}}\textcolor{red}{-0.082}\\(0.133)\end{tabular} & \begin{tabular}{@{}c@{}}\textcolor{red}{-0.080}\\(0.226)\end{tabular} & \begin{tabular}{@{}c@{}}\textcolor{red}{-0.098}\\(0.058)\end{tabular} & \begin{tabular}{@{}c@{}}\textcolor{red}{-0.090}\\(0.077)\end{tabular} & \begin{tabular}{@{}c@{}}\textcolor{red}{-0.069}\\(0.131)\end{tabular} & \begin{tabular}{@{}c@{}}\textbf{\textcolor{red}{-0.036}}\\(0.203)\end{tabular} \\
\textbf{} & \textbf{SVI} & \begin{tabular}{@{}c@{}}\textcolor{red}{-0.011}\\(0.053)\end{tabular} & \begin{tabular}{@{}c@{}}\textbf{\textcolor{red}{-0.003}}\\(0.072)\end{tabular} & \begin{tabular}{@{}c@{}}\textbf{\textcolor{darkgreen}{0.023}}\\(0.140)\end{tabular} & \begin{tabular}{@{}c@{}}\textcolor{darkgreen}{0.071}\\(0.238)\end{tabular} & \begin{tabular}{@{}c@{}}\textcolor{red}{-0.020}\\(0.056)\end{tabular} & \begin{tabular}{@{}c@{}}\textbf{\textcolor{red}{-0.011}}\\(0.076)\end{tabular} & \begin{tabular}{@{}c@{}}\textbf{\textcolor{darkgreen}{0.019}}\\(0.147)\end{tabular} & \begin{tabular}{@{}c@{}}\textcolor{darkgreen}{0.068}\\(0.256)\end{tabular} \\
\textbf{} & \textbf{linear} & \begin{tabular}{@{}c@{}}\textbf{\textcolor{darkgreen}{0.007}}\\(0.053)\end{tabular} & \begin{tabular}{@{}c@{}}\textcolor{darkgreen}{0.017}\\(0.080)\end{tabular} & \begin{tabular}{@{}c@{}}\textcolor{darkgreen}{0.048}\\(0.155)\end{tabular} & \begin{tabular}{@{}c@{}}\textcolor{darkgreen}{0.097}\\(0.258)\end{tabular} & \begin{tabular}{@{}c@{}}\textbf{\textcolor{darkgreen}{0.001}}\\(0.033)\end{tabular} & \begin{tabular}{@{}c@{}}\textcolor{darkgreen}{0.012}\\(0.046)\end{tabular} & \begin{tabular}{@{}c@{}}\textcolor{darkgreen}{0.048}\\(0.087)\end{tabular} & \begin{tabular}{@{}c@{}}\textcolor{darkgreen}{0.112}\\(0.147)\end{tabular} \\
\cline{1-10}
\multirow[t]{3}{*}{\textbf{Random Walk}} & \textbf{AHBS} & \textcolor{red}{-0.064} & \textcolor{red}{-0.069} & \textcolor{red}{-0.085} & \textcolor{red}{-0.091} & \textcolor{red}{-0.125} & \textcolor{red}{-0.128} & \textcolor{red}{-0.138} & \textcolor{red}{-0.141} \\
\textbf{} & \textbf{SVI} & \textcolor{red}{-0.036} & \textcolor{red}{-0.042} & \textcolor{red}{-0.057} & \textcolor{red}{-0.070} & \textcolor{red}{-0.048} & \textcolor{red}{-0.053} & \textcolor{red}{-0.059} & \textcolor{red}{-0.068} \\
\textbf{} & \textbf{linear} & \textcolor{red}{-0.018} & \textcolor{red}{-0.022} & \textcolor{red}{-0.040} & \textbf{\textcolor{red}{-0.051}} & \textcolor{red}{-0.020} & \textcolor{red}{-0.025} & \textcolor{red}{-0.036} & \textcolor{red}{-0.043} \\
\cline{1-10}
\bottomrule
\end{tabular}
\end{table}

    Secondly, we check the ability of convolutional LSTM model to fit the shock related to FOMC conferences, 
    which will be done by introducing a dummy variable $X$ denoting the date of the announcement and one day after according to the methodology laid out in \Cref{sec:ML_arch}. 

    When it comes to the comparison of out-of-sample RMSE between the ML model and the benchamark within one construction method the pattern is the same as in case of the model
    without the announcement dummy as shown in \Cref{tab:rmse_iv_surface_convXLSTM}. The SVI still performs the best. Most notably however, there seems to be a minor gain from 
    including the information regarding FOMC conference.
    Most model errors are slightly smaller than their dummy-less counterpart. 
    The global impact is however limited due to a small number of FOMC announcements in the whole sample.
    
    \begin{table}[H]
\caption{Comparison of out-of-sample RMSE for ConvLSTM and benchmarks.}
\label{tab:rmse_iv_surface}
\begin{tabular}{llllllllll}
\toprule
 & Type & \multicolumn{4}{c}{Call} & \multicolumn{4}{c}{Put} \\
 & Horizon & 1 & 2 & 5 & 10 & 1 & 2 & 5 & 10 \\
Model type & Model &  &  &  &  &  &  &  &  \\
\midrule
\multirow[t]{3}{*}{\textbf{ConvLSTM}} & \textbf{AHBS} & \begin{tabular}{@{}c@{}}0.206\\(0.001)\end{tabular} & \begin{tabular}{@{}c@{}}0.208\\(0.001)\end{tabular} & \begin{tabular}{@{}c@{}}0.210\\(0.001)\end{tabular} & \begin{tabular}{@{}c@{}}0.213\\(0.003)\end{tabular} & \begin{tabular}{@{}c@{}}0.154\\(0.001)\end{tabular} & \begin{tabular}{@{}c@{}}0.156\\(0.001)\end{tabular} & \begin{tabular}{@{}c@{}}0.161\\(0.002)\end{tabular} & \begin{tabular}{@{}c@{}}0.167\\(0.005)\end{tabular} \\
\textbf{} & \textbf{SVI} & \begin{tabular}{@{}c@{}}\textbf{0.084}\\(0.004)\end{tabular} & \begin{tabular}{@{}c@{}}\textbf{0.086}\\(0.004)\end{tabular} & \begin{tabular}{@{}c@{}}\textbf{0.101}\\(0.003)\end{tabular} & \begin{tabular}{@{}c@{}}\textbf{0.115}\\(0.005)\end{tabular} & \begin{tabular}{@{}c@{}}\textbf{0.075}\\(0.004)\end{tabular} & \begin{tabular}{@{}c@{}}\textbf{0.079}\\(0.004)\end{tabular} & \begin{tabular}{@{}c@{}}0.094\\(0.007)\end{tabular} & \begin{tabular}{@{}c@{}}0.112\\(0.012)\end{tabular} \\
\textbf{} & \textbf{linear} & \begin{tabular}{@{}c@{}}0.115\\(0.005)\end{tabular} & \begin{tabular}{@{}c@{}}0.116\\(0.005)\end{tabular} & \begin{tabular}{@{}c@{}}0.131\\(0.006)\end{tabular} & \begin{tabular}{@{}c@{}}0.146\\(0.011)\end{tabular} & \begin{tabular}{@{}c@{}}0.087\\(0.003)\end{tabular} & \begin{tabular}{@{}c@{}}0.091\\(0.004)\end{tabular} & \begin{tabular}{@{}c@{}}0.106\\(0.005)\end{tabular} & \begin{tabular}{@{}c@{}}0.120\\(0.008)\end{tabular} \\
\cline{1-10}
\multirow[t]{3}{*}{\textbf{Random Walk}} & \textbf{AHBS} & 0.208 & 0.211 & 0.215 & 0.219 & 0.155 & 0.156 & 0.162 & 0.167 \\
\textbf{} & \textbf{SVI} & 0.092 & 0.105 & 0.107 & 0.116 & 0.084 & 0.088 & \textbf{0.093} & \textbf{0.107} \\
\textbf{} & \textbf{linear} & 0.124 & 0.134 & 0.137 & 0.146 & 0.102 & 0.112 & 0.111 & 0.126 \\
\cline{1-10}
\bottomrule
\end{tabular}
\end{table}

    In terms of MPE as shown in \Cref{tab:MPE_iv_surface_convXLSTM} the inclusion of the dummy variable $X$ has also caused a minor improvement except for AHBS.
    Nevertheless, RMSE and MPE both seem to confirm the addition of the announcement dummy usually improves error metrics both in short and long horizons.

    \begin{table}[H] \centering
\caption{Comparison of out-of-sample mean MPE for ConvLSTM and benchmarks. Standard deviation is displayed inside the brackets.}
\label{tab:MPE_iv_surface_convXLSTM}
\fontsize{10}{14}\selectfont
\begin{tabular}{llllllllll}
\toprule
 & Type & \multicolumn{4}{c}{Call} & \multicolumn{4}{c}{Put} \\
 & Horizon & 1 & 2 & 5 & 10 & 1 & 2 & 5 & 10 \\
Model type & Model &  &  &  &  &  &  &  &  \\
\midrule
\multirow[t]{3}{*}{\textbf{ConvLSTM}} & \textbf{AHBS} & \begin{tabular}{@{}c@{}}\textcolor{red}{-0.087}\\(0.055)\end{tabular} & \begin{tabular}{@{}c@{}}\textcolor{red}{-0.098}\\(0.073)\end{tabular} & \begin{tabular}{@{}c@{}}\textcolor{red}{-0.125}\\(0.124)\end{tabular} & \begin{tabular}{@{}c@{}}\textcolor{red}{-0.154}\\(0.204)\end{tabular} & \begin{tabular}{@{}c@{}}\textcolor{red}{-0.098}\\(0.065)\end{tabular} & \begin{tabular}{@{}c@{}}\textcolor{red}{-0.090}\\(0.087)\end{tabular} & \begin{tabular}{@{}c@{}}\textcolor{red}{-0.070}\\(0.149)\end{tabular} & \begin{tabular}{@{}c@{}}\textbf{\textcolor{red}{-0.039}}\\(0.235)\end{tabular} \\
\textbf{} & \textbf{SVI} & \begin{tabular}{@{}c@{}}\textbf{\textcolor{red}{-0.004}}\\(0.052)\end{tabular} & \begin{tabular}{@{}c@{}}\textbf{\textcolor{darkgreen}{0.005}}\\(0.071)\end{tabular} & \begin{tabular}{@{}c@{}}\textbf{\textcolor{darkgreen}{0.037}}\\(0.133)\end{tabular} & \begin{tabular}{@{}c@{}}\textcolor{darkgreen}{0.094}\\(0.222)\end{tabular} & \begin{tabular}{@{}c@{}}\textcolor{red}{-0.017}\\(0.057)\end{tabular} & \begin{tabular}{@{}c@{}}\textbf{\textcolor{red}{-0.009}}\\(0.079)\end{tabular} & \begin{tabular}{@{}c@{}}\textbf{\textcolor{darkgreen}{0.018}}\\(0.149)\end{tabular} & \begin{tabular}{@{}c@{}}\textcolor{darkgreen}{0.064}\\(0.258)\end{tabular} \\
\textbf{} & \textbf{linear} & \begin{tabular}{@{}c@{}}\textcolor{darkgreen}{0.025}\\(0.056)\end{tabular} & \begin{tabular}{@{}c@{}}\textcolor{darkgreen}{0.042}\\(0.083)\end{tabular} & \begin{tabular}{@{}c@{}}\textcolor{darkgreen}{0.089}\\(0.159)\end{tabular} & \begin{tabular}{@{}c@{}}\textcolor{darkgreen}{0.165}\\(0.266)\end{tabular} & \begin{tabular}{@{}c@{}}\textbf{\textcolor{darkgreen}{0.002}}\\(0.046)\end{tabular} & \begin{tabular}{@{}c@{}}\textcolor{darkgreen}{0.009}\\(0.063)\end{tabular} & \begin{tabular}{@{}c@{}}\textcolor{darkgreen}{0.032}\\(0.112)\end{tabular} & \begin{tabular}{@{}c@{}}\textcolor{darkgreen}{0.073}\\(0.184)\end{tabular} \\
\cline{1-10}
\multirow[t]{3}{*}{\textbf{Random Walk}} & \textbf{AHBS} & \textcolor{red}{-0.064} & \textcolor{red}{-0.069} & \textcolor{red}{-0.085} & \textcolor{red}{-0.091} & \textcolor{red}{-0.125} & \textcolor{red}{-0.128} & \textcolor{red}{-0.138} & \textcolor{red}{-0.141} \\
\textbf{} & \textbf{SVI} & \textcolor{red}{-0.036} & \textcolor{red}{-0.042} & \textcolor{red}{-0.057} & \textcolor{red}{-0.070} & \textcolor{red}{-0.048} & \textcolor{red}{-0.053} & \textcolor{red}{-0.059} & \textcolor{red}{-0.068} \\
\textbf{} & \textbf{linear} & \textcolor{red}{-0.018} & \textcolor{red}{-0.022} & \textcolor{red}{-0.040} & \textbf{\textcolor{red}{-0.051}} & \textcolor{red}{-0.020} & \textcolor{red}{-0.025} & \textcolor{red}{-0.036} & \textcolor{red}{-0.043} \\
\cline{1-10}
\bottomrule
\end{tabular}
\end{table}

    To assess the statistical significance of these findings, we employ the Diebold-Mariano test introduced in \parencite{Diebold} 
    with HAC errors to account for overlapping multi-horizon forecast windows, 
    a widely used method for evaluating differences in predictive accuracy across competing forecasting models. 
    Specifically, we test for a difference between the forecasts generated by the ML models and their benchmark counterparts in favour of ML models. 
    In addition, we compare models estimated with and without the announcement dummy $X$, both over the full sample and during 
    FOMC conferences days. This allows us to examine whether the inclusion of announcement effects improves model performance overall, 
    as well as to assess the robustness of the ML models during periods characterized by heightened market uncertainty and informational shocks.

    First we proceed with comparing the models without an announcement dummy and their benchmarks.
    Therefore, we verify the following hypotheses:

    \begin{equation*}
        \begin{aligned}
            \begin{cases}
                H_0:\text{The forecasting accuracy of the convolutional LSTM model without the announcement } \\
                \text{dummy is the same the benchmark random walk.} \\
                H_1:\text{The forecasting accuracy of the convolutional LSTM model without the announcement } \\
                \text{dummy is better than the benchmark random walk.}
            \end{cases}
        \end{aligned}
    \end{equation*}

    As we can observe in \Cref{tab:dm_iv_surface_conv_bench} except for a few isolated cases
    we do not have enough evidence to reject the null hypothesis. This means that although the ML model improves the basic error metrics it does not always systematically beat the benchmark.
    It does not however come as a surprise. Most financial time series have a very low signal-to-noise ratio and as a result, even though the ML model is associated with a smaller aggregated RMSE,
    the outperformance is not always statistically significant. 
    
    \begin{table}[H]
\caption{Comparison of out-of-sample Diebold-Mariano mean p-value for ConvLSTM and benchmarks. Standard deviation is displayed inside the brackets.}
\label{tab:dm_iv_surface_conv_bench}
\fontsize{10}{14}\selectfont
\begin{tabular}{llllllllll}
\toprule
 & Type & \multicolumn{4}{c}{Call} & \multicolumn{4}{c}{Put} \\
 & Horizon & 1 & 2 & 5 & 10 & 1 & 2 & 5 & 10 \\
Model type & Model &  &  &  &  &  &  &  &  \\
\midrule
\multirow[t]{3}{*}{\textbf{ConvLSTM}} & \textbf{AHBS} & \begin{tabular}{@{}c@{}}0.509\\(0.273)\end{tabular} & \begin{tabular}{@{}c@{}}0.229\\(0.258)\end{tabular} & \begin{tabular}{@{}c@{}}0.114\\(0.233)\end{tabular} & \begin{tabular}{@{}c@{}}0.179\\(0.335)\end{tabular} & \begin{tabular}{@{}c@{}}0.447\\(0.206)\end{tabular} & \begin{tabular}{@{}c@{}}0.415\\(0.231)\end{tabular} & \begin{tabular}{@{}c@{}}0.385\\(0.285)\end{tabular} & \begin{tabular}{@{}c@{}}0.398\\(0.361)\end{tabular} \\
\textbf{} & \textbf{SVI} & \begin{tabular}{@{}c@{}}0.063\\(0.062)\end{tabular} & \begin{tabular}{@{}c@{}}\textbf{\textcolor{UBCred}{0.002}}\\(0.001)\end{tabular} & \begin{tabular}{@{}c@{}}0.269\\(0.089)\end{tabular} & \begin{tabular}{@{}c@{}}0.508\\(0.213)\end{tabular} & \begin{tabular}{@{}c@{}}0.069\\(0.079)\end{tabular} & \begin{tabular}{@{}c@{}}\textbf{\textcolor{UBCred}{0.021}}\\(0.034)\end{tabular} & \begin{tabular}{@{}c@{}}0.684\\(0.324)\end{tabular} & \begin{tabular}{@{}c@{}}0.751\\(0.364)\end{tabular} \\
\textbf{} & \textbf{linear} & \begin{tabular}{@{}c@{}}0.112\\(0.146)\end{tabular} & \begin{tabular}{@{}c@{}}\textbf{\textcolor{UBCred}{0.014}}\\(0.031)\end{tabular} & \begin{tabular}{@{}c@{}}0.348\\(0.296)\end{tabular} & \begin{tabular}{@{}c@{}}0.551\\(0.319)\end{tabular} & \begin{tabular}{@{}c@{}}\textbf{\textcolor{UBCred}{0.000}}\\(0.000)\end{tabular} & \begin{tabular}{@{}c@{}}\textbf{\textcolor{UBCred}{0.000}}\\(0.000)\end{tabular} & \begin{tabular}{@{}c@{}}0.386\\(0.193)\end{tabular} & \begin{tabular}{@{}c@{}}0.599\\(0.259)\end{tabular} \\
\cline{1-10}
\bottomrule
\end{tabular}
\end{table}

    In case of announcement days only the case is much clearer. 
    In \Cref{tab:dm_iv_surface_conv_bench_ann} we collect the p-values of the same Diebold-Mariano test
    but performed on forecasts on days of FOMC conferences. Therefore, the hypotheses being verified are:

    \begin{equation*}
        \begin{aligned}
            \begin{cases}
                H_0:\text{The forecasting accuracy during announcement days of the convolutional LSTM } \\
                \text{model without the announcement dummy is the same the benchmark random walk.} \\
                H_1:\text{The forecasting accuracy during announcement days of the convolutional LSTM model without the announcement } \\
                \text{model without the announcement dummy is better than the benchmark random walk.}
            \end{cases}
        \end{aligned}
    \end{equation*}

    As it can be seen, we cannot confidently reject the null hypotheses in all cases, which is to be expected
    as the model does not have access to the information regarding the upcoming announcement.

    \begin{table}[H]
\caption{Comparison of out-of-sample Diebold-Mariano mean p-value for ConvLSTM and benchmarks on announcement days. Standard deviation is displayed inside the brackets.}
\label{tab:dm_iv_surface_conv_bench_ann}
\fontsize{10}{14}\selectfont
\begin{tabular}{llllllllll}
\toprule
 & Type & \multicolumn{4}{c}{Call} & \multicolumn{4}{c}{Put} \\
 & Horizon & 1 & 2 & 5 & 10 & 1 & 2 & 5 & 10 \\
Model type & Model &  &  &  &  &  &  &  &  \\
\midrule
\multirow[t]{3}{*}{\textbf{ConvLSTM}} & \textbf{AHBS} & \begin{tabular}{@{}c@{}}0.445\\(0.173)\end{tabular} & \begin{tabular}{@{}c@{}}0.600\\(0.271)\end{tabular} & \begin{tabular}{@{}c@{}}0.470\\(0.413)\end{tabular} & \begin{tabular}{@{}c@{}}0.497\\(0.309)\end{tabular} & \begin{tabular}{@{}c@{}}0.524\\(0.462)\end{tabular} & \begin{tabular}{@{}c@{}}0.215\\(0.383)\end{tabular} & \begin{tabular}{@{}c@{}}0.400\\(0.384)\end{tabular} & \begin{tabular}{@{}c@{}}0.889\\(0.144)\end{tabular} \\
\textbf{} & \textbf{SVI} & \begin{tabular}{@{}c@{}}0.678\\(0.107)\end{tabular} & \begin{tabular}{@{}c@{}}0.507\\(0.135)\end{tabular} & \begin{tabular}{@{}c@{}}0.920\\(0.080)\end{tabular} & \begin{tabular}{@{}c@{}}0.785\\(0.214)\end{tabular} & \begin{tabular}{@{}c@{}}0.666\\(0.066)\end{tabular} & \begin{tabular}{@{}c@{}}0.479\\(0.109)\end{tabular} & \begin{tabular}{@{}c@{}}0.716\\(0.310)\end{tabular} & \begin{tabular}{@{}c@{}}0.926\\(0.095)\end{tabular} \\
\textbf{} & \textbf{linear} & \begin{tabular}{@{}c@{}}0.598\\(0.076)\end{tabular} & \begin{tabular}{@{}c@{}}0.719\\(0.086)\end{tabular} & \begin{tabular}{@{}c@{}}0.946\\(0.042)\end{tabular} & \begin{tabular}{@{}c@{}}0.911\\(0.082)\end{tabular} & \begin{tabular}{@{}c@{}}0.718\\(0.031)\end{tabular} & \begin{tabular}{@{}c@{}}0.274\\(0.043)\end{tabular} & \begin{tabular}{@{}c@{}}0.961\\(0.015)\end{tabular} & \begin{tabular}{@{}c@{}}0.954\\(0.022)\end{tabular} \\
\cline{1-10}
\bottomrule
\end{tabular}
\end{table}

    However, in \Cref{tab:dm_comp_global} we display the results of Diebold-Mariano test between the ML models themselves - with and without the information about 
    the announcement of interest rate policy occuring at a given date. Therefore, we check the following hypotheses:

    \begin{equation*}
        \begin{aligned}
            \begin{cases}
                H_0:\text{The forecasting accuracy of the convolutional LSTM model with the announcement } \\
                \text{dummy is the same as the convolutional LSTM model without the announcement dummy.} \\
                H_1:\text{The forecasting accuracy of the convolutional LSTM model with the announcement } \\
                \text{dummy is better than the convolutional LSTM model without the announcement dummy.}
            \end{cases}
        \end{aligned}
    \end{equation*}

    Globally except for AHBS call options, at 5\% significance level we do not have enough evidence to reject null hypothesis. 
    This can be mostly attributed to the instability of Diebold-Mariano test across seeds as the standard deviations are quite high.

    \begin{table}[H]
\caption{Comparison of out-of-sample Diebold-Mariano mean p-value for ConvLSTM with and without announcement dummy. Standard deviation is displayed inside the brackets.}
\label{tab:dm_comp_global}
\fontsize{10}{14}\selectfont
\begin{tabular}{llllllllll}
\toprule
 & Type & \multicolumn{4}{c}{Call} & \multicolumn{4}{c}{Put} \\
 & Horizon & 1 & 2 & 5 & 10 & 1 & 2 & 5 & 10 \\
Model type & Model &  &  &  &  &  &  &  &  \\
\midrule
\multirow[t]{3}{*}{\textbf{ConvLSTM}} & \textbf{AHBS} & \begin{tabular}{@{}c@{}}\textbf{\textcolor{UBCred}{0.000}}\\(0.000)\end{tabular} & \begin{tabular}{@{}c@{}}\textbf{\textcolor{UBCred}{0.000}}\\(0.000)\end{tabular} & \begin{tabular}{@{}c@{}}\textbf{\textcolor{UBCred}{0.000}}\\(0.000)\end{tabular} & \begin{tabular}{@{}c@{}}\textbf{\textcolor{UBCred}{0.040}}\\(0.097)\end{tabular} & \begin{tabular}{@{}c@{}}0.311\\(0.476)\end{tabular} & \begin{tabular}{@{}c@{}}0.293\\(0.441)\end{tabular} & \begin{tabular}{@{}c@{}}0.311\\(0.383)\end{tabular} & \begin{tabular}{@{}c@{}}0.424\\(0.346)\end{tabular} \\
\textbf{} & \textbf{SVI} & \begin{tabular}{@{}c@{}}0.310\\(0.450)\end{tabular} & \begin{tabular}{@{}c@{}}0.336\\(0.407)\end{tabular} & \begin{tabular}{@{}c@{}}0.349\\(0.423)\end{tabular} & \begin{tabular}{@{}c@{}}0.287\\(0.375)\end{tabular} & \begin{tabular}{@{}c@{}}0.173\\(0.360)\end{tabular} & \begin{tabular}{@{}c@{}}0.204\\(0.323)\end{tabular} & \begin{tabular}{@{}c@{}}0.254\\(0.392)\end{tabular} & \begin{tabular}{@{}c@{}}0.302\\(0.413)\end{tabular} \\
\textbf{} & \textbf{linear} & \begin{tabular}{@{}c@{}}0.460\\(0.427)\end{tabular} & \begin{tabular}{@{}c@{}}0.267\\(0.302)\end{tabular} & \begin{tabular}{@{}c@{}}0.140\\(0.159)\end{tabular} & \begin{tabular}{@{}c@{}}0.175\\(0.182)\end{tabular} & \begin{tabular}{@{}c@{}}0.624\\(0.490)\end{tabular} & \begin{tabular}{@{}c@{}}0.520\\(0.500)\end{tabular} & \begin{tabular}{@{}c@{}}0.431\\(0.481)\end{tabular} & \begin{tabular}{@{}c@{}}0.329\\(0.500)\end{tabular} \\
\cline{1-10}
\bottomrule
\end{tabular}
\end{table}

    Much stronger result pertains to FOMC conferences days only. We verify almost the same hypotheses as above:

    \begin{equation*}
        \begin{aligned}
            \begin{cases}
                H_0:\text{The forecasting accuracy during announcement days of the convolutional LSTM model } \\
                \text{with the announcement dummy is the same as the model without the announcement dummy.} \\
                H_1:\text{The forecasting accuracy during announcement days of the convolutional LSTM model } \\
                \text{with the announcement dummy is better than the model without the announcement dummy.}
            \end{cases}
        \end{aligned}
    \end{equation*}

    We conclude that during policy announcement days almost all results are statistically significant at 5\% level 
    especially for call options as presented in \Cref{tab:dm_comp_ann} with relatively small standard deviations.
    Thus, ML models augmented with the announcement dummy forecast IV during FOMC conferences usually more accurately than models without this additional information.

    \begin{table}[H]
\caption{Comparison of out-of-sample Diebold-Mariano mean p-value for ConvLSTM with and without announcement dummy. Standard deviation is displayed inside the brackets.}
\label{tab:dm_comp_ann}
\fontsize{10}{14}\selectfont
\begin{tabular}{llllllllll}
\toprule
 & Type & \multicolumn{4}{c}{Call} & \multicolumn{4}{c}{Put} \\
 & Horizon & 1 & 2 & 5 & 10 & 1 & 2 & 5 & 10 \\
Model type & Model &  &  &  &  &  &  &  &  \\
\midrule
\multirow[t]{3}{*}{\textbf{ConvLSTM}} & \textbf{AHBS} & \begin{tabular}{@{}c@{}}\textbf{\textcolor{UBCred}{0.001}}\\(0.001)\end{tabular} & \begin{tabular}{@{}c@{}}\textbf{\textcolor{UBCred}{0.003}}\\(0.002)\end{tabular} & \begin{tabular}{@{}c@{}}\textbf{\textcolor{UBCred}{0.011}}\\(0.009)\end{tabular} & \begin{tabular}{@{}c@{}}0.087\\(0.118)\end{tabular} & \begin{tabular}{@{}c@{}}\textbf{\textcolor{UBCred}{0.019}}\\(0.041)\end{tabular} & \begin{tabular}{@{}c@{}}\textbf{\textcolor{UBCred}{0.029}}\\(0.067)\end{tabular} & \begin{tabular}{@{}c@{}}0.164\\(0.319)\end{tabular} & \begin{tabular}{@{}c@{}}0.199\\(0.391)\end{tabular} \\
\textbf{} & \textbf{SVI} & \begin{tabular}{@{}c@{}}\textbf{\textcolor{UBCred}{0.001}}\\(0.000)\end{tabular} & \begin{tabular}{@{}c@{}}\textbf{\textcolor{UBCred}{0.001}}\\(0.000)\end{tabular} & \begin{tabular}{@{}c@{}}\textbf{\textcolor{UBCred}{0.002}}\\(0.001)\end{tabular} & \begin{tabular}{@{}c@{}}\textbf{\textcolor{UBCred}{0.004}}\\(0.004)\end{tabular} & \begin{tabular}{@{}c@{}}0.053\\(0.029)\end{tabular} & \begin{tabular}{@{}c@{}}\textbf{\textcolor{UBCred}{0.039}}\\(0.021)\end{tabular} & \begin{tabular}{@{}c@{}}0.071\\(0.077)\end{tabular} & \begin{tabular}{@{}c@{}}0.176\\(0.392)\end{tabular} \\
\textbf{} & \textbf{linear} & \begin{tabular}{@{}c@{}}\textbf{\textcolor{UBCred}{0.004}}\\(0.001)\end{tabular} & \begin{tabular}{@{}c@{}}\textbf{\textcolor{UBCred}{0.003}}\\(0.001)\end{tabular} & \begin{tabular}{@{}c@{}}\textbf{\textcolor{UBCred}{0.003}}\\(0.000)\end{tabular} & \begin{tabular}{@{}c@{}}\textbf{\textcolor{UBCred}{0.005}}\\(0.002)\end{tabular} & \begin{tabular}{@{}c@{}}\textbf{\textcolor{UBCred}{0.011}}\\(0.005)\end{tabular} & \begin{tabular}{@{}c@{}}\textbf{\textcolor{UBCred}{0.007}}\\(0.004)\end{tabular} & \begin{tabular}{@{}c@{}}\textbf{\textcolor{UBCred}{0.022}}\\(0.032)\end{tabular} & \begin{tabular}{@{}c@{}}0.154\\(0.240)\end{tabular} \\
\cline{1-10}
\bottomrule
\end{tabular}
\end{table}

    Therefore, summing up the results above, we can reject the hypothesis that models with and without additional exogeneous dummy variable have equal predictive performance,
    but the difference is only clearly visible during announcement days. 
    The aforementioned results support our view that adding the date of FOMC meetings 
    as an exogeneous feature improves ML model performance, during policy annoucement days. However, globally the improvement is not as much pronounced.

\subsubsection{Model diagnostics}\label{sec:model_diag}

    Even though we have shown that ML framework, especially when augmented with exogeneous information gives a significant edge in forecasting IV surface, in what follows
    we also attempt to perform a more in-depth analysis and diagnostics of the models. 
    
    First of all, the natural extension of the tables with global error metrics is taking a closer look at the local error, 
    which means the average error across log-moneyness and maturity. We stick to the model which learns using SVI interpolated data augmented with the announcement dummy where applicable, 
    as it seemed to achieve the best performance. We also focus on horizon 1. 
    Analogous graphs for other construction methods and horizons 
    are available in \Cref{sec:models_diag_errors_RMSE}, \Cref{sec:models_diag_errors_MPE} and \Cref{sec:models_diag_errors_DM}.

\begin{figure}[H]
        \centering
        \includegraphics[width=\linewidth]{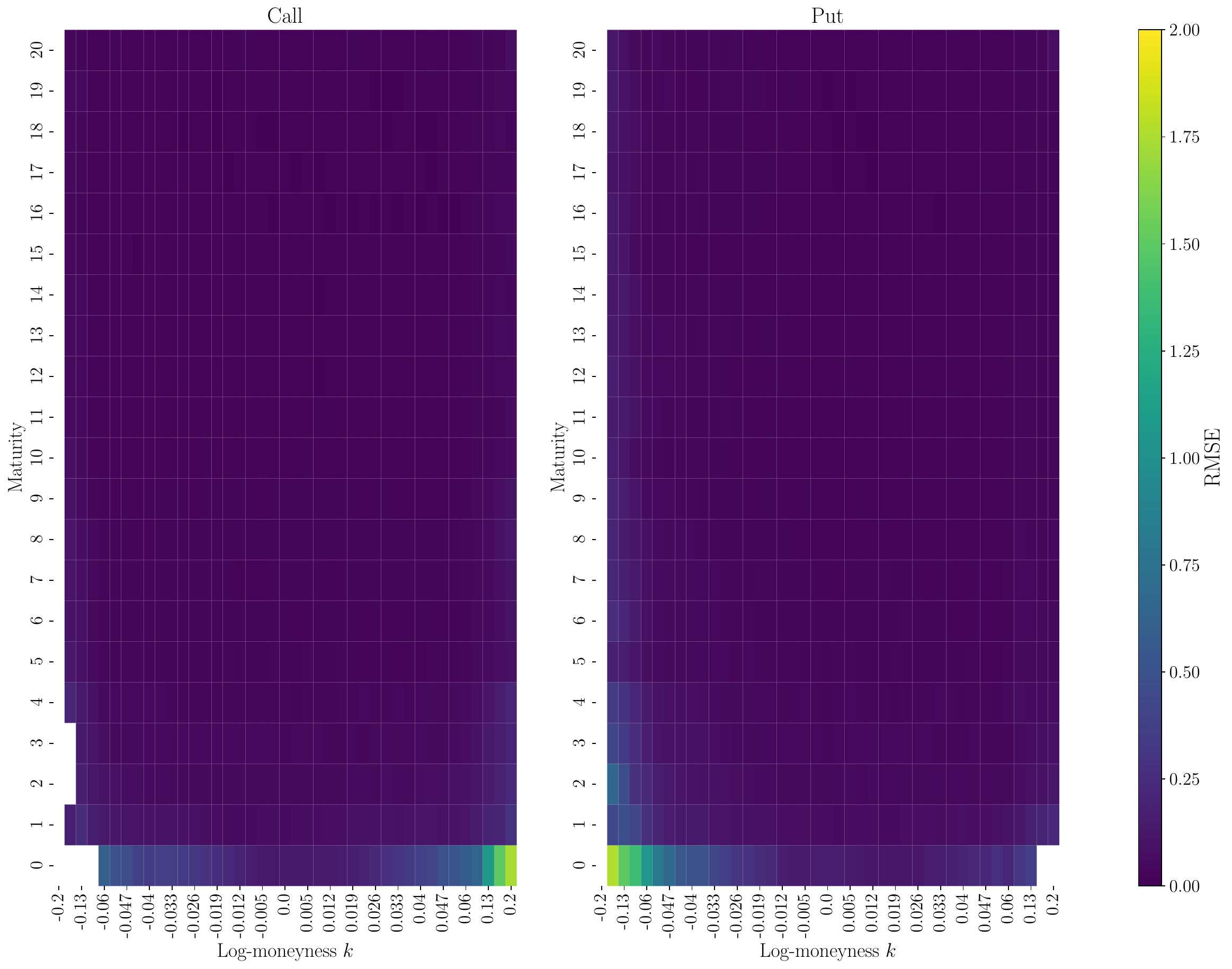}
        \caption{Out-of-sample RMSE for predictions from the model with announcement dummy - surface interpolated by SVI model, prediction horizon 1.}
        \label{fig:RMSE_grid_convLSTM}
    \end{figure}

\begin{figure}[H]
        \centering
        \includegraphics[width=\linewidth]{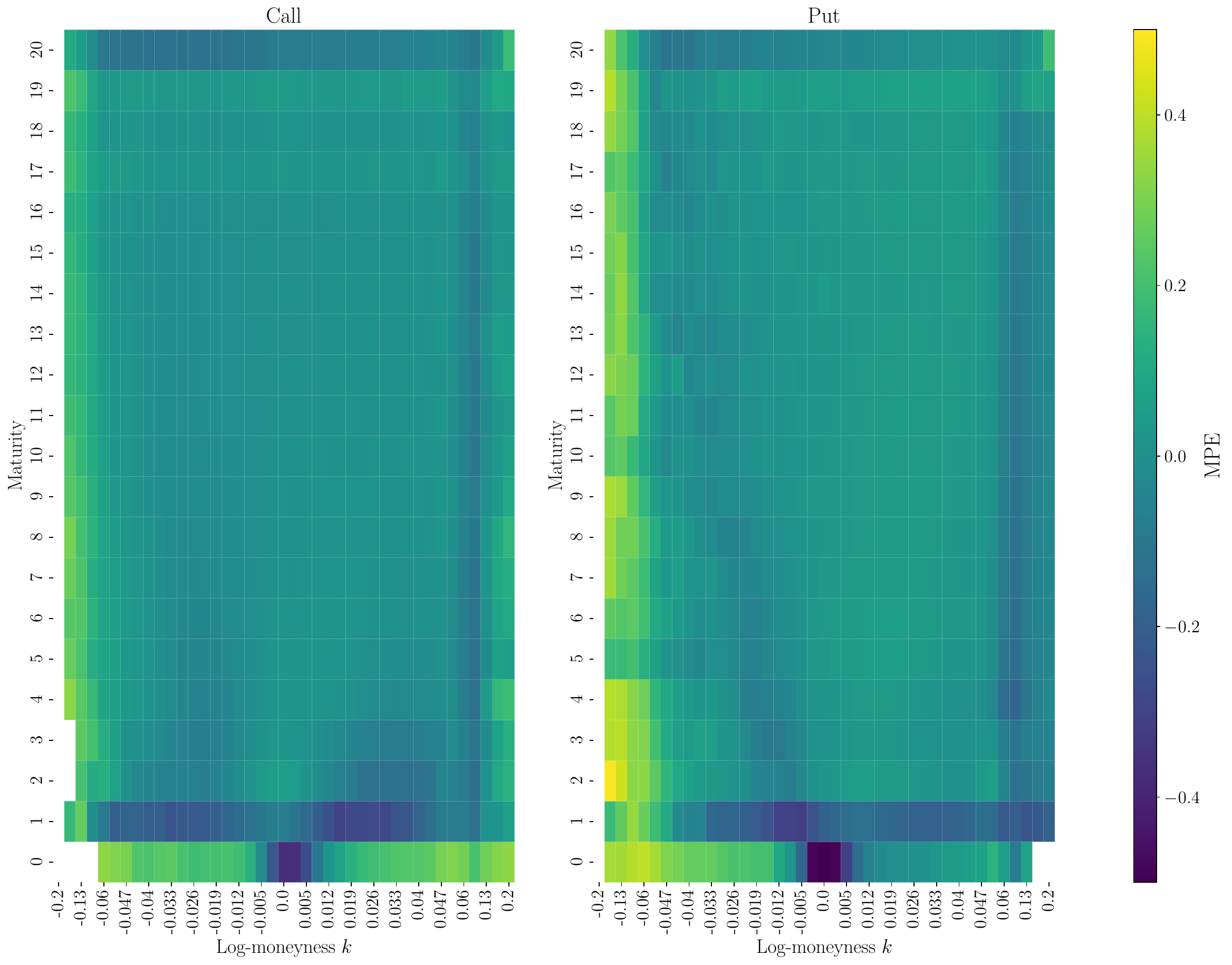}
        \caption{Out-of-sample MPE for predictions from the model with announcement dummy - surface interpolated by SVI model, prediction horizon 1.}
        \label{fig:MPE_grid_convLSTM}
    \end{figure}

    In the following graphs \Cref{fig:RMSE_grid_convLSTM} and \Cref{fig:MPE_grid_convLSTM} we break down the RMSE and MPE, respectively per moneyness and maturity. As it can be observed
    high values of both error metrics are mainly concentrated around the wings of short-maturity options. In addition, MPE pattern
    shows the model on average underestimates the IV especially in the wings, while slightly overestimates around ATM.
    This can be attributed to much lower variation in that region of the surface.
    Moreover, longer-maturity options react much more slowly to any shocks because the highest impact on options price comes from its time value. 
    Therefore, we could confirm the expectation that a better model fit can be achieved for ATM implied volatility and longer maturities. 
    This has profound implicatons for forecasting. Implied volatility for long maturities are much more predictable and can be more easily captured by the model.
    Also ATM IV seems to be forecasted with a relatively high precision. 
    At the same time using the model to predict IV in the wings of shorter maturities requires much more caution. Therefore, these results would contradict
    the hypothesis that the model performance is the same across the entire IV surface. As predicted, the error is concentrated in the short-maturity wings.

    In addition, previously discussed Diebold-Mariano test aggregated on the whole surface tells little about where exactly the shortcomings of the model are concentrated.
    The grid-level analysis reveals that we can reject the null hypothesis that the ML model has the same forecasting accuracy
    as its benchmark at $5\%$ significance level for short time horizons (except for 0 DTE options) and in the ITM wing of call options and OTM wing of put options. 
    It could be explained by a relatively lower volatility of longer maturities which implies the model cannot achieve a significant edge over the benchmark. 
    On the other hand, 0 DTE options are notoriously difficult to predict and price due to very short maturity and quickly dropping time value.

    \begin{figure}[H]
        \centering
        \includegraphics[width=\linewidth]{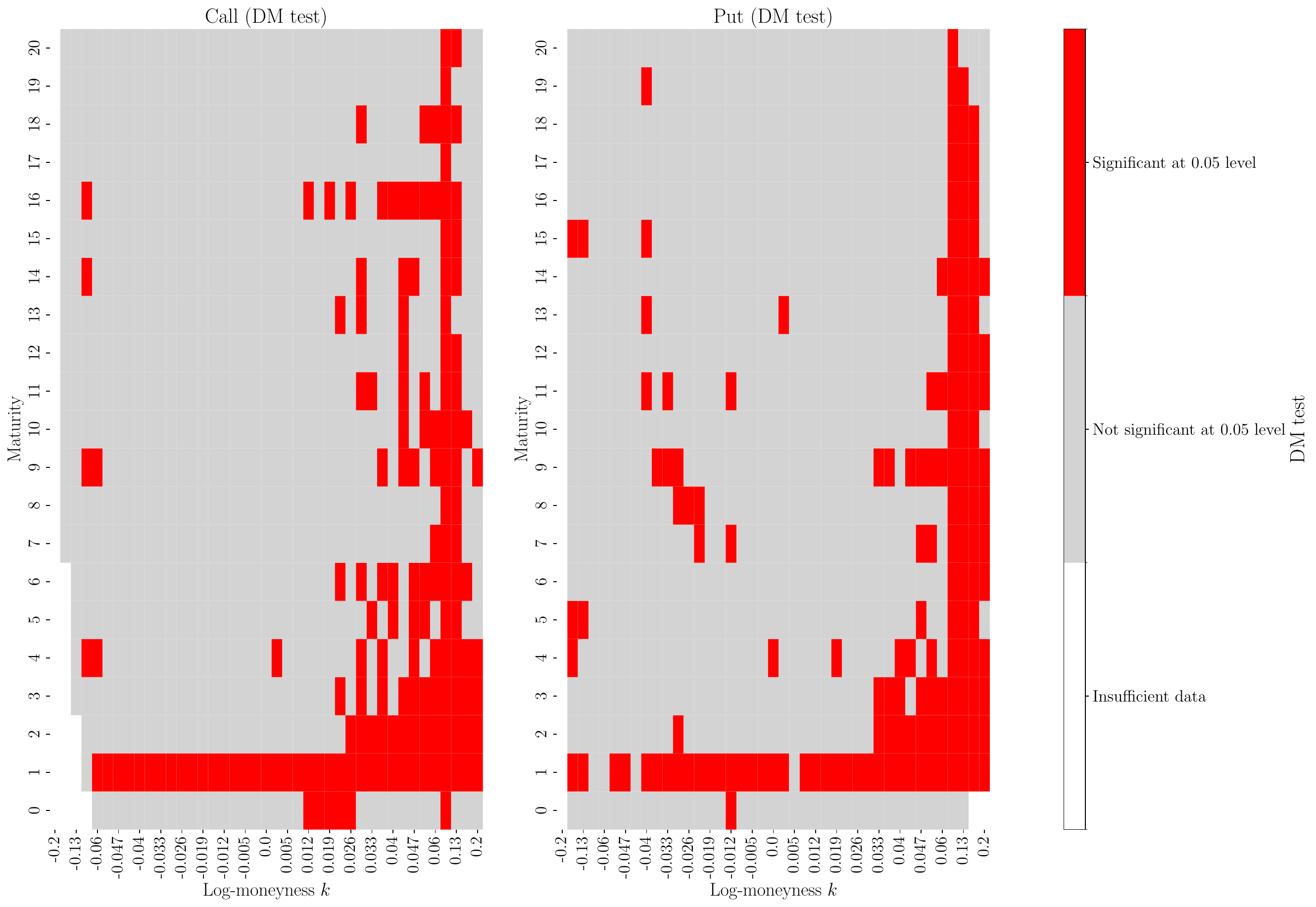}
        \caption{Diebold-Mariano test comparing predictions from the model without announcement dummy to the benchmark - surface interpolated by SVI model, prediction horizon 1.}
        \label{fig:DM_grid_convLSTM}
    \end{figure}

    These observatons and error metrics being smaller for ML models yield mixed conclusions. The performance of the ML framework is definitely not uniform across the IV surface
    and therefore, we cannot with full confidence reject our original hypothesis that ML framework beats the random walk in prediction power both globally and during US monetary policy announcement days.

\begin{figure}[H]
        \centering
        \includegraphics[width=\linewidth]{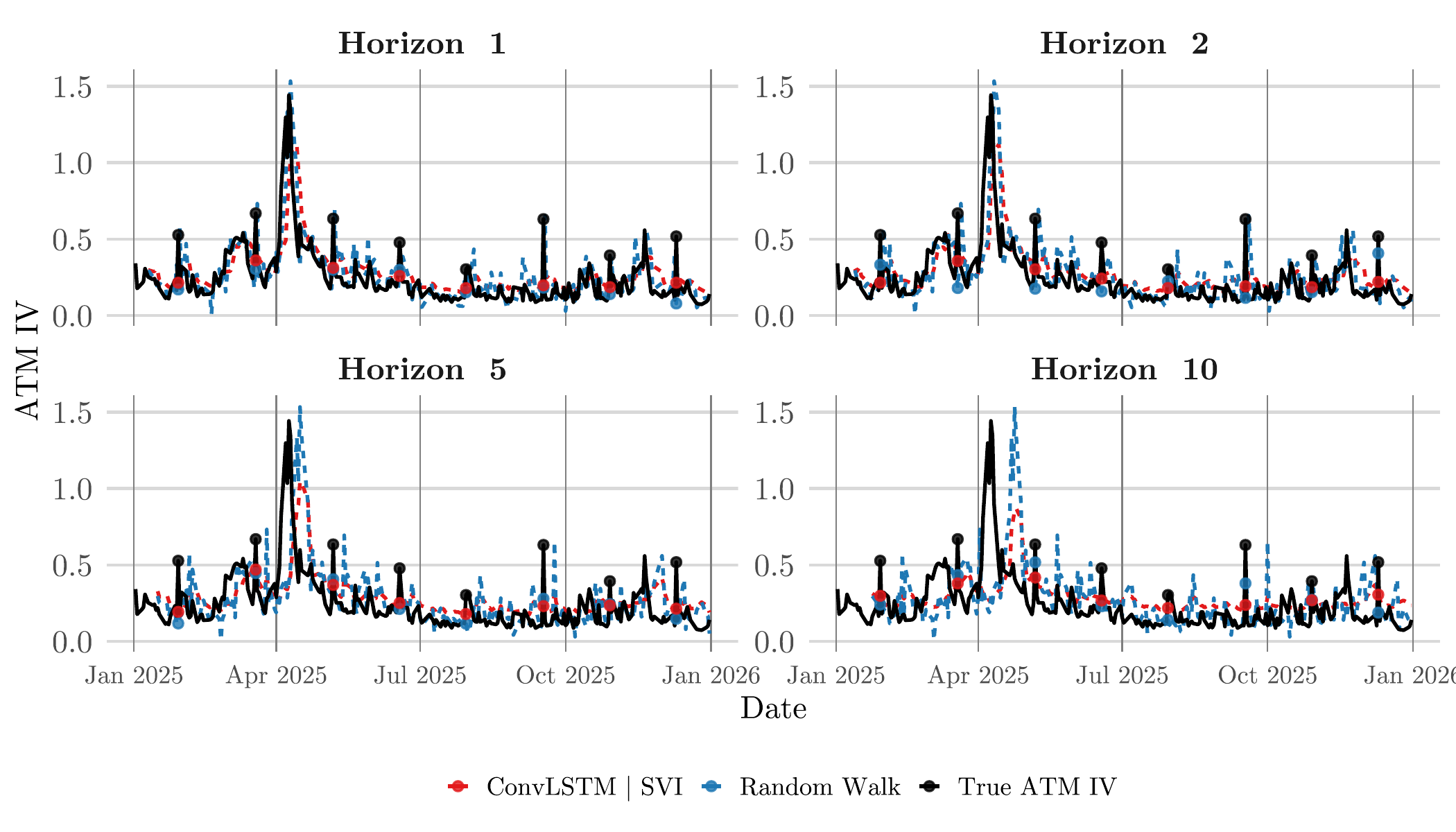}
        \caption{Comparison between Random Walk and ConvLSTM model without announcement dummy out-of-sample predictions of ATM IV paired with SVI model, call options. 
        Dots represent an annoucement date.}\label{fig:ATM_IV_preds_convLSTM}
    \end{figure}

    Secondly, as the first part of our analysis focused on the ATM IV, we turn to visualising the out-of-sample predictions depicted in \Cref{fig:ATM_IV_preds_convLSTM}.
    The graph makes it clear the performance of the convolutional LSTM model trained on SVI interpolated data is quite accurate 
    especially in shorter forecasting horizons. The pattern of forecats suggests the model has learnt the strong persistence of IV 
    which is one of its main characteristics as discussed in \parencite{Cont}. The model seem to have also learnt the return to the average volatility
    in longer horizons, although the persistent element is much more pronounced. 

\begin{figure}[H]
        \centering
        \includegraphics[width=\linewidth]{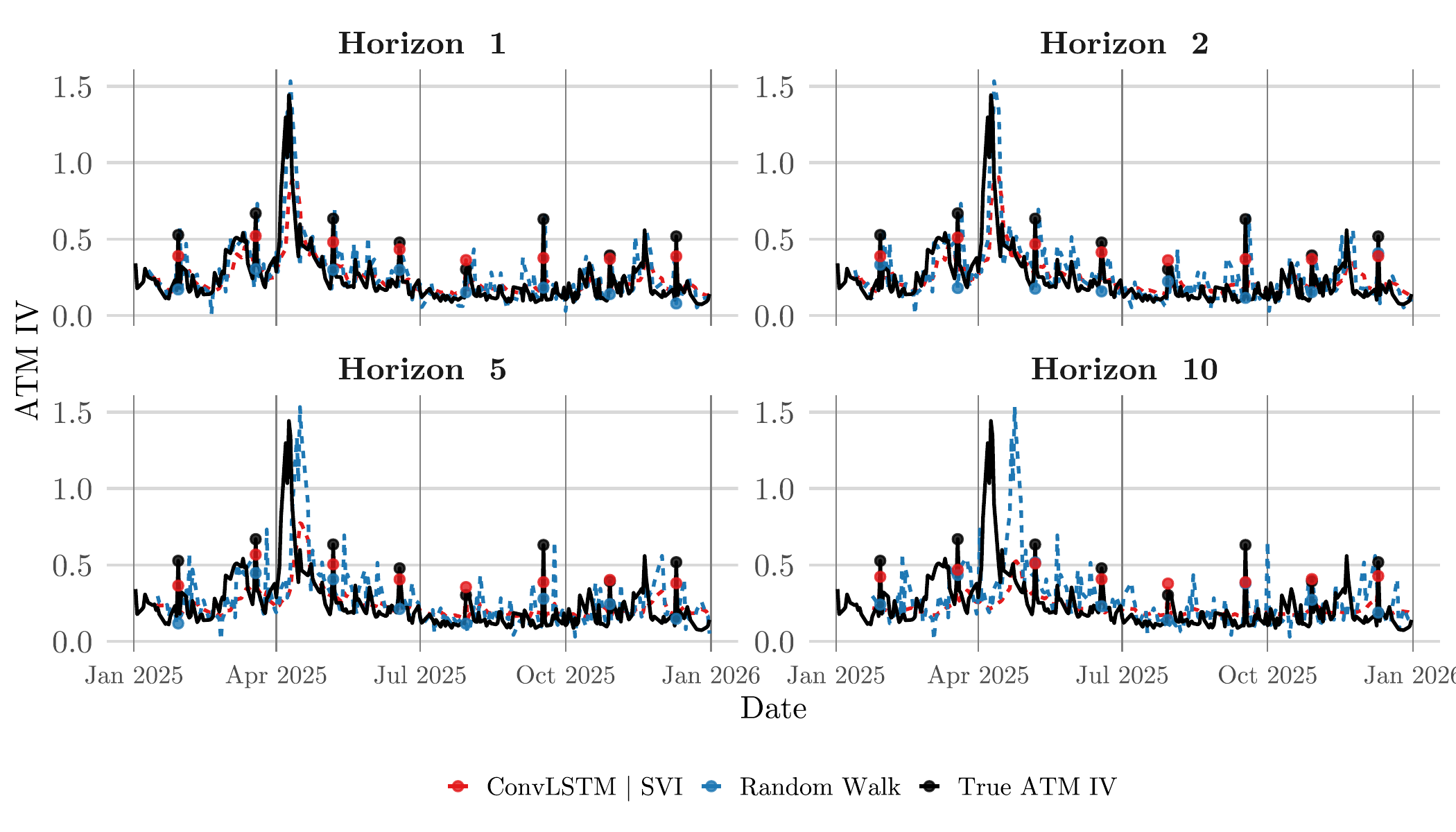}
        \caption{Comparison between Random Walk and ConvLSTM model with announcement dummy out-of-sample predictions of ATM IV paired with SVI model, call options. 
        Dots represent an annoucement date}\label{fig:ATM_IV_preds_convXLSTM}
    \end{figure}

    Nevertheless, what the model seems to have missed is the impact of scheduled FOMC meetings, which are highlighted on the graph by black dots.
    Naturally the model, since it learns high persistence and has no way of knowing about the upcoming significant announcement, systematically underestimates IV on those days.
    However, the model augmented with the announcement dummy variable, whose predictions of ATM IV are shown in \Cref{fig:ATM_IV_preds_convXLSTM},
    takes the associated effect into account and raises the IV whenever the meeting is scheduled.
    Still, the fit is not perfect because not all meetings result in such a tangible shock, yet the difference with the baseline model is noticable.

\begin{figure}[H]
        \centering
        \begin{subfigure}[b]{0.49\linewidth}
            \centering
            \includegraphics[width=\linewidth]{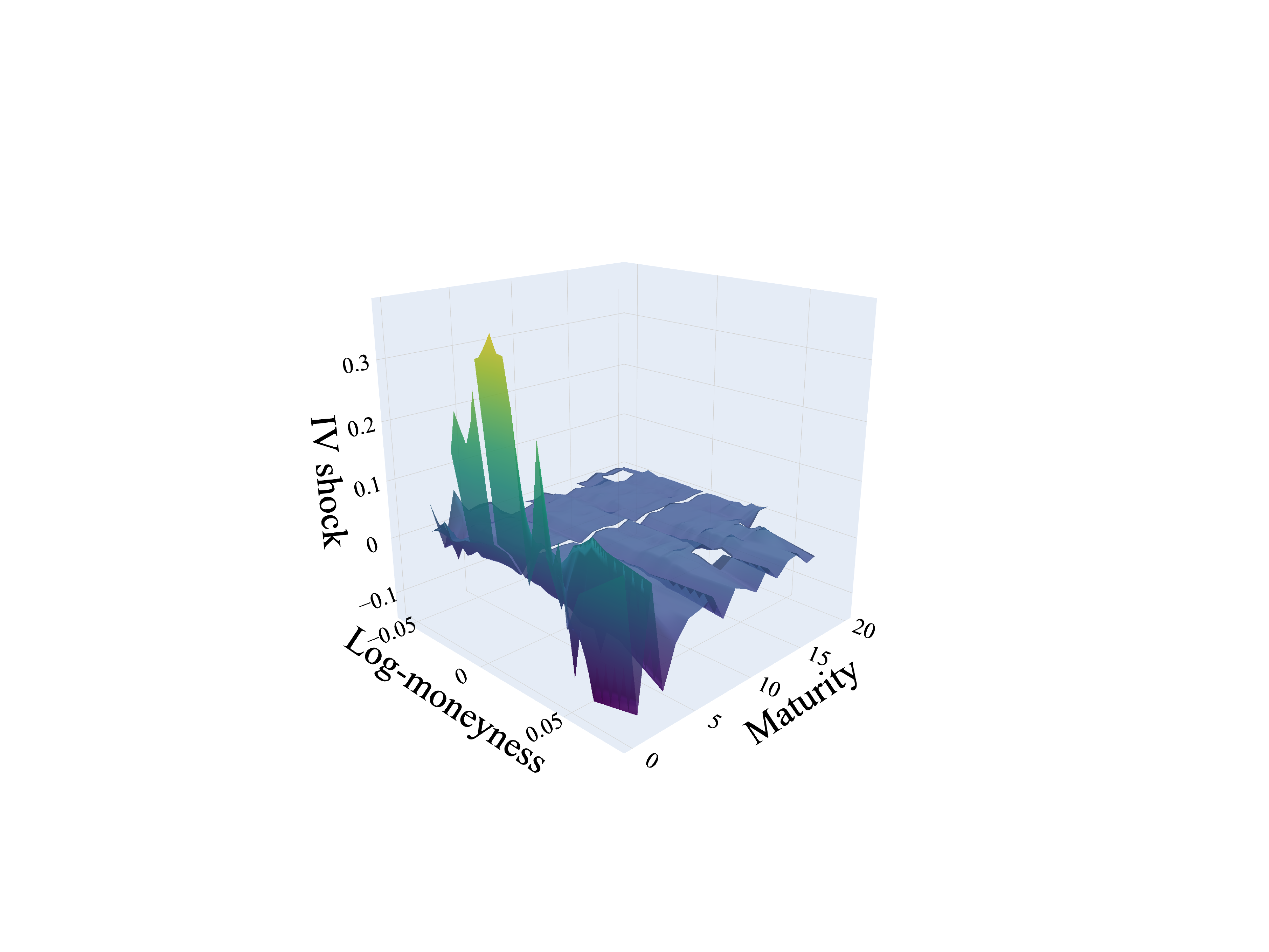}
            \caption{Average difference between the implied volatility of call options at the time of the FOMC conference and the day before.}\label{fig:shock_call_true}
        \end{subfigure}
        \hfill
        \begin{subfigure}[b]{0.49\linewidth}
            \centering
            \includegraphics[width=\linewidth]{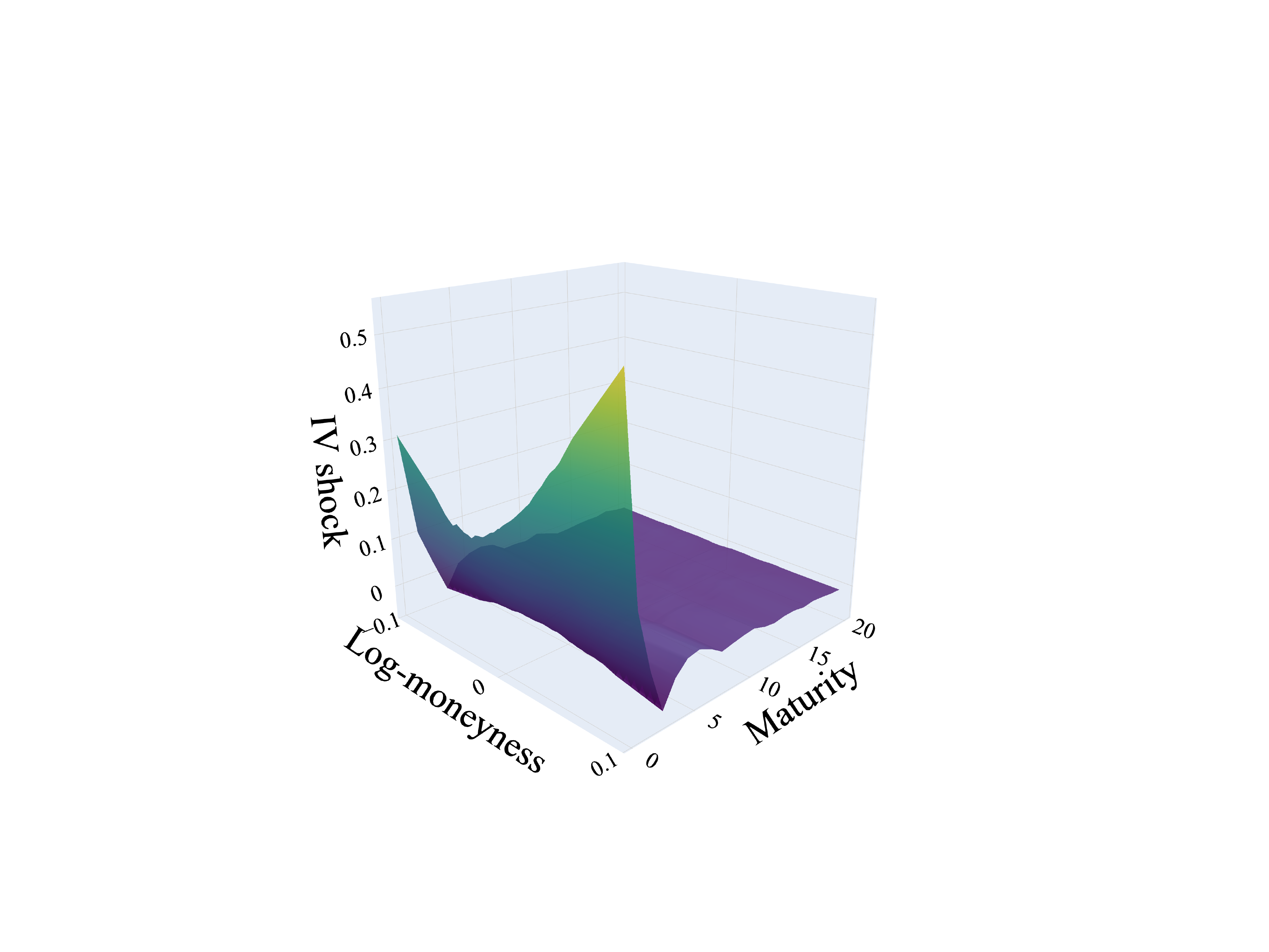}
            \caption{Average difference between the fitted (with the ConvLSTM model with an announcement dummy and interpolated by SVI) implied volatility $\hat{\sigma}^{\text{IV}}$ 
                of call options at the time of the FOMC conference and the day before.}\label{fig:shock_call_model}
        \end{subfigure}
    \end{figure}

\begin{figure}[H]
        \centering
        \begin{subfigure}[b]{0.49\linewidth}
            \centering
            \includegraphics[width=\linewidth]{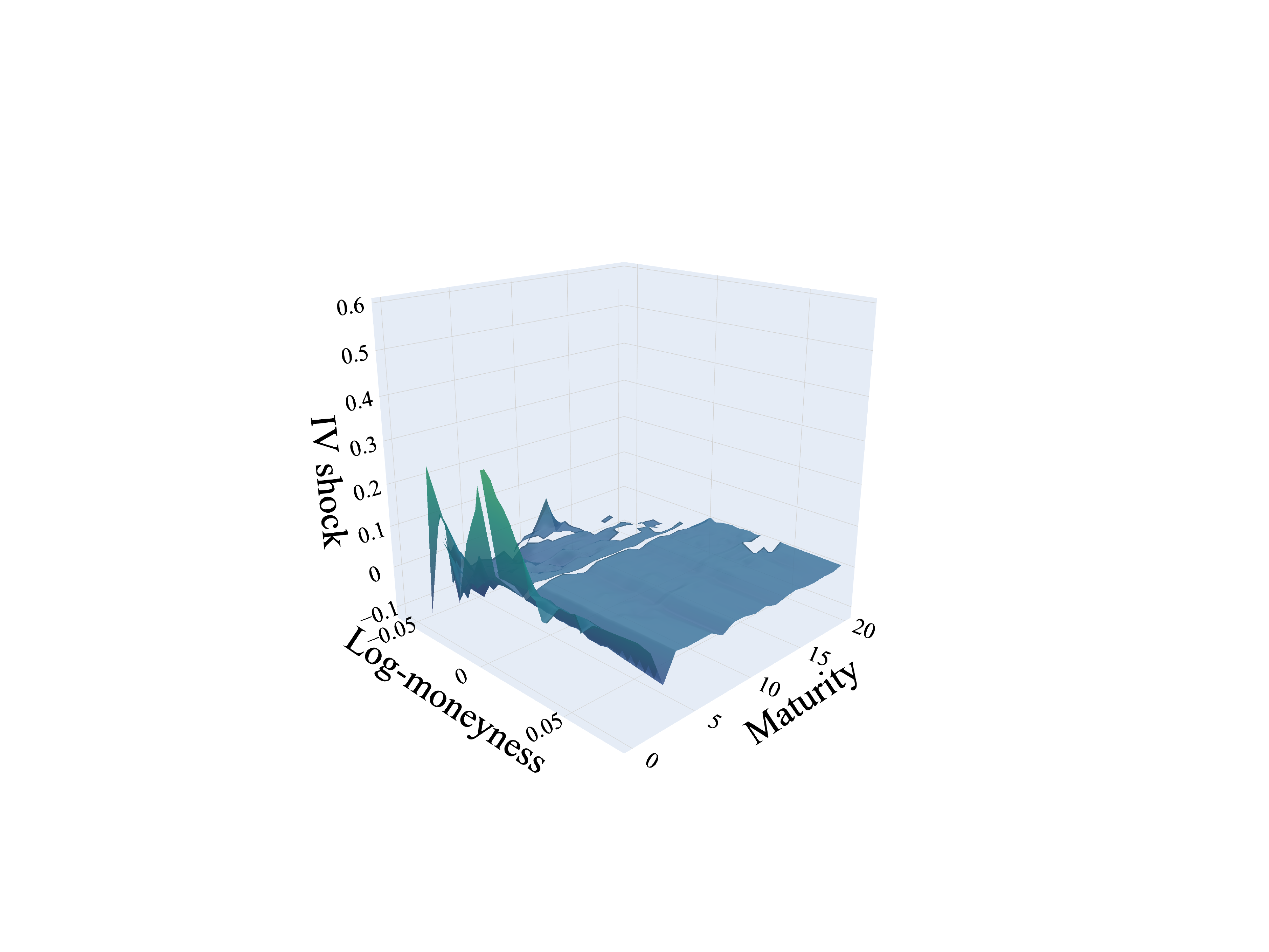}
            \caption{Average difference between the implied volatility of put options at the time of the FOMC conference and the day before.}\label{fig:shock_put_true}
        \end{subfigure}
        \hfill
        \begin{subfigure}[b]{0.49\linewidth}
            \centering
            \includegraphics[width=\linewidth]{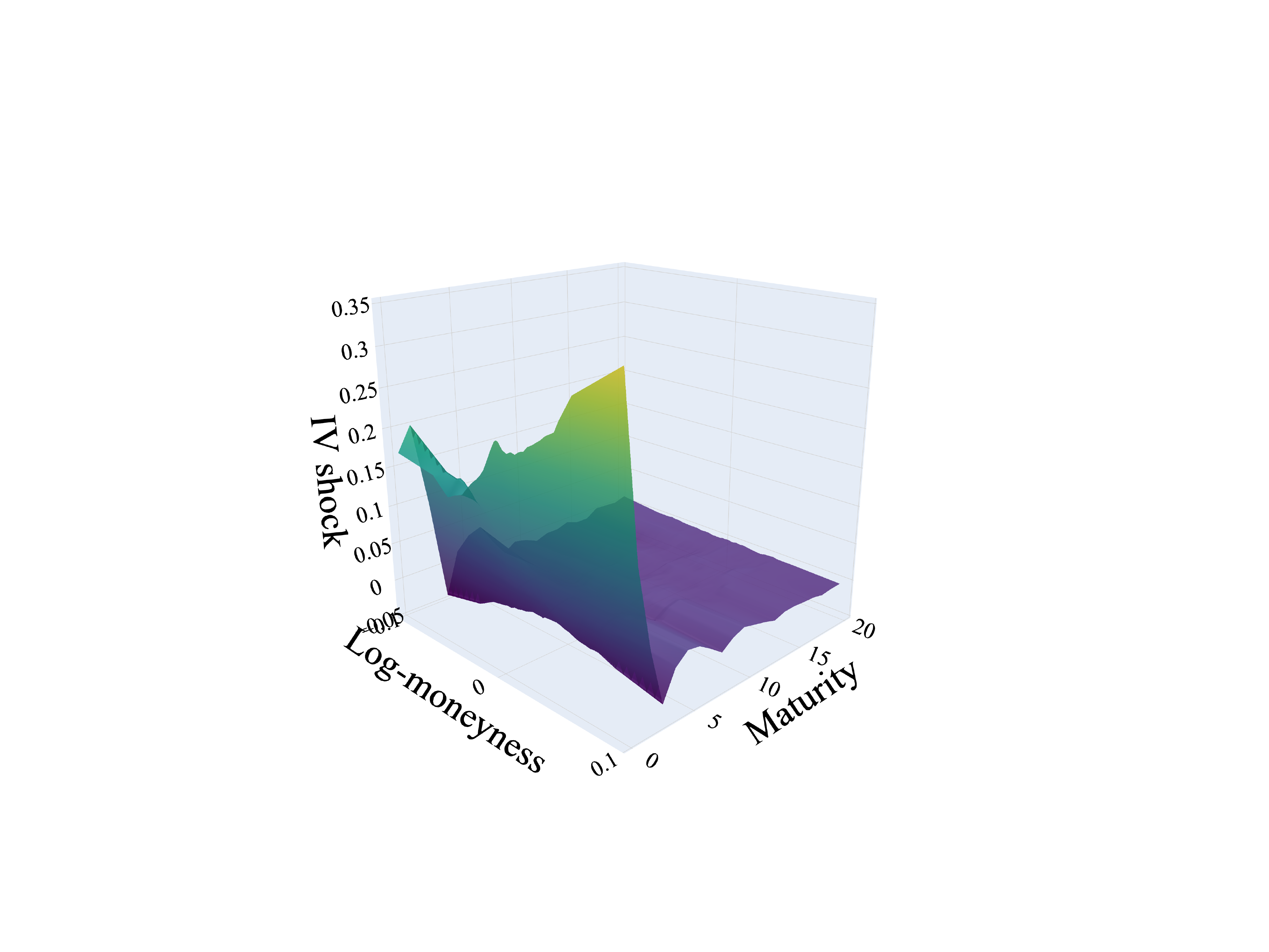}
            \caption{Average difference between the fitted (with the ConvLSTM model with an announcement dummy and interpolated by SVI) implied volatility $\hat{\sigma}^{\text{IV}}$ 
                of put options at the time of the FOMC conference and the day before.}\label{fig:shock_put_model}
        \end{subfigure}
    \end{figure}

    In order to further investigate the extent to which the ML model has learnt the effect of pre-announcement uncertainty we plot the average difference in forecasted IV
    between the day of the policy announcement and the day before presented in \Cref{fig:shock_call_model} for call options and \Cref{fig:shock_put_model} for put options. 
    We compare it to the actual average shock across the surface plotted in \Cref{fig:shock_call_true} for call options and \Cref{fig:shock_put_true} for put options.

    First of all, it needs to be noted that our previous observations regarding first two hypotheses based on the plots of a true shock still hold. 
    Not only can we observe the increase in IV both for call and put options, this increase is also the highest for short dated options and subsides with time to maturity.
    The positive shock in IV before the annoncement is also the strongest for ATM IV.

    When it comes to the ML model it seems to have correctly learnt the level and shape of the shock, especially for call options, albeit the effect is smoothed out.
    In case of put options the OTM (right wing) is not consistent with the true pattern possibly due to interpolation error - the SVI could not perfectly fit this wing
    if it was not as convex.
    Nevertheless, these results reinforce the evidence in favour of the hypothesis that adding the policy announcement as an exogeneous variables improves the model performance.
    The ML model has generally correctly identified the pattern of elevated volatility and the level on average is also consistent with the observed true market data which
    results in a better fit on those days.

    \subsubsection{Impulse Response}\label{sec:imp_res}

    As a last element of deriving insights from the fitted model we calculate and plot the impulse response (IRF) from a pre-announcement shock.
    Doing this analysis on raw data would be quite hard, yet training the models allows us to simulate the (model implied) reaction of the IV to the FOMC conference.
    To calculate the IRF, assuming the shock happens at time $\tau$, we follow the steps outlined below:
    
    \begin{enumerate}
        \item Compute the average pre-announcement surfaces ($\bar{\sigma}(K^*,T^*) = \bar{\hat{\sigma}}^{\text{IV}}_{\tau-1}(K^*,T^*),\dots,\bar{\hat{\sigma}}^{\text{IV}}_{\tau-l}(K^*,T^*)$);
        \item Predict the surface $\E\left[\hat{\sigma}^{\text{IV}}_{\tau}(K^*,T^*)\right]$ setting the dummy to 1 or 0;
        \item Predict subsequent surfaces using the two versions from step 1 and 2 and calculate the difference between them.
    \end{enumerate}

    We thus predict the announcement and non-annoucement impulse response for horizon $h$ for all pairs $(K^*_i,T^*_j)$:

    \begin{equation*}
        IRF_h(K^*_i,T^*_j) = \E\left[\sigma^{\text{IV}}_{\tau + h}(K^*_i,T^*_j)|X_\tau=1, \bar{\sigma}(K^*_i,T^*_j)\right] - 
        \E\left[\hat{\sigma}^{\text{IV}}_{\tau + h}(K^*_i,T^*_j) | X_\tau=0, \bar{\sigma}(K^*_i,T^*_j)\right]
    \end{equation*}

\begin{figure}[H]
        \centering
        \includegraphics[width=0.9\linewidth]{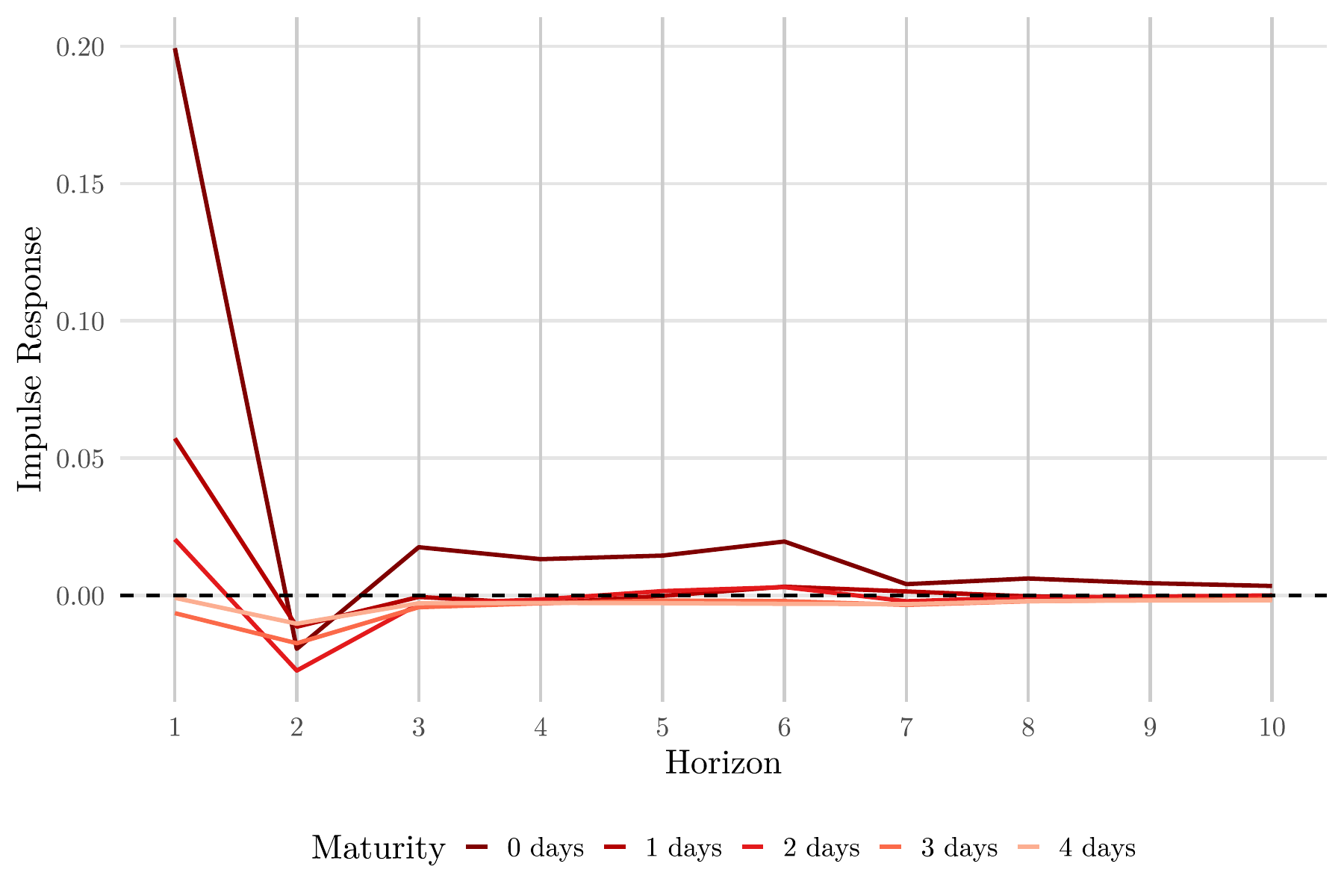}
        \caption{The difference between the announcement and non-announcement IRF for ATM IV 0 DTE of call options for the ConvLSTM model trained on the surface interpolated by SVI.}
        \label{fig:IRF_call}
    \end{figure}

    As our previous analyses have shown the most interesting results pertain to ATM short dated options, thus we focus on this one particular point on the surface.
    The results are shown in \Cref{fig:IRF_call}, which makes it clear the pre-announcement effect is quite substantial in size even up to 20\% on average for 0 DTE options. 
    The values of IRF at horizon $1$ confirm that the annoucement results in the resolution of uncertainty and the resulting drop in IV, as well. 
    Moreover, again the plot implies the longer the maturity, the smaller the impulse response.
    The plot also corroborates our previous observations regarding the model. Apparently, it has learnt high persistence of IV, which in the absence of other shocks and data results in an indefinitely elevated IV.   
 
\section{Concluding remarks}\label{sec:conc}

In this section we gather all the results and review them in terms of stated hypotheses. 
We also discuss the limitations of our approach and lay out the potential for further studies.

\subsection{Limitations}\label{sec:conc_lim}

    Primiraly, the main limitation of our approach is the scope of ML architecture. The trained model is rather small in terms of ML framework. 
    Furthermore, only one type of architecture finally made it to our analysis. However, it has to be noted that this choice of architecture has a profound 
    theoretical justification and was the result of many trials and errors. In addition, we deem the performance of the model satisfactory given its size and low cost of training.
    
    Second of all, as the analysis is centered around FOMC meetings, we have only introduced one exogeneous variable, which however managed to significantly 
    improve the performance of the model. 

    In addition, we found out the data in the wings, of especially short-maturity options is of poor quality due to low liquidity and arbitrage violations.
    It singificantly affects our ability to derive meaningful conclusions regarding the pre-announcement effect in these regions of implied volatility surface.

    Lastly, our results lack the verification in an actual trading environment. We considered only statistical significance, which we believe to be a proper starting point.

\subsection{Future research}\label{sec:conc_fut_research}

    Nevertheless, all these limitations imply a huge potential for future research.

    First of all, we could include many more exogeneous variables, for instance the return and volatility of the underlying, VIX, as well as economic variables and
    naturally more economic events and scheduled data and policy announcements.

    Having added these variables, we could consider bigger models and alternative architectures like attention mechanisms, 
    especially when it comes to connecting exogeneous features with endogeneous past observations of IV surface.

    Last but not least, we would like to apply the results of our study to algorithmic trading, leveraging our model especially around policy announcement days.
    It would allow us to check the economic significance of our results, complementing the statistical significance.

\subsection{Conclusion}\label{sec:conc_conc}

    In conclusion, the pre-announcement effect of elevated uncertainty translates prominently into options market, where investors price the policy and systemic risk.
    We explored the evolution of IV derived from options on S\&P 500 around the FOMC conferences in high frequency and rejected the hypothesis that 
    IV does not change before nor during the FOMC conference. 
    We have discovered it monotonically increases prior to the announcement in expectation of higher realised volatility. 
    However, we noticed there is a timing mismatch between the increase 
    of IV, which for short dated options starts up to two days before the announcement and the jump in realised volatility which occurs only after the conference 
    as the markets react to the announcement. At the same time this effect seems to be stronger for short dated options, which confirms our second research question.
    Due to poorer data quality we could not find enough evidence to back or refute the expectation that the pre-announcement effect is stronger for OTM options.
    Nevertheless, our analysis implies the increase in IV is centered around ATM point and the ITM wing.

    Moreover, we have effectively leveraged ML framework to forecast the entire IV surface both during regular and abnormal days. 
    We have concluded that the convolutional 2D LSTM paired with SVI interpolation/extrapolation generally achieves the best performance in terms of both RMSE and MPE. 
    Nevertheless, we could not confirm a full statistical significance of this better performance. 
    The edge in terms of predictive accuracy is not uniform across IV surface and is concentrated mainly around short maturities and ATM point.
    
    Furthermore, we have managed to succeed in the first attempt to quantitatively forecast the pre-announcement effect.
    We achieved this by including exogeneous information inside ML framework in the form of a dummy variable indicating the days of FOMC conferences.
    We have rejected the hypothesis that models having access to this information have equal predictive performance to models which are not augmented with the announcement dummy
    during policy announcement days.
    
    Finally, we have conducted a much deeper analysis of the fitted model. Our study has revealed significant insights into the ML framework.
    We have discovered that the model has mainly learnt the persistence in the IV surface and the return to the mean in the long-term. 
    We have uncovered evidence suggesting that the closest fit can be attained across ATM and long-maturity options.

    Even though our study reinforces the view that machine learning models can effectively forecast the implied volatility surface, 
    it also highlights the importance of careful feature engineering. 
    In particular, incorporating scheduled policy announcements such as FOMC conferences can materially improve predictive performance on days with such announcements. 
    Without accounting for such event-driven dynamics, the advantage of ML models remains limited, especially when dealing with noisy financial time series. 
    More broadly, our findings suggest that understanding the economic mechanisms underlying market behavior is just 
    as important as the sophistication of the forecasting model itself.

\newpage

\section{Data and Code availability}\label{sec:data_code_availability}

The data used for this analysis was obtained from Chicago Board Options Exchange (CBOE) market data service \parencite{CBOE}. 
It is paid (subscription based) and therefore, cannot be shared directly.

However, code is publicly available on \href{https://github.com/lukada13/When-the-FED-speaks}{GitHub}.


\printbibliography

@article{BlackScholes,
  author  = "Fischer Black and Myron Scholes",
  title   = "The Pricing of Options and Corporate Liabilities",
  journal = "The Journal of Political Economy",
  year    = 1973,
  volume  = "81",
  number  = "3",
  pages   = "637--654"
}

@article{Beaver,
  author  = "William H. Beaver",
  title   = "The Information Content of Annual Earnings Announcements",
  journal = "Journal of Accounting Research",
  year    = 1968,
  volume  = "6",
  pages   = "67--92"
}

@article{Cohen,
  author  = "Daniel A. Cohen and Aiyesha Dey and Thomas Z. Lys and Shyam V. Sunder",
  title   = "Earnings announcement premia and the limits to arbitrage",
  journal = "The Journal of Accounting and Economics",
  year    = 2007,
  volume  = "43",
  pages   = "153--180"
}

@article{Cont,
  author  = "Rama Cont and José da Fonseca",
  title   = "Dynamics of Implied Volatility Surfaces",
  journal = "Quantitative Finance",
  year    = 2002,
  volume  = "2",
  number  = "2",
  pages   = "45--60"
}

@article{Dubinsky,
  author  = "Andrew Dubinsky and Michael Johannes and Andreas Kaeck and Norman J. Seeger",
  title   = "Option Pricing of Earnings Announcement Risks",
  journal = "The Review of Financial Studies ",
  year    = 2019,
  volume  = "32",
  number  = "2",
  pages   = "646--687"
}

@article{EderingtonLee,
  author  = "Louis H. Ederington and Jae Ha Lee",
  title   = "The Creation and Resolution of Market Uncertainty: The Impact of Information Releases on Implied Volatility",
  journal = "The Journal of Financial and Quantitative Analysis",
  year    = 1996,
  volume  = "31",
  number  = "4",
  pages   = "513--539"
}

@article{Kelly,
  author  = "Bryan Kelly and Luboš Pástor and Pietro Veronesi",
  title   = "The Price of Political Uncertainty: Theory and Evidence from the Option Market",
  journal = "The Journal of Finance",
  year    = 2016,
  volume  = "71",
  number  = "5",
  pages   = "2417--2480"
}

@article{LuccaMoench,
  author  = "David O. Lucca and Emanuel Moench",
  title   = "The Pre-FOMC Announcement Drift",
  journal = "The Journal of Finance",
  year    = 2015,
  volume  = "70",
  number  = "1",
  pages   = "329--371"
}

@article{Sharpe,
author = {William F. Sharpe},
title = {Capital Asset Prices: A Theory of Market Equilibrium under Conditions of Risk},
journal = {Journal of Finance},
volume = {19},
number = {3},
pages = {425--442},
year = {1964}}

@article{PattelWolfson1979,
  author  = "James M. Patell and Mark A. Wolfson",
  title   = "Anticipated Information Reflected in Call Option Prices",
  journal = "The Journal of Accounting and Economics",
  year    = 1979,
  volume  = "1",
  number  = "2",
  pages   = "117--140"
}

@article{PattelWolfson1981,
  author  = "James M. Patell and Mark A. Wolfson",
  title   = "The Ex Ante and Ex Post Price Effects of Quarterly Earnings Announcements Reflected in Option and Stock Prices",
  journal = "The Journal of Accounting Research",
  year    = 1981,
  volume  = "19",
  number  = "2",
  pages   = "434--458"
}

@article{SavorWilson,
  author  = "Pavel G. Savor and Mungo I. Wilson",
  title   = "How Much Do Investors Care About Macroeconomic Risk? Evidence from Scheduled Economic Announcements",
  journal = "Journal of Financial and Quantitative Analysis",
  year    = 2013,
  volume  = "48",
  number  = "2",
  pages   = "343--372"
}

@article{LamontFrazzini,
  author  = "Owen Lamont and Andrea Frazzini",
  title   = "The Earnings Announcement Premium and Trading Volume",
  journal = "NBER Working Papers",
  year = 2007,
  note = "Working Paper 13090"
}

@article{FlemingRemolona,
  author  = "Michael J. Fleming and Eli M. Remolona",
  title   = "Price Formation and Liquidity in the U.S. Treasury Market: The Response to Public Information",
  journal = "The Journal of Finance",
  year = 1999,
  volume = "54",
  number = "5",
  pages = "1901--1915"
}

@article{Andersen,
  author  = "Torben G. Andersen and Tim Bollerslev and Francis X. Diebold and Clara Vega",
  title   = "Micro Effects of Macro Announcements: Real-Time Price Discovery in Foreign Exchange",
  journal = "The American Economic Review",
  year = 2003,
  volume = "93",
  number = "1",
  pages = "38--62"
}

@article{Kurov,
  author  = "Alexander Kurov and Marketa H. Wolfe and Thomas Gilbert",
  title   = "The disappearing pre-FOMC announcement drift",
  journal = "Finance Research Letters",
  year = 2021,
  volume = "40"
}

@article{Plihal,
  author  = "Tomáš Plíhal",
  title   = "Scheduled macroeconomic news announcements and Forex volatility forecasting",
  journal = "Journal of Forecasting",
  year = 2021,
  volume = "40",
  number = "8",
  pages = "1379--1397"
}

@article{McQueenRoley,
  author  = "Grant McQueen and V. Vance Roley",
  title   = "Stock Prices, News, and Business Conditions",
  journal = "The Review of Financial Studies",
  year = 1993,
  volume = "6",
  number = "3",
  pages = "683--707"
}

@article{ChanGrey,
  author  = "Kam Fong Chan and Philip Gray",
  title   = "Volatility Jumps and Macroeconomic News Announcement",
  journal = "Journal of Futures Markets",
  year = 2018,
  volume = "38",
  number = "8",
  pages = "881--897"
}

@article{JonesLamont,
  author  = "Charles M. Jones, Owen Lamont, Robin L. Lumsdaine",
  title   = "Macroeconomic news and bond market volatility",
  journal = "Journal of Financial Economics",
  year = 1998,
  volume = "47",
  number = "3",
  pages = "315--337"
}

@article{Lee,
  author  = "Suzanne S. Lee",
  title   = "Jumps and Information Flow in Financial Markets",
  journal = "The Review of Financial Studies",
  year = 2011,
  volume = "25",
  number = "2",
  pages = "439--479"
}

@article{JiangLo,
  author  = "George J. Jiang and Ingrid Lo and Adrien Verdelhan",
  title   = "Information Shocks, Liquidity Shocks, Jumps, and Price Discovery: Evidence from the U.S. Treasury Market",
  journal = "The Journal of Financial and Quantitative Analysis",
  year = 2011,
  volume = "46",
  number = "2",
  pages = "527--551"
}

@article{LahayeLaurent,
  author  = "Jérôme Lahaye and Sébastien Laurent and Christopher J. Neely",
  title   = "Jumps, cojumps and macro announcements",
  journal = "The Journal of Applied Econometrics",
  year = 2011,
  volume = "26",
  number = "6",
  pages = "893--921"
}

@article{Dungey,
  author  = "Mardi Dungey and Michael McKenzie and L. Vanessa Smith",
  title   = "Empirical evidence on jumps in the term structure of the US Treasury Market",
  journal = "The Journal of Empirical Finance",
  year = 2009,
  volume = "16",
  number = "3",
  pages = "430--445"
}

@article{ChenGrith,
author = {Ying Chen and Maria Grith and Hannah L. H. Lai},
title = {Neural Tangent Kernel in Implied Volatility Forecasting: A Nonlinear Functional Autoregression Approach},
journal = {Journal of Business \& Economic Statistics},
volume = {44},
number = {1},
pages = {24--38},
year = {2026}}

@article{Zhang,
  author  = "Wenyong Zhang and Lingfei Li and Gongqiu Zhang",
  title   = "A two-step framework for arbitrage-free prediction of the implied volatility surface",
  journal = "Quantitative Finance",
  year = 2023,
  volume = "23",
  number = "1",
  pages = "21--34"
}

@article{FenglerHardle,
    author = {Fengler, Matthias R. and Härdle, Wolfgang K. and Mammen, Enno},
    title = {A semiparametric factor model for implied volatility surface dynamics},
    journal = {Journal of Financial Econometrics},
    volume = {5},
    number = {2},
    pages = {189-218},
    year = {2007}
}

@article{Fengler,
author = {Matthias R. Fengler},
title = {Arbitrage-free smoothing of the implied volatility surface},
journal = {Quantitative Finance},
volume = {9},
number = {4},
pages = {417--428},
year = {2009}
}

@article{Orosi,
author = {Orosi, Greg},
title = {Arbitrage-free call option surface construction using regression splines},
journal = {Applied Stochastic Models in Business and Industry},
volume = {31},
number = {4},
pages = {515-527},
year = {2015}
}

@article{Gatheral,
author = {Jim Gatheral and Antoine Jacquier},
title = {Arbitrage-free SVI volatility surfaces},
journal = {Quantitative Finance},
volume = {14},
number = {1},
pages = {59--71},
year = {2014}}

@article{Horvath,
author = {Blanka Horvath and Aitor Muguruza and Mehdi Tomas},
title = {Deep learning volatility: a deep neural network perspective on pricing and calibration in (rough) volatility models},
journal = {Quantitative Finance},
volume = {21},
number = {1},
pages = {11--27},
year = {2021}}

@article{Almeida,
author = {Caio Almeida and Jianqing Fan and Gustavo Freire and Francesca Tang},
title = {Can a Machine Correct Option Pricing Models?},
journal = {Journal of Business \& Economic Statistics},
volume = {41},
number = {3},
pages = {995--1009},
year = {2023}}

@article{Colangelo,
	author = {Francesco Audrino and Dominik Colangelo},
	title = {Semi-parametric forecasts of the implied volatility surface using regression trees},
	volume = {20},
	pages = {421-434},
	year = {2010},
	journal = {Statistics and Computing},
}

@TechReport{Chen,
type={Papers},
institution={arXiv.org},
author={Shengli Chen and Zili Zhang},
title={Forecasting Implied Volatility Smile Surface via Deep Learning and Attention Mechanism},
year={2019},
month={Dec},
number={1912.11059},
url={https://ideas.repec.org/p/arx/papers/1912.11059.html},
urldate = {2026-04-13}
}

@TechReport{Bloch,
type={Papers},
institution={SSRN},
author={Daniel A. Bloch and Arthur Böök},
title={Deep Learning Based Dynamic Implied Volatility Surface},
year={2019},
url={https://papers.ssrn.com/sol3/papers.cfm?abstract_id=3952842},
urldate = {2026-04-13}
}

@article{LeeMoment,
author = {Lee, Roger W.},
title = {The Moment Formula for Implied Volatility at Extreme Strikes},
journal = {Mathematical Finance},
volume = {14},
number = {3},
pages = {469-480},
year = {2004}
}

@inproceedings{Shi,
  author    = {Shi, Xingjian and Chen, Zhourong and Wang, Hao and Yeung, Dit-Yan and Wong, Wai-Kin and Woo, Wang-chun},
  title     = {Convolutional LSTM Network: A Machine Learning Approach for Precipitation Nowcasting},
  booktitle = {Advances in Neural Information Processing Systems (NeurIPS)},
  year      = {2015}
}

@article{Hochreiter,
  author  = {Hochreiter, Sepp and Schmidhuber, J{\"u}rgen},
  title   = {Long Short-Term Memory},
  journal = {Neural Computation},
  year    = {1997},
  volume  = {9},
  number  = {8},
  pages   = {1735--1780}
}

@article{Bergeron,
	author = {Maxime Bergeron and Nicholas Fung and John Hull and Zissis Poulos and Andreas Veneris},
	title = {Variational Autoencoders: A Hands-Off Approach to Volatility},
	volume = {4},
	number = {2},
	pages = {125-138},
	year = {2022},
	journal = {The Journal of Financial Data Science},
}

@article{Goncalves,
author = {Sílvia Gonçalves and Massimo Guidolin},
 journal = {The Journal of Business},
 number = {3},
 pages = {1591--1635},
title = {Predictable Dynamics in the S\&P 500 Index Options Implied Volatility Surface},
 urldate = {2026-04-13},
 volume = {79},
 year = {2006}
}

@article{Dumas,
 author = {Bernard Dumas and Jeff Fleming and Robert E. Whaley},
 journal = {The Journal of Finance},
 number = {6},
 pages = {2059--2106},
 title = {Implied Volatility Functions: Empirical Tests},
 urldate = {2026-04-13},
 volume = {53},
 year = {1998}
}

@article{Diebold,
  author  = {Francis X. Diebold and Roberto S. Mariano},
  title   = {Comparing Predictive Accuracy},
  journal = {Journal of Business \& Economic Statistics},
  volume  = {13},
  number  = {3},
  pages   = {253--263},
  year    = {1995}
}

@article{GhadariLSTMcrypto,
title = {Designing a cryptocurrency trading system with deep reinforcement learning utilizing LSTM neural networks and XGBoost feature selection},
journal = {Applied Soft Computing},
volume = {175},
pages = {113029},
year = {2025},
author = {Hamidreza Ghadiri and Ehsan Hajizadeh}}

@article{ParkLSTMcrypto,
title = {Intelligent cryptocurrency trading system using integrated AdaBoost-LSTM with market turbulence knowledge},
journal = {Applied Soft Computing},
volume = {145},
pages = {110568},
year = {2023},
author = {Sangjin Park and Jae-Suk Yang},}

@article{FiszederVolatility,
title = {Identification of Bitcoin volatility drivers using statistical and machine learning methods},
journal = {Applied Soft Computing},
volume = {188},
pages = {114384},
year = {2026},
author = {Piotr Fiszeder and Witold Orzeszko and Radosław Pietrzyk and Grzegorz Dudek},}

@article{SongVolatility,
title = {Forecasting realized volatility using deep learning quantile function},
journal = {Applied Soft Computing},
volume = {175},
pages = {113016},
year = {2025},
author = {Jungyoon Song and Hyunju Lee and Jongu Lee and Woojin Chang},}

@article{MoShipping,
title = {Annual dilated convolutional LSTM network for time charter rate forecasting},
journal = {Applied Soft Computing},
volume = {126},
pages = {109259},
year = {2022},
author = {Jixian Mo and Ruobin Gao and Jiahui Liu and Liang Du and Kum Fai Yuen}}

@article{YangCyclones,
title = {A spatio-temporal graph-guided convolutional LSTM for tropical cyclones precipitation nowcasting},
journal = {Applied Soft Computing},
volume = {124},
pages = {109003},
year = {2022},
author = {Xuying Yang and Feng Zhang and Peng Sun and Xiaofan Li and Zhenhong Du and Renyi Liu}}

@article{SumitroCOVID,
title = {Learning where to look for COVID-19 growth: Multivariate analysis of COVID-19 cases over time using explainable convolution–LSTM},
journal = {Applied Soft Computing},
volume = {109},
pages = {107469},
year = {2021},
author = {Novanto Yudistira and Sutiman Bambang Sumitro and Alberth Nahas and Nelly Florida Riama}}

@online{fred,
  author = {{Federal Reserve Bank of St. Louis}},
  title = "FRED Data",
  url = "https://fred.stlouisfed.org",
  urldate = {2025-12-30},
  year = ""
}

@online{FED,
  author = {{Board of Governors of the Federal Reserve System}},
  title = "Federal Open Market Committee - Historical Materials by Year",
  url = "https://www.federalreserve.gov/monetarypolicy/fomc_historical_year.htm",
  urldate = {2025-12-30},
  year = ""
}

@online{CBOE,
  author = {{Chicago Board Options Exchange}},
  title = "CBOE datashop",
  url = "https://datashop.cboe.com/option-quote-intervals",
  urldate = {2026-03-11},  
  year = ""
}


\newpage

\appendix
\renewcommand{\theequation}{\Alph{section}\hspace{0.15em}-\hspace{0.05em}\arabic{equation}}
\numberwithin{equation}{section}

\section{figures}\label{sec:app_figures}

\subsection{Call options}

\begin{figure}[H]
    \centering
    \includegraphics[width=\linewidth]{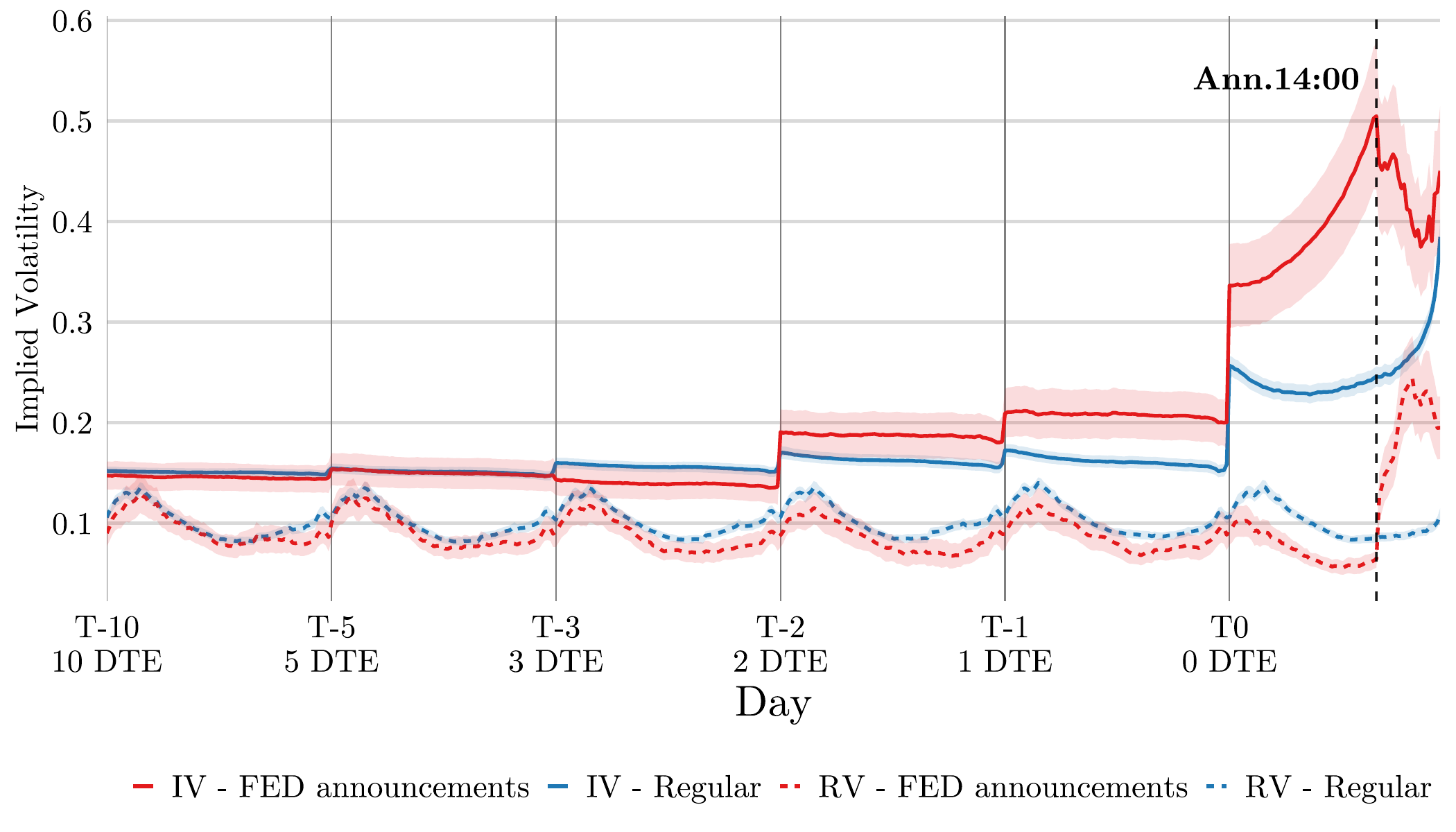}
    \caption{Average ATM Implied and Realised Volatility in the period 2016-2025 for call options expiring on the day of announcement.}
\end{figure}

\begin{figure}[H]
    \centering
    \includegraphics[width=\linewidth]{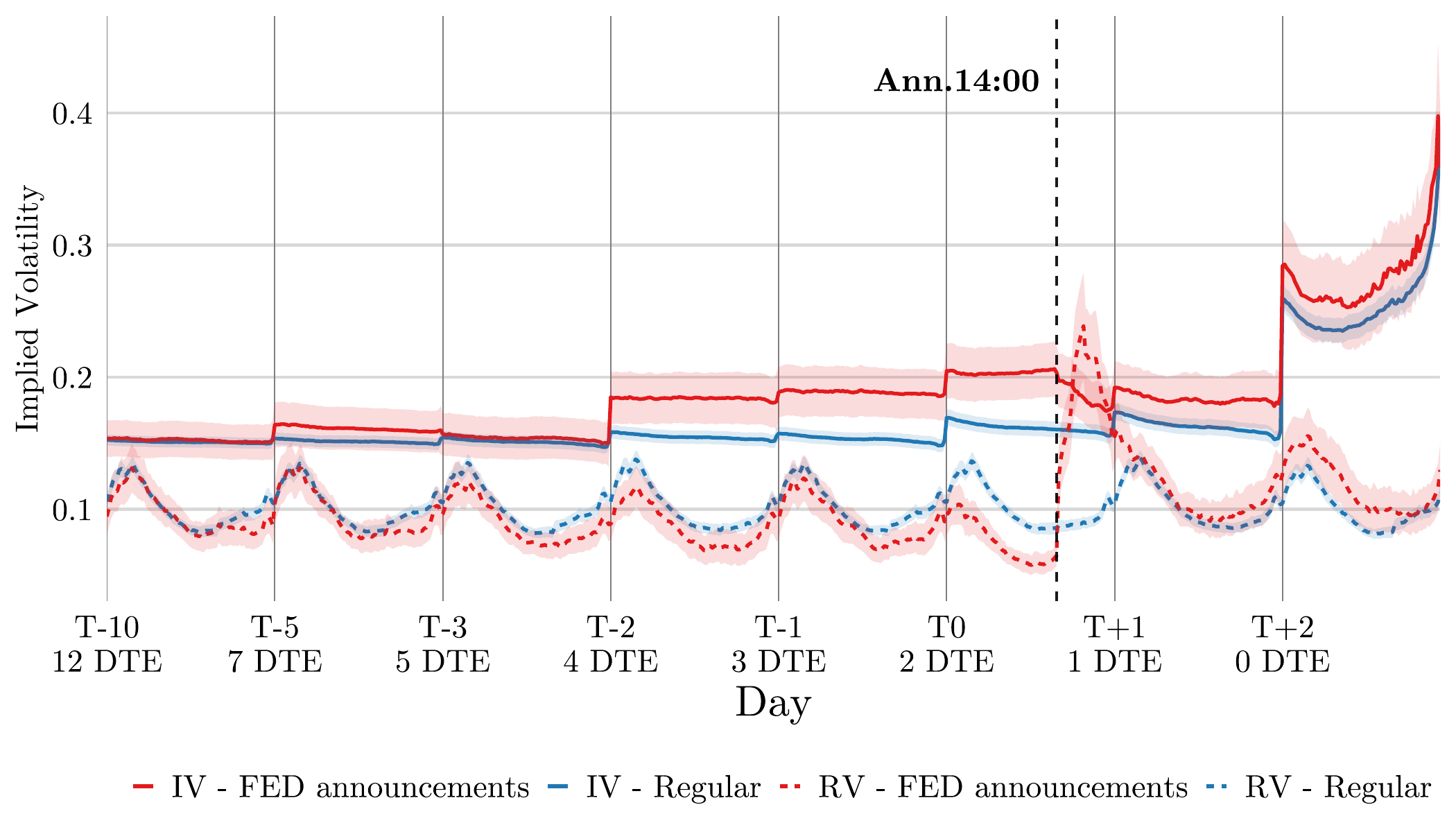}
    \caption{Average ATM Implied and Realised Volatility in the period 2016-2025 for call options expiring 2 trading days after the announcement.}
\end{figure}

\begin{figure}[H]
    \centering
    \includegraphics[width=\linewidth]{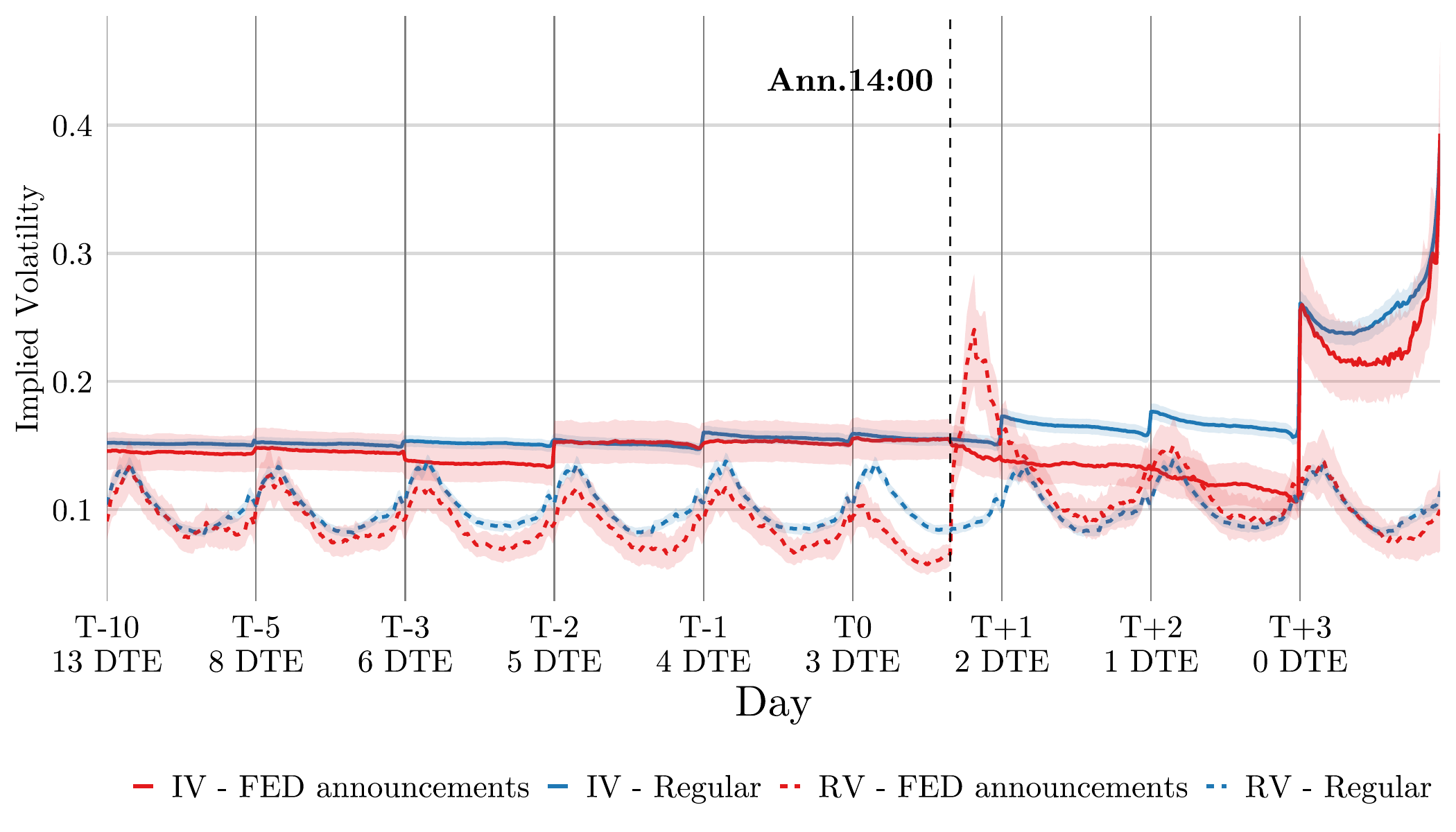}
    \caption{Average ATM Implied and Realised Volatility in the period 2016-2025 for call options expiring 3 trading days after the announcement.}
\end{figure}

\begin{figure}[H]
    \centering
    \includegraphics[width=\linewidth]{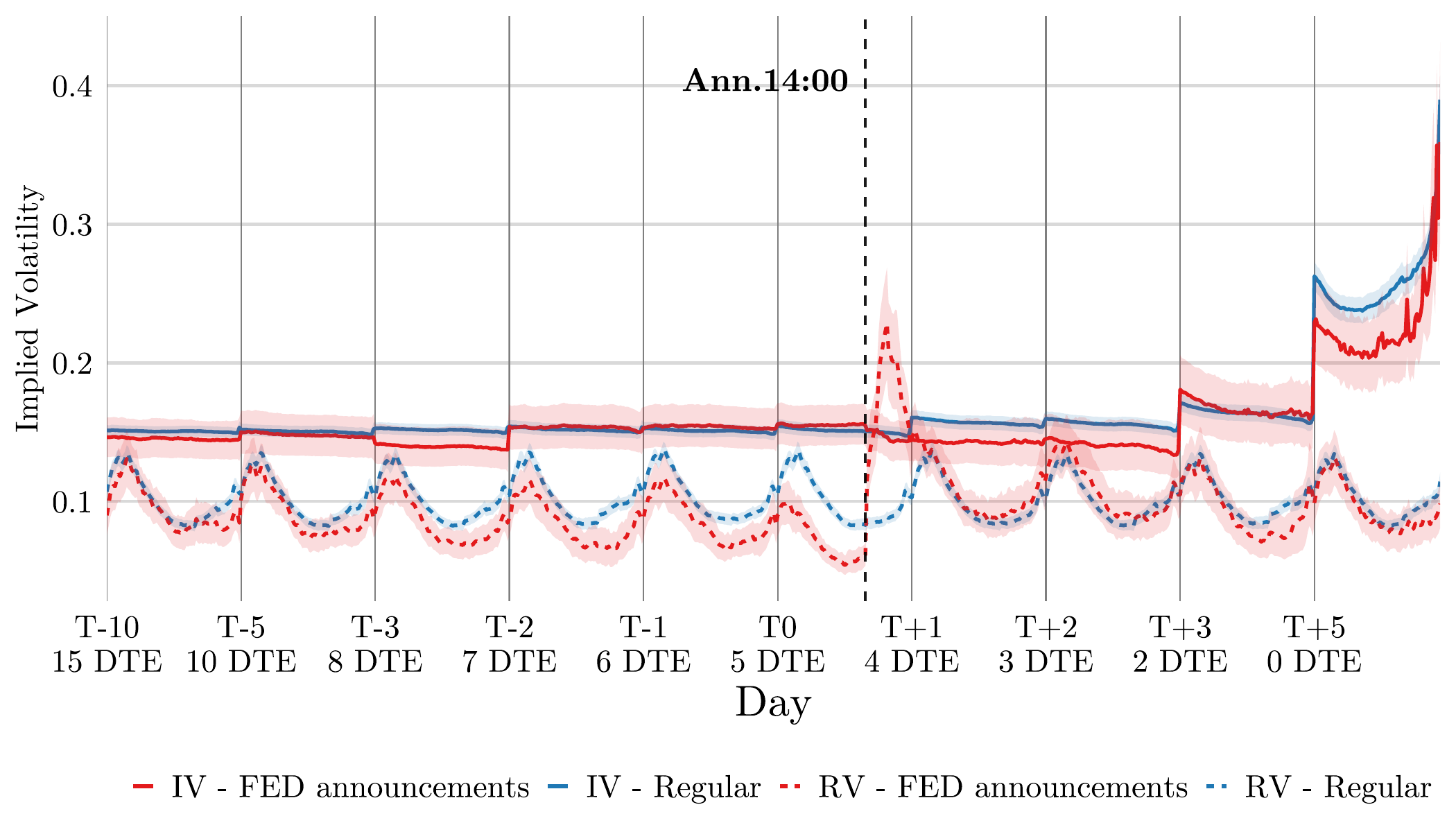}
    \caption{Average ATM Implied and Realised Volatility in the period 2016-2025 for call options expiring 5 trading days after the announcement.}
\end{figure}

\begin{figure}[H]
    \centering
    \includegraphics[width=\linewidth]{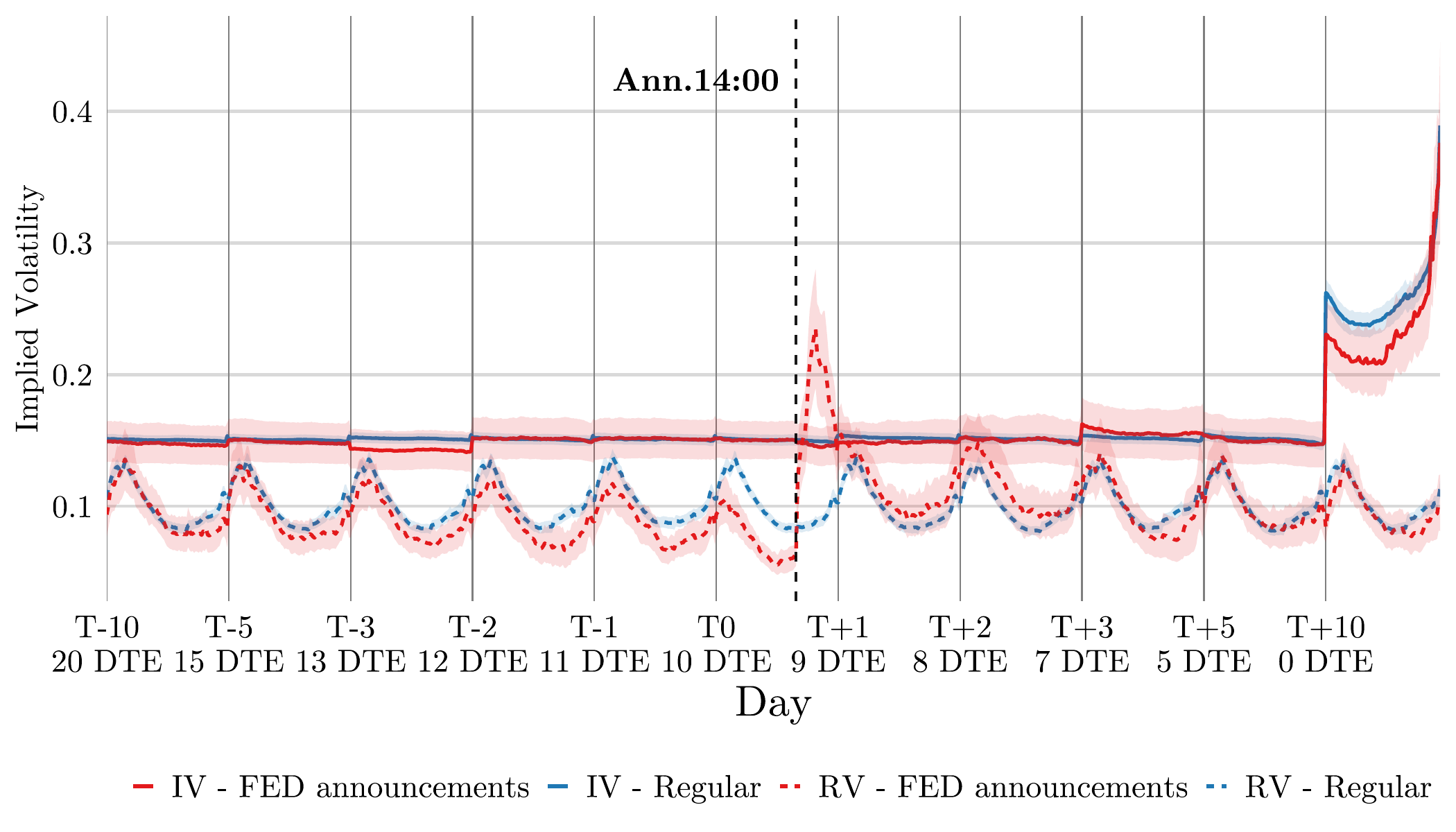}
    \caption{Average ATM Implied and Realised Volatility in the period 2016-2025 for call options expiring 10 trading days after the announcement.}
\end{figure}

\subsection{Put options}

\begin{figure}[H]
    \centering
    \includegraphics[width=\linewidth]{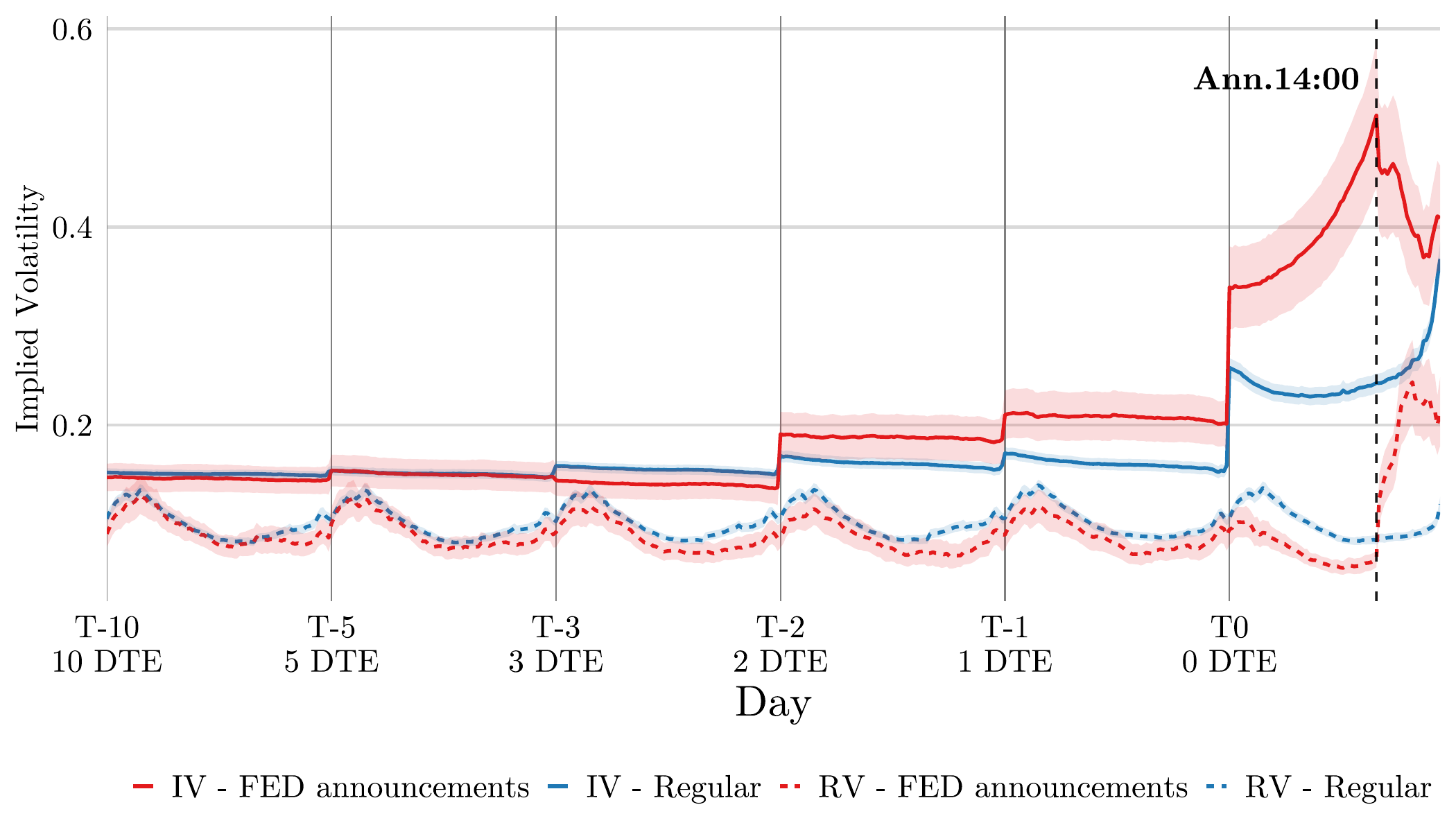}
    \caption{Average ATM Implied and Realised Volatility in the period 2016-2025 for put options expiring on the day of announcement.}
\end{figure}

\begin{figure}[H]
    \centering
    \includegraphics[width=\linewidth]{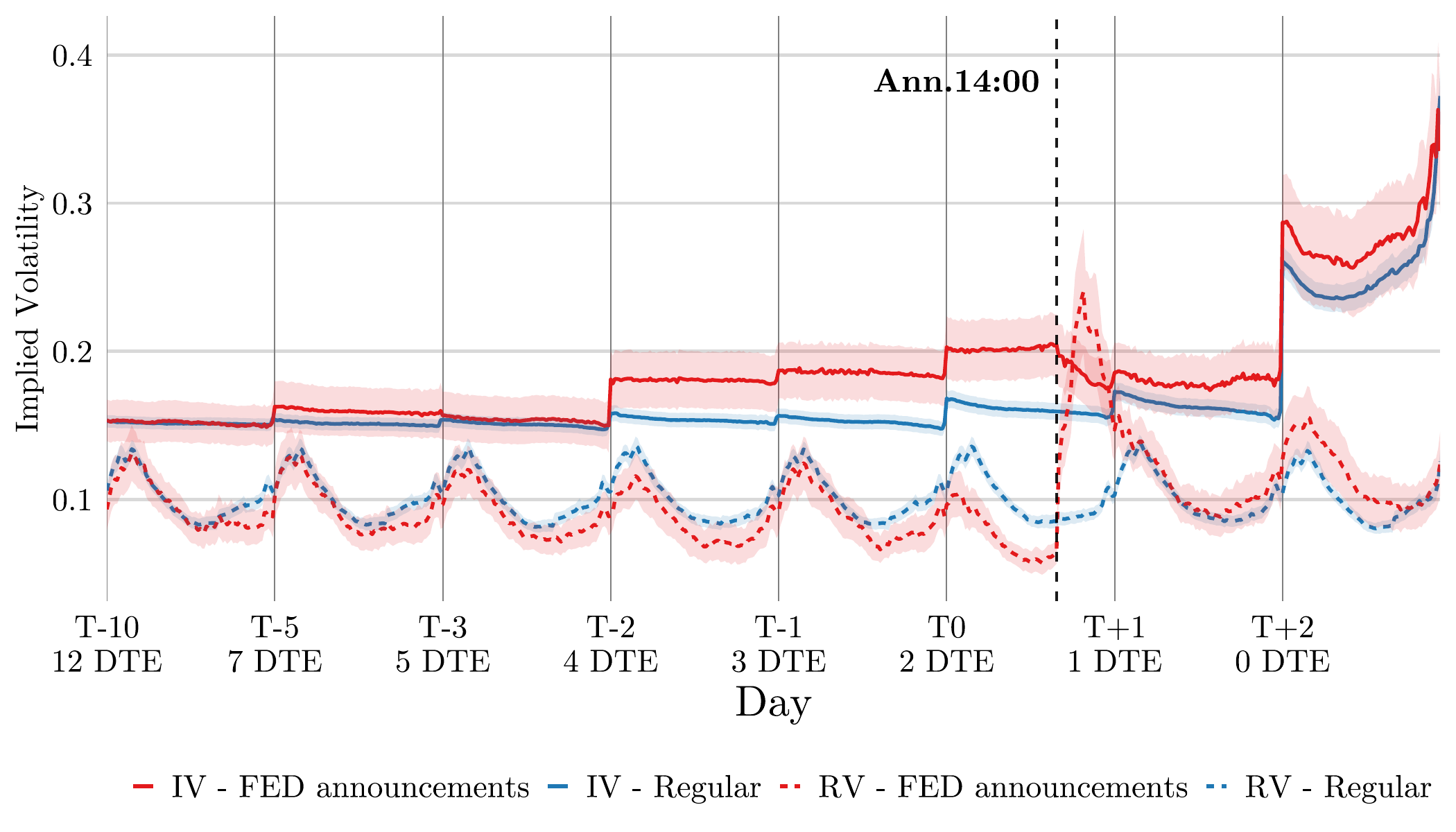}
    \caption{Average ATM Implied and Realised Volatility in the period 2016-2025 for put options expiring 2 trading days after the announcement.}
\end{figure}

\begin{figure}[H]
    \centering
    \includegraphics[width=\linewidth]{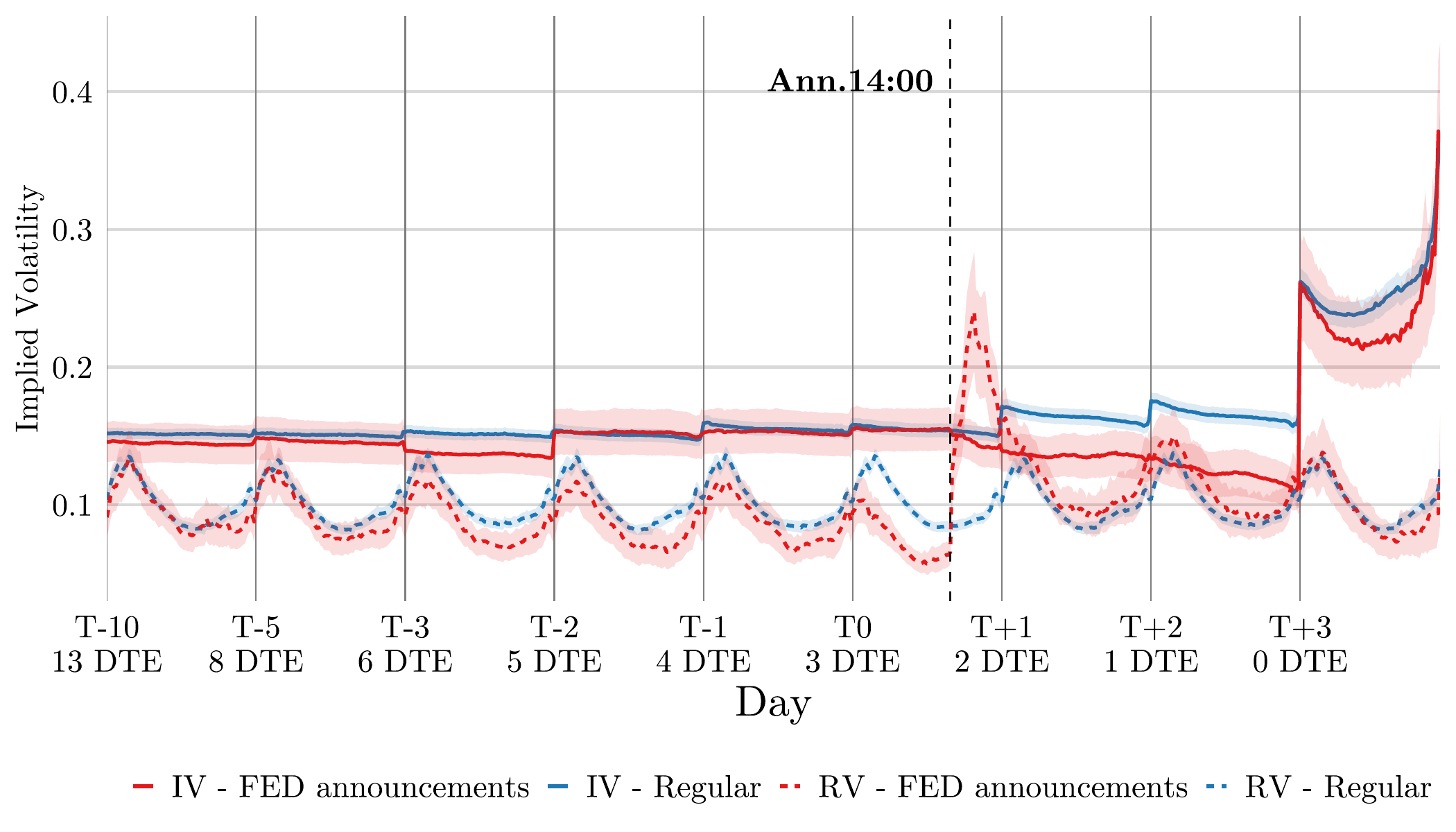}
    \caption{Average ATM Implied and Realised Volatility in the period 2016-2025 for put options expiring 3 trading days after the announcement.}
\end{figure}

\begin{figure}[H]
    \centering
    \includegraphics[width=\linewidth]{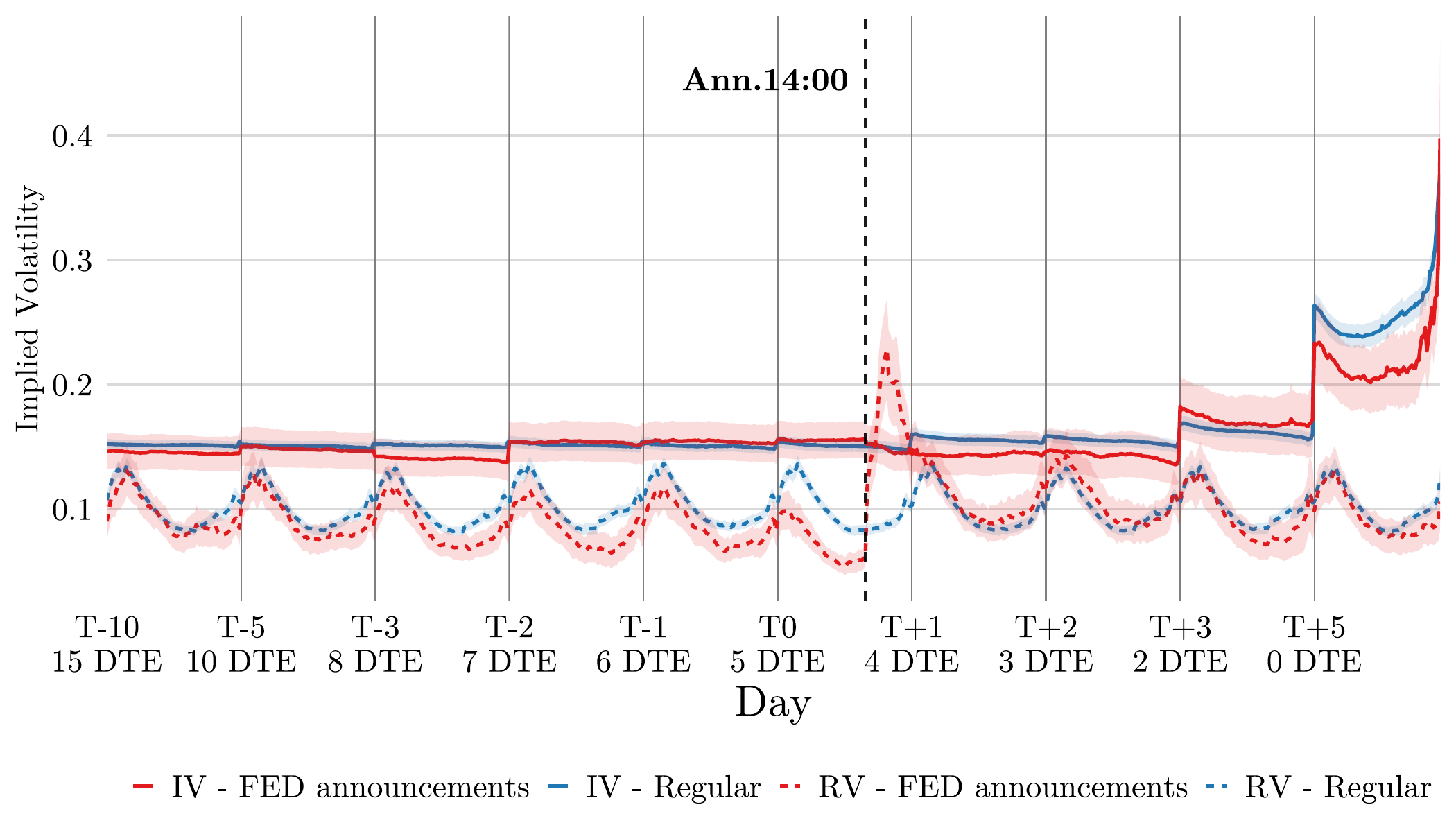}
    \caption{Average ATM Implied and Realised Volatility in the period 2016-2025 for put options expiring 5 trading days after the announcement.}
\end{figure}

\begin{figure}[H]
    \centering
    \includegraphics[width=\linewidth]{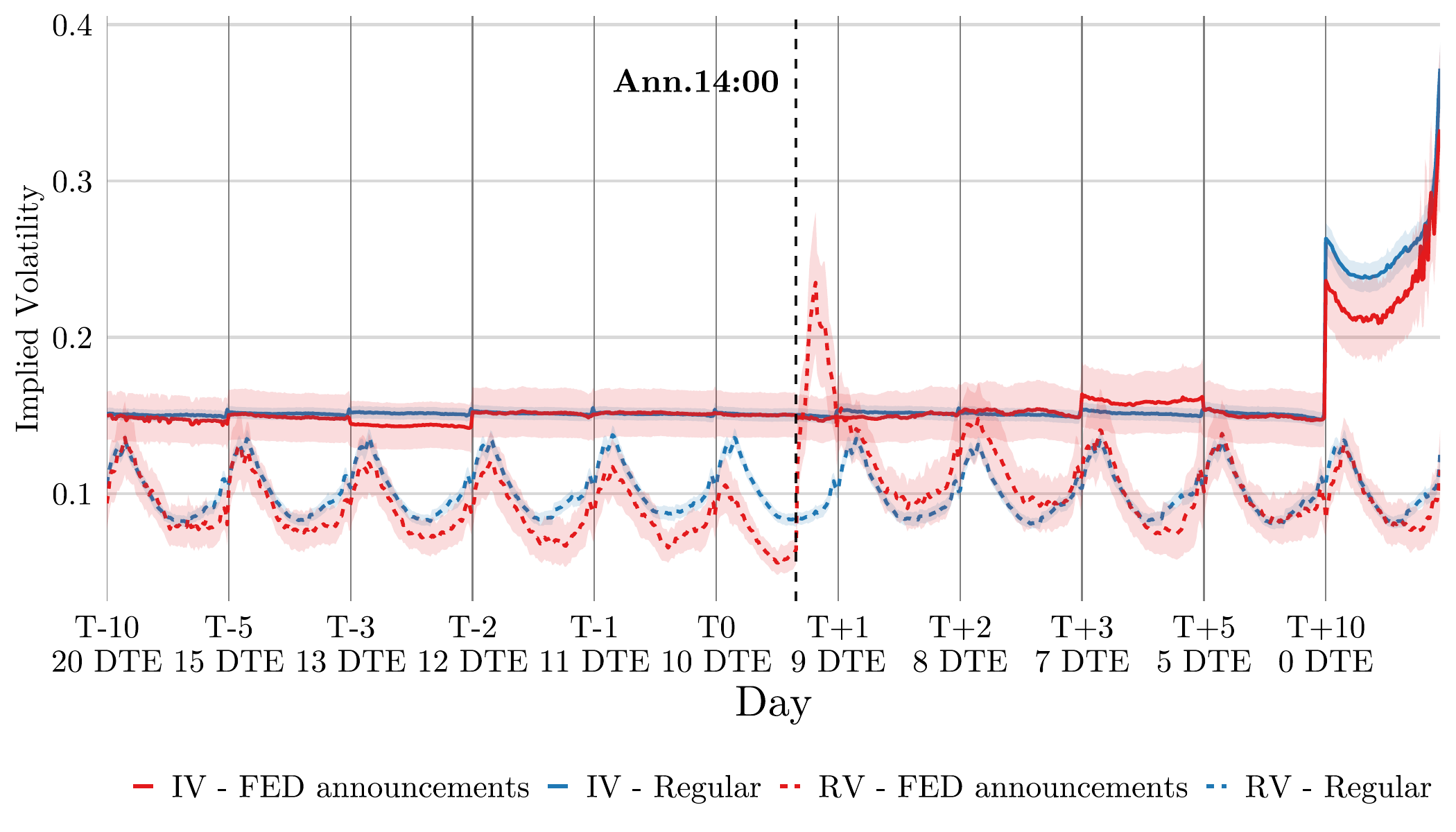}
    \caption{Average ATM Implied and Realised Volatility in the period 2016-2025 for put options expiring 10 trading days after the announcement.}
\end{figure}

\section{figures - IV}\label{sec:app_figures_subsamples}

\subsection{Call options}

\begin{figure}[H]
    \centering
    \includegraphics[width=\linewidth]{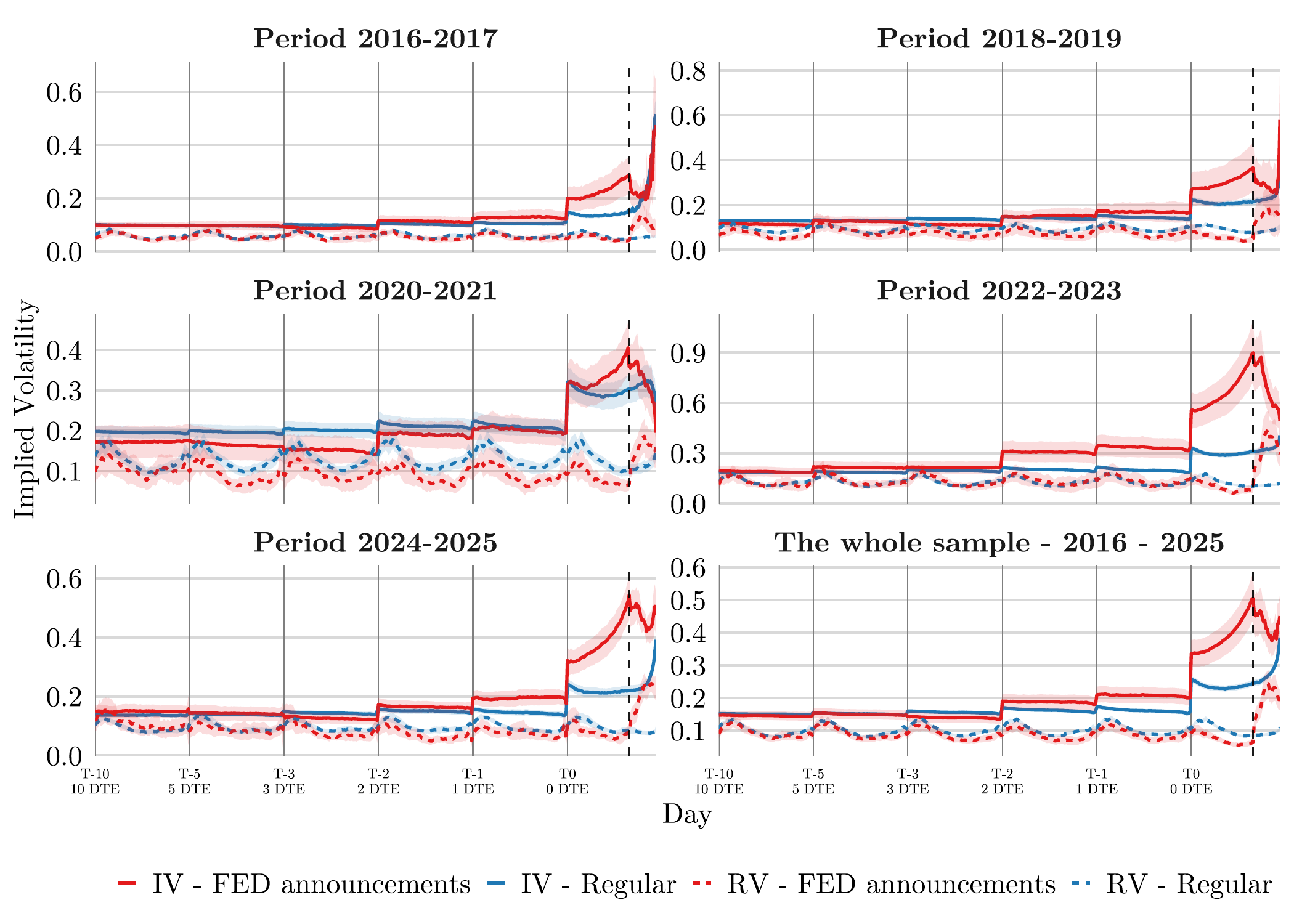}
    \caption{Average ATM Implied and Realised Volatility for call options expiring on the day of announcement.}
\end{figure}

\begin{figure}[H]
    \centering
    \includegraphics[width=\linewidth]{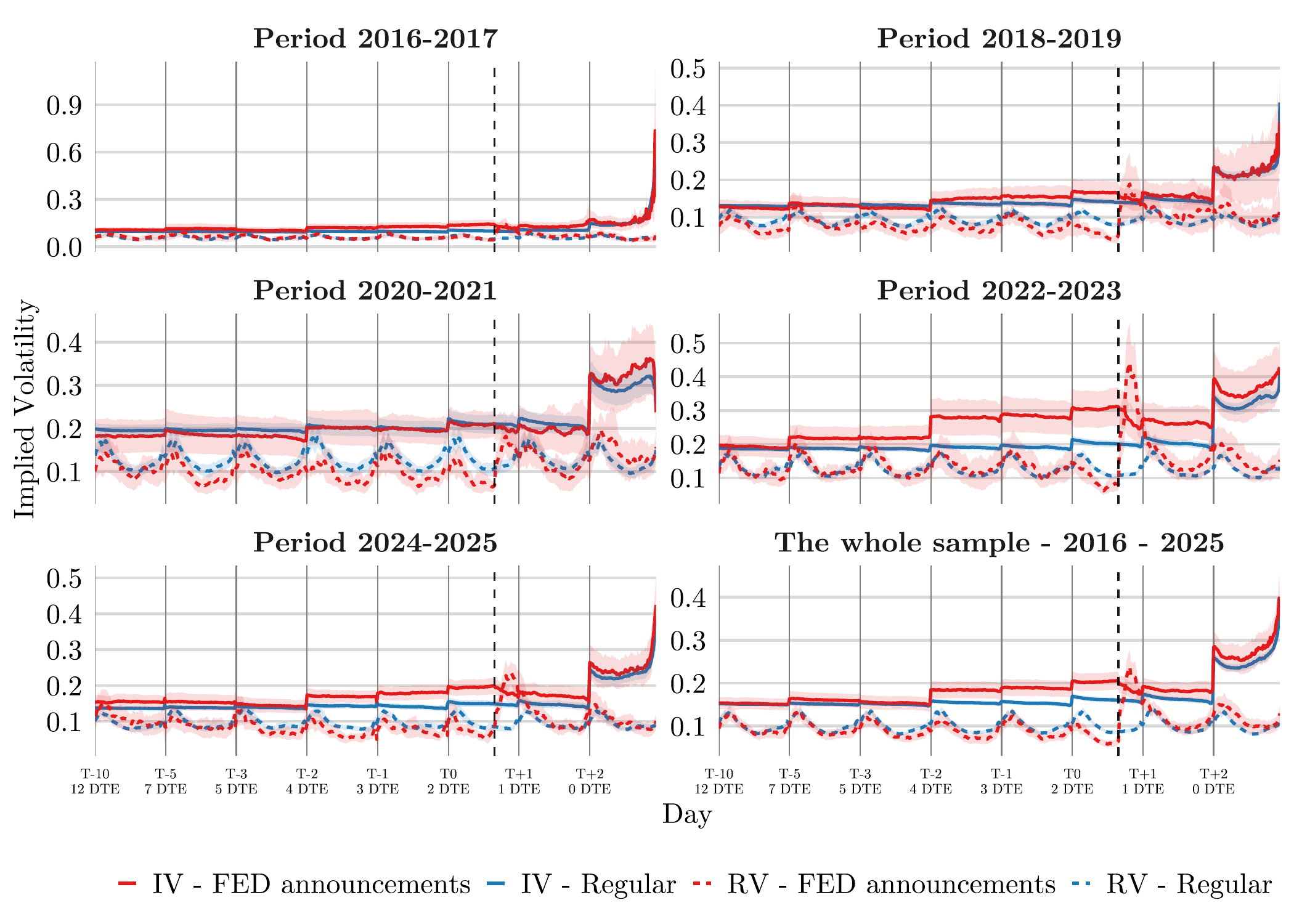}
    \caption{Average ATM Implied and Realised Volatility for call options expiring 2 trading days after the announcement.}
\end{figure}

\begin{figure}[H]
    \centering
    \includegraphics[width=\linewidth]{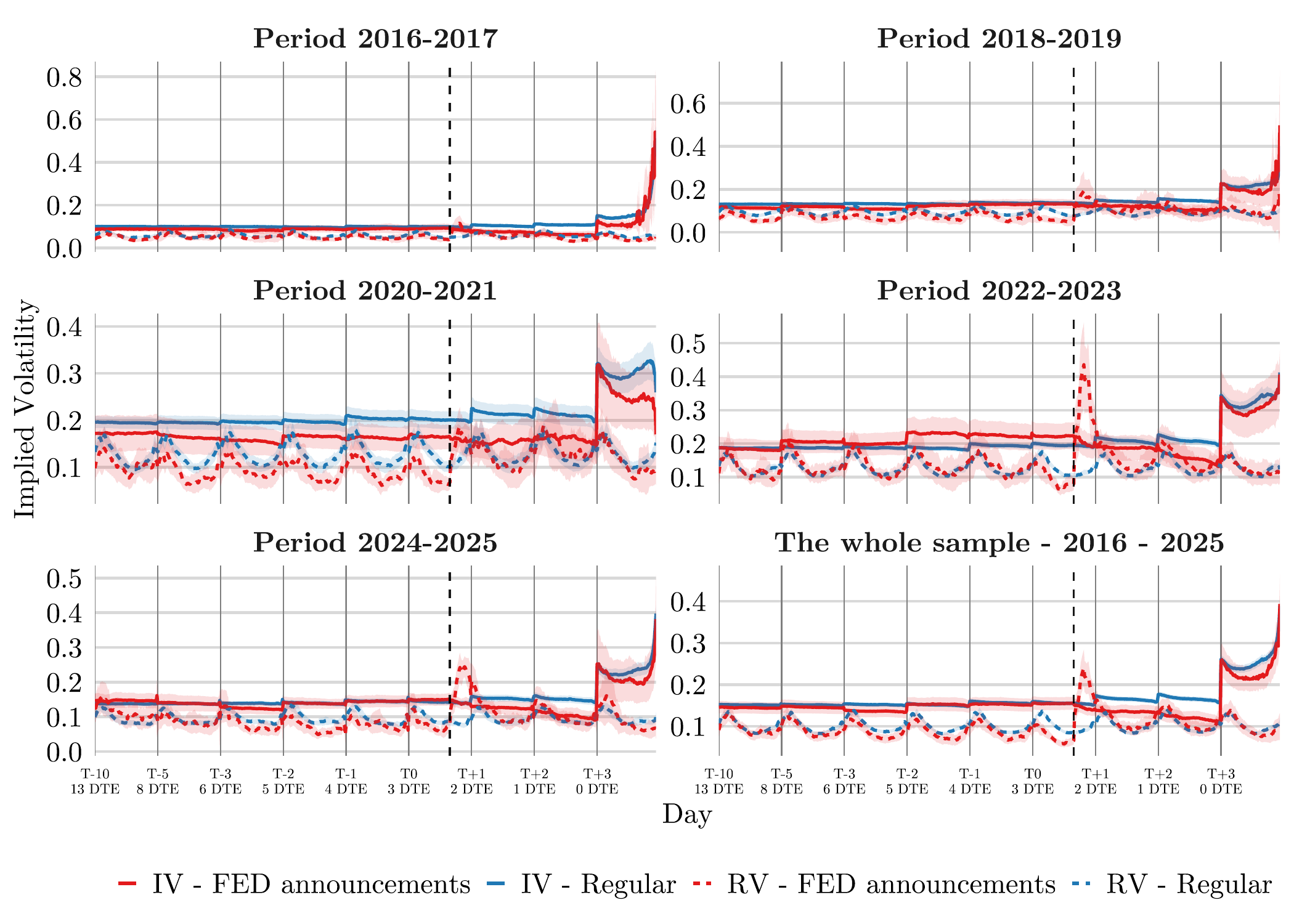}
    \caption{Average ATM Implied and Realised Volatility for call options expiring 3 trading days after the announcement.}
\end{figure}

\begin{figure}[H]
    \centering
    \includegraphics[width=\linewidth]{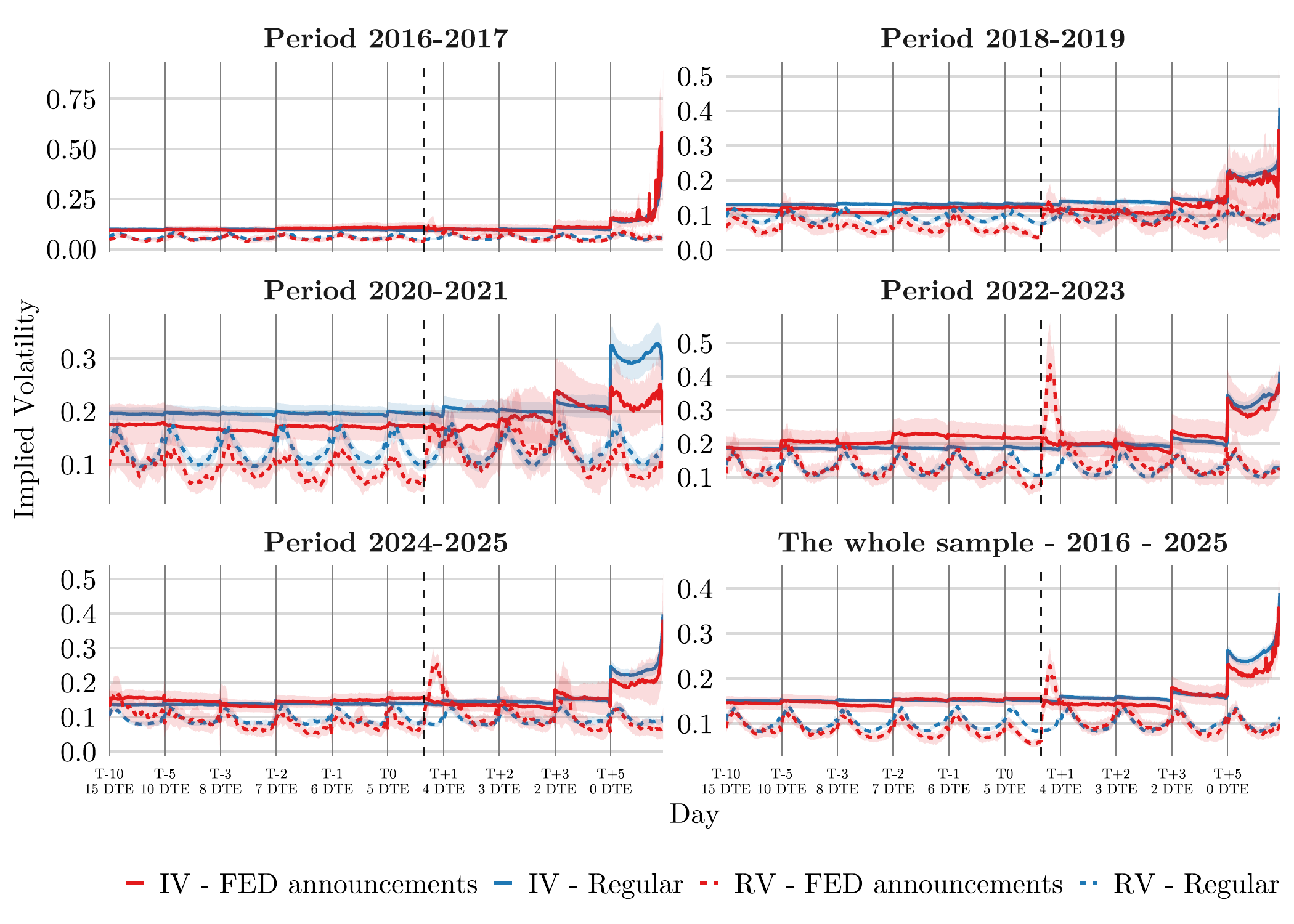}
    \caption{Average ATM Implied and Realised Volatility for call options expiring 5 trading days after the announcement.}
\end{figure}

\begin{figure}[H]
    \centering
    \includegraphics[width=\linewidth]{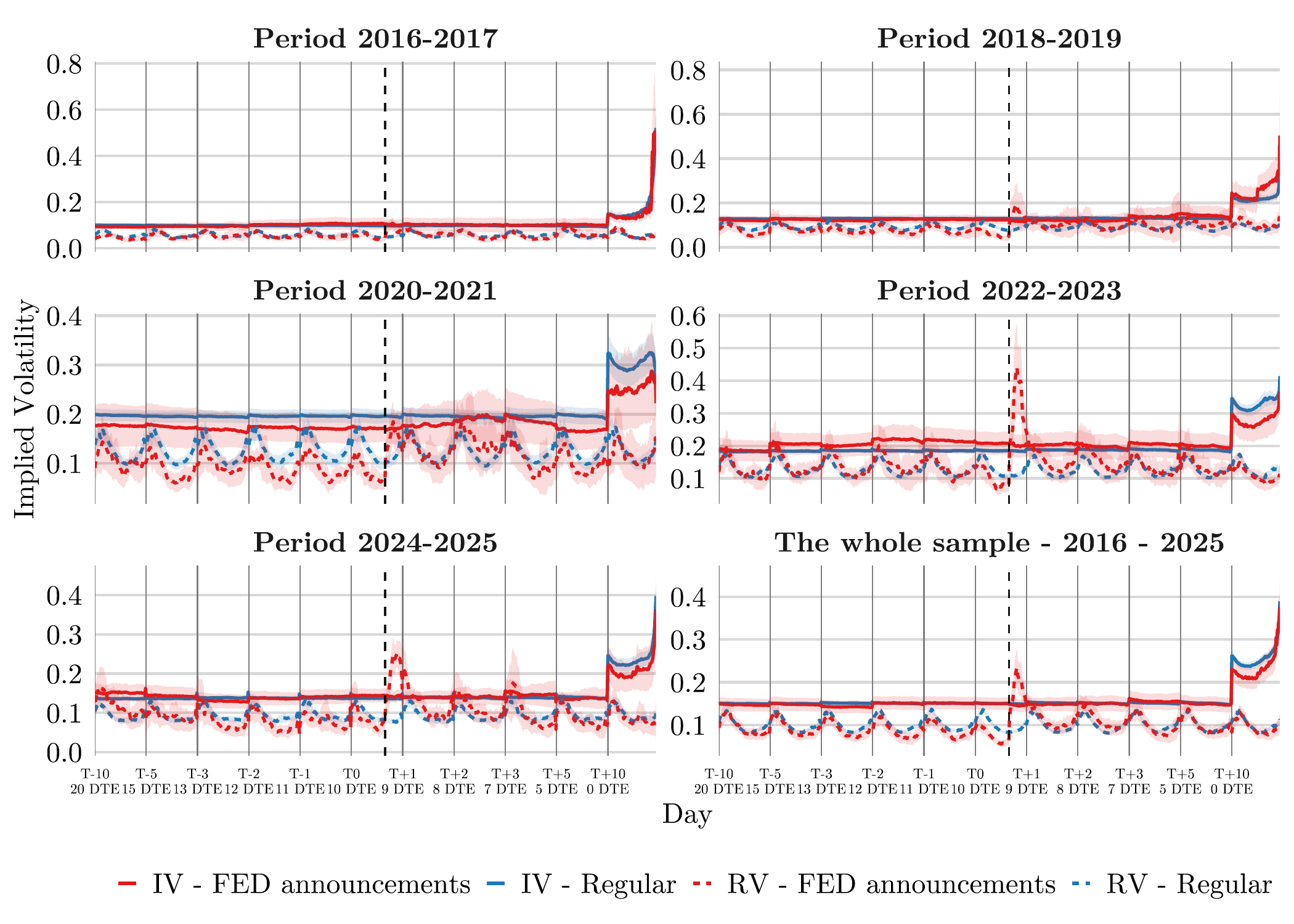}
    \caption{Average ATM Implied and Realised Volatility for call options expiring 10 trading days after the announcement.}
\end{figure}

\subsection{Put options}

\begin{figure}[H]
    \centering
    \includegraphics[width=\linewidth]{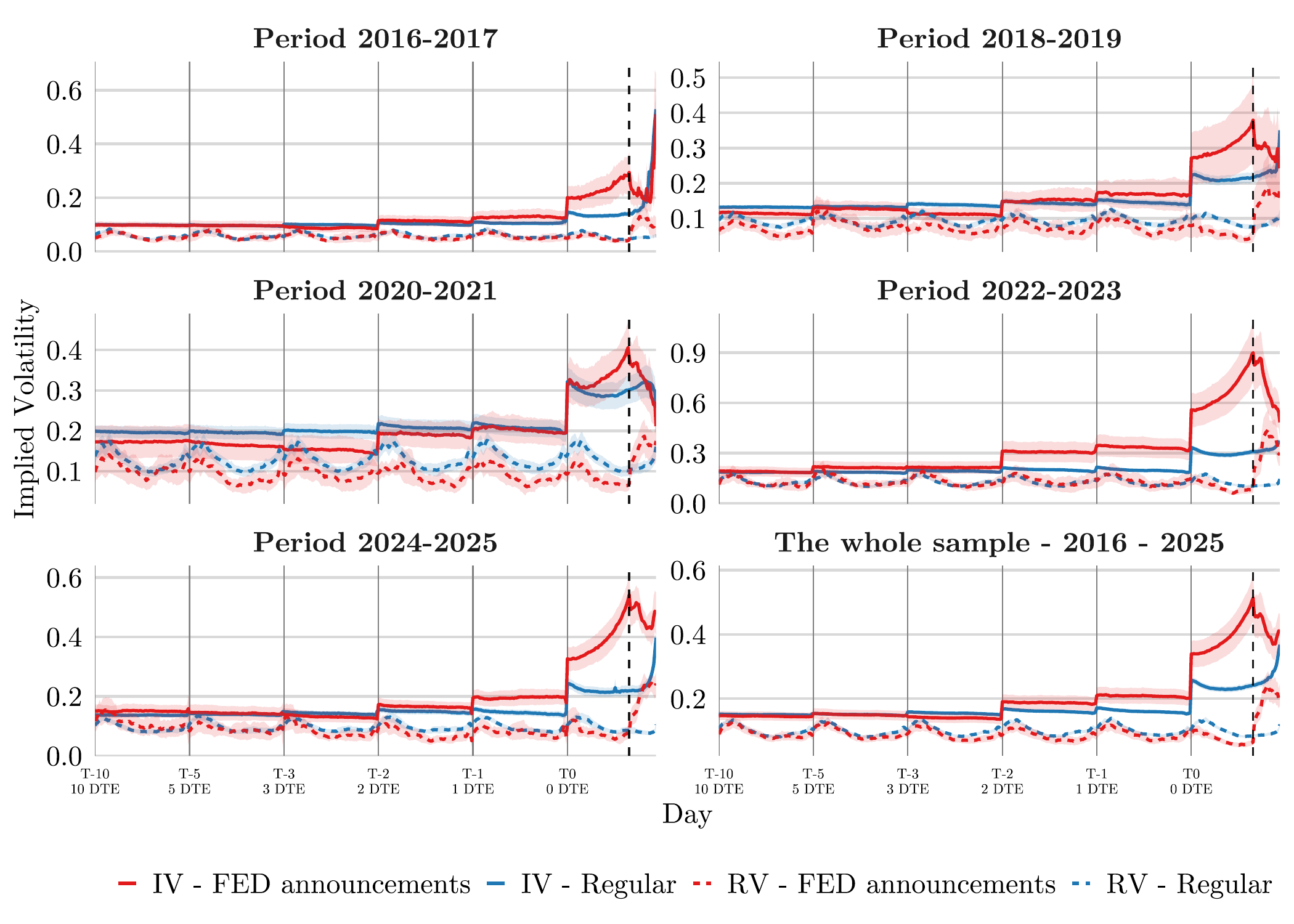}
    \caption{Average ATM Implied and Realised Volatility for put options expiring on the day of announcement.}
\end{figure}

\begin{figure}[H]
    \centering
    \includegraphics[width=\linewidth]{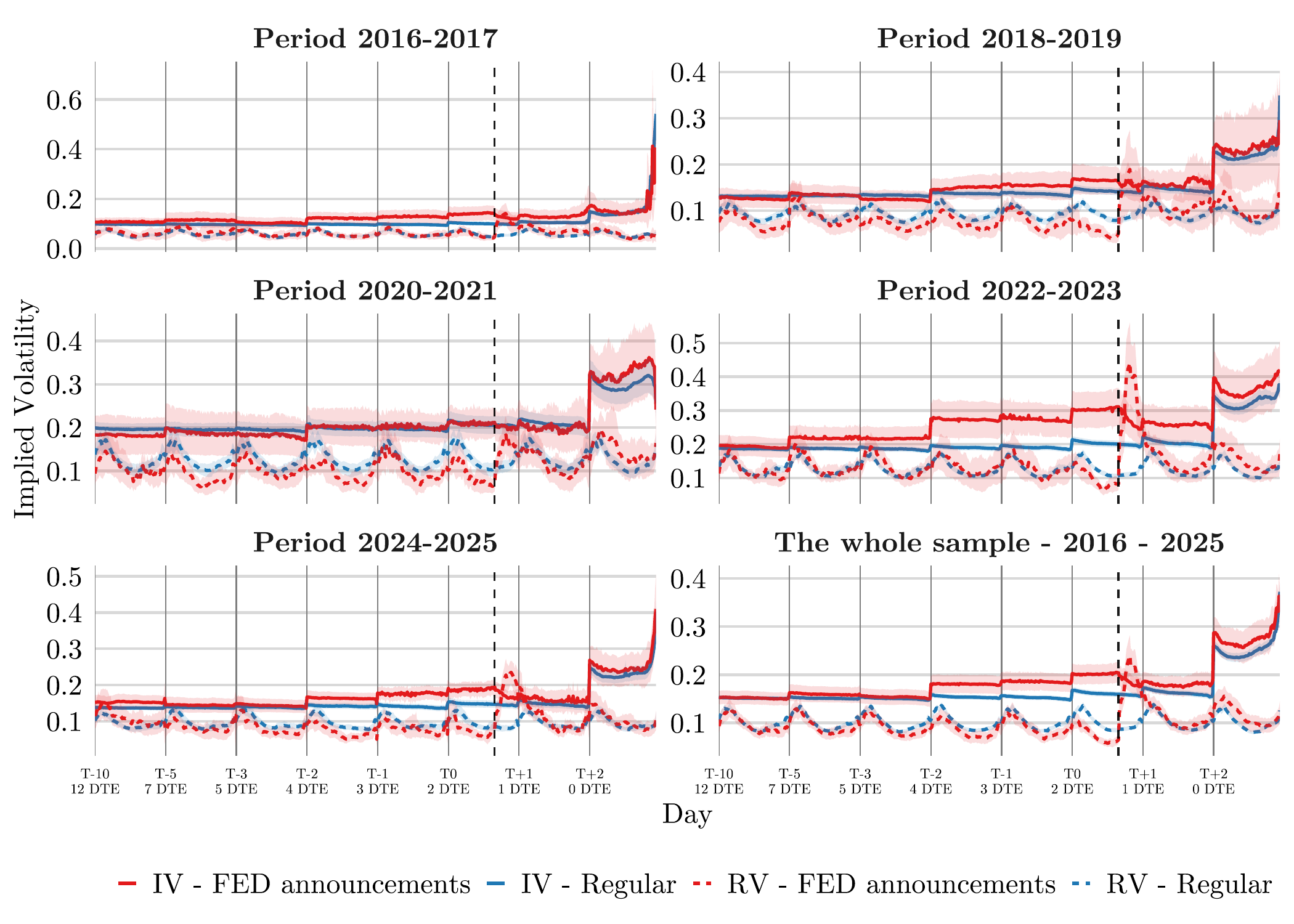}
    \caption{Average ATM Implied and Realised Volatility for put options expiring 2 trading days after the announcement.}
\end{figure}

\begin{figure}[H]
    \centering
    \includegraphics[width=\linewidth]{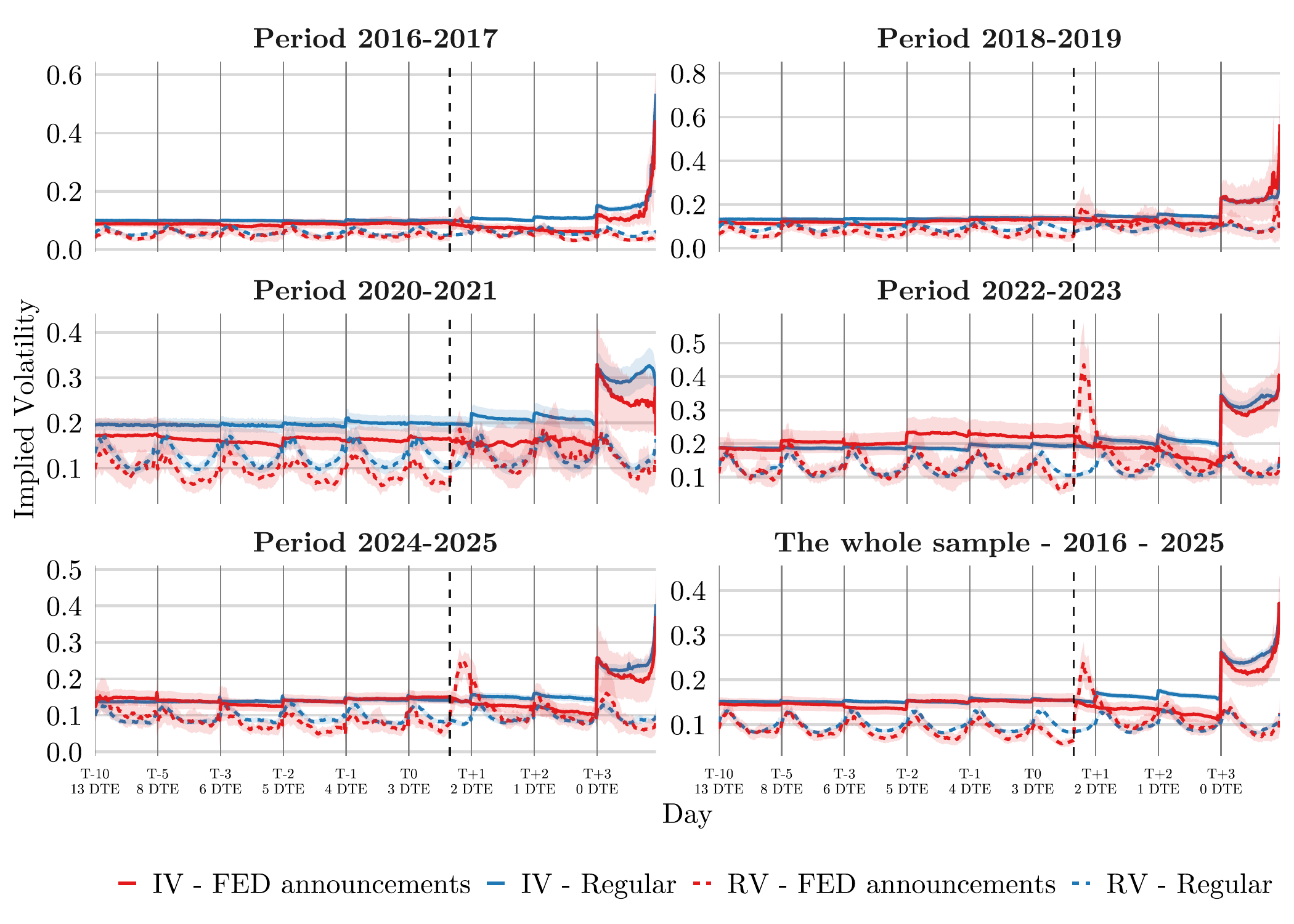}
    \caption{Average ATM Implied and Realised Volatility for put options expiring 3 trading days after the announcement.}
\end{figure}

\begin{figure}[H]
    \centering
    \includegraphics[width=\linewidth]{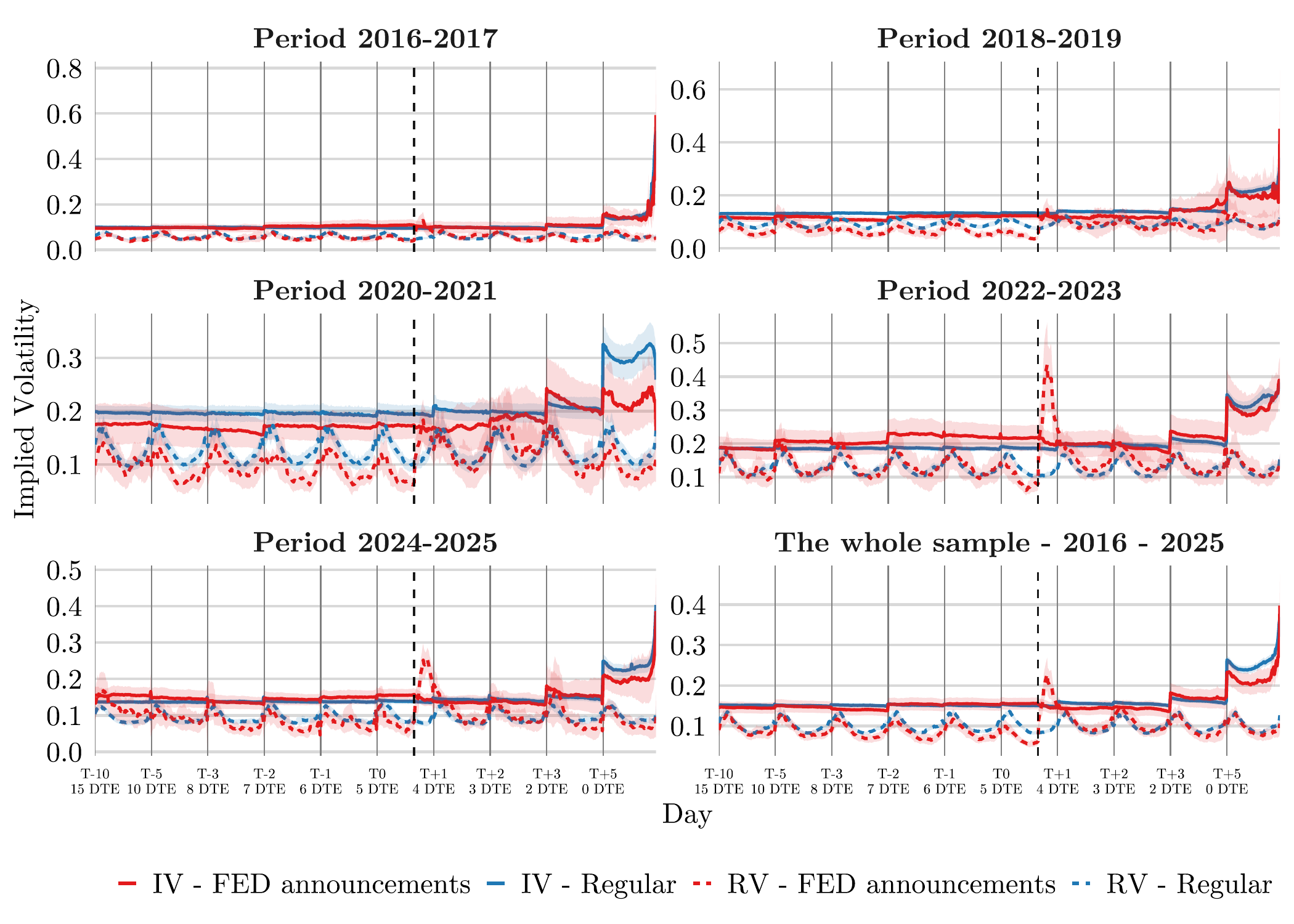}
    \caption{Average ATM Implied and Realised Volatility for put options expiring 5 trading days after the announcement.}
\end{figure}

\begin{figure}[H]
    \centering
    \includegraphics[width=\linewidth]{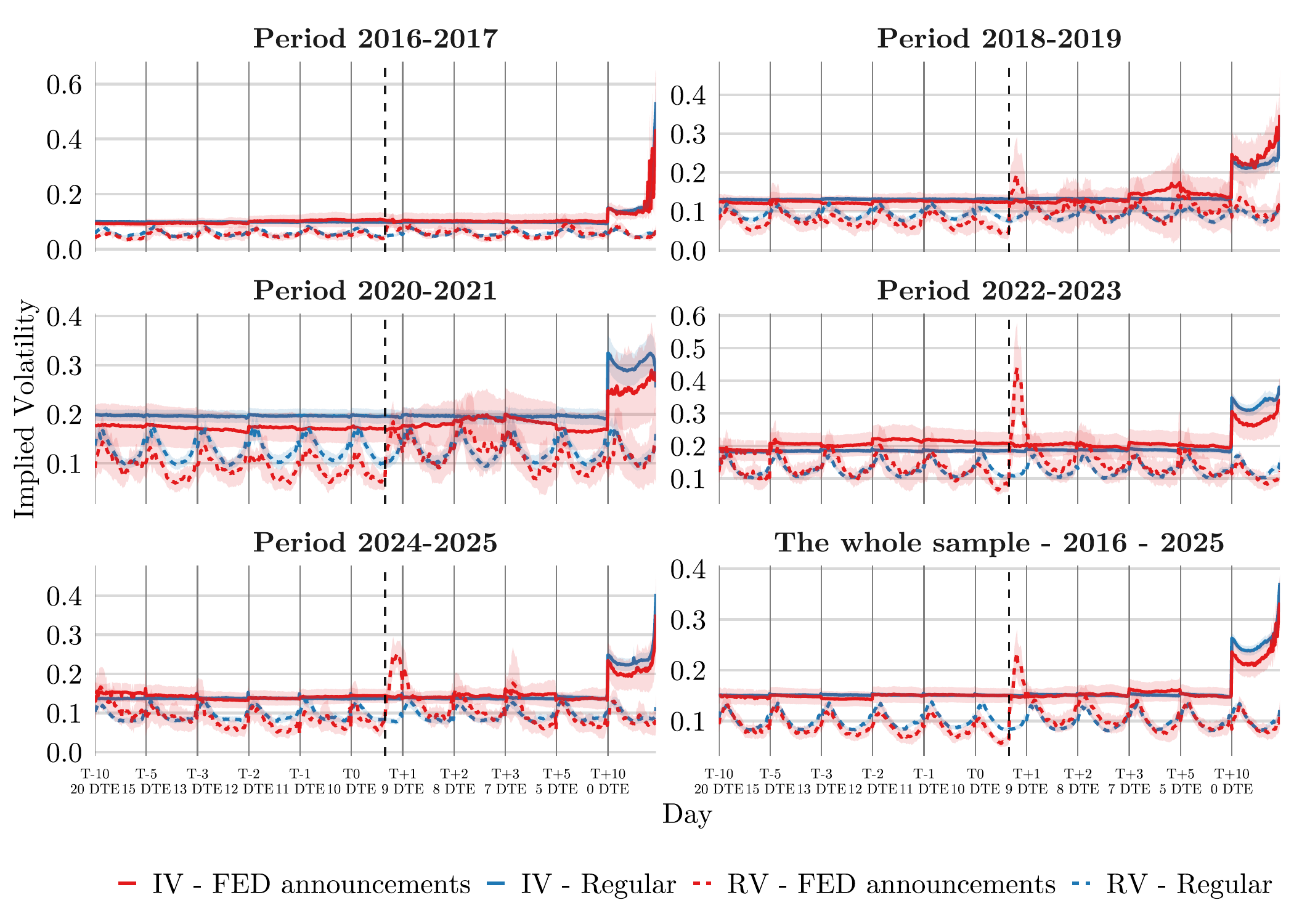}
    \caption{Average ATM Implied and Realised Volatility for put options expiring 10 trading days after the announcement.}
\end{figure}

\section{Tests}\label{sec:app_tests}

\subsection{Calls}

The IV drift is defined as:

\begin{equation*}
    \begin{aligned}
        &\bar{IV}\mathbbm{1}_{\{t=\tau\}} - \bar{IV}\mathbbm{1}_{\{t\neq\tau\}}\\
    \end{aligned}
\end{equation*}

where $\tau$ denotes the time of the announcement (FOMC conference at 2pm) and $\bar{IV}$ is the average implied volatility.

Hypotheses of Mann-Whitney U test (applied as the distibutions of IV are not normally distibuted):

\begin{equation*}
    \begin{aligned}
        \begin{cases}
            H_0:\text{The distibution of IV at a given time is the same as the distibution of IV during all other periods.} \\
            H_1:\text{The distibution of IV at a given time differs from the distibution of IV during all other periods.}
        \end{cases}
    \end{aligned}
\end{equation*}

\begin{table}[H]
\centering
\caption{The Mann-Whitney U test around FED interest rates announcements across log-moneyness $k$ for call options expiring on the day of the announcement. \\$H_0$: \textit{The distibution of IV at a given time around the announcement is the same as the distibution of IV during all other periods.}\\*, **, and *** denote significance at 5\%, 1\%, and 0.1\% levels, respectively.}
\centering
\fontsize{8}{10}\selectfont
\begin{tabular}[t]{>{}l>{}lccllccllc}
\toprule
\textbf{DTE} & \textbf{Time} & $k$ = -0.02 & $k$ = -0.015 & $k$ = -0.01 & $k$ = -0.005 & $k$ = 0 & $k$ = 0.005 & $k$ = 0.01 & $k$ = 0.015 & $k$ = 0.02\\
\cmidrule{1-11}
\midrule
 & \textbf{10:00} & 0.397 & 0.441 & 0.454 & 0.496 & 0.931 & 0.457 & 0.417 & 0.410 & 0.372\\

 & \textbf{12:00} & 0.460 & 0.467 & 0.477 & 0.480 & 0.924 & 0.447 & 0.445 & 0.445 & 0.437\\

 & \textbf{14:00} & 0.401 & 0.372 & 0.402 & 0.398 & 0.904 & 0.363 & 0.326 & 0.348 & 0.318\\

\multirow{-4}{*}{\raggedright\arraybackslash \textbf{3}} & \textbf{16:00} & 0.489 & 0.465 & 0.496 & 0.497 & 0.930 & 0.481 & 0.506 & 0.538 & 0.610\\
\cmidrule{1-11}
 & \textbf{10:00} & 0.358 & 0.288 & 0.276 & 0.298 & \textbf{\textcolor{UBCred}{0.004**}} & 0.231 & 0.236 & 0.234 & 0.217\\

 & \textbf{12:00} & 0.324 & 0.228 & 0.183 & 0.169 & \textbf{\textcolor{UBCred}{0.001***}} & 0.120 & 0.104 & 0.094 & 0.108\\

 & \textbf{14:00} & 0.415 & 0.265 & 0.195 & 0.173 & \textbf{\textcolor{UBCred}{0.001***}} & 0.119 & 0.120 & 0.132 & 0.186\\

\multirow{-4}{*}{\raggedright\arraybackslash \textbf{2}} & \textbf{16:00} & 0.184 & 0.108 & 0.092 & 0.095 & \textbf{\textcolor{UBCred}{0.000***}} & 0.089 & 0.068 & 0.104 & 0.112\\
\cmidrule{1-11}
 & \textbf{10:00} & 0.080 & \textbf{\textcolor{UBCred}{0.029*}} & \textbf{\textcolor{UBCred}{0.018*}} & \textbf{\textcolor{UBCred}{0.012*}} & \textbf{\textcolor{UBCred}{0.000***}} & \textbf{\textcolor{UBCred}{0.009**}} & \textbf{\textcolor{UBCred}{0.013*}} & \textbf{\textcolor{UBCred}{0.009**}} & \textbf{\textcolor{UBCred}{0.028*}}\\

 & \textbf{12:00} & \textbf{\textcolor{UBCred}{0.049*}} & \textbf{\textcolor{UBCred}{0.013*}} & \textbf{\textcolor{UBCred}{0.004**}} & \textbf{\textcolor{UBCred}{0.003**}} & \textbf{\textcolor{UBCred}{0.000***}} & \textbf{\textcolor{UBCred}{0.002**}} & \textbf{\textcolor{UBCred}{0.002**}} & \textbf{\textcolor{UBCred}{0.006**}} & \textbf{\textcolor{UBCred}{0.018*}}\\

 & \textbf{14:00} & \textbf{\textcolor{UBCred}{0.031*}} & \textbf{\textcolor{UBCred}{0.008**}} & \textbf{\textcolor{UBCred}{0.003**}} & \textbf{\textcolor{UBCred}{0.002**}} & \textbf{\textcolor{UBCred}{0.000***}} & \textbf{\textcolor{UBCred}{0.003**}} & \textbf{\textcolor{UBCred}{0.003**}} & \textbf{\textcolor{UBCred}{0.011*}} & 0.059\\

\multirow{-4}{*}{\raggedright\arraybackslash \textbf{1}} & \textbf{16:00} & \textbf{\textcolor{UBCred}{0.011*}} & \textbf{\textcolor{UBCred}{0.002**}} & \textbf{\textcolor{UBCred}{0.000***}} & \textbf{\textcolor{UBCred}{0.000***}} & \textbf{\textcolor{UBCred}{0.000***}} & \textbf{\textcolor{UBCred}{0.000***}} & \textbf{\textcolor{UBCred}{0.002**}} & \textbf{\textcolor{UBCred}{0.018*}} & 0.078\\
\cmidrule{1-11}
 & \textbf{10:00} & \textbf{\textcolor{UBCred}{0.000***}} & \textbf{\textcolor{UBCred}{0.000***}} & \textbf{\textcolor{UBCred}{0.000***}} & \textbf{\textcolor{UBCred}{0.000***}} & \textbf{\textcolor{UBCred}{0.000***}} & \textbf{\textcolor{UBCred}{0.000***}} & \textbf{\textcolor{UBCred}{0.000***}} & \textbf{\textcolor{UBCred}{0.000***}} & \textbf{\textcolor{UBCred}{0.000***}}\\

 & \textbf{12:00} & \textbf{\textcolor{UBCred}{0.000***}} & \textbf{\textcolor{UBCred}{0.000***}} & \textbf{\textcolor{UBCred}{0.000***}} & \textbf{\textcolor{UBCred}{0.000***}} & \textbf{\textcolor{UBCred}{0.000***}} & \textbf{\textcolor{UBCred}{0.000***}} & \textbf{\textcolor{UBCred}{0.000***}} & \textbf{\textcolor{UBCred}{0.000***}} & \textbf{\textcolor{UBCred}{0.000***}}\\

\multirow{-3}{*}{\raggedright\arraybackslash \textbf{0}} & \textbf{14:00} & \textbf{\textcolor{UBCred}{0.000***}} & \textbf{\textcolor{UBCred}{0.000***}} & \textbf{\textcolor{UBCred}{0.000***}} & \textbf{\textcolor{UBCred}{0.000***}} & \textbf{\textcolor{UBCred}{0.000***}} & \textbf{\textcolor{UBCred}{0.000***}} & \textbf{\textcolor{UBCred}{0.000***}} & \textbf{\textcolor{UBCred}{0.000***}} & \textbf{\textcolor{UBCred}{0.000***}}\\
\bottomrule
\end{tabular}
\end{table}

\begin{table}[H]
\centering
\caption{The Mann-Whitney U test around FED interest rates announcements across log-moneyness $k$ for call options expiring 2 trading days after the announcement. \\$H_0$: \textit{The distibution of IV at a given time around the announcement is the same as the distibution of IV during all other periods.}\\*, **, and *** denote significance at 5\%, 1\%, and 0.1\% levels, respectively.}
\centering
\fontsize{8}{10}\selectfont
\begin{tabular}[t]{>{}l>{}lccllccllc}
\toprule
\textbf{DTE} & \textbf{Time} & $k$ = -0.02 & $k$ = -0.015 & $k$ = -0.01 & $k$ = -0.005 & $k$ = 0 & $k$ = 0.005 & $k$ = 0.01 & $k$ = 0.015 & $k$ = 0.02\\
\cmidrule{1-11}
\midrule
 & \textbf{10:00} & 0.471 & 0.473 & 0.459 & 0.474 & 0.328 & 0.464 & 0.466 & 0.422 & 0.367\\

 & \textbf{12:00} & 0.471 & 0.464 & 0.463 & 0.470 & 0.303 & 0.422 & 0.416 & 0.430 & 0.420\\

 & \textbf{14:00} & 0.408 & 0.415 & 0.413 & 0.408 & 0.247 & 0.375 & 0.375 & 0.346 & 0.292\\

\multirow{-4}{*}{\raggedright\arraybackslash \textbf{5}} & \textbf{16:00} & 0.490 & 0.511 & 0.508 & 0.500 & 0.343 & 0.511 & 0.476 & 0.513 & 0.505\\
\cmidrule{1-11}
 & \textbf{10:00} & 0.359 & 0.366 & 0.365 & 0.379 & \textbf{\textcolor{UBCred}{0.000***}} & 0.286 & 0.281 & 0.230 & 0.169\\

 & \textbf{12:00} & 0.269 & 0.265 & 0.242 & 0.226 & \textbf{\textcolor{UBCred}{0.000***}} & 0.190 & 0.163 & 0.141 & 0.107\\

 & \textbf{14:00} & 0.274 & 0.282 & 0.276 & 0.263 & \textbf{\textcolor{UBCred}{0.000***}} & 0.203 & 0.161 & 0.158 & 0.154\\

\multirow{-4}{*}{\raggedright\arraybackslash \textbf{4}} & \textbf{16:00} & 0.156 & 0.186 & 0.180 & 0.154 & \textbf{\textcolor{UBCred}{0.000***}} & 0.130 & 0.118 & 0.069 & 0.069\\
\cmidrule{1-11}
 & \textbf{10:00} & 0.078 & 0.102 & 0.113 & 0.105 & \textbf{\textcolor{UBCred}{0.000***}} & 0.088 & 0.083 & \textbf{\textcolor{UBCred}{0.047*}} & \textbf{\textcolor{UBCred}{0.040*}}\\

 & \textbf{12:00} & 0.055 & \textbf{\textcolor{UBCred}{0.047*}} & 0.050 & 0.052 & \textbf{\textcolor{UBCred}{0.000***}} & \textbf{\textcolor{UBCred}{0.046*}} & \textbf{\textcolor{UBCred}{0.045*}} & \textbf{\textcolor{UBCred}{0.040*}} & \textbf{\textcolor{UBCred}{0.041*}}\\

 & \textbf{14:00} & \textbf{\textcolor{UBCred}{0.037*}} & \textbf{\textcolor{UBCred}{0.044*}} & \textbf{\textcolor{UBCred}{0.050*}} & 0.054 & \textbf{\textcolor{UBCred}{0.000***}} & 0.058 & 0.059 & 0.062 & 0.085\\

\multirow{-4}{*}{\raggedright\arraybackslash \textbf{3}} & \textbf{16:00} & \textbf{\textcolor{UBCred}{0.012*}} & \textbf{\textcolor{UBCred}{0.018*}} & \textbf{\textcolor{UBCred}{0.026*}} & \textbf{\textcolor{UBCred}{0.038*}} & \textbf{\textcolor{UBCred}{0.000***}} & \textbf{\textcolor{UBCred}{0.033*}} & \textbf{\textcolor{UBCred}{0.024*}} & \textbf{\textcolor{UBCred}{0.023*}} & \textbf{\textcolor{UBCred}{0.042*}}\\
\cmidrule{1-11}
 & \textbf{10:00} & \textbf{\textcolor{UBCred}{0.005**}} & \textbf{\textcolor{UBCred}{0.004**}} & \textbf{\textcolor{UBCred}{0.006**}} & \textbf{\textcolor{UBCred}{0.010**}} & \textbf{\textcolor{UBCred}{0.000***}} & \textbf{\textcolor{UBCred}{0.008**}} & \textbf{\textcolor{UBCred}{0.007**}} & \textbf{\textcolor{UBCred}{0.004**}} & \textbf{\textcolor{UBCred}{0.002**}}\\

 & \textbf{12:00} & \textbf{\textcolor{UBCred}{0.001**}} & \textbf{\textcolor{UBCred}{0.001***}} & \textbf{\textcolor{UBCred}{0.001**}} & \textbf{\textcolor{UBCred}{0.002**}} & \textbf{\textcolor{UBCred}{0.000***}} & \textbf{\textcolor{UBCred}{0.002**}} & \textbf{\textcolor{UBCred}{0.001**}} & \textbf{\textcolor{UBCred}{0.001***}} & \textbf{\textcolor{UBCred}{0.001***}}\\

 & \textbf{14:00} & \textbf{\textcolor{UBCred}{0.000***}} & \textbf{\textcolor{UBCred}{0.000***}} & \textbf{\textcolor{UBCred}{0.000***}} & \textbf{\textcolor{UBCred}{0.001**}} & \textbf{\textcolor{UBCred}{0.000***}} & \textbf{\textcolor{UBCred}{0.001***}} & \textbf{\textcolor{UBCred}{0.001***}} & \textbf{\textcolor{UBCred}{0.000***}} & \textbf{\textcolor{UBCred}{0.000***}}\\

\multirow{-4}{*}{\raggedright\arraybackslash \textbf{2}} & \textbf{16:00} & 0.095 & 0.079 & 0.107 & 0.146 & \textbf{\textcolor{UBCred}{0.000***}} & 0.170 & 0.131 & 0.092 & 0.115\\
\bottomrule
\end{tabular}
\end{table}

\begin{table}[H]
\centering
\caption{The Mann-Whitney U test around FED interest rates announcements across log-moneyness $k$ for call options expiring 3 trading days after the announcement. \\$H_0$: \textit{The distibution of IV at a given time around the announcement is the same as the distibution of IV during all other periods.}\\*, **, and *** denote significance at 5\%, 1\%, and 0.1\% levels, respectively.}
\centering
\fontsize{8}{10}\selectfont
\begin{tabular}[t]{>{}l>{}lccllccllc}
\toprule
\textbf{DTE} & \textbf{Time} & $k$ = -0.02 & $k$ = -0.015 & $k$ = -0.01 & $k$ = -0.005 & $k$ = 0 & $k$ = 0.005 & $k$ = 0.01 & $k$ = 0.015 & $k$ = 0.02\\
\cmidrule{1-11}
\midrule
 & \textbf{10:00} & 0.503 & 0.510 & 0.507 & 0.535 & 0.930 & 0.477 & 0.453 & 0.410 & 0.399\\

 & \textbf{12:00} & 0.521 & 0.535 & 0.535 & 0.558 & 0.940 & 0.504 & 0.481 & 0.443 & 0.407\\

 & \textbf{14:00} & 0.458 & 0.474 & 0.484 & 0.482 & 0.924 & 0.446 & 0.409 & 0.380 & 0.330\\

\multirow{-4}{*}{\raggedright\arraybackslash \textbf{6}} & \textbf{16:00} & 0.558 & 0.569 & 0.581 & 0.572 & 0.944 & 0.549 & 0.537 & 0.518 & 0.448\\
\cmidrule{1-11}
 & \textbf{10:00} & 0.397 & 0.411 & 0.394 & 0.398 & 0.320 & 0.337 & 0.305 & 0.243 & 0.233\\

 & \textbf{12:00} & 0.334 & 0.316 & 0.307 & 0.282 & 0.216 & 0.239 & 0.198 & 0.169 & 0.151\\

 & \textbf{14:00} & 0.302 & 0.299 & 0.268 & 0.253 & 0.193 & 0.222 & 0.178 & 0.152 & 0.126\\

\multirow{-4}{*}{\raggedright\arraybackslash \textbf{5}} & \textbf{16:00} & 0.189 & 0.194 & 0.180 & 0.173 & 0.151 & 0.144 & 0.134 & 0.106 & 0.081\\
\cmidrule{1-11}
 & \textbf{10:00} & 0.116 & 0.113 & 0.113 & 0.098 & 0.279 & 0.089 & 0.064 & \textbf{\textcolor{UBCred}{0.048*}} & \textbf{\textcolor{UBCred}{0.034*}}\\

 & \textbf{12:00} & 0.054 & 0.054 & 0.056 & 0.053 & 0.197 & \textbf{\textcolor{UBCred}{0.045*}} & \textbf{\textcolor{UBCred}{0.038*}} & \textbf{\textcolor{UBCred}{0.033*}} & \textbf{\textcolor{UBCred}{0.026*}}\\

 & \textbf{14:00} & \textbf{\textcolor{UBCred}{0.046*}} & 0.051 & 0.051 & 0.055 & 0.194 & 0.051 & \textbf{\textcolor{UBCred}{0.039*}} & \textbf{\textcolor{UBCred}{0.033*}} & \textbf{\textcolor{UBCred}{0.025*}}\\

\multirow{-4}{*}{\raggedright\arraybackslash \textbf{4}} & \textbf{16:00} & \textbf{\textcolor{UBCred}{0.021*}} & \textbf{\textcolor{UBCred}{0.027*}} & \textbf{\textcolor{UBCred}{0.031*}} & \textbf{\textcolor{UBCred}{0.035*}} & 0.129 & \textbf{\textcolor{UBCred}{0.032*}} & \textbf{\textcolor{UBCred}{0.025*}} & \textbf{\textcolor{UBCred}{0.019*}} & \textbf{\textcolor{UBCred}{0.017*}}\\
\cmidrule{1-11}
 & \textbf{10:00} & \textbf{\textcolor{UBCred}{0.008**}} & \textbf{\textcolor{UBCred}{0.010**}} & \textbf{\textcolor{UBCred}{0.012*}} & \textbf{\textcolor{UBCred}{0.014*}} & 0.265 & \textbf{\textcolor{UBCred}{0.014*}} & \textbf{\textcolor{UBCred}{0.008**}} & \textbf{\textcolor{UBCred}{0.006**}} & \textbf{\textcolor{UBCred}{0.003**}}\\

 & \textbf{12:00} & \textbf{\textcolor{UBCred}{0.002**}} & \textbf{\textcolor{UBCred}{0.003**}} & \textbf{\textcolor{UBCred}{0.003**}} & \textbf{\textcolor{UBCred}{0.004**}} & 0.154 & \textbf{\textcolor{UBCred}{0.003**}} & \textbf{\textcolor{UBCred}{0.002**}} & \textbf{\textcolor{UBCred}{0.001**}} & \textbf{\textcolor{UBCred}{0.001***}}\\

 & \textbf{14:00} & \textbf{\textcolor{UBCred}{0.001***}} & \textbf{\textcolor{UBCred}{0.001**}} & \textbf{\textcolor{UBCred}{0.002**}} & \textbf{\textcolor{UBCred}{0.003**}} & 0.179 & \textbf{\textcolor{UBCred}{0.002**}} & \textbf{\textcolor{UBCred}{0.001**}} & \textbf{\textcolor{UBCred}{0.000***}} & \textbf{\textcolor{UBCred}{0.000***}}\\

\multirow{-4}{*}{\raggedright\arraybackslash \textbf{3}} & \textbf{16:00} & \textbf{\textcolor{UBCred}{0.045*}} & \textbf{\textcolor{UBCred}{0.050*}} & 0.069 & 0.075 & 0.658 & 0.060 & \textbf{\textcolor{UBCred}{0.042*}} & \textbf{\textcolor{UBCred}{0.046*}} & \textbf{\textcolor{UBCred}{0.028*}}\\
\bottomrule
\end{tabular}
\end{table}

\begin{table}[H]
\centering
\caption{The Mann-Whitney U test around FED interest rates announcements across log-moneyness $k$ for call options expiring 5 trading days after the announcement. \\$H_0$: \textit{The distibution of IV at a given time around the announcement is the same as the distibution of IV during all other periods.}\\*, **, and *** denote significance at 5\%, 1\%, and 0.1\% levels, respectively.}
\centering
\fontsize{8}{10}\selectfont
\begin{tabular}[t]{>{}l>{}lccllccllc}
\toprule
\textbf{DTE} & \textbf{Time} & $k$ = -0.02 & $k$ = -0.015 & $k$ = -0.01 & $k$ = -0.005 & $k$ = 0 & $k$ = 0.005 & $k$ = 0.01 & $k$ = 0.015 & $k$ = 0.02\\
\cmidrule{1-11}
\midrule
 & \textbf{10:00} & 0.755 & 0.759 & 0.764 & 0.772 & 0.892 & 0.747 & 0.737 & 0.706 & 0.699\\

 & \textbf{12:00} & 0.783 & 0.785 & 0.780 & 0.793 & 0.904 & 0.786 & 0.768 & 0.754 & 0.733\\

 & \textbf{14:00} & 0.762 & 0.759 & 0.762 & 0.762 & 0.892 & 0.732 & 0.708 & 0.720 & 0.677\\

\multirow{-4}{*}{\raggedright\arraybackslash \textbf{8}} & \textbf{16:00} & 0.840 & 0.847 & 0.837 & 0.832 & 0.913 & 0.810 & 0.803 & 0.815 & 0.792\\
\cmidrule{1-11}
 & \textbf{10:00} & 0.723 & 0.716 & 0.702 & 0.708 & 0.317 & 0.680 & 0.642 & 0.622 & 0.572\\

 & \textbf{12:00} & 0.629 & 0.615 & 0.585 & 0.579 & 0.211 & 0.551 & 0.513 & 0.497 & 0.445\\

 & \textbf{14:00} & 0.645 & 0.632 & 0.604 & 0.590 & 0.222 & 0.535 & 0.504 & 0.514 & 0.491\\

\multirow{-4}{*}{\raggedright\arraybackslash \textbf{7}} & \textbf{16:00} & 0.541 & 0.564 & 0.511 & 0.503 & 0.185 & 0.467 & 0.465 & 0.454 & 0.417\\
\cmidrule{1-11}
 & \textbf{10:00} & 0.458 & 0.446 & 0.438 & 0.407 & 0.197 & 0.398 & 0.353 & 0.317 & 0.257\\

 & \textbf{12:00} & 0.334 & 0.332 & 0.320 & 0.314 & 0.132 & 0.289 & 0.267 & 0.220 & 0.206\\

 & \textbf{14:00} & 0.324 & 0.330 & 0.323 & 0.319 & 0.146 & 0.310 & 0.328 & 0.264 & 0.259\\

\multirow{-4}{*}{\raggedright\arraybackslash \textbf{6}} & \textbf{16:00} & 0.238 & 0.270 & 0.281 & 0.295 & 0.115 & 0.246 & 0.249 & 0.253 & 0.223\\
\cmidrule{1-11}
 & \textbf{10:00} & 0.173 & 0.187 & 0.200 & 0.189 & 0.153 & 0.193 & 0.175 & 0.146 & 0.118\\

 & \textbf{12:00} & 0.096 & 0.117 & 0.120 & 0.116 & 0.091 & 0.106 & 0.086 & 0.079 & 0.060\\

 & \textbf{14:00} & 0.065 & 0.081 & 0.113 & 0.112 & 0.081 & 0.103 & 0.085 & 0.063 & \textbf{\textcolor{UBCred}{0.043*}}\\

\multirow{-4}{*}{\raggedright\arraybackslash \textbf{5}} & \textbf{16:00} & 0.307 & 0.351 & 0.376 & 0.406 & 0.346 & 0.406 & 0.360 & 0.356 & 0.314\\
\bottomrule
\end{tabular}
\end{table}

\begin{table}[H]
\centering
\caption{The Mann-Whitney U test around FED interest rates announcements across log-moneyness $k$ for call options expiring 10 trading days after the announcement. \\$H_0$: \textit{The distibution of IV at a given time around the announcement is the same as the distibution of IV during all other periods.}\\*, **, and *** denote significance at 5\%, 1\%, and 0.1\% levels, respectively.}
\centering
\fontsize{8}{10}\selectfont
\begin{tabular}[t]{>{}l>{}lccllccllc}
\toprule
\textbf{DTE} & \textbf{Time} & $k$ = -0.02 & $k$ = -0.015 & $k$ = -0.01 & $k$ = -0.005 & $k$ = 0 & $k$ = 0.005 & $k$ = 0.01 & $k$ = 0.015 & $k$ = 0.02\\
\cmidrule{1-11}
\midrule
 & \textbf{10:00} & 0.630 & 0.636 & 0.627 & 0.642 & 0.716 & 0.629 & 0.643 & 0.637 & 0.610\\

 & \textbf{12:00} & 0.641 & 0.641 & 0.637 & 0.653 & 0.720 & 0.657 & 0.659 & 0.662 & 0.648\\

 & \textbf{14:00} & 0.598 & 0.598 & 0.613 & 0.621 & 0.684 & 0.634 & 0.613 & 0.621 & 0.583\\

\multirow{-4}{*}{\raggedright\arraybackslash \textbf{13}} & \textbf{16:00} & 0.691 & 0.690 & 0.673 & 0.702 & 0.736 & 0.705 & 0.710 & 0.706 & 0.717\\
\cmidrule{1-11}
 & \textbf{10:00} & 0.605 & 0.589 & 0.575 & 0.602 & 0.297 & 0.583 & 0.585 & 0.596 & 0.569\\

 & \textbf{12:00} & 0.550 & 0.545 & 0.529 & 0.520 & 0.232 & 0.517 & 0.497 & 0.513 & 0.500\\

 & \textbf{14:00} & 0.575 & 0.568 & 0.551 & 0.550 & 0.242 & 0.531 & 0.524 & 0.510 & 0.520\\

\multirow{-4}{*}{\raggedright\arraybackslash \textbf{12}} & \textbf{16:00} & 0.508 & 0.494 & 0.487 & 0.496 & 0.199 & 0.490 & 0.499 & 0.487 & 0.488\\
\cmidrule{1-11}
 & \textbf{10:00} & 0.510 & 0.481 & 0.471 & 0.480 & 0.275 & 0.458 & 0.466 & 0.491 & 0.477\\

 & \textbf{12:00} & 0.424 & 0.414 & 0.404 & 0.406 & 0.208 & 0.397 & 0.408 & 0.423 & 0.415\\

 & \textbf{14:00} & 0.407 & 0.423 & 0.428 & 0.418 & 0.226 & 0.430 & 0.442 & 0.468 & 0.462\\

\multirow{-4}{*}{\raggedright\arraybackslash \textbf{11}} & \textbf{16:00} & 0.376 & 0.394 & 0.397 & 0.409 & 0.229 & 0.410 & 0.440 & 0.437 & 0.436\\
\cmidrule{1-11}
 & \textbf{10:00} & 0.435 & 0.419 & 0.414 & 0.429 & 0.305 & 0.431 & 0.420 & 0.441 & 0.440\\

 & \textbf{12:00} & 0.349 & 0.360 & 0.357 & 0.348 & 0.265 & 0.353 & 0.366 & 0.374 & 0.367\\

 & \textbf{14:00} & 0.316 & 0.330 & 0.305 & 0.324 & 0.269 & 0.345 & 0.341 & 0.360 & 0.412\\

\multirow{-4}{*}{\raggedright\arraybackslash \textbf{10}} & \textbf{16:00} & 0.438 & 0.468 & 0.463 & 0.487 & 0.478 & 0.536 & 0.553 & 0.579 & 0.585\\
\bottomrule
\end{tabular}
\end{table}

\subsection{Puts}
\begin{table}[H]
\centering
\caption{The Mann-Whitney U test around FED interest rates announcements across log-moneyness $k$ for put options expiring on the day of the announcement. \\$H_0$: \textit{The distibution of IV at a given time around the announcement is the same as the distibution of IV during all other periods.}\\*, **, and *** denote significance at 5\%, 1\%, and 0.1\% levels, respectively.}
\centering
\fontsize{8}{10}\selectfont
\begin{tabular}[t]{>{}l>{}lccllccllc}
\toprule
\textbf{DTE} & \textbf{Time} & $k$ = -0.02 & $k$ = -0.015 & $k$ = -0.01 & $k$ = -0.005 & $k$ = 0 & $k$ = 0.005 & $k$ = 0.01 & $k$ = 0.015 & $k$ = 0.02\\
\cmidrule{1-11}
\midrule
 & \textbf{10:00} & 0.623 & 0.586 & 0.508 & 0.469 & 0.906 & 0.420 & 0.374 & 0.314 & 0.228\\

 & \textbf{12:00} & 0.518 & 0.532 & 0.486 & 0.453 & 0.907 & 0.421 & 0.376 & 0.318 & 0.282\\

 & \textbf{14:00} & 0.534 & 0.443 & 0.390 & 0.366 & 0.857 & 0.305 & 0.258 & 0.228 & 0.159\\

\multirow{-4}{*}{\raggedright\arraybackslash \textbf{3}} & \textbf{16:00} & 0.317 & 0.322 & 0.361 & 0.404 & 0.903 & 0.382 & 0.308 & 0.267 & 0.214\\
\cmidrule{1-11}
 & \textbf{10:00} & 0.744 & 0.607 & 0.463 & 0.351 & \textbf{\textcolor{UBCred}{0.003**}} & 0.256 & 0.227 & 0.208 & 0.206\\

 & \textbf{12:00} & 0.749 & 0.496 & 0.348 & 0.219 & \textbf{\textcolor{UBCred}{0.001***}} & 0.140 & 0.112 & 0.096 & 0.121\\

 & \textbf{14:00} & 0.754 & 0.509 & 0.266 & 0.215 & \textbf{\textcolor{UBCred}{0.001***}} & 0.151 & 0.118 & 0.140 & 0.148\\

\multirow{-4}{*}{\raggedright\arraybackslash \textbf{2}} & \textbf{16:00} & 0.905 & 0.693 & 0.356 & 0.167 & \textbf{\textcolor{UBCred}{0.000***}} & 0.115 & 0.092 & 0.097 & 0.125\\
\cmidrule{1-11}
 & \textbf{10:00} & 0.866 & 0.369 & 0.077 & \textbf{\textcolor{UBCred}{0.027*}} & \textbf{\textcolor{UBCred}{0.000***}} & \textbf{\textcolor{UBCred}{0.012*}} & \textbf{\textcolor{UBCred}{0.011*}} & \textbf{\textcolor{UBCred}{0.013*}} & \textbf{\textcolor{UBCred}{0.033*}}\\

 & \textbf{12:00} & 0.183 & \textbf{\textcolor{UBCred}{0.050*}} & \textbf{\textcolor{UBCred}{0.004**}} & \textbf{\textcolor{UBCred}{0.005**}} & \textbf{\textcolor{UBCred}{0.000***}} & \textbf{\textcolor{UBCred}{0.002**}} & \textbf{\textcolor{UBCred}{0.003**}} & \textbf{\textcolor{UBCred}{0.004**}} & \textbf{\textcolor{UBCred}{0.008**}}\\

 & \textbf{14:00} & 0.070 & \textbf{\textcolor{UBCred}{0.008**}} & \textbf{\textcolor{UBCred}{0.002**}} & \textbf{\textcolor{UBCred}{0.002**}} & \textbf{\textcolor{UBCred}{0.000***}} & \textbf{\textcolor{UBCred}{0.002**}} & \textbf{\textcolor{UBCred}{0.002**}} & \textbf{\textcolor{UBCred}{0.003**}} & \textbf{\textcolor{UBCred}{0.012*}}\\

\multirow{-4}{*}{\raggedright\arraybackslash \textbf{1}} & \textbf{16:00} & 0.075 & \textbf{\textcolor{UBCred}{0.049*}} & \textbf{\textcolor{UBCred}{0.004**}} & \textbf{\textcolor{UBCred}{0.001**}} & \textbf{\textcolor{UBCred}{0.000***}} & \textbf{\textcolor{UBCred}{0.000***}} & \textbf{\textcolor{UBCred}{0.000***}} & \textbf{\textcolor{UBCred}{0.001**}} & \textbf{\textcolor{UBCred}{0.003**}}\\
\cmidrule{1-11}
 & \textbf{10:00} & 0.683 & 0.168 & \textbf{\textcolor{UBCred}{0.001**}} & \textbf{\textcolor{UBCred}{0.000***}} & \textbf{\textcolor{UBCred}{0.000***}} & \textbf{\textcolor{UBCred}{0.000***}} & \textbf{\textcolor{UBCred}{0.000***}} & \textbf{\textcolor{UBCred}{0.000***}} & \textbf{\textcolor{UBCred}{0.000***}}\\

 & \textbf{12:00} & 0.089 & \textbf{\textcolor{UBCred}{0.004**}} & \textbf{\textcolor{UBCred}{0.000***}} & \textbf{\textcolor{UBCred}{0.000***}} & \textbf{\textcolor{UBCred}{0.000***}} & \textbf{\textcolor{UBCred}{0.000***}} & \textbf{\textcolor{UBCred}{0.000***}} & \textbf{\textcolor{UBCred}{0.000***}} & \textbf{\textcolor{UBCred}{0.000***}}\\

\multirow{-3}{*}{\raggedright\arraybackslash \textbf{0}} & \textbf{14:00} & \textbf{\textcolor{UBCred}{0.000***}} & \textbf{\textcolor{UBCred}{0.000***}} & \textbf{\textcolor{UBCred}{0.000***}} & \textbf{\textcolor{UBCred}{0.000***}} & \textbf{\textcolor{UBCred}{0.000***}} & \textbf{\textcolor{UBCred}{0.000***}} & \textbf{\textcolor{UBCred}{0.000***}} & \textbf{\textcolor{UBCred}{0.000***}} & \textbf{\textcolor{UBCred}{0.000***}}\\
\bottomrule
\end{tabular}
\end{table}

\begin{table}[H]
\centering
\caption{The Mann-Whitney U test around FED interest rates announcements across log-moneyness $k$ for put options expiring 2 trading days after the announcement. \\$H_0$: \textit{The distibution of IV at a given time around the announcement is the same as the distibution of IV during all other periods.}\\*, **, and *** denote significance at 5\%, 1\%, and 0.1\% levels, respectively.}
\centering
\fontsize{8}{10}\selectfont
\begin{tabular}[t]{>{}l>{}lccllccllc}
\toprule
\textbf{DTE} & \textbf{Time} & $k$ = -0.02 & $k$ = -0.015 & $k$ = -0.01 & $k$ = -0.005 & $k$ = 0 & $k$ = 0.005 & $k$ = 0.01 & $k$ = 0.015 & $k$ = 0.02\\
\cmidrule{1-11}
\midrule
 & \textbf{10:00} & 0.579 & 0.424 & 0.444 & 0.515 & 0.330 & 0.522 & 0.573 & 0.602 & 0.633\\

 & \textbf{12:00} & 0.567 & 0.511 & 0.424 & 0.445 & 0.308 & 0.487 & 0.522 & 0.571 & 0.697\\

 & \textbf{14:00} & 0.381 & 0.368 & 0.334 & 0.374 & 0.234 & 0.407 & 0.425 & 0.484 & 0.567\\

\multirow{-4}{*}{\raggedright\arraybackslash \textbf{5}} & \textbf{16:00} & 0.460 & 0.478 & 0.389 & 0.422 & 0.326 & 0.448 & 0.523 & 0.589 & 0.658\\
\cmidrule{1-11}
 & \textbf{10:00} & 0.766 & 0.521 & 0.478 & 0.459 & \textbf{\textcolor{UBCred}{0.001***}} & 0.468 & 0.512 & 0.588 & 0.744\\

 & \textbf{12:00} & 0.630 & 0.400 & 0.367 & 0.357 & \textbf{\textcolor{UBCred}{0.000***}} & 0.335 & 0.380 & 0.454 & 0.561\\

 & \textbf{14:00} & 0.508 & 0.406 & 0.301 & 0.299 & \textbf{\textcolor{UBCred}{0.000***}} & 0.339 & 0.392 & 0.508 & 0.684\\

\multirow{-4}{*}{\raggedright\arraybackslash \textbf{4}} & \textbf{16:00} & 0.728 & 0.592 & 0.417 & 0.311 & \textbf{\textcolor{UBCred}{0.000***}} & 0.359 & 0.367 & 0.419 & 0.539\\
\cmidrule{1-11}
 & \textbf{10:00} & 0.366 & 0.130 & 0.178 & 0.156 & \textbf{\textcolor{UBCred}{0.000***}} & 0.158 & 0.145 & 0.215 & 0.297\\

 & \textbf{12:00} & 0.207 & 0.079 & 0.099 & 0.087 & \textbf{\textcolor{UBCred}{0.000***}} & 0.096 & 0.091 & 0.143 & 0.224\\

 & \textbf{14:00} & 0.097 & \textbf{\textcolor{UBCred}{0.047*}} & \textbf{\textcolor{UBCred}{0.029*}} & 0.051 & \textbf{\textcolor{UBCred}{0.000***}} & 0.069 & 0.063 & 0.099 & 0.173\\

\multirow{-4}{*}{\raggedright\arraybackslash \textbf{3}} & \textbf{16:00} & 0.160 & 0.090 & \textbf{\textcolor{UBCred}{0.041*}} & 0.053 & \textbf{\textcolor{UBCred}{0.000***}} & 0.073 & 0.106 & 0.180 & 0.273\\
\cmidrule{1-11}
 & \textbf{10:00} & 0.318 & 0.100 & \textbf{\textcolor{UBCred}{0.012*}} & \textbf{\textcolor{UBCred}{0.020*}} & \textbf{\textcolor{UBCred}{0.000***}} & \textbf{\textcolor{UBCred}{0.025*}} & \textbf{\textcolor{UBCred}{0.039*}} & \textbf{\textcolor{UBCred}{0.037*}} & 0.070\\

 & \textbf{12:00} & 0.054 & \textbf{\textcolor{UBCred}{0.003**}} & \textbf{\textcolor{UBCred}{0.003**}} & \textbf{\textcolor{UBCred}{0.003**}} & \textbf{\textcolor{UBCred}{0.000***}} & \textbf{\textcolor{UBCred}{0.004**}} & \textbf{\textcolor{UBCred}{0.005**}} & \textbf{\textcolor{UBCred}{0.007**}} & \textbf{\textcolor{UBCred}{0.012*}}\\

 & \textbf{14:00} & \textbf{\textcolor{UBCred}{0.040*}} & \textbf{\textcolor{UBCred}{0.002**}} & \textbf{\textcolor{UBCred}{0.001**}} & \textbf{\textcolor{UBCred}{0.003**}} & \textbf{\textcolor{UBCred}{0.000***}} & \textbf{\textcolor{UBCred}{0.001***}} & \textbf{\textcolor{UBCred}{0.001**}} & \textbf{\textcolor{UBCred}{0.001***}} & \textbf{\textcolor{UBCred}{0.000***}}\\

\multirow{-4}{*}{\raggedright\arraybackslash \textbf{2}} & \textbf{16:00} & 0.559 & 0.478 & 0.414 & 0.314 & \textbf{\textcolor{UBCred}{0.001***}} & 0.323 & 0.411 & 0.544 & 0.733\\
\bottomrule
\end{tabular}
\end{table}

\begin{table}[H]
\centering
\caption{The Mann-Whitney U test around FED interest rates announcements across log-moneyness $k$ for put options expiring 3 trading days after the announcement. \\$H_0$: \textit{The distibution of IV at a given time around the announcement is the same as the distibution of IV during all other periods.}\\*, **, and *** denote significance at 5\%, 1\%, and 0.1\% levels, respectively.}
\centering
\fontsize{8}{10}\selectfont
\begin{tabular}[t]{>{}l>{}lccllccllc}
\toprule
\textbf{DTE} & \textbf{Time} & $k$ = -0.02 & $k$ = -0.015 & $k$ = -0.01 & $k$ = -0.005 & $k$ = 0 & $k$ = 0.005 & $k$ = 0.01 & $k$ = 0.015 & $k$ = 0.02\\
\cmidrule{1-11}
\midrule
 & \textbf{10:00} & 0.659 & 0.596 & 0.575 & 0.563 & 0.906 & 0.528 & 0.495 & 0.442 & 0.389\\

 & \textbf{12:00} & 0.681 & 0.594 & 0.586 & 0.567 & 0.922 & 0.527 & 0.489 & 0.449 & 0.420\\

 & \textbf{14:00} & 0.573 & 0.553 & 0.527 & 0.511 & 0.893 & 0.458 & 0.427 & 0.372 & 0.334\\

\multirow{-4}{*}{\raggedright\arraybackslash \textbf{6}} & \textbf{16:00} & 0.539 & 0.500 & 0.545 & 0.558 & 0.913 & 0.494 & 0.462 & 0.451 & 0.398\\
\cmidrule{1-11}
 & \textbf{10:00} & 0.542 & 0.535 & 0.482 & 0.461 & 0.295 & 0.393 & 0.333 & 0.326 & 0.285\\

 & \textbf{12:00} & 0.448 & 0.435 & 0.391 & 0.348 & 0.213 & 0.304 & 0.250 & 0.199 & 0.186\\

 & \textbf{14:00} & 0.486 & 0.433 & 0.368 & 0.320 & 0.185 & 0.267 & 0.221 & 0.184 & 0.167\\

\multirow{-4}{*}{\raggedright\arraybackslash \textbf{5}} & \textbf{16:00} & 0.526 & 0.380 & 0.322 & 0.266 & 0.144 & 0.207 & 0.190 & 0.155 & 0.135\\
\cmidrule{1-11}
 & \textbf{10:00} & 0.483 & 0.297 & 0.189 & 0.134 & 0.270 & 0.093 & 0.090 & 0.056 & 0.064\\

 & \textbf{12:00} & 0.154 & 0.091 & 0.088 & 0.084 & 0.177 & 0.062 & \textbf{\textcolor{UBCred}{0.044*}} & \textbf{\textcolor{UBCred}{0.038*}} & \textbf{\textcolor{UBCred}{0.025*}}\\

 & \textbf{14:00} & 0.179 & 0.111 & 0.084 & 0.073 & 0.176 & 0.060 & \textbf{\textcolor{UBCred}{0.046*}} & \textbf{\textcolor{UBCred}{0.033*}} & \textbf{\textcolor{UBCred}{0.026*}}\\

\multirow{-4}{*}{\raggedright\arraybackslash \textbf{4}} & \textbf{16:00} & 0.082 & 0.067 & \textbf{\textcolor{UBCred}{0.047*}} & 0.052 & 0.133 & \textbf{\textcolor{UBCred}{0.037*}} & \textbf{\textcolor{UBCred}{0.027*}} & \textbf{\textcolor{UBCred}{0.019*}} & \textbf{\textcolor{UBCred}{0.013*}}\\
\cmidrule{1-11}
 & \textbf{10:00} & 0.232 & 0.051 & \textbf{\textcolor{UBCred}{0.025*}} & \textbf{\textcolor{UBCred}{0.025*}} & 0.237 & \textbf{\textcolor{UBCred}{0.016*}} & \textbf{\textcolor{UBCred}{0.012*}} & \textbf{\textcolor{UBCred}{0.006**}} & \textbf{\textcolor{UBCred}{0.005**}}\\

 & \textbf{12:00} & \textbf{\textcolor{UBCred}{0.036*}} & \textbf{\textcolor{UBCred}{0.016*}} & \textbf{\textcolor{UBCred}{0.012*}} & \textbf{\textcolor{UBCred}{0.010**}} & 0.144 & \textbf{\textcolor{UBCred}{0.005**}} & \textbf{\textcolor{UBCred}{0.003**}} & \textbf{\textcolor{UBCred}{0.002**}} & \textbf{\textcolor{UBCred}{0.001**}}\\

 & \textbf{14:00} & \textbf{\textcolor{UBCred}{0.049*}} & \textbf{\textcolor{UBCred}{0.016*}} & \textbf{\textcolor{UBCred}{0.005**}} & \textbf{\textcolor{UBCred}{0.006**}} & 0.136 & \textbf{\textcolor{UBCred}{0.003**}} & \textbf{\textcolor{UBCred}{0.002**}} & \textbf{\textcolor{UBCred}{0.001***}} & \textbf{\textcolor{UBCred}{0.001**}}\\

\multirow{-4}{*}{\raggedright\arraybackslash \textbf{3}} & \textbf{16:00} & 0.235 & 0.112 & 0.156 & 0.130 & 0.533 & 0.093 & 0.061 & \textbf{\textcolor{UBCred}{0.035*}} & \textbf{\textcolor{UBCred}{0.021*}}\\
\bottomrule
\end{tabular}
\end{table}

\begin{table}[H]
\centering
\caption{The Mann-Whitney U test around FED interest rates announcements across log-moneyness $k$ for put options expiring 5 trading days after the announcement. \\$H_0$: \textit{The distibution of IV at a given time around the announcement is the same as the distibution of IV during all other periods.}\\*, **, and *** denote significance at 5\%, 1\%, and 0.1\% levels, respectively.}
\centering
\fontsize{8}{10}\selectfont
\begin{tabular}[t]{>{}l>{}lccllccllc}
\toprule
\textbf{DTE} & \textbf{Time} & $k$ = -0.02 & $k$ = -0.015 & $k$ = -0.01 & $k$ = -0.005 & $k$ = 0 & $k$ = 0.005 & $k$ = 0.01 & $k$ = 0.015 & $k$ = 0.02\\
\cmidrule{1-11}
\midrule
 & \textbf{10:00} & 0.811 & 0.753 & 0.748 & 0.772 & 0.845 & 0.736 & 0.731 & 0.705 & 0.688\\

 & \textbf{12:00} & 0.801 & 0.785 & 0.779 & 0.785 & 0.870 & 0.781 & 0.750 & 0.758 & 0.726\\

 & \textbf{14:00} & 0.799 & 0.773 & 0.752 & 0.761 & 0.849 & 0.742 & 0.732 & 0.700 & 0.664\\

\multirow{-4}{*}{\raggedright\arraybackslash \textbf{8}} & \textbf{16:00} & 0.817 & 0.773 & 0.806 & 0.795 & 0.890 & 0.787 & 0.776 & 0.746 & 0.716\\
\cmidrule{1-11}
 & \textbf{10:00} & 0.817 & 0.772 & 0.755 & 0.732 & 0.302 & 0.680 & 0.682 & 0.658 & 0.630\\

 & \textbf{12:00} & 0.703 & 0.661 & 0.637 & 0.629 & 0.205 & 0.594 & 0.551 & 0.521 & 0.501\\

 & \textbf{14:00} & 0.716 & 0.674 & 0.644 & 0.635 & 0.211 & 0.598 & 0.548 & 0.544 & 0.490\\

\multirow{-4}{*}{\raggedright\arraybackslash \textbf{7}} & \textbf{16:00} & 0.694 & 0.675 & 0.635 & 0.593 & 0.186 & 0.574 & 0.558 & 0.516 & 0.480\\
\cmidrule{1-11}
 & \textbf{10:00} & 0.636 & 0.561 & 0.503 & 0.480 & 0.185 & 0.422 & 0.409 & 0.382 & 0.346\\

 & \textbf{12:00} & 0.486 & 0.421 & 0.388 & 0.384 & 0.132 & 0.337 & 0.319 & 0.275 & 0.277\\

 & \textbf{14:00} & 0.477 & 0.390 & 0.359 & 0.359 & 0.132 & 0.347 & 0.321 & 0.306 & 0.297\\

\multirow{-4}{*}{\raggedright\arraybackslash \textbf{6}} & \textbf{16:00} & 0.364 & 0.318 & 0.338 & 0.310 & 0.105 & 0.279 & 0.267 & 0.248 & 0.229\\
\cmidrule{1-11}
 & \textbf{10:00} & 0.510 & 0.312 & 0.279 & 0.250 & 0.139 & 0.221 & 0.202 & 0.176 & 0.156\\

 & \textbf{12:00} & 0.165 & 0.204 & 0.192 & 0.170 & 0.086 & 0.134 & 0.118 & 0.097 & 0.080\\

 & \textbf{14:00} & 0.153 & 0.141 & 0.129 & 0.129 & 0.070 & 0.103 & 0.090 & 0.081 & 0.066\\

\multirow{-4}{*}{\raggedright\arraybackslash \textbf{5}} & \textbf{16:00} & 0.388 & 0.489 & 0.513 & 0.509 & 0.302 & 0.479 & 0.389 & 0.348 & 0.306\\
\bottomrule
\end{tabular}
\end{table}

\begin{table}[H]
\centering
\caption{The Mann-Whitney U test around FED interest rates announcements across log-moneyness $k$ for put options expiring 10 trading days after the announcement. \\$H_0$: \textit{The distibution of IV at a given time around the announcement is the same as the distibution of IV during all other periods.}\\*, **, and *** denote significance at 5\%, 1\%, and 0.1\% levels, respectively.}
\centering
\fontsize{8}{10}\selectfont
\begin{tabular}[t]{>{}l>{}lccllccllc}
\toprule
\textbf{DTE} & \textbf{Time} & $k$ = -0.02 & $k$ = -0.015 & $k$ = -0.01 & $k$ = -0.005 & $k$ = 0 & $k$ = 0.005 & $k$ = 0.01 & $k$ = 0.015 & $k$ = 0.02\\
\cmidrule{1-11}
\midrule
 & \textbf{10:00} & 0.542 & 0.536 & 0.562 & 0.572 & 0.669 & 0.595 & 0.578 & 0.594 & 0.573\\

 & \textbf{12:00} & 0.523 & 0.552 & 0.562 & 0.598 & 0.687 & 0.613 & 0.607 & 0.601 & 0.579\\

 & \textbf{14:00} & 0.484 & 0.512 & 0.509 & 0.539 & 0.625 & 0.553 & 0.550 & 0.555 & 0.520\\

\multirow{-4}{*}{\raggedright\arraybackslash \textbf{13}} & \textbf{16:00} & 0.469 & 0.516 & 0.541 & 0.576 & 0.682 & 0.587 & 0.597 & 0.564 & 0.573\\
\cmidrule{1-11}
 & \textbf{10:00} & 0.487 & 0.512 & 0.532 & 0.553 & 0.294 & 0.563 & 0.544 & 0.557 & 0.538\\

 & \textbf{12:00} & 0.402 & 0.459 & 0.465 & 0.494 & 0.257 & 0.495 & 0.499 & 0.464 & 0.447\\

 & \textbf{14:00} & 0.425 & 0.465 & 0.487 & 0.483 & 0.246 & 0.506 & 0.501 & 0.481 & 0.476\\

\multirow{-4}{*}{\raggedright\arraybackslash \textbf{12}} & \textbf{16:00} & 0.393 & 0.432 & 0.467 & 0.480 & 0.244 & 0.489 & 0.471 & 0.465 & 0.428\\
\cmidrule{1-11}
 & \textbf{10:00} & 0.467 & 0.434 & 0.450 & 0.453 & 0.261 & 0.428 & 0.428 & 0.420 & 0.419\\

 & \textbf{12:00} & 0.295 & 0.333 & 0.350 & 0.366 & 0.211 & 0.359 & 0.361 & 0.344 & 0.331\\

 & \textbf{14:00} & 0.276 & 0.318 & 0.342 & 0.346 & 0.235 & 0.362 & 0.374 & 0.366 & 0.365\\

\multirow{-4}{*}{\raggedright\arraybackslash \textbf{11}} & \textbf{16:00} & 0.246 & 0.284 & 0.319 & 0.322 & 0.207 & 0.344 & 0.356 & 0.351 & 0.325\\
\cmidrule{1-11}
 & \textbf{10:00} & 0.293 & 0.331 & 0.360 & 0.363 & 0.317 & 0.358 & 0.365 & 0.343 & 0.327\\

 & \textbf{12:00} & 0.211 & 0.291 & 0.315 & 0.314 & 0.251 & 0.321 & 0.319 & 0.313 & 0.301\\

 & \textbf{14:00} & 0.201 & 0.263 & 0.281 & 0.296 & 0.245 & 0.299 & 0.295 & 0.288 & 0.274\\

\multirow{-4}{*}{\raggedright\arraybackslash \textbf{10}} & \textbf{16:00} & 0.353 & 0.414 & 0.444 & 0.460 & 0.335 & 0.479 & 0.477 & 0.440 & 0.436\\
\bottomrule
\end{tabular}
\end{table}




\section{Model diagnostics}\label{sec:models_diag}

\subsection{Surface errors}\label{sec:models_diag_errors}

\subsubsection{Call options}

\begin{figure}[H]
    \centering

    \begin{subfigure}{0.49\linewidth}
        \centering
        \includegraphics[width=\linewidth]{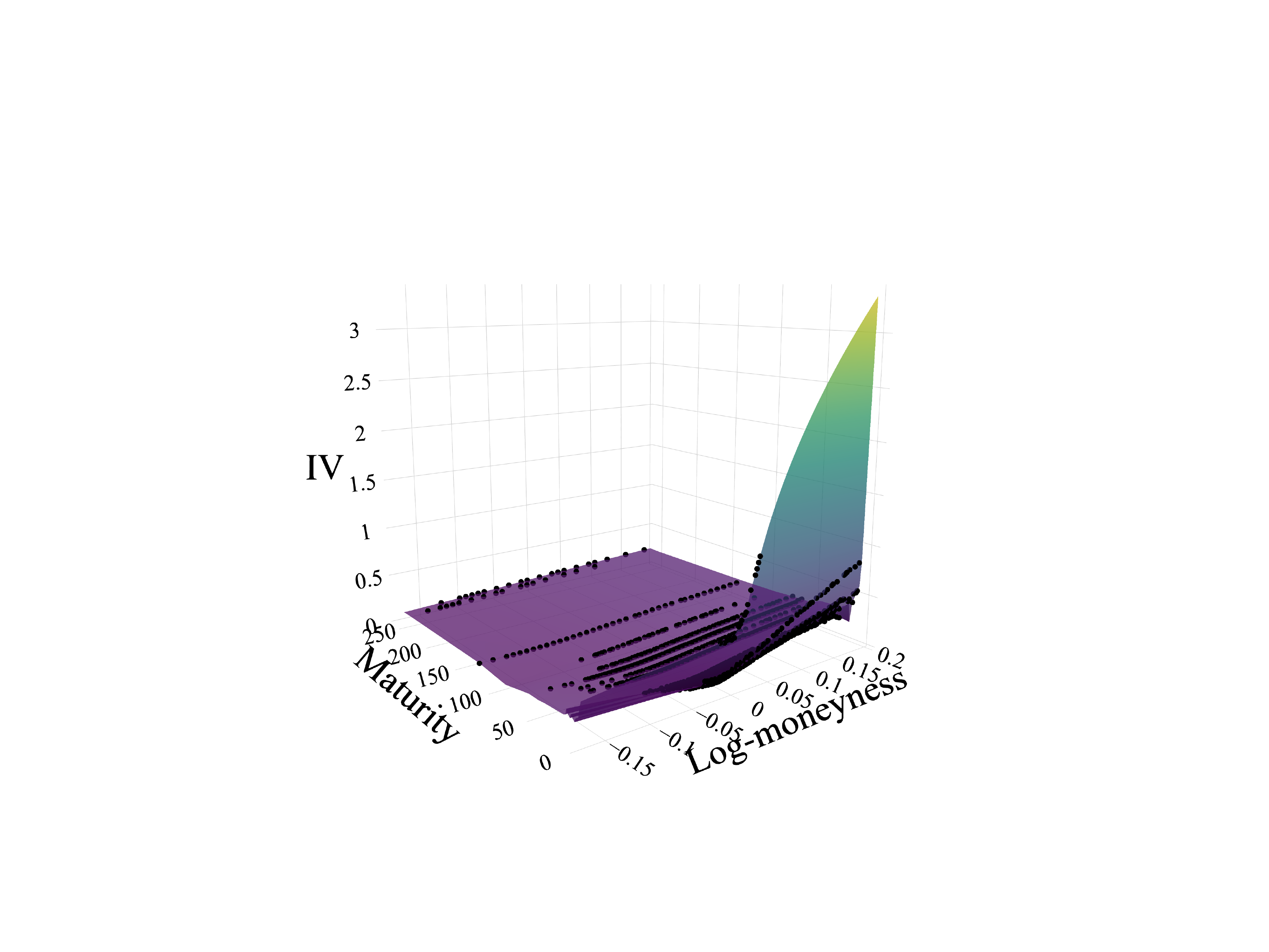}
        \caption{Linear interpolation}
    \end{subfigure}
    \hfill
    \begin{subfigure}{0.49\linewidth}
        \centering
        \includegraphics[width=\linewidth]{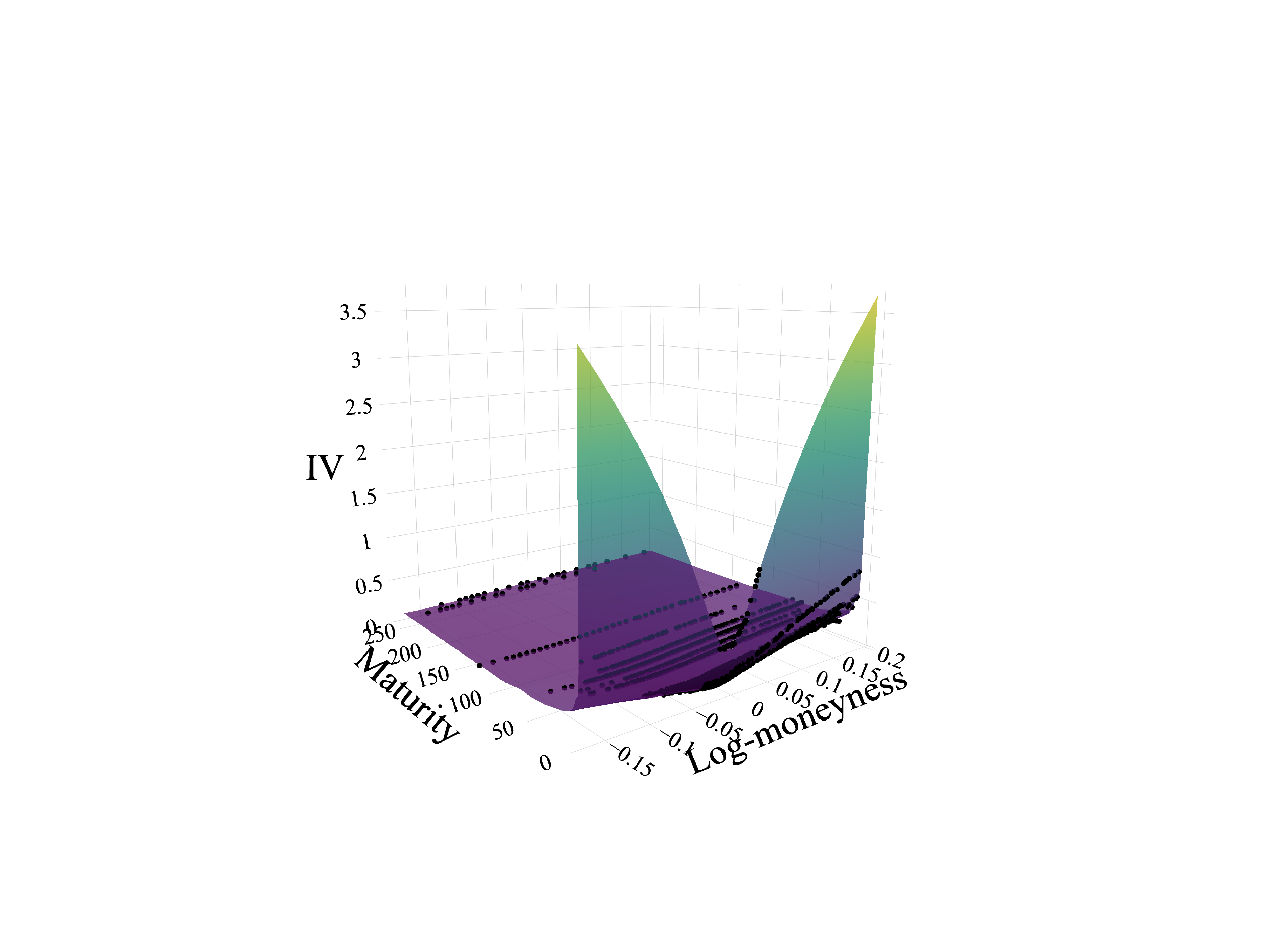}
        \caption{SVI}
    \end{subfigure}

    \vspace{0.5em}
    \hfill
    \begin{subfigure}{0.49\linewidth}
        \centering
        \includegraphics[width=\linewidth]{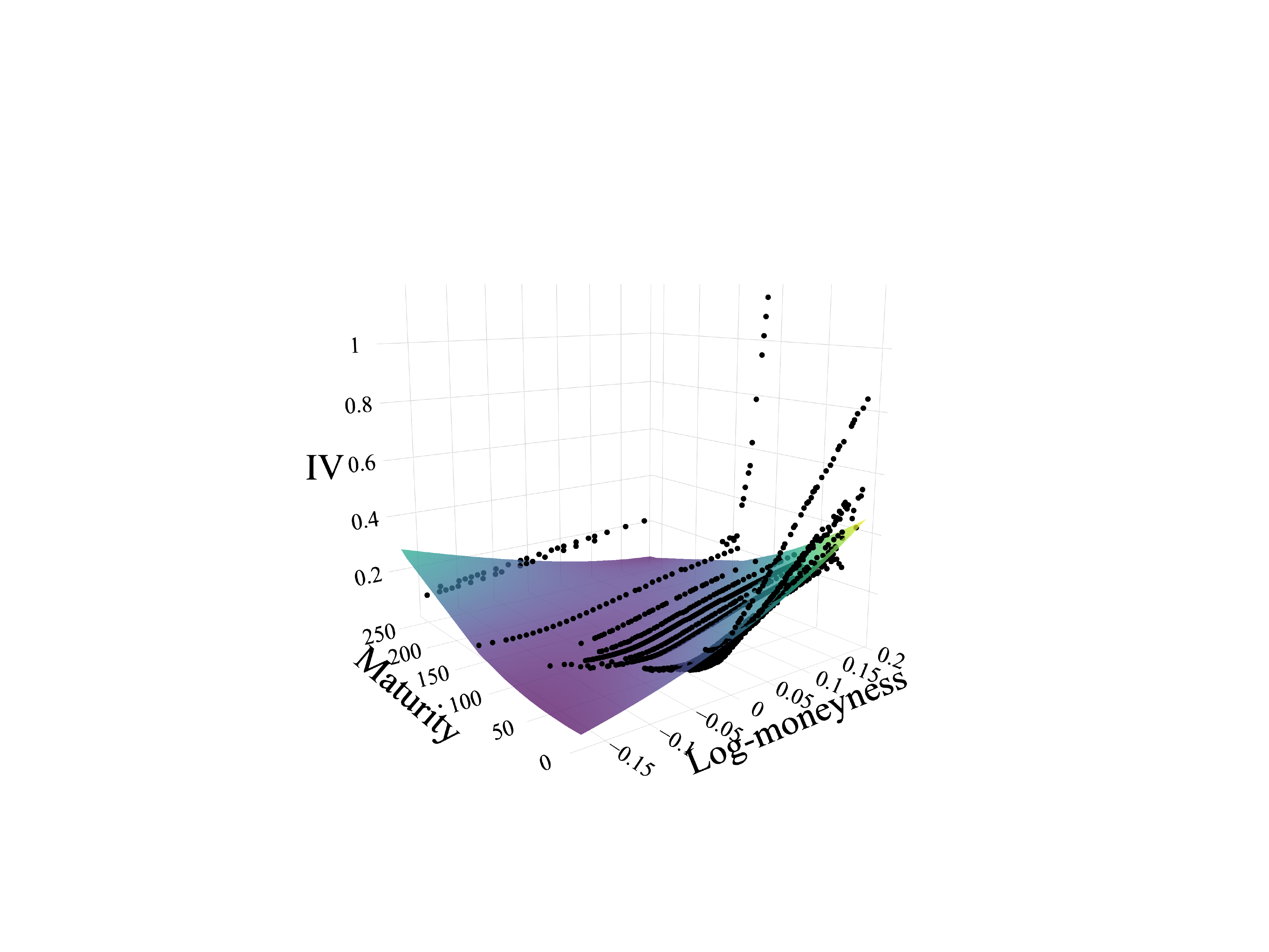}
        \caption{AHBS}
    \end{subfigure}

    \caption{Models fitted to the implied volatility surface of call options on 21 March 2018.}
\end{figure}

\subsubsection{Put options}

\begin{figure}[H]
    \centering

    \begin{subfigure}{0.49\linewidth}
        \centering
        \includegraphics[width=\linewidth]{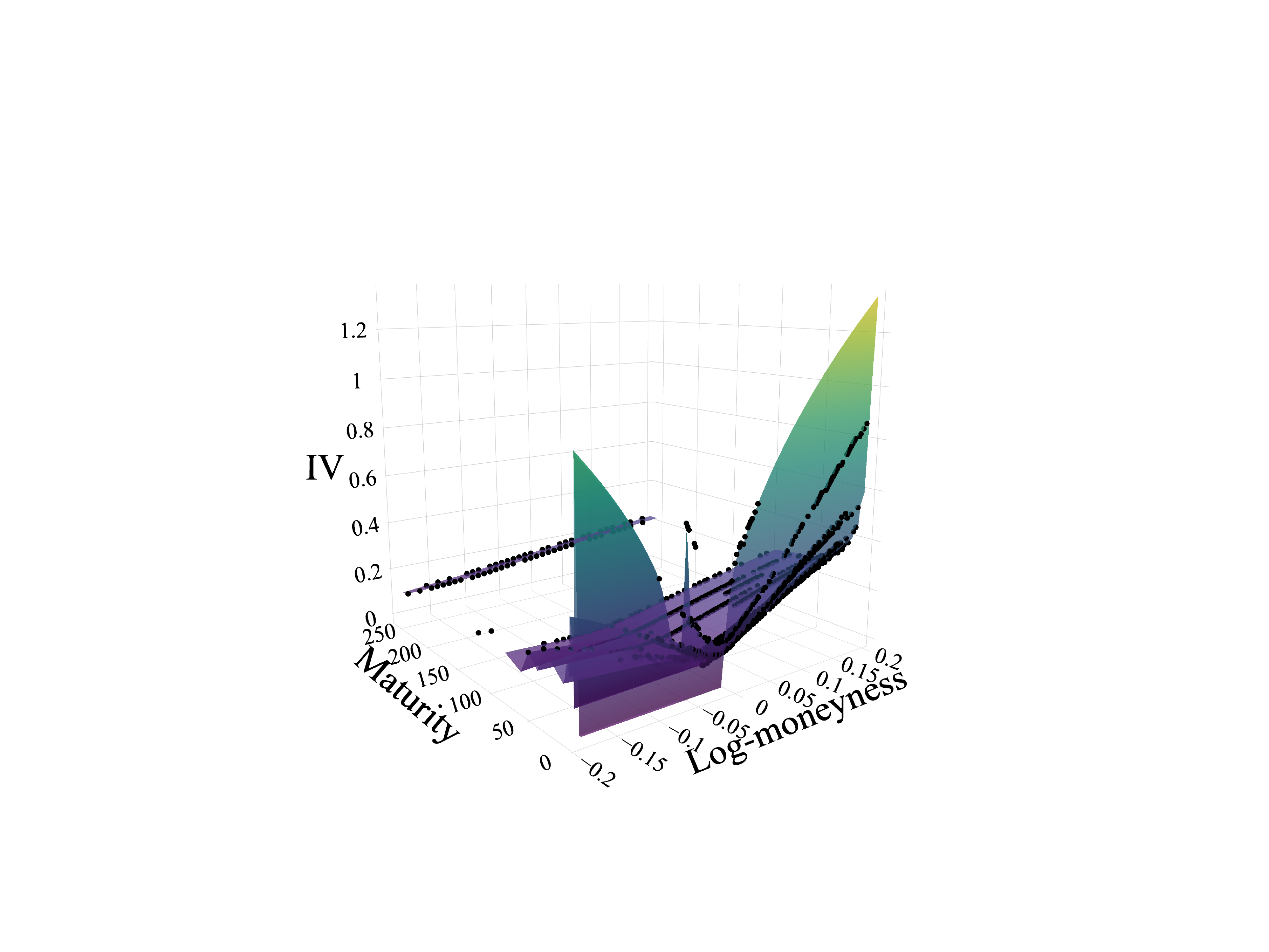}
        \caption{Linear interpolation}
    \end{subfigure}
    \hfill
    \begin{subfigure}{0.49\linewidth}
        \centering
        \includegraphics[width=\linewidth]{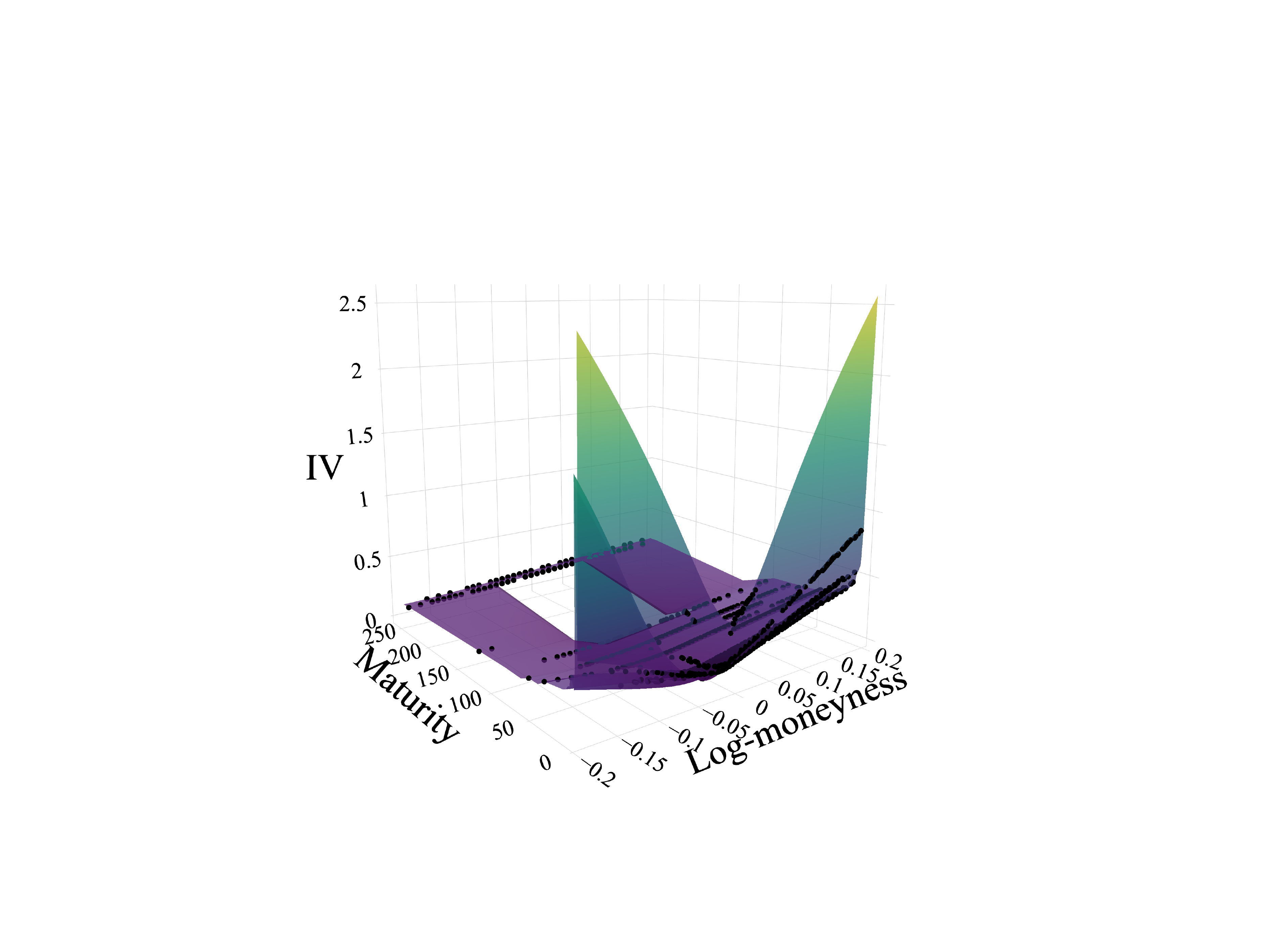}
        \caption{SVI}
    \end{subfigure}

    \vspace{0.5em}

    \hfill
    \begin{subfigure}{0.49\linewidth}
        \centering
        \includegraphics[width=\linewidth]{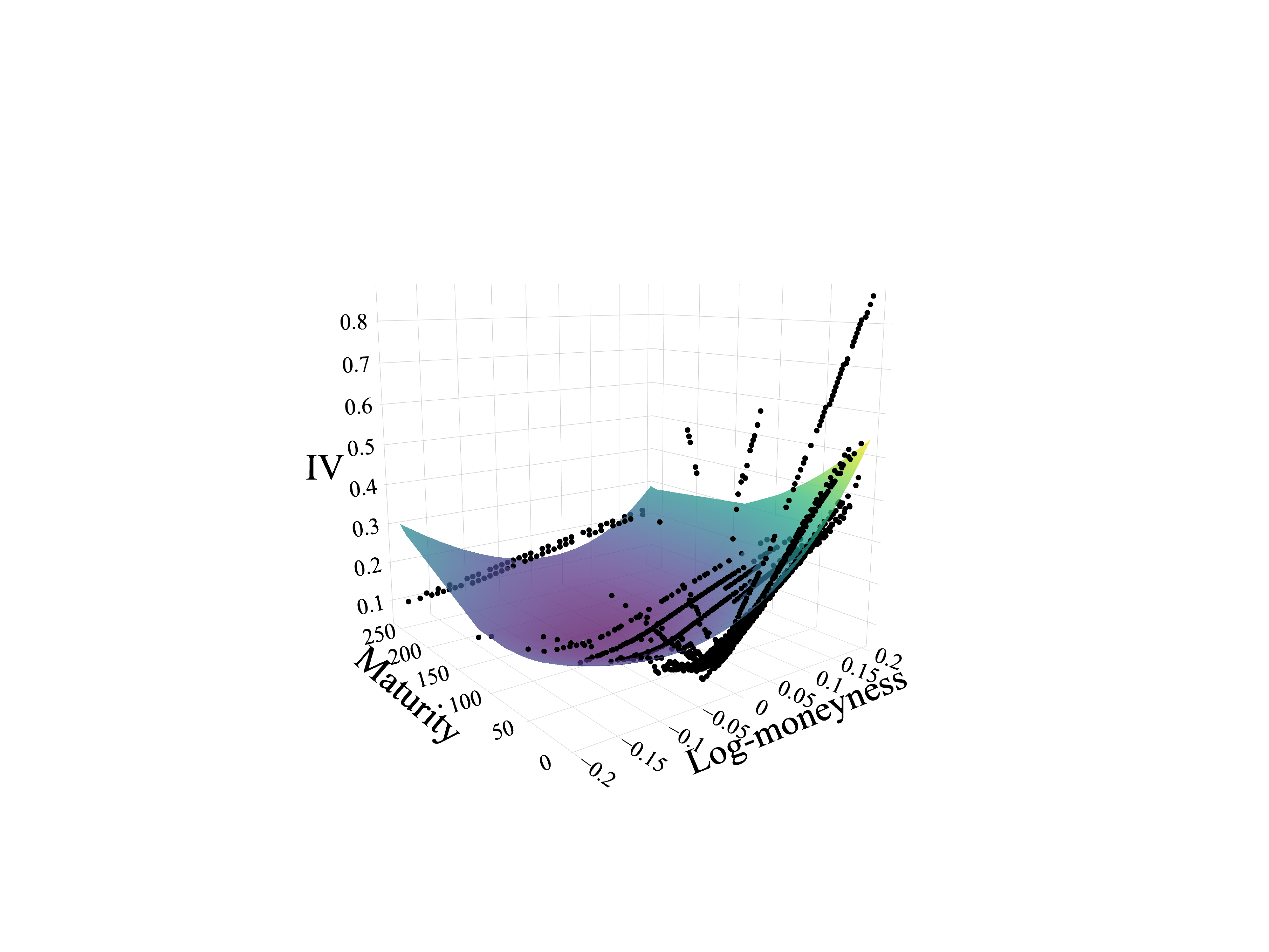}
        \caption{AHBS}
    \end{subfigure}

    \caption{Models fitted to the implied volatility surface of put options on 21 March 2018.}
\end{figure}

\subsection{ML models global errors}\label{sec:ml_models_metrics_global}

Many tables here are exact copies from the text for the sake of completeness.

These tables display the fit of models without the information regarding the announcement.

\begin{table}[H]
\caption{Comparison of out-of-sample RMSE for ConvLSTM and benchmarks.}
\label{tab:rmse_iv_surface}
\begin{tabular}{llllllllll}
\toprule
 & Type & \multicolumn{4}{c}{Call} & \multicolumn{4}{c}{Put} \\
 & Horizon & 1 & 2 & 5 & 10 & 1 & 2 & 5 & 10 \\
Model type & Model &  &  &  &  &  &  &  &  \\
\midrule
\multirow[t]{3}{*}{\textbf{ConvLSTM}} & \textbf{AHBS} & \begin{tabular}{@{}c@{}}0.208\\(0.001)\end{tabular} & \begin{tabular}{@{}c@{}}0.210\\(0.001)\end{tabular} & \begin{tabular}{@{}c@{}}0.212\\(0.002)\end{tabular} & \begin{tabular}{@{}c@{}}0.216\\(0.003)\end{tabular} & \begin{tabular}{@{}c@{}}0.155\\(0.001)\end{tabular} & \begin{tabular}{@{}c@{}}0.156\\(0.001)\end{tabular} & \begin{tabular}{@{}c@{}}0.161\\(0.003)\end{tabular} & \begin{tabular}{@{}c@{}}0.167\\(0.006)\end{tabular} \\
\textbf{} & \textbf{SVI} & \begin{tabular}{@{}c@{}}\textbf{0.085}\\(0.003)\end{tabular} & \begin{tabular}{@{}c@{}}\textbf{0.087}\\(0.002)\end{tabular} & \begin{tabular}{@{}c@{}}\textbf{0.102}\\(0.001)\end{tabular} & \begin{tabular}{@{}c@{}}0.117\\(0.004)\end{tabular} & \begin{tabular}{@{}c@{}}\textbf{0.077}\\(0.003)\end{tabular} & \begin{tabular}{@{}c@{}}\textbf{0.081}\\(0.003)\end{tabular} & \begin{tabular}{@{}c@{}}0.097\\(0.005)\end{tabular} & \begin{tabular}{@{}c@{}}0.114\\(0.009)\end{tabular} \\
\textbf{} & \textbf{linear} & \begin{tabular}{@{}c@{}}0.116\\(0.004)\end{tabular} & \begin{tabular}{@{}c@{}}0.118\\(0.004)\end{tabular} & \begin{tabular}{@{}c@{}}0.133\\(0.005)\end{tabular} & \begin{tabular}{@{}c@{}}0.149\\(0.009)\end{tabular} & \begin{tabular}{@{}c@{}}0.087\\(0.002)\end{tabular} & \begin{tabular}{@{}c@{}}0.091\\(0.002)\end{tabular} & \begin{tabular}{@{}c@{}}0.109\\(0.003)\end{tabular} & \begin{tabular}{@{}c@{}}0.127\\(0.006)\end{tabular} \\
\cline{1-10}
\multirow[t]{3}{*}{\textbf{Random Walk}} & \textbf{AHBS} & 0.208 & 0.211 & 0.215 & 0.219 & 0.155 & 0.156 & 0.162 & 0.167 \\
\textbf{} & \textbf{SVI} & 0.092 & 0.105 & 0.107 & \textbf{0.116} & 0.084 & 0.088 & \textbf{0.093} & \textbf{0.107} \\
\textbf{} & \textbf{linear} & 0.124 & 0.134 & 0.137 & 0.146 & 0.102 & 0.112 & 0.111 & 0.126 \\
\cline{1-10}
\bottomrule
\end{tabular}
\end{table}

\begin{table}[H] \centering
\caption{Comparison of out-of-sample MPE for ConvLSTM and benchmarks.}
\label{tab:MPE_iv_surface}
\begin{tabular}{llllllllll}
\toprule
 & Type & \multicolumn{4}{c}{Call} & \multicolumn{4}{c}{Put} \\
 & Horizon & 1 & 2 & 5 & 10 & 1 & 2 & 5 & 10 \\
Model type & Model &  &  &  &  &  &  &  &  \\
\midrule
\multirow[t]{3}{*}{\textbf{ConvLSTM}} & \textbf{AHBS} & \begin{tabular}{@{}c@{}}\textcolor{red}{-0.068}\\(0.057)\end{tabular} & \begin{tabular}{@{}c@{}}\textcolor{red}{-0.072}\\(0.076)\end{tabular} & \begin{tabular}{@{}c@{}}\textcolor{red}{-0.082}\\(0.133)\end{tabular} & \begin{tabular}{@{}c@{}}\textcolor{red}{-0.080}\\(0.226)\end{tabular} & \begin{tabular}{@{}c@{}}\textcolor{red}{-0.098}\\(0.058)\end{tabular} & \begin{tabular}{@{}c@{}}\textcolor{red}{-0.090}\\(0.077)\end{tabular} & \begin{tabular}{@{}c@{}}\textcolor{red}{-0.069}\\(0.131)\end{tabular} & \begin{tabular}{@{}c@{}}\textbf{\textcolor{red}{-0.036}}\\(0.203)\end{tabular} \\
\textbf{} & \textbf{SVI} & \begin{tabular}{@{}c@{}}\textcolor{red}{-0.011}\\(0.053)\end{tabular} & \begin{tabular}{@{}c@{}}\textbf{\textcolor{red}{-0.003}}\\(0.072)\end{tabular} & \begin{tabular}{@{}c@{}}\textbf{\textcolor{darkgreen}{0.023}}\\(0.140)\end{tabular} & \begin{tabular}{@{}c@{}}\textcolor{darkgreen}{0.071}\\(0.238)\end{tabular} & \begin{tabular}{@{}c@{}}\textcolor{red}{-0.020}\\(0.056)\end{tabular} & \begin{tabular}{@{}c@{}}\textbf{\textcolor{red}{-0.011}}\\(0.076)\end{tabular} & \begin{tabular}{@{}c@{}}\textbf{\textcolor{darkgreen}{0.019}}\\(0.147)\end{tabular} & \begin{tabular}{@{}c@{}}\textcolor{darkgreen}{0.068}\\(0.256)\end{tabular} \\
\textbf{} & \textbf{linear} & \begin{tabular}{@{}c@{}}\textbf{\textcolor{darkgreen}{0.007}}\\(0.053)\end{tabular} & \begin{tabular}{@{}c@{}}\textcolor{darkgreen}{0.017}\\(0.080)\end{tabular} & \begin{tabular}{@{}c@{}}\textcolor{darkgreen}{0.048}\\(0.155)\end{tabular} & \begin{tabular}{@{}c@{}}\textcolor{darkgreen}{0.097}\\(0.258)\end{tabular} & \begin{tabular}{@{}c@{}}\textbf{\textcolor{darkgreen}{0.001}}\\(0.033)\end{tabular} & \begin{tabular}{@{}c@{}}\textcolor{darkgreen}{0.012}\\(0.046)\end{tabular} & \begin{tabular}{@{}c@{}}\textcolor{darkgreen}{0.048}\\(0.087)\end{tabular} & \begin{tabular}{@{}c@{}}\textcolor{darkgreen}{0.112}\\(0.147)\end{tabular} \\
\cline{1-10}
\multirow[t]{3}{*}{\textbf{Random Walk}} & \textbf{AHBS} & \textcolor{red}{-0.064} & \textcolor{red}{-0.069} & \textcolor{red}{-0.085} & \textcolor{red}{-0.091} & \textcolor{red}{-0.125} & \textcolor{red}{-0.128} & \textcolor{red}{-0.138} & \textcolor{red}{-0.141} \\
\textbf{} & \textbf{SVI} & \textcolor{red}{-0.036} & \textcolor{red}{-0.042} & \textcolor{red}{-0.057} & \textcolor{red}{-0.070} & \textcolor{red}{-0.048} & \textcolor{red}{-0.053} & \textcolor{red}{-0.059} & \textcolor{red}{-0.068} \\
\textbf{} & \textbf{linear} & \textcolor{red}{-0.018} & \textcolor{red}{-0.022} & \textcolor{red}{-0.040} & \textbf{\textcolor{red}{-0.051}} & \textcolor{red}{-0.020} & \textcolor{red}{-0.025} & \textcolor{red}{-0.036} & \textcolor{red}{-0.043} \\
\cline{1-10}
\bottomrule
\end{tabular}
\end{table}

These tables display the fit of models augmented with the information regarding the announcement as a dummy variable.

\begin{table}[H] \centering
\caption{Comparison of out-of-sample MPE for ConvLSTM and benchmarks.}
\label{tab:MPE_iv_surface}
\begin{tabular}{llllllllll}
\toprule
 & Type & \multicolumn{4}{c}{Call} & \multicolumn{4}{c}{Put} \\
 & Horizon & 1 & 2 & 5 & 10 & 1 & 2 & 5 & 10 \\
Model type & Model &  &  &  &  &  &  &  &  \\
\midrule
\multirow[t]{3}{*}{\textbf{ConvLSTM}} & \textbf{AHBS} & \begin{tabular}{@{}c@{}}\textcolor{red}{-0.087}\\(0.055)\end{tabular} & \begin{tabular}{@{}c@{}}\textcolor{red}{-0.098}\\(0.073)\end{tabular} & \begin{tabular}{@{}c@{}}\textcolor{red}{-0.125}\\(0.124)\end{tabular} & \begin{tabular}{@{}c@{}}\textcolor{red}{-0.154}\\(0.204)\end{tabular} & \begin{tabular}{@{}c@{}}\textcolor{red}{-0.098}\\(0.065)\end{tabular} & \begin{tabular}{@{}c@{}}\textcolor{red}{-0.090}\\(0.087)\end{tabular} & \begin{tabular}{@{}c@{}}\textcolor{red}{-0.070}\\(0.149)\end{tabular} & \begin{tabular}{@{}c@{}}\textbf{\textcolor{red}{-0.039}}\\(0.235)\end{tabular} \\
\textbf{} & \textbf{SVI} & \begin{tabular}{@{}c@{}}\textbf{\textcolor{red}{-0.004}}\\(0.052)\end{tabular} & \begin{tabular}{@{}c@{}}\textbf{\textcolor{darkgreen}{0.005}}\\(0.071)\end{tabular} & \begin{tabular}{@{}c@{}}\textbf{\textcolor{darkgreen}{0.037}}\\(0.133)\end{tabular} & \begin{tabular}{@{}c@{}}\textcolor{darkgreen}{0.094}\\(0.222)\end{tabular} & \begin{tabular}{@{}c@{}}\textcolor{red}{-0.017}\\(0.057)\end{tabular} & \begin{tabular}{@{}c@{}}\textbf{\textcolor{red}{-0.009}}\\(0.079)\end{tabular} & \begin{tabular}{@{}c@{}}\textbf{\textcolor{darkgreen}{0.018}}\\(0.149)\end{tabular} & \begin{tabular}{@{}c@{}}\textcolor{darkgreen}{0.064}\\(0.258)\end{tabular} \\
\textbf{} & \textbf{linear} & \begin{tabular}{@{}c@{}}\textcolor{darkgreen}{0.025}\\(0.056)\end{tabular} & \begin{tabular}{@{}c@{}}\textcolor{darkgreen}{0.042}\\(0.083)\end{tabular} & \begin{tabular}{@{}c@{}}\textcolor{darkgreen}{0.089}\\(0.159)\end{tabular} & \begin{tabular}{@{}c@{}}\textcolor{darkgreen}{0.165}\\(0.266)\end{tabular} & \begin{tabular}{@{}c@{}}\textbf{\textcolor{darkgreen}{0.002}}\\(0.046)\end{tabular} & \begin{tabular}{@{}c@{}}\textcolor{darkgreen}{0.009}\\(0.063)\end{tabular} & \begin{tabular}{@{}c@{}}\textcolor{darkgreen}{0.032}\\(0.112)\end{tabular} & \begin{tabular}{@{}c@{}}\textcolor{darkgreen}{0.073}\\(0.184)\end{tabular} \\
\cline{1-10}
\multirow[t]{3}{*}{\textbf{Random Walk}} & \textbf{AHBS} & \textcolor{red}{-0.064} & \textcolor{red}{-0.069} & \textcolor{red}{-0.085} & \textcolor{red}{-0.091} & \textcolor{red}{-0.125} & \textcolor{red}{-0.128} & \textcolor{red}{-0.138} & \textcolor{red}{-0.141} \\
\textbf{} & \textbf{SVI} & \textcolor{red}{-0.036} & \textcolor{red}{-0.042} & \textcolor{red}{-0.057} & \textcolor{red}{-0.070} & \textcolor{red}{-0.048} & \textcolor{red}{-0.053} & \textcolor{red}{-0.059} & \textcolor{red}{-0.068} \\
\textbf{} & \textbf{linear} & \textcolor{red}{-0.018} & \textcolor{red}{-0.022} & \textcolor{red}{-0.040} & \textbf{\textcolor{red}{-0.051}} & \textcolor{red}{-0.020} & \textcolor{red}{-0.025} & \textcolor{red}{-0.036} & \textcolor{red}{-0.043} \\
\cline{1-10}
\bottomrule
\end{tabular}
\end{table}

These tables compare the fit only on the days of announcements - models without the information regarding the announcement.

\begin{table}[H]
\caption{Comparison of out-of-sample RMSE for ConvLSTM and benchmarks.}
\label{tab:rmse_iv_surface}
\begin{tabular}{llllllllll}
\toprule
 & Type & \multicolumn{4}{c}{Call} & \multicolumn{4}{c}{Put} \\
 & Horizon & 1 & 2 & 5 & 10 & 1 & 2 & 5 & 10 \\
Model type & Model &  &  &  &  &  &  &  &  \\
\midrule
\multirow[t]{3}{*}{\textbf{ConvLSTM}} & \textbf{AHBS} & \begin{tabular}{@{}c@{}}0.229\\(0.002)\end{tabular} & \begin{tabular}{@{}c@{}}0.229\\(0.002)\end{tabular} & \begin{tabular}{@{}c@{}}0.231\\(0.003)\end{tabular} & \begin{tabular}{@{}c@{}}0.230\\(0.005)\end{tabular} & \begin{tabular}{@{}c@{}}0.163\\(0.002)\end{tabular} & \begin{tabular}{@{}c@{}}0.164\\(0.002)\end{tabular} & \begin{tabular}{@{}c@{}}0.164\\(0.004)\end{tabular} & \begin{tabular}{@{}c@{}}0.168\\(0.008)\end{tabular} \\
\textbf{} & \textbf{SVI} & \begin{tabular}{@{}c@{}}0.099\\(0.004)\end{tabular} & \begin{tabular}{@{}c@{}}\textbf{0.100}\\(0.004)\end{tabular} & \begin{tabular}{@{}c@{}}0.112\\(0.005)\end{tabular} & \begin{tabular}{@{}c@{}}0.125\\(0.009)\end{tabular} & \begin{tabular}{@{}c@{}}0.076\\(0.002)\end{tabular} & \begin{tabular}{@{}c@{}}0.081\\(0.003)\end{tabular} & \begin{tabular}{@{}c@{}}0.088\\(0.005)\end{tabular} & \begin{tabular}{@{}c@{}}0.093\\(0.013)\end{tabular} \\
\textbf{} & \textbf{linear} & \begin{tabular}{@{}c@{}}0.123\\(0.005)\end{tabular} & \begin{tabular}{@{}c@{}}0.125\\(0.006)\end{tabular} & \begin{tabular}{@{}c@{}}0.141\\(0.009)\end{tabular} & \begin{tabular}{@{}c@{}}0.157\\(0.014)\end{tabular} & \begin{tabular}{@{}c@{}}0.082\\(0.002)\end{tabular} & \begin{tabular}{@{}c@{}}0.088\\(0.003)\end{tabular} & \begin{tabular}{@{}c@{}}0.094\\(0.004)\end{tabular} & \begin{tabular}{@{}c@{}}0.102\\(0.007)\end{tabular} \\
\cline{1-10}
\multirow[t]{3}{*}{\textbf{Random Walk}} & \textbf{AHBS} & 0.229 & 0.229 & 0.230 & 0.229 & 0.163 & 0.166 & 0.163 & 0.162 \\
\textbf{} & \textbf{SVI} & \textbf{0.093} & 0.100 & \textbf{0.097} & \textbf{0.114} & 0.071 & \textbf{0.080} & 0.084 & \textbf{0.064} \\
\textbf{} & \textbf{linear} & 0.116 & 0.113 & 0.111 & 0.126 & \textbf{0.071} & 0.100 & \textbf{0.080} & 0.064 \\
\cline{1-10}
\bottomrule
\end{tabular}
\end{table}

\begin{table}[H] \centering
\caption{Comparison of out-of-sample MPE for ConvLSTM and benchmarks.}
\label{tab:MPE_iv_surface}
\begin{tabular}{llllllllll}
\toprule
 & Type & \multicolumn{4}{c}{Call} & \multicolumn{4}{c}{Put} \\
 & Horizon & 1 & 2 & 5 & 10 & 1 & 2 & 5 & 10 \\
Model type & Model &  &  &  &  &  &  &  &  \\
\midrule
\multirow[t]{3}{*}{\textbf{ConvLSTM}} & \textbf{AHBS} & \begin{tabular}{@{}c@{}}\textcolor{darkgreen}{0.049}\\(0.050)\end{tabular} & \begin{tabular}{@{}c@{}}\textcolor{darkgreen}{0.059}\\(0.067)\end{tabular} & \begin{tabular}{@{}c@{}}\textcolor{darkgreen}{0.041}\\(0.118)\end{tabular} & \begin{tabular}{@{}c@{}}\textbf{\textcolor{red}{-0.031}}\\(0.207)\end{tabular} & \begin{tabular}{@{}c@{}}\textcolor{red}{-0.012}\\(0.053)\end{tabular} & \begin{tabular}{@{}c@{}}\textcolor{darkgreen}{0.003}\\(0.071)\end{tabular} & \begin{tabular}{@{}c@{}}\textcolor{red}{-0.006}\\(0.125)\end{tabular} & \begin{tabular}{@{}c@{}}\textbf{\textcolor{red}{-0.002}}\\(0.202)\end{tabular} \\
\textbf{} & \textbf{SVI} & \begin{tabular}{@{}c@{}}\textcolor{darkgreen}{0.062}\\(0.047)\end{tabular} & \begin{tabular}{@{}c@{}}\textcolor{darkgreen}{0.093}\\(0.065)\end{tabular} & \begin{tabular}{@{}c@{}}\textcolor{darkgreen}{0.094}\\(0.128)\end{tabular} & \begin{tabular}{@{}c@{}}\textcolor{darkgreen}{0.088}\\(0.230)\end{tabular} & \begin{tabular}{@{}c@{}}\textcolor{darkgreen}{0.050}\\(0.051)\end{tabular} & \begin{tabular}{@{}c@{}}\textcolor{darkgreen}{0.076}\\(0.071)\end{tabular} & \begin{tabular}{@{}c@{}}\textcolor{darkgreen}{0.075}\\(0.139)\end{tabular} & \begin{tabular}{@{}c@{}}\textcolor{darkgreen}{0.089}\\(0.249)\end{tabular} \\
\textbf{} & \textbf{linear} & \begin{tabular}{@{}c@{}}\textcolor{darkgreen}{0.069}\\(0.047)\end{tabular} & \begin{tabular}{@{}c@{}}\textcolor{darkgreen}{0.107}\\(0.072)\end{tabular} & \begin{tabular}{@{}c@{}}\textcolor{darkgreen}{0.122}\\(0.142)\end{tabular} & \begin{tabular}{@{}c@{}}\textcolor{darkgreen}{0.123}\\(0.248)\end{tabular} & \begin{tabular}{@{}c@{}}\textcolor{darkgreen}{0.067}\\(0.030)\end{tabular} & \begin{tabular}{@{}c@{}}\textcolor{darkgreen}{0.096}\\(0.042)\end{tabular} & \begin{tabular}{@{}c@{}}\textcolor{darkgreen}{0.101}\\(0.081)\end{tabular} & \begin{tabular}{@{}c@{}}\textcolor{darkgreen}{0.118}\\(0.139)\end{tabular} \\
\cline{1-10}
\multirow[t]{3}{*}{\textbf{Random Walk}} & \textbf{AHBS} & \textcolor{darkgreen}{0.034} & \textcolor{darkgreen}{0.050} & \textcolor{darkgreen}{0.056} & \textcolor{red}{-0.036} & \textcolor{red}{-0.052} & \textbf{\textcolor{red}{-0.003}} & \textcolor{red}{-0.084} & \textcolor{red}{-0.120} \\
\textbf{} & \textbf{SVI} & \textbf{\textcolor{red}{-0.000}} & \textbf{\textcolor{darkgreen}{0.021}} & \textbf{\textcolor{darkgreen}{0.021}} & \textcolor{red}{-0.068} & \textbf{\textcolor{red}{-0.008}} & \textcolor{darkgreen}{0.019} & \textbf{\textcolor{red}{-0.001}} & \textcolor{red}{-0.066} \\
\textbf{} & \textbf{linear} & \textcolor{darkgreen}{0.014} & \textcolor{darkgreen}{0.042} & \textcolor{darkgreen}{0.039} & \textcolor{red}{-0.047} & \textcolor{darkgreen}{0.013} & \textcolor{darkgreen}{0.041} & \textcolor{darkgreen}{0.015} & \textcolor{red}{-0.066} \\
\cline{1-10}
\bottomrule
\end{tabular}
\end{table}

These tables display the fit of models augmented with the information regarding the announcement as a dummy variable only on the days of announcement.

\begin{table}[H]
\caption{Comparison of out-of-sample RMSE for ConvLSTM and benchmarks.}
\label{tab:rmse_iv_surface}
\begin{tabular}{llllllllll}
\toprule
 & Type & \multicolumn{4}{c}{Call} & \multicolumn{4}{c}{Put} \\
 & Horizon & 1 & 2 & 5 & 10 & 1 & 2 & 5 & 10 \\
Model type & Model &  &  &  &  &  &  &  &  \\
\midrule
\multirow[t]{3}{*}{\textbf{ConvLSTM}} & \textbf{AHBS} & \begin{tabular}{@{}c@{}}0.222\\(0.001)\end{tabular} & \begin{tabular}{@{}c@{}}0.222\\(0.001)\end{tabular} & \begin{tabular}{@{}c@{}}0.222\\(0.002)\end{tabular} & \begin{tabular}{@{}c@{}}0.223\\(0.002)\end{tabular} & \begin{tabular}{@{}c@{}}0.161\\(0.001)\end{tabular} & \begin{tabular}{@{}c@{}}0.162\\(0.001)\end{tabular} & \begin{tabular}{@{}c@{}}0.162\\(0.003)\end{tabular} & \begin{tabular}{@{}c@{}}0.166\\(0.006)\end{tabular} \\
\textbf{} & \textbf{SVI} & \begin{tabular}{@{}c@{}}\textbf{0.071}\\(0.003)\end{tabular} & \begin{tabular}{@{}c@{}}\textbf{0.072}\\(0.003)\end{tabular} & \begin{tabular}{@{}c@{}}\textbf{0.083}\\(0.004)\end{tabular} & \begin{tabular}{@{}c@{}}\textbf{0.093}\\(0.006)\end{tabular} & \begin{tabular}{@{}c@{}}\textbf{0.061}\\(0.003)\end{tabular} & \begin{tabular}{@{}c@{}}\textbf{0.063}\\(0.003)\end{tabular} & \begin{tabular}{@{}c@{}}\textbf{0.072}\\(0.008)\end{tabular} & \begin{tabular}{@{}c@{}}0.081\\(0.018)\end{tabular} \\
\textbf{} & \textbf{linear} & \begin{tabular}{@{}c@{}}0.084\\(0.007)\end{tabular} & \begin{tabular}{@{}c@{}}0.084\\(0.007)\end{tabular} & \begin{tabular}{@{}c@{}}0.099\\(0.009)\end{tabular} & \begin{tabular}{@{}c@{}}0.113\\(0.013)\end{tabular} & \begin{tabular}{@{}c@{}}0.066\\(0.002)\end{tabular} & \begin{tabular}{@{}c@{}}0.070\\(0.003)\end{tabular} & \begin{tabular}{@{}c@{}}0.075\\(0.006)\end{tabular} & \begin{tabular}{@{}c@{}}0.082\\(0.012)\end{tabular} \\
\cline{1-10}
\multirow[t]{3}{*}{\textbf{Random Walk}} & \textbf{AHBS} & 0.229 & 0.229 & 0.230 & 0.229 & 0.163 & 0.166 & 0.163 & 0.162 \\
\textbf{} & \textbf{SVI} & 0.093 & 0.100 & 0.097 & 0.114 & 0.071 & 0.080 & 0.084 & \textbf{0.064} \\
\textbf{} & \textbf{linear} & 0.116 & 0.113 & 0.111 & 0.126 & 0.071 & 0.100 & 0.080 & 0.064 \\
\cline{1-10}
\bottomrule
\end{tabular}
\end{table}

\begin{table}[H] \centering
\caption{Comparison of out-of-sample MPE for ConvLSTM and benchmarks.}
\label{tab:MPE_iv_surface}
\begin{tabular}{llllllllll}
\toprule
 & Type & \multicolumn{4}{c}{Call} & \multicolumn{4}{c}{Put} \\
 & Horizon & 1 & 2 & 5 & 10 & 1 & 2 & 5 & 10 \\
Model type & Model &  &  &  &  &  &  &  &  \\
\midrule
\multirow[t]{3}{*}{\textbf{ConvLSTM}} & \textbf{AHBS} & \begin{tabular}{@{}c@{}}\textcolor{red}{-0.040}\\(0.052)\end{tabular} & \begin{tabular}{@{}c@{}}\textcolor{red}{-0.037}\\(0.068)\end{tabular} & \begin{tabular}{@{}c@{}}\textcolor{red}{-0.073}\\(0.113)\end{tabular} & \begin{tabular}{@{}c@{}}\textcolor{red}{-0.166}\\(0.190)\end{tabular} & \begin{tabular}{@{}c@{}}\textcolor{red}{-0.047}\\(0.060)\end{tabular} & \begin{tabular}{@{}c@{}}\textcolor{red}{-0.031}\\(0.080)\end{tabular} & \begin{tabular}{@{}c@{}}\textcolor{red}{-0.039}\\(0.142)\end{tabular} & \begin{tabular}{@{}c@{}}\textbf{\textcolor{red}{-0.036}}\\(0.234)\end{tabular} \\
\textbf{} & \textbf{SVI} & \begin{tabular}{@{}c@{}}\textcolor{darkgreen}{0.049}\\(0.060)\end{tabular} & \begin{tabular}{@{}c@{}}\textcolor{darkgreen}{0.079}\\(0.077)\end{tabular} & \begin{tabular}{@{}c@{}}\textcolor{darkgreen}{0.082}\\(0.134)\end{tabular} & \begin{tabular}{@{}c@{}}\textcolor{darkgreen}{0.082}\\(0.226)\end{tabular} & \begin{tabular}{@{}c@{}}\textcolor{darkgreen}{0.027}\\(0.062)\end{tabular} & \begin{tabular}{@{}c@{}}\textcolor{darkgreen}{0.050}\\(0.082)\end{tabular} & \begin{tabular}{@{}c@{}}\textcolor{darkgreen}{0.045}\\(0.149)\end{tabular} & \begin{tabular}{@{}c@{}}\textcolor{darkgreen}{0.054}\\(0.258)\end{tabular} \\
\textbf{} & \textbf{linear} & \begin{tabular}{@{}c@{}}\textcolor{darkgreen}{0.074}\\(0.062)\end{tabular} & \begin{tabular}{@{}c@{}}\textcolor{darkgreen}{0.116}\\(0.086)\end{tabular} & \begin{tabular}{@{}c@{}}\textcolor{darkgreen}{0.140}\\(0.157)\end{tabular} & \begin{tabular}{@{}c@{}}\textcolor{darkgreen}{0.156}\\(0.269)\end{tabular} & \begin{tabular}{@{}c@{}}\textcolor{darkgreen}{0.048}\\(0.047)\end{tabular} & \begin{tabular}{@{}c@{}}\textcolor{darkgreen}{0.070}\\(0.062)\end{tabular} & \begin{tabular}{@{}c@{}}\textcolor{darkgreen}{0.063}\\(0.110)\end{tabular} & \begin{tabular}{@{}c@{}}\textcolor{darkgreen}{0.060}\\(0.185)\end{tabular} \\
\cline{1-10}
\multirow[t]{3}{*}{\textbf{Random Walk}} & \textbf{AHBS} & \textcolor{darkgreen}{0.034} & \textcolor{darkgreen}{0.050} & \textcolor{darkgreen}{0.056} & \textbf{\textcolor{red}{-0.036}} & \textcolor{red}{-0.052} & \textbf{\textcolor{red}{-0.003}} & \textcolor{red}{-0.084} & \textcolor{red}{-0.120} \\
\textbf{} & \textbf{SVI} & \textbf{\textcolor{red}{-0.000}} & \textbf{\textcolor{darkgreen}{0.021}} & \textbf{\textcolor{darkgreen}{0.021}} & \textcolor{red}{-0.068} & \textbf{\textcolor{red}{-0.008}} & \textcolor{darkgreen}{0.019} & \textbf{\textcolor{red}{-0.001}} & \textcolor{red}{-0.066} \\
\textbf{} & \textbf{linear} & \textcolor{darkgreen}{0.014} & \textcolor{darkgreen}{0.042} & \textcolor{darkgreen}{0.039} & \textcolor{red}{-0.047} & \textcolor{darkgreen}{0.013} & \textcolor{darkgreen}{0.041} & \textcolor{darkgreen}{0.015} & \textcolor{red}{-0.066} \\
\cline{1-10}
\bottomrule
\end{tabular}
\end{table}

\subsection{Surface errors - RMSE}\label{sec:models_diag_errors_RMSE}


These graphs display the fit of models without the information regarding the announcement per maturity and moneyness in terms of RMSE.

\begin{figure}[H]
    \centering
    \includegraphics[width=\linewidth]{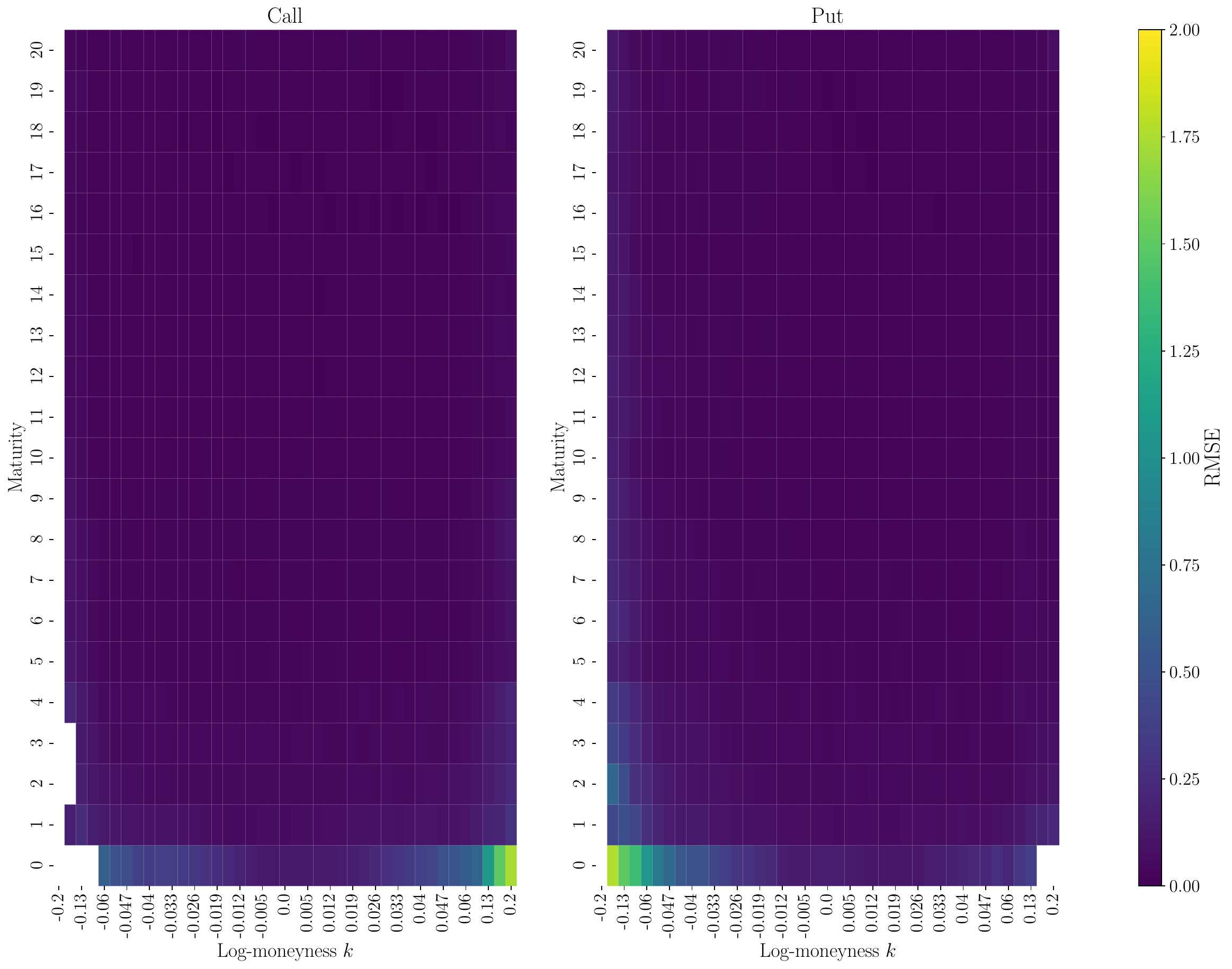}
    \caption{Out-of-sample RMSE comparison between naive Random Walk predictions and convLSTM predictions for SVI model, prediction horizon 1.}
\end{figure}

\begin{figure}[H]
    \centering
    \includegraphics[width=\linewidth]{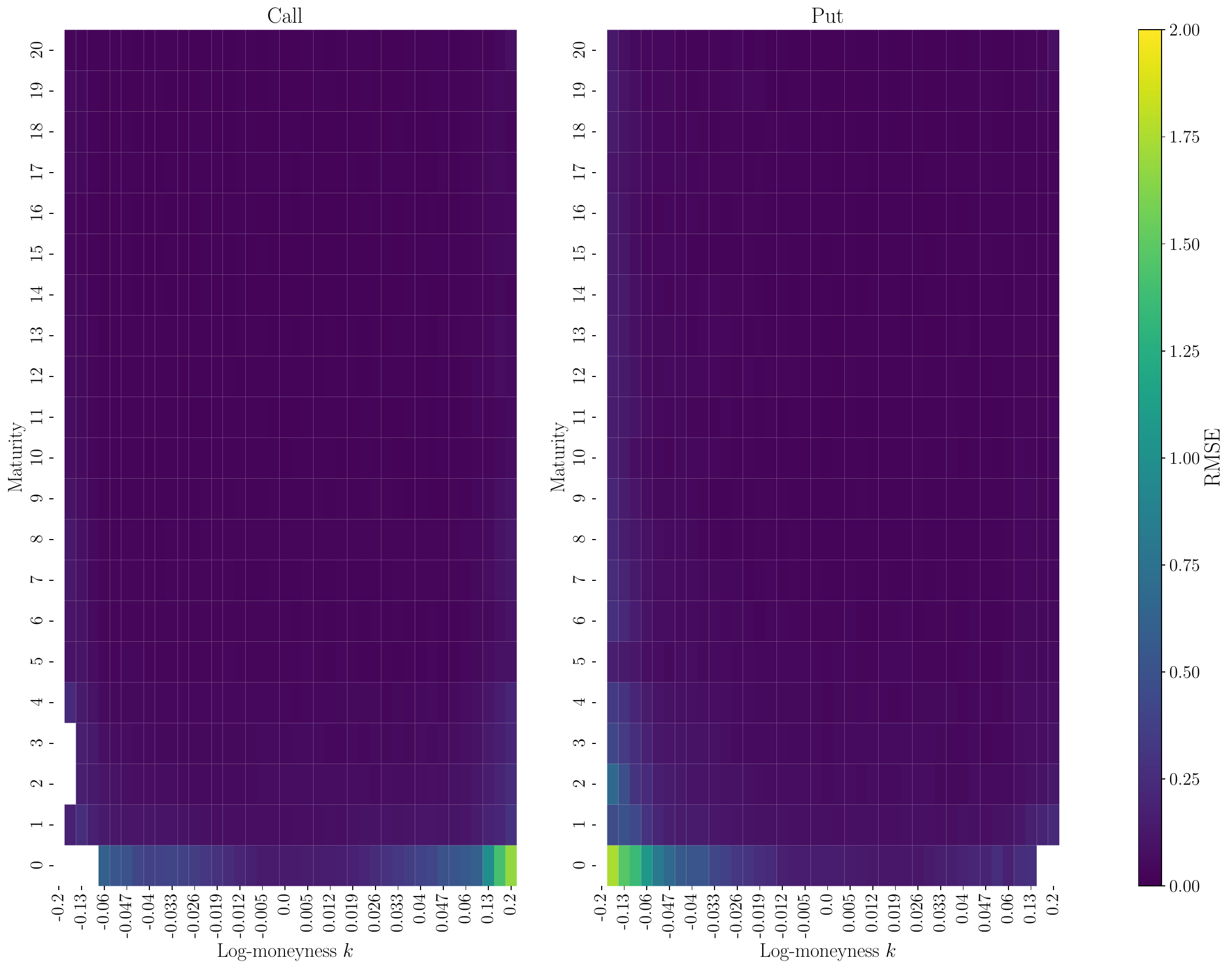}
    \caption{Out-of-sample RMSE comparison between naive Random Walk predictions and convLSTM predictions for SVI model, prediction horizon 2.}
\end{figure}

\begin{figure}[H]
    \centering
    \includegraphics[width=\linewidth]{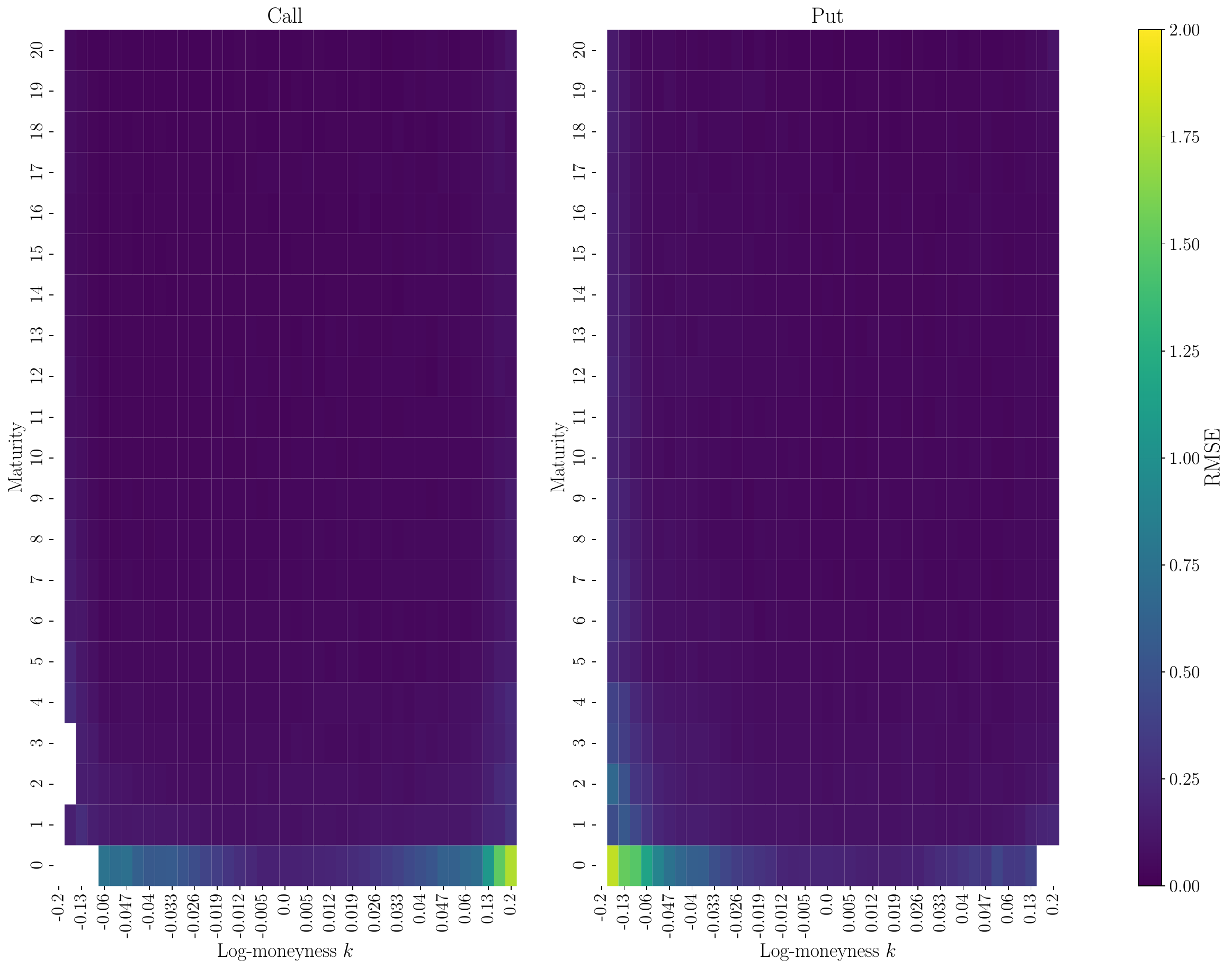}
    \caption{Out-of-sample RMSE comparison between naive Random Walk predictions and convLSTM predictions for SVI model, prediction horizon 5.}
\end{figure}

\begin{figure}[H]
    \centering
    \includegraphics[width=\linewidth]{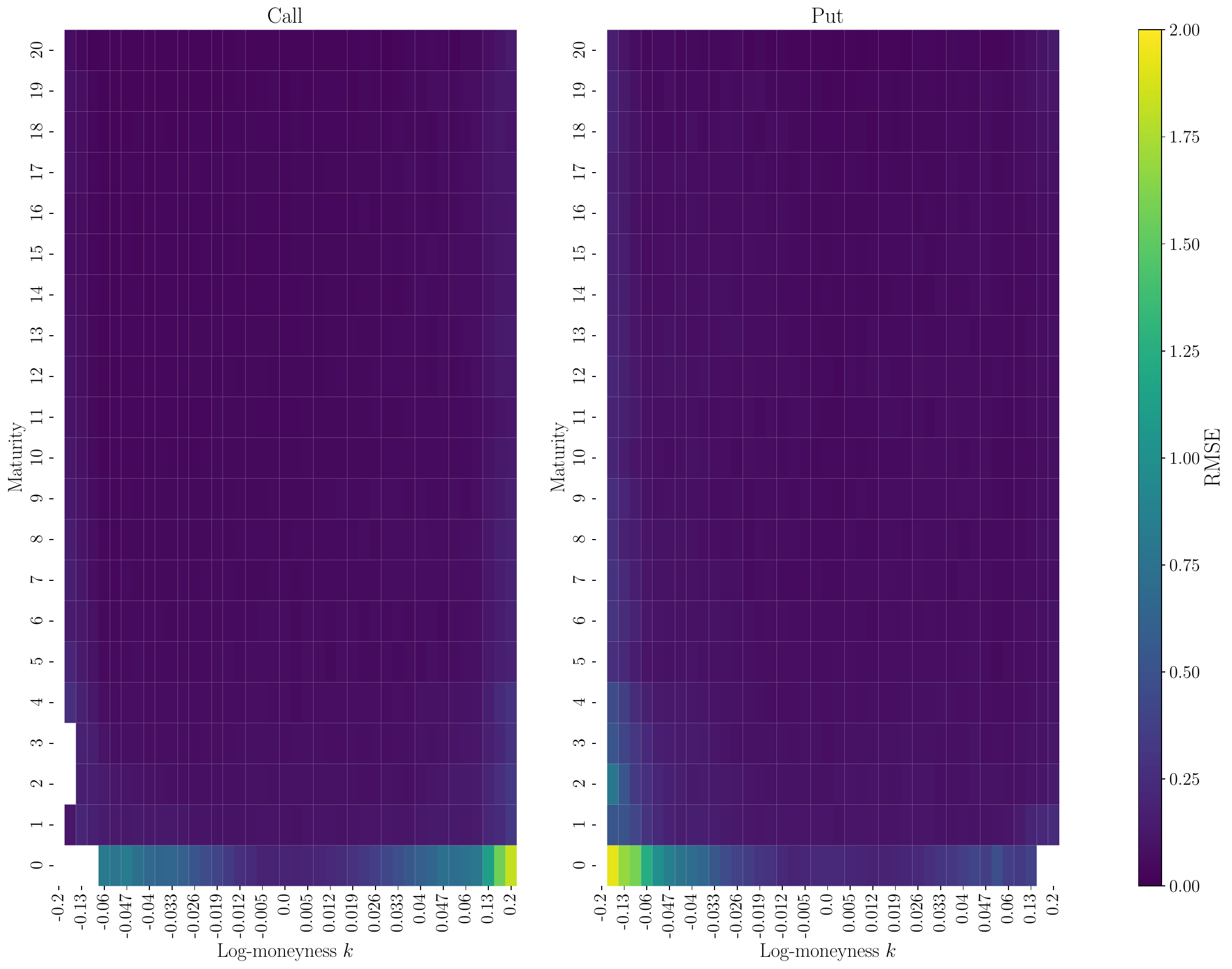}
    \caption{Out-of-sample RMSE comparison between naive Random Walk predictions and convLSTM predictions for SVI model, prediction horizon 10.}
\end{figure}


\begin{figure}[H]
    \centering
    \includegraphics[width=\linewidth]{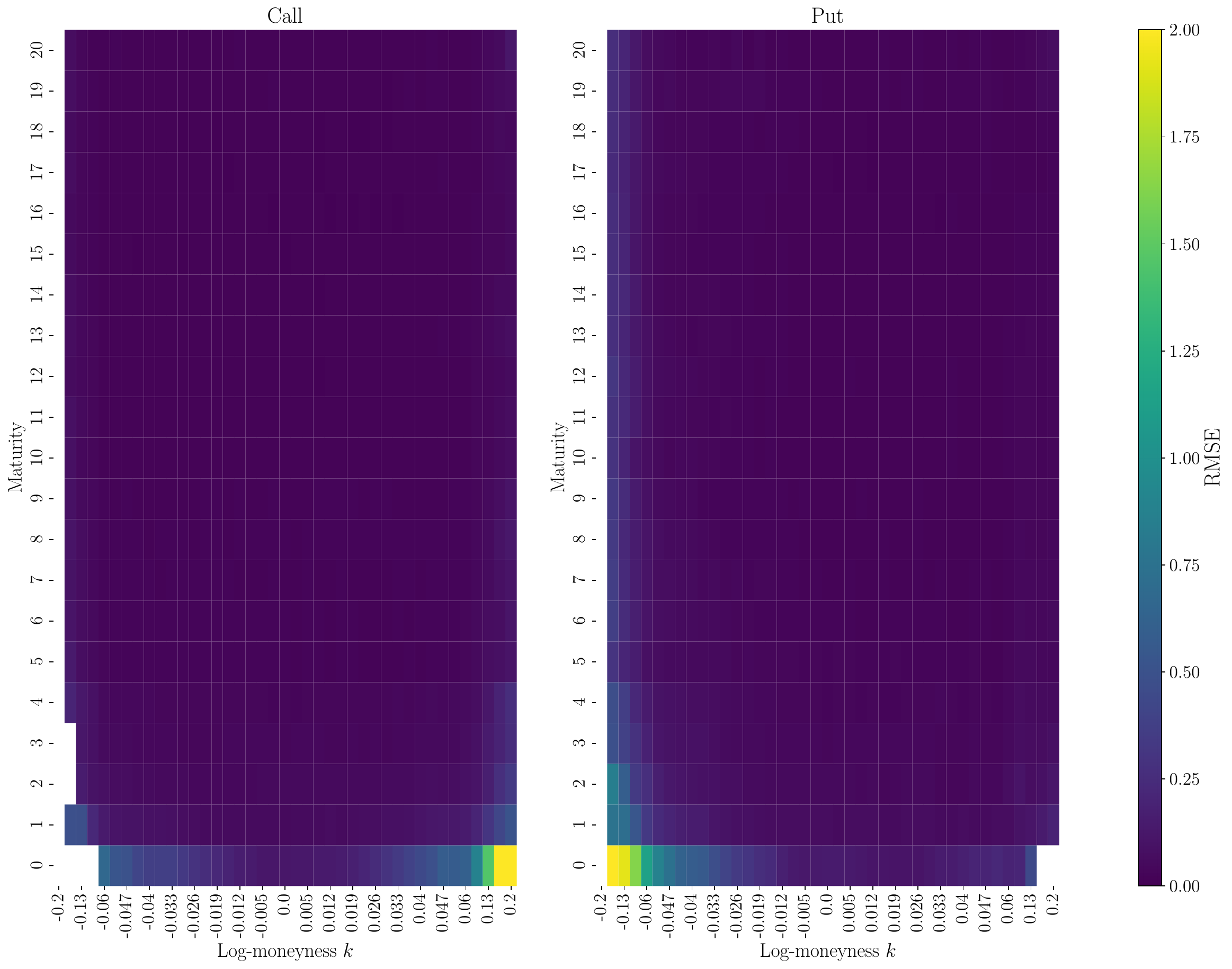}
    \caption{Out-of-sample RMSE comparison between naive Random Walk predictions and convLSTM predictions for linear interpolation, prediction horizon 1.}
\end{figure}

\begin{figure}[H]
    \centering
    \includegraphics[width=\linewidth]{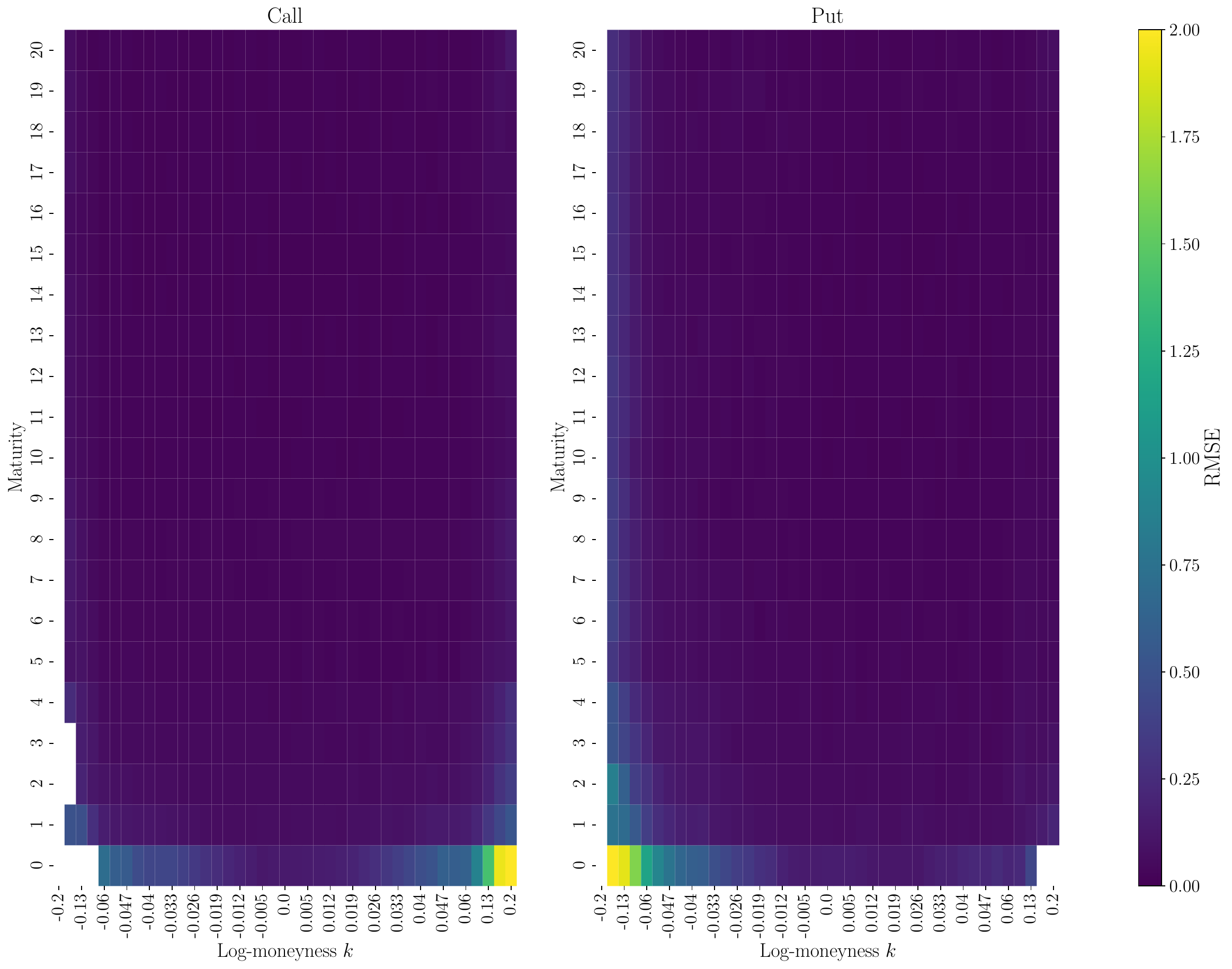}
    \caption{Out-of-sample RMSE comparison between naive Random Walk predictions and convLSTM predictions for linear interpolation, prediction horizon 2.}
\end{figure}

\begin{figure}[H]
    \centering
    \includegraphics[width=\linewidth]{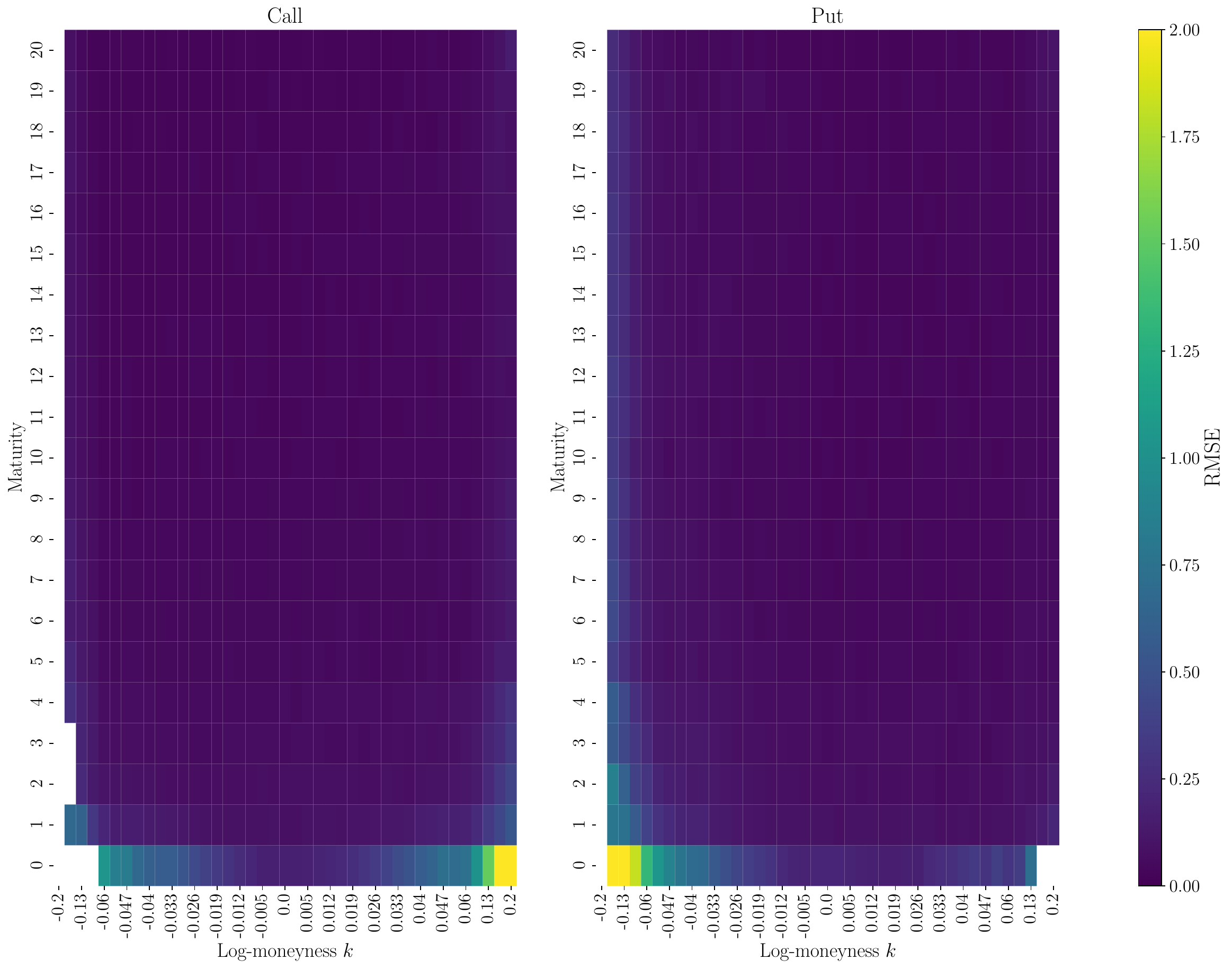}
    \caption{Out-of-sample RMSE comparison between naive Random Walk predictions and convLSTM predictions for linear interpolation, prediction horizon 5.}
\end{figure}

\begin{figure}[H]
    \centering
    \includegraphics[width=\linewidth]{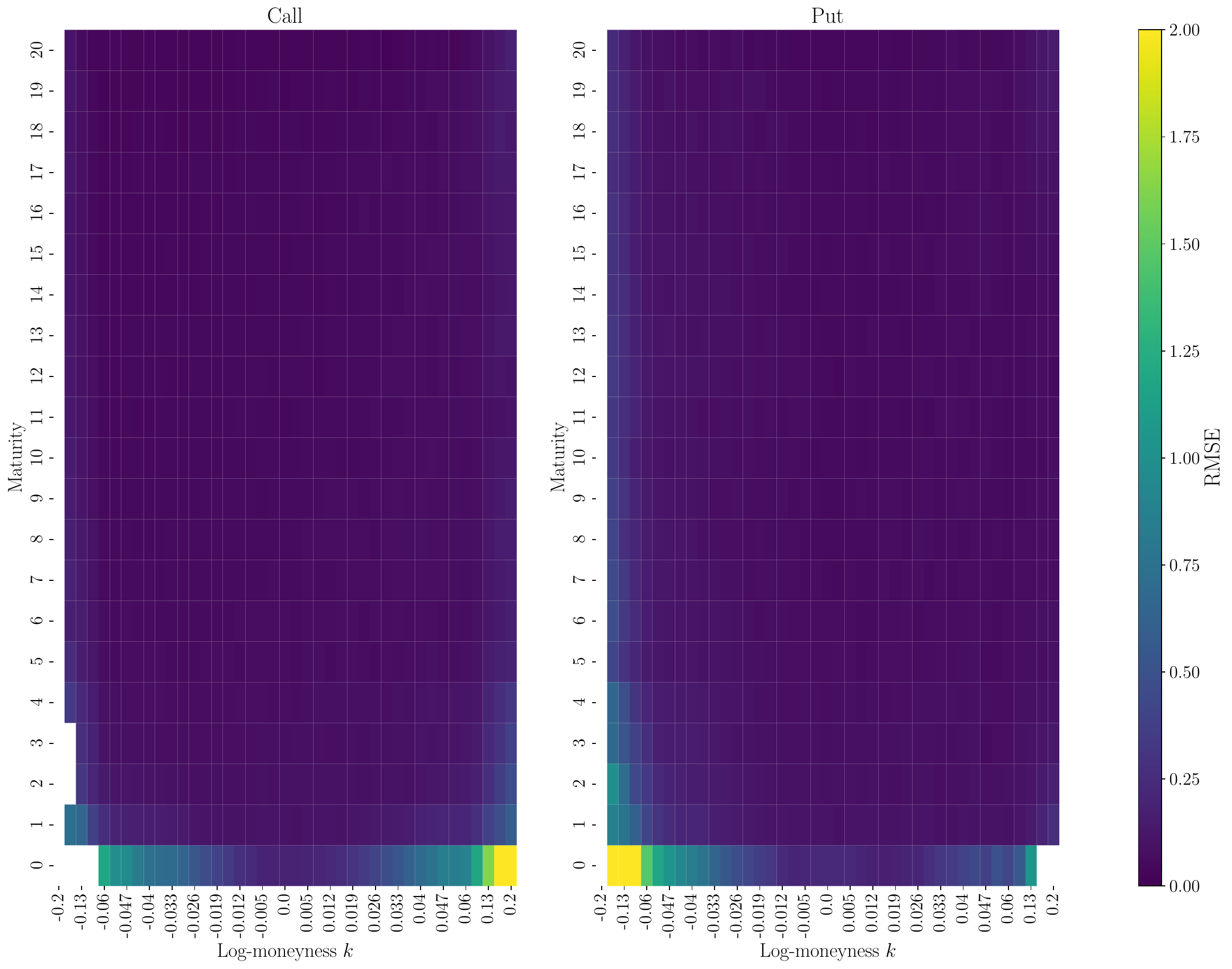}
    \caption{Out-of-sample RMSE comparison between naive Random Walk predictions and convLSTM predictions for linear interpolation, prediction horizon 10.}
\end{figure}


\begin{figure}[H]
    \centering
    \includegraphics[width=\linewidth]{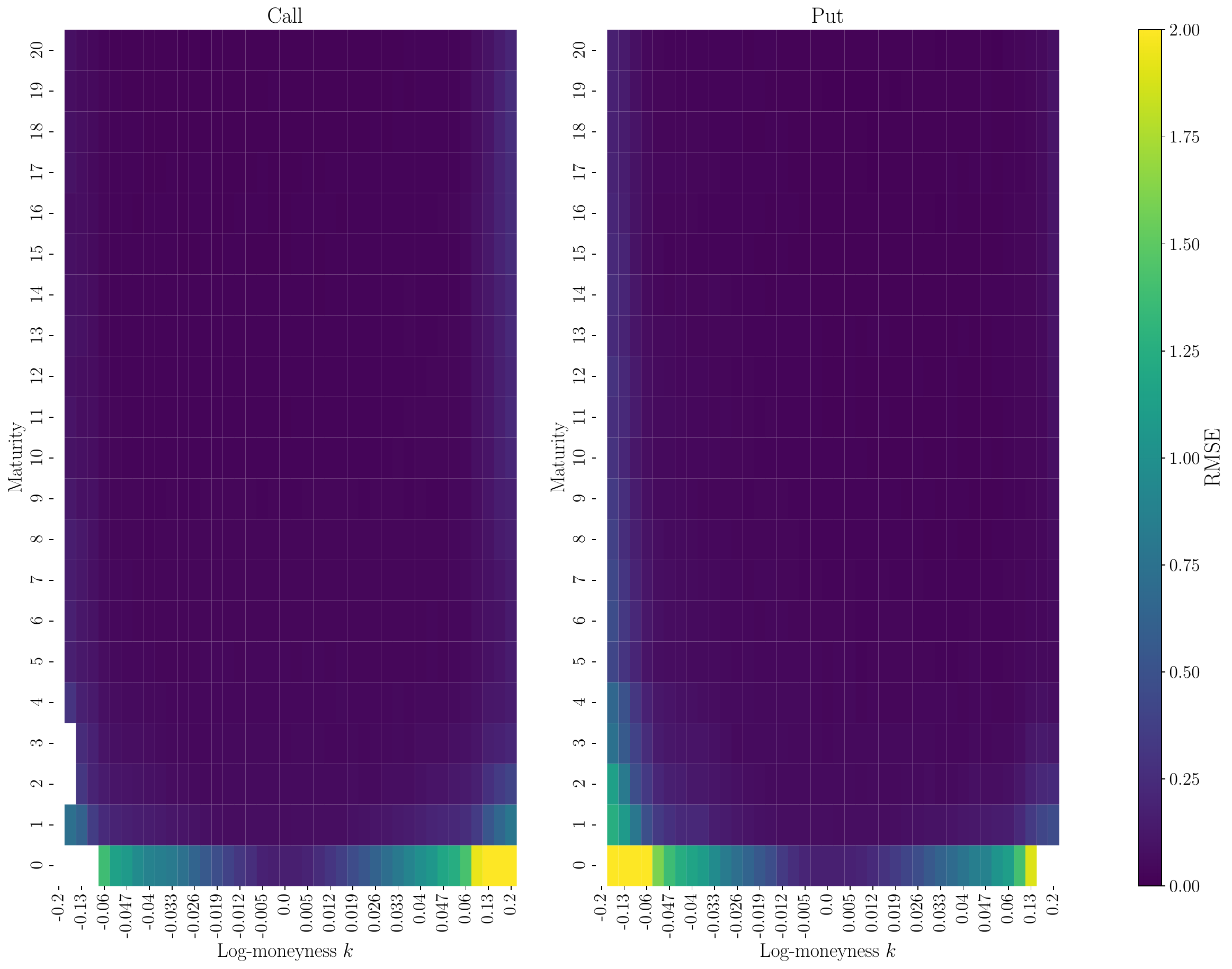}
    \caption{Out-of-sample RMSE comparison between naive Random Walk predictions and convLSTM predictions for AHBS model, prediction horizon 1.}
\end{figure}

\begin{figure}[H]
    \centering
    \includegraphics[width=\linewidth]{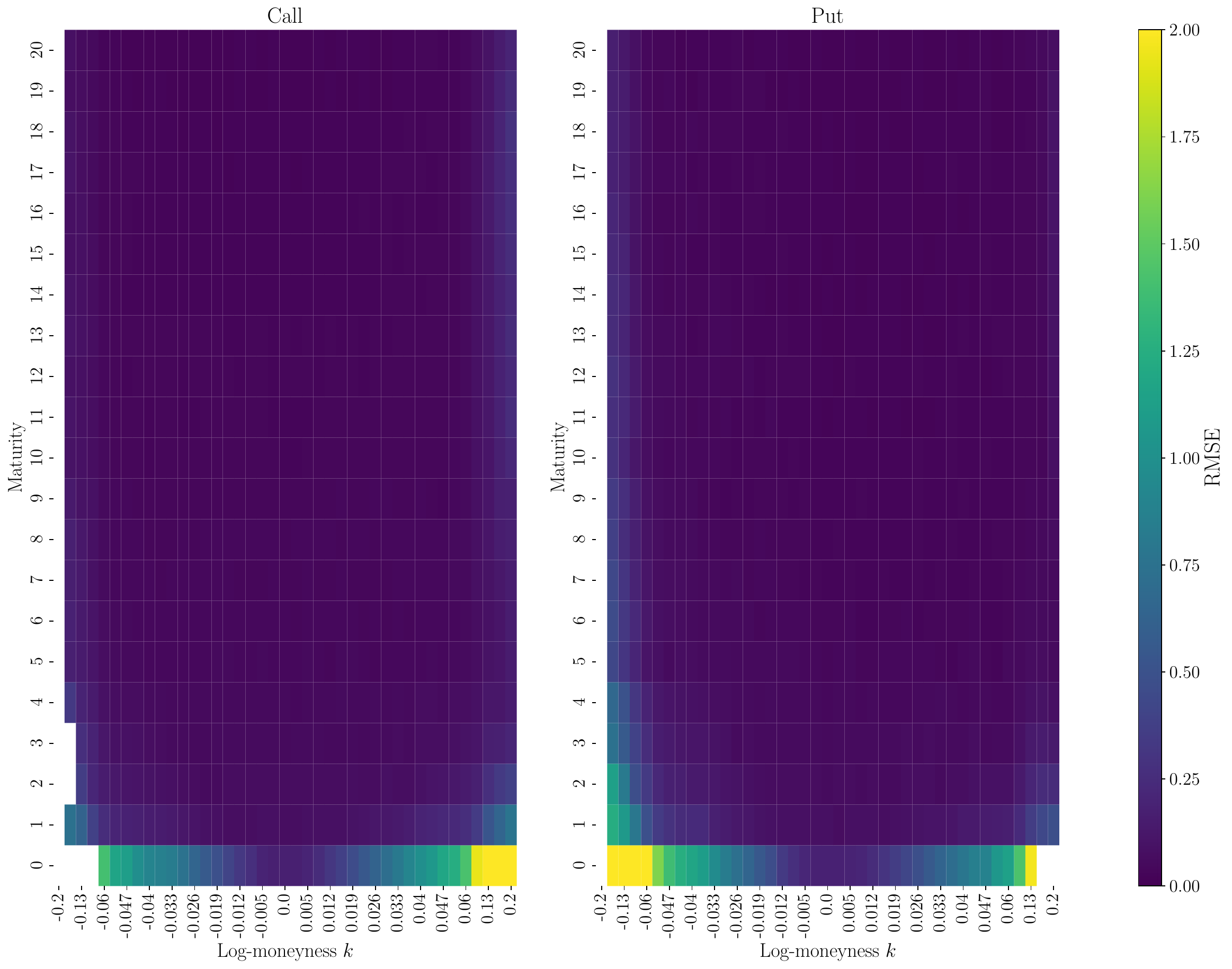}
    \caption{Out-of-sample RMSE comparison between naive Random Walk predictions and convLSTM predictions for AHBS model, prediction horizon 2.}
\end{figure}

\begin{figure}[H]
    \centering
    \includegraphics[width=\linewidth]{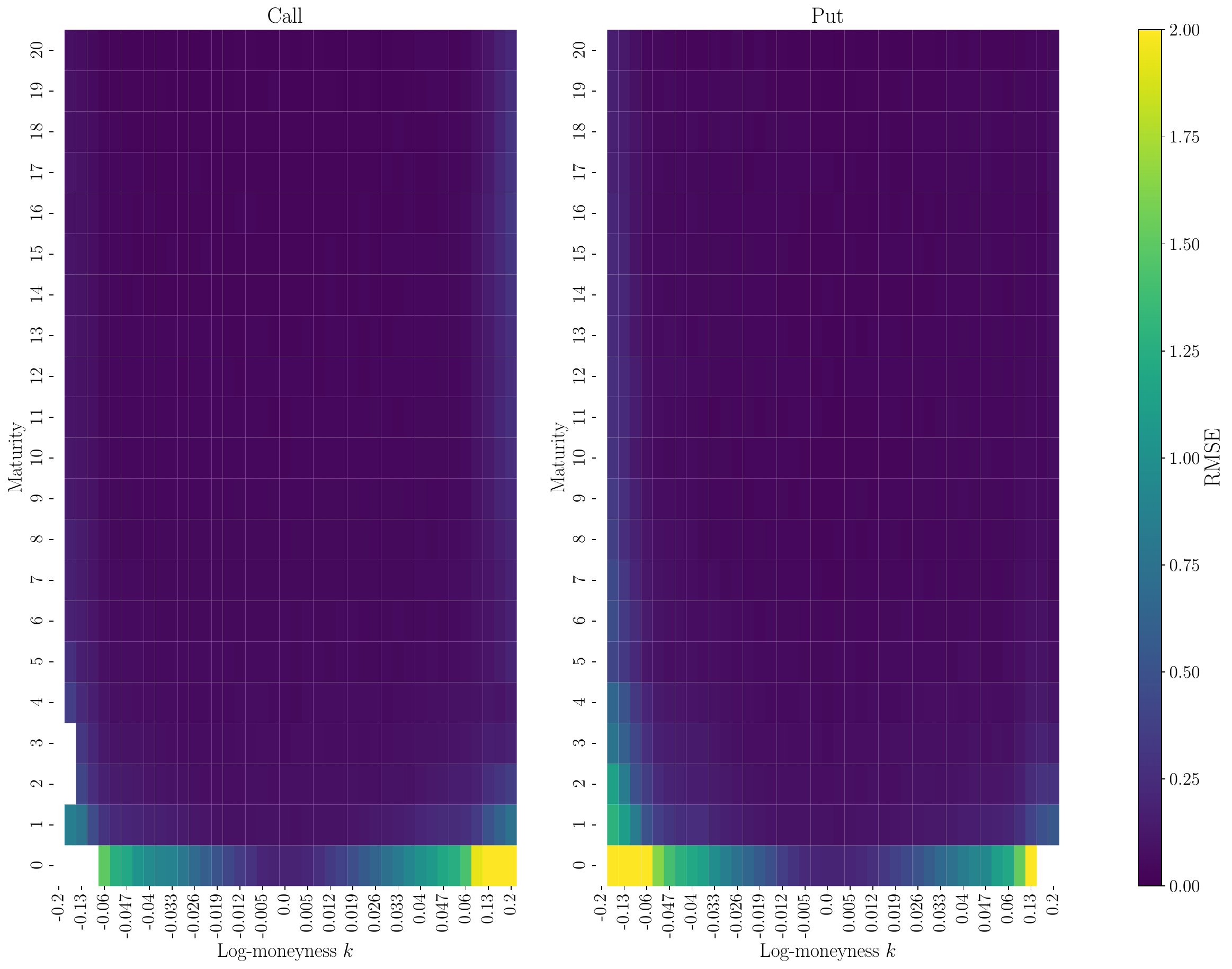}
    \caption{Out-of-sample RMSE comparison between naive Random Walk predictions and convLSTM predictions for AHBS model, prediction horizon 5.}
\end{figure}

\begin{figure}[H]
    \centering
    \includegraphics[width=\linewidth]{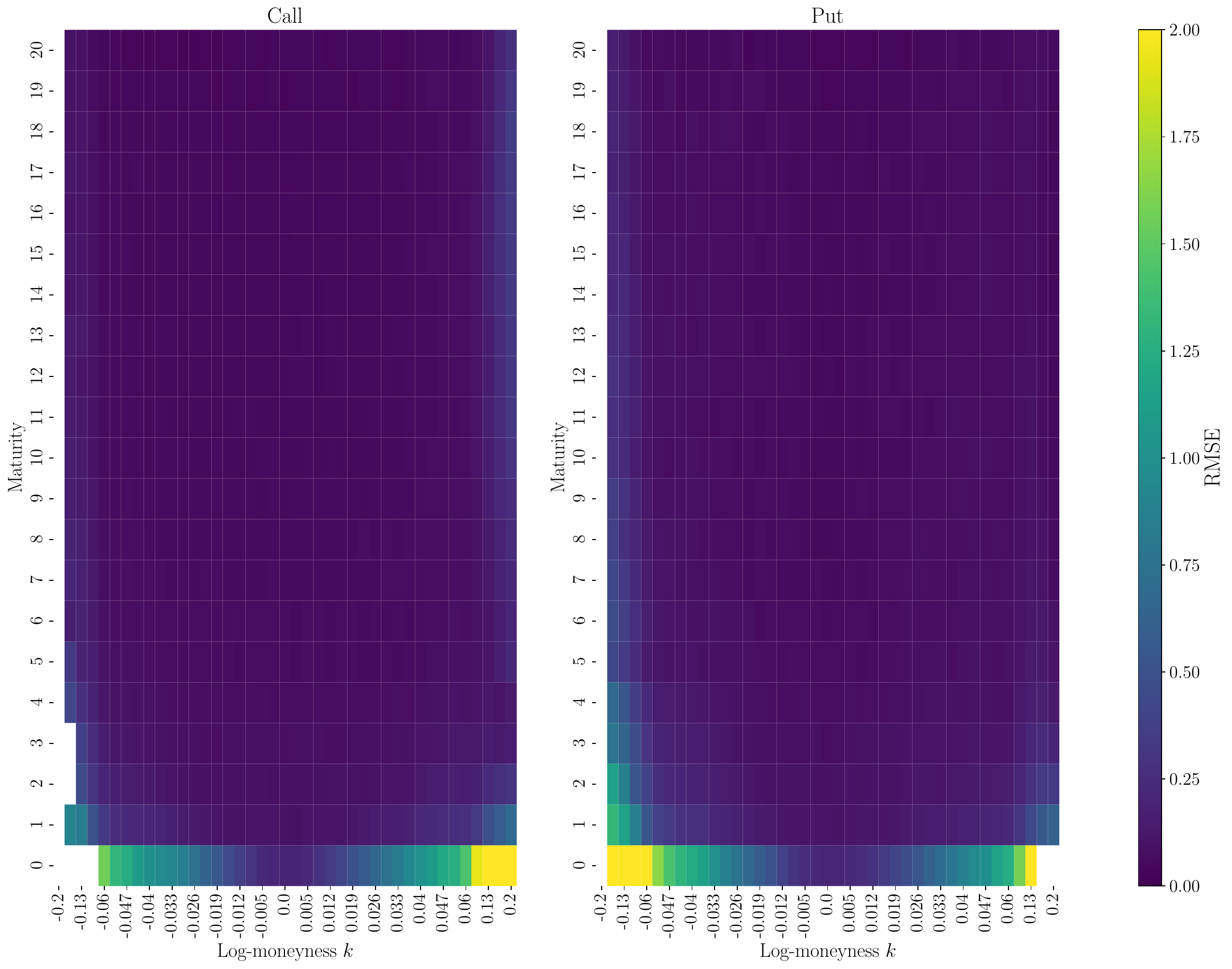}
    \caption{Out-of-sample RMSE comparison between naive Random Walk predictions and convLSTM predictions for AHBS model, prediction horizon 10.}
\end{figure}

\subsection{Surface errors - MPE}\label{sec:models_diag_errors_MPE}

These graphs display the fit of models without the information regarding the announcement per maturity and moneyness in terms of MPE.


\begin{figure}[H]
    \centering
    \includegraphics[width=\linewidth]{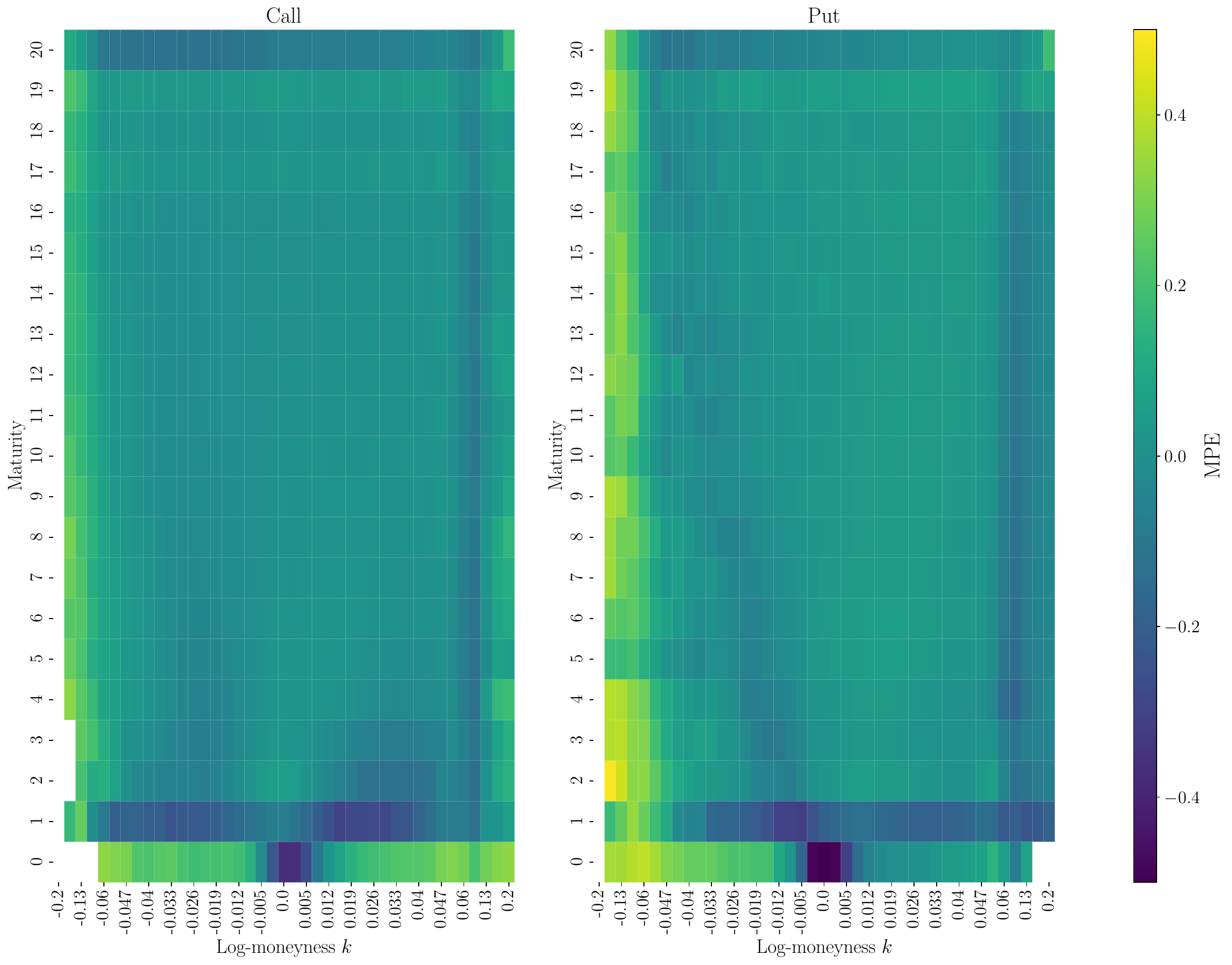}
    \caption{Out-of-sample MPE comparison between naive Random Walk predictions and convLSTM predictions for SSVI model, prediction horizon 1.}
\end{figure}

\begin{figure}[H]
    \centering
    \includegraphics[width=\linewidth]{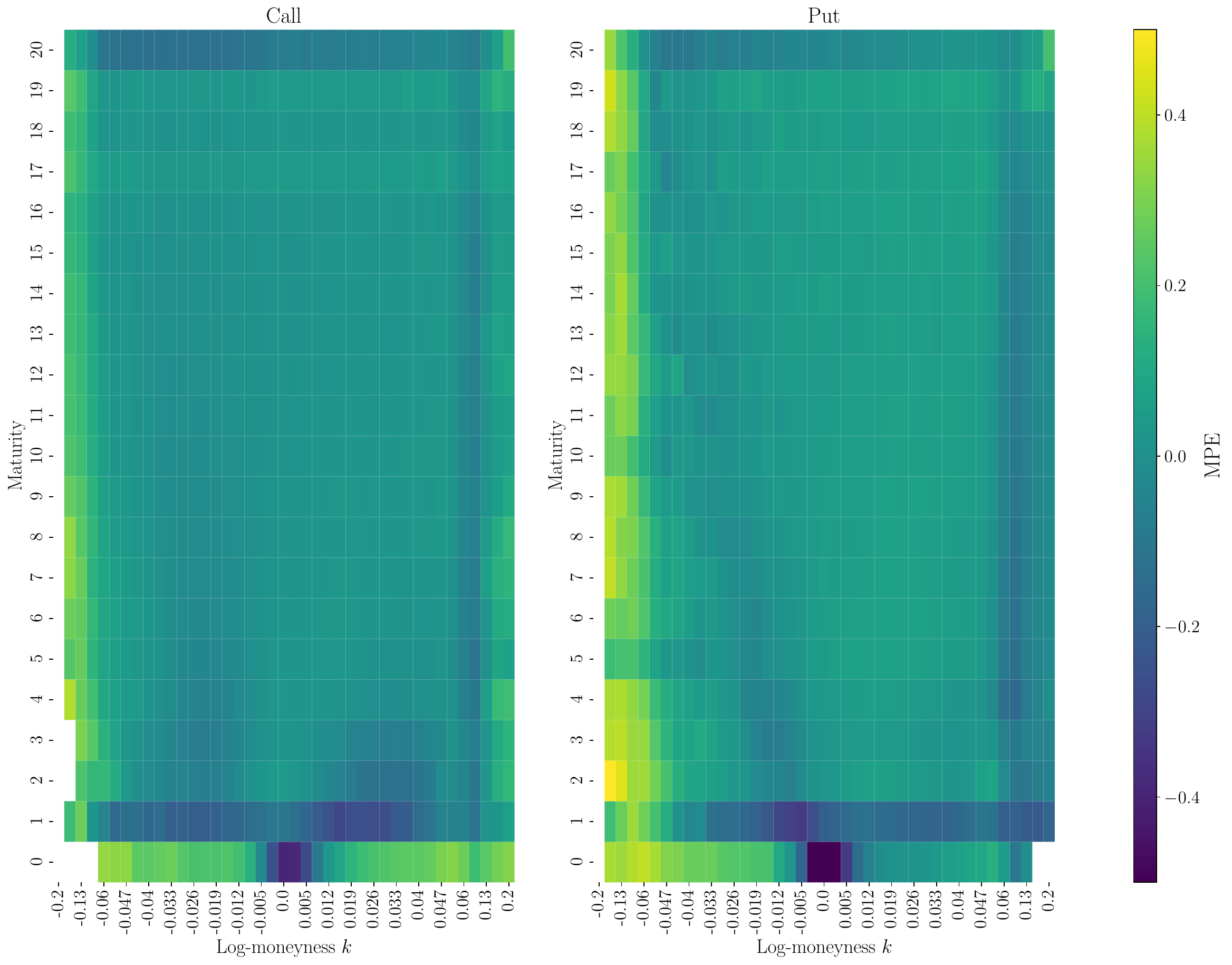}
    \caption{Out-of-sample MPE comparison between naive Random Walk predictions and convLSTM predictions for SVI model, prediction horizon 2.}
\end{figure}

\begin{figure}[H]
    \centering
    \includegraphics[width=\linewidth]{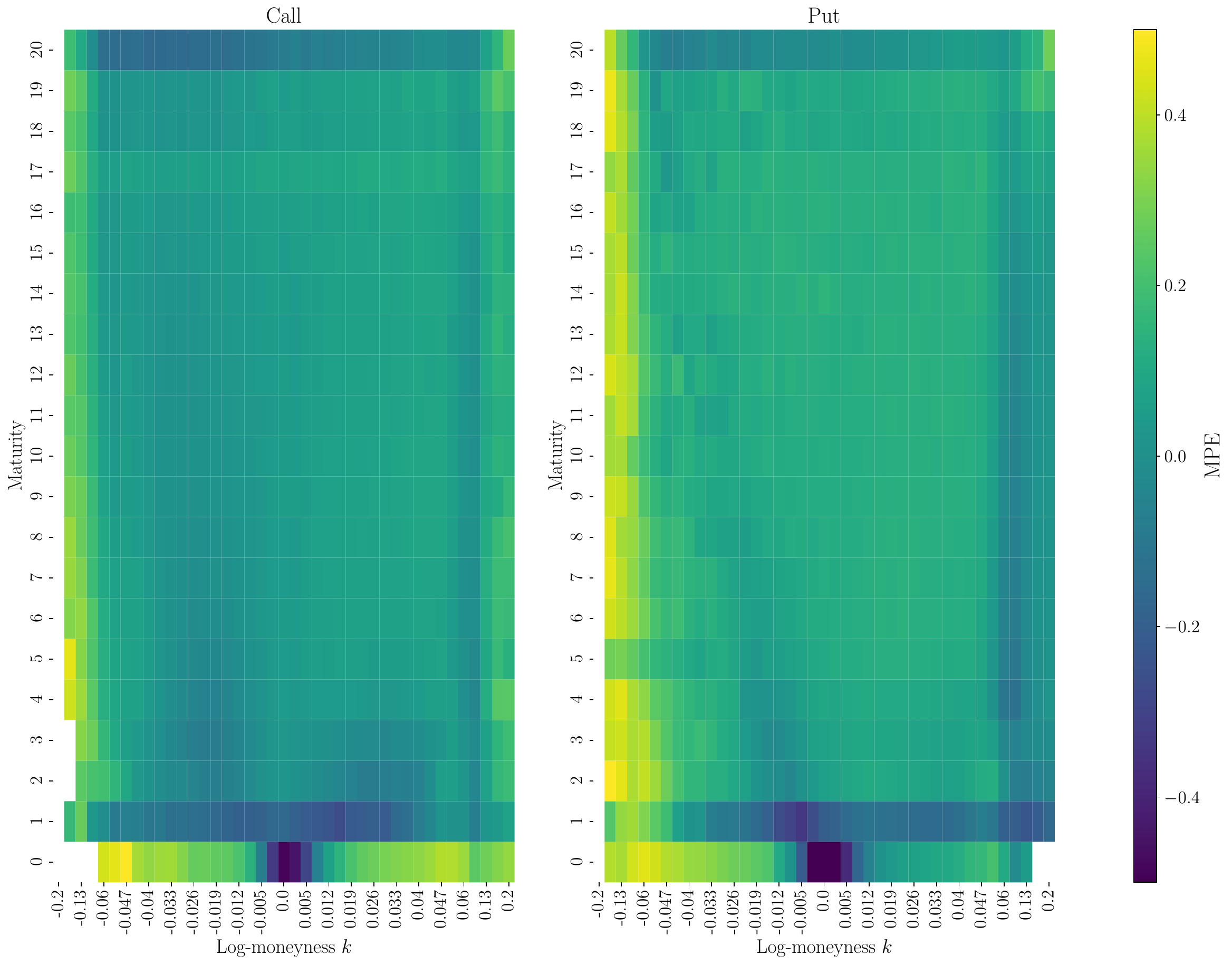}
    \caption{Out-of-sample MPE comparison between naive Random Walk predictions and convLSTM predictions for SVI model, prediction horizon 5.}
\end{figure}

\begin{figure}[H]
    \centering
    \includegraphics[width=\linewidth]{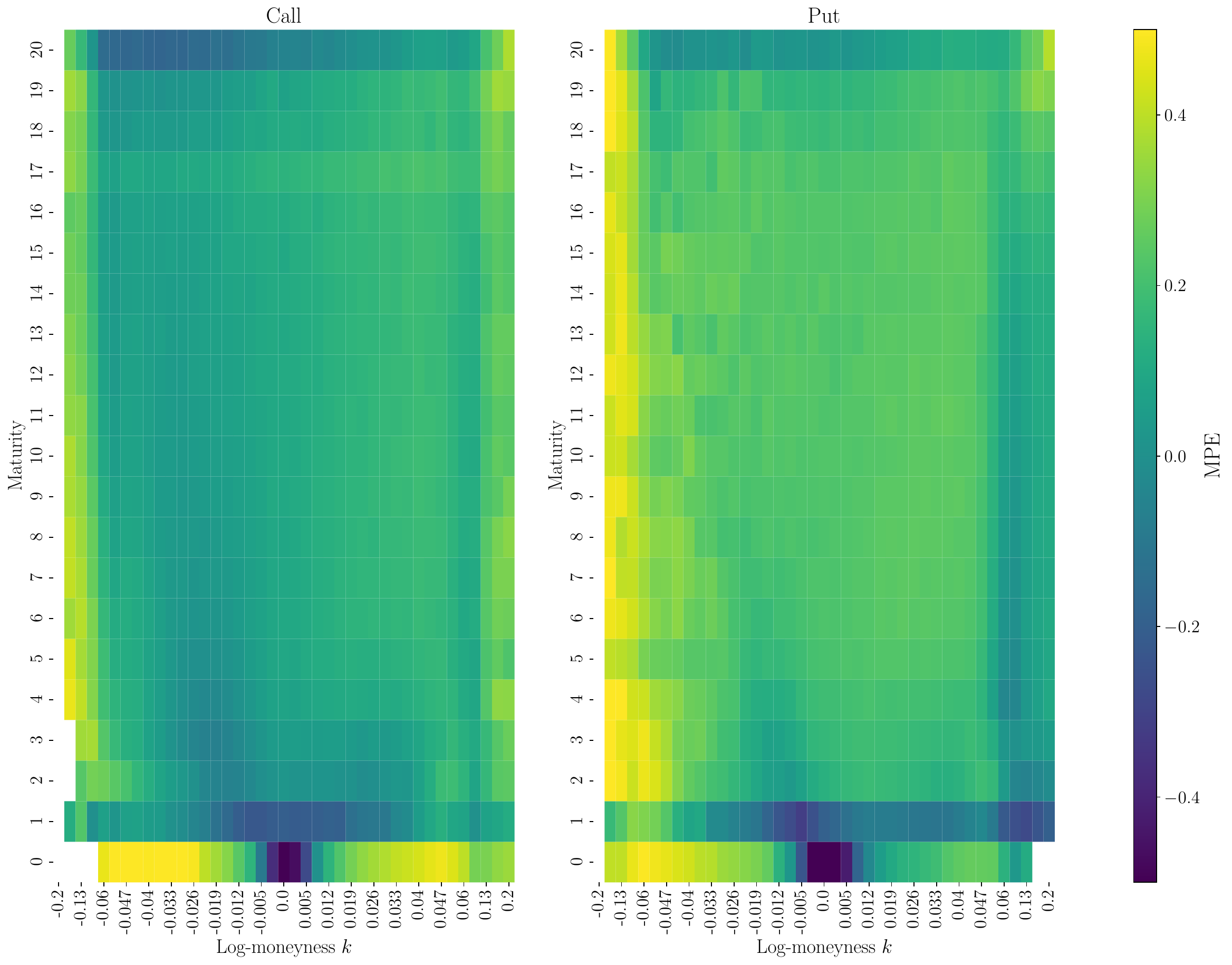}
    \caption{Out-of-sample MPE comparison between naive Random Walk predictions and convLSTM predictions for SVI model, prediction horizon 10.}
\end{figure}


\begin{figure}[H]
    \centering
    \includegraphics[width=\linewidth]{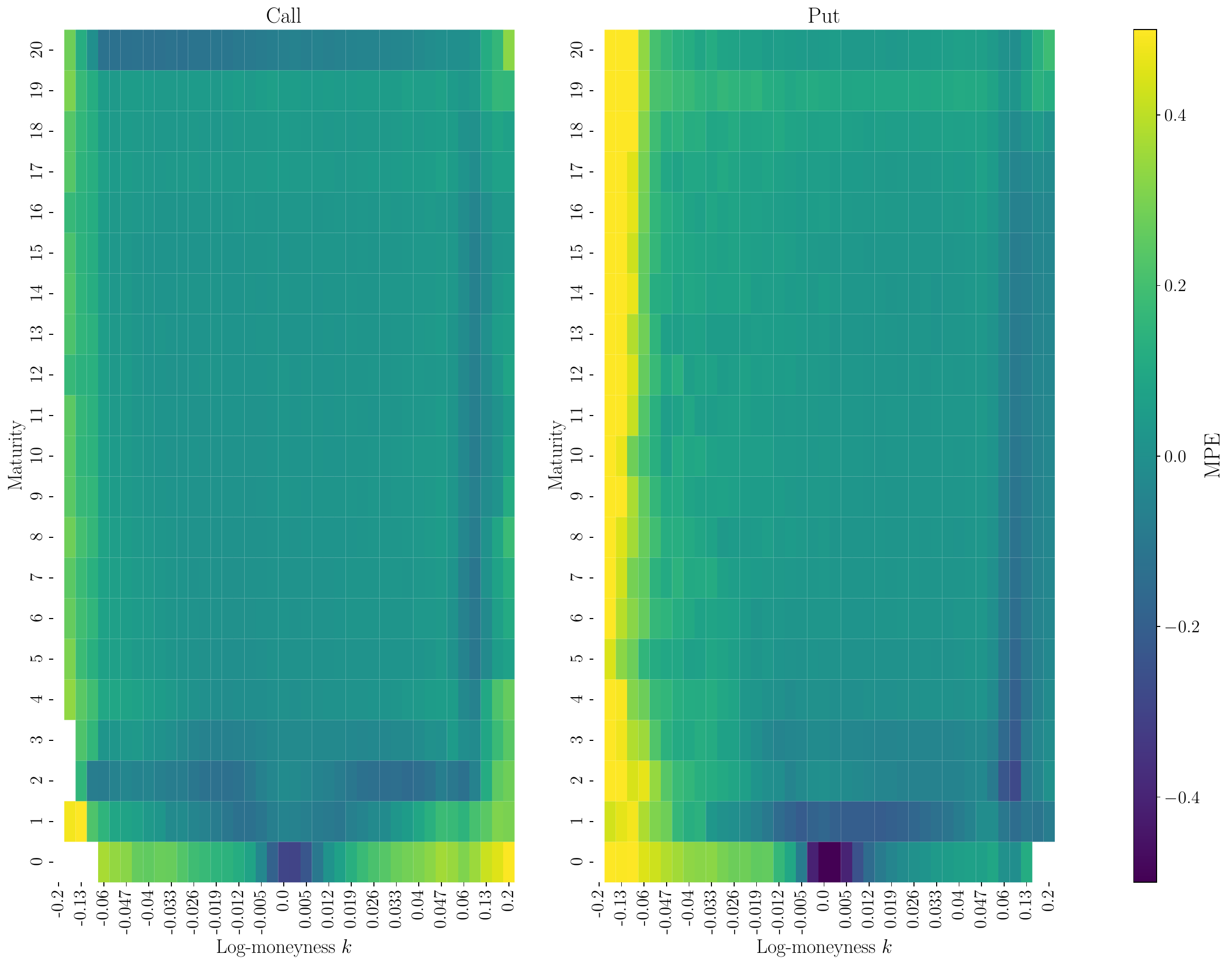}
    \caption{Out-of-sample MPE comparison between naive Random Walk predictions and convLSTM predictions for linear interpolation, prediction horizon 1.}
\end{figure}

\begin{figure}[H]
    \centering
    \includegraphics[width=\linewidth]{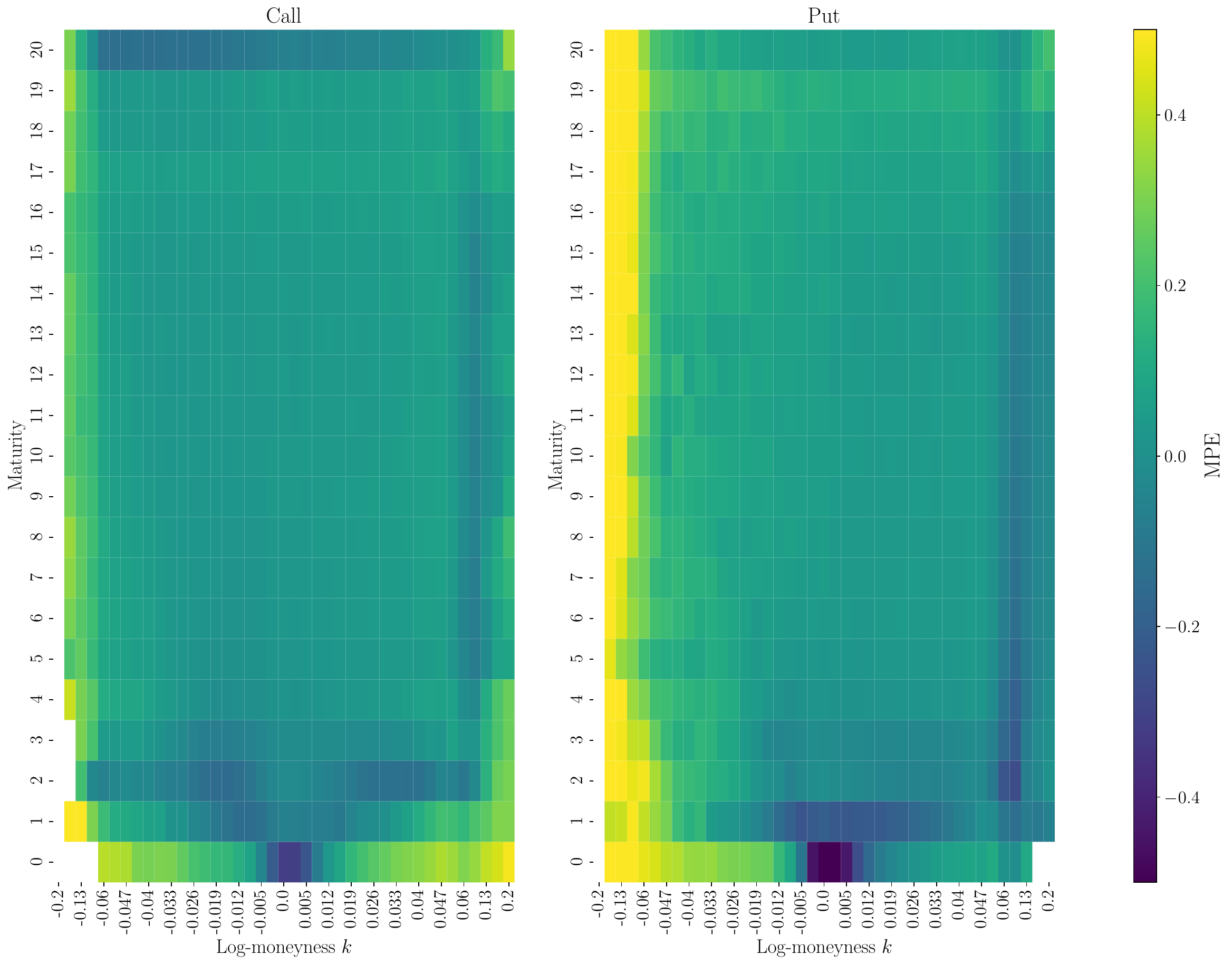}
    \caption{Out-of-sample MPE comparison between naive Random Walk predictions and convLSTM predictions for linear interpolation, prediction horizon 2.}
\end{figure}

\begin{figure}[H]
    \centering
    \includegraphics[width=\linewidth]{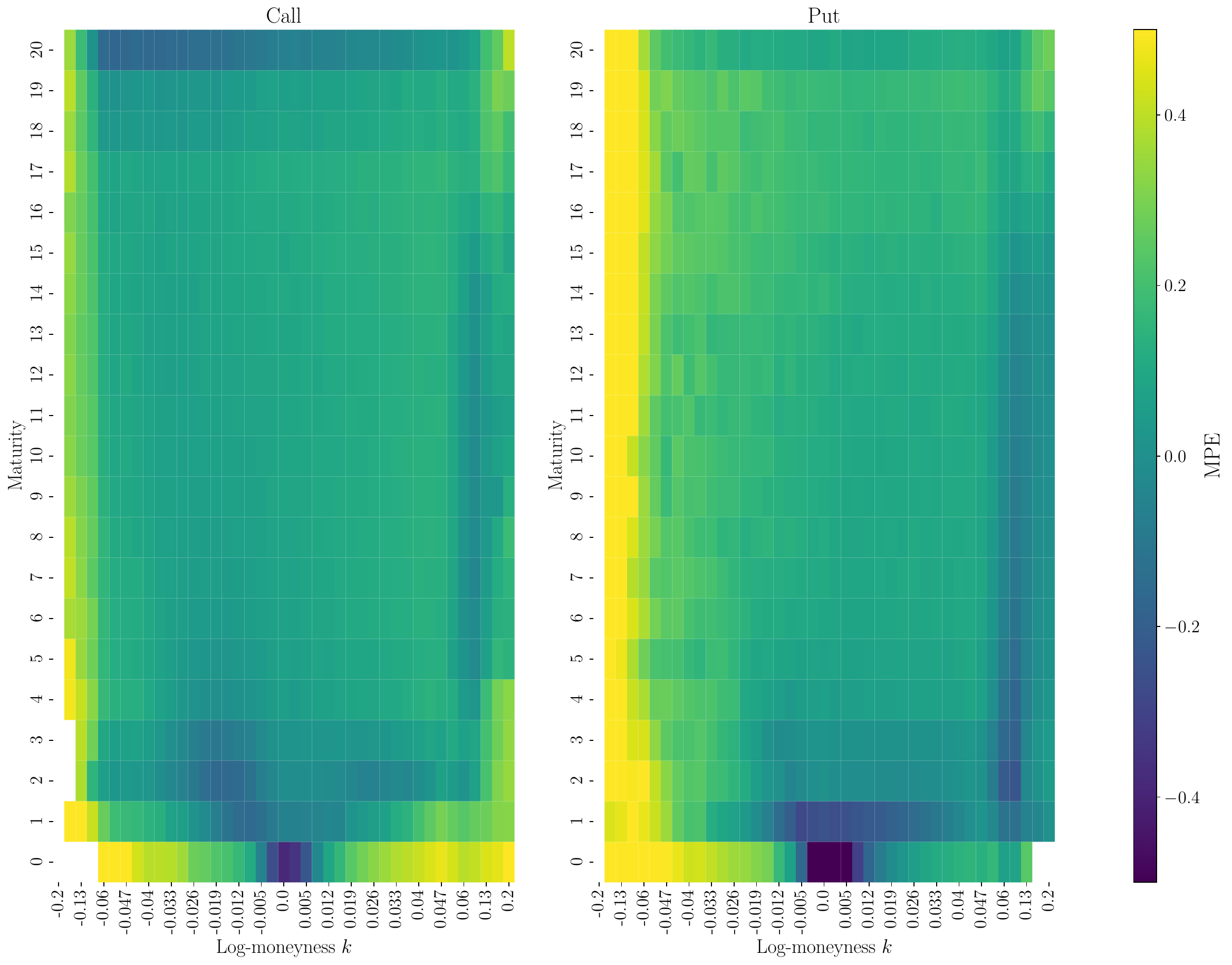}
    \caption{Out-of-sample MPE comparison between naive Random Walk predictions and convLSTM predictions for linear interpolation, prediction horizon 5.}
\end{figure}

\begin{figure}[H]
    \centering
    \includegraphics[width=\linewidth]{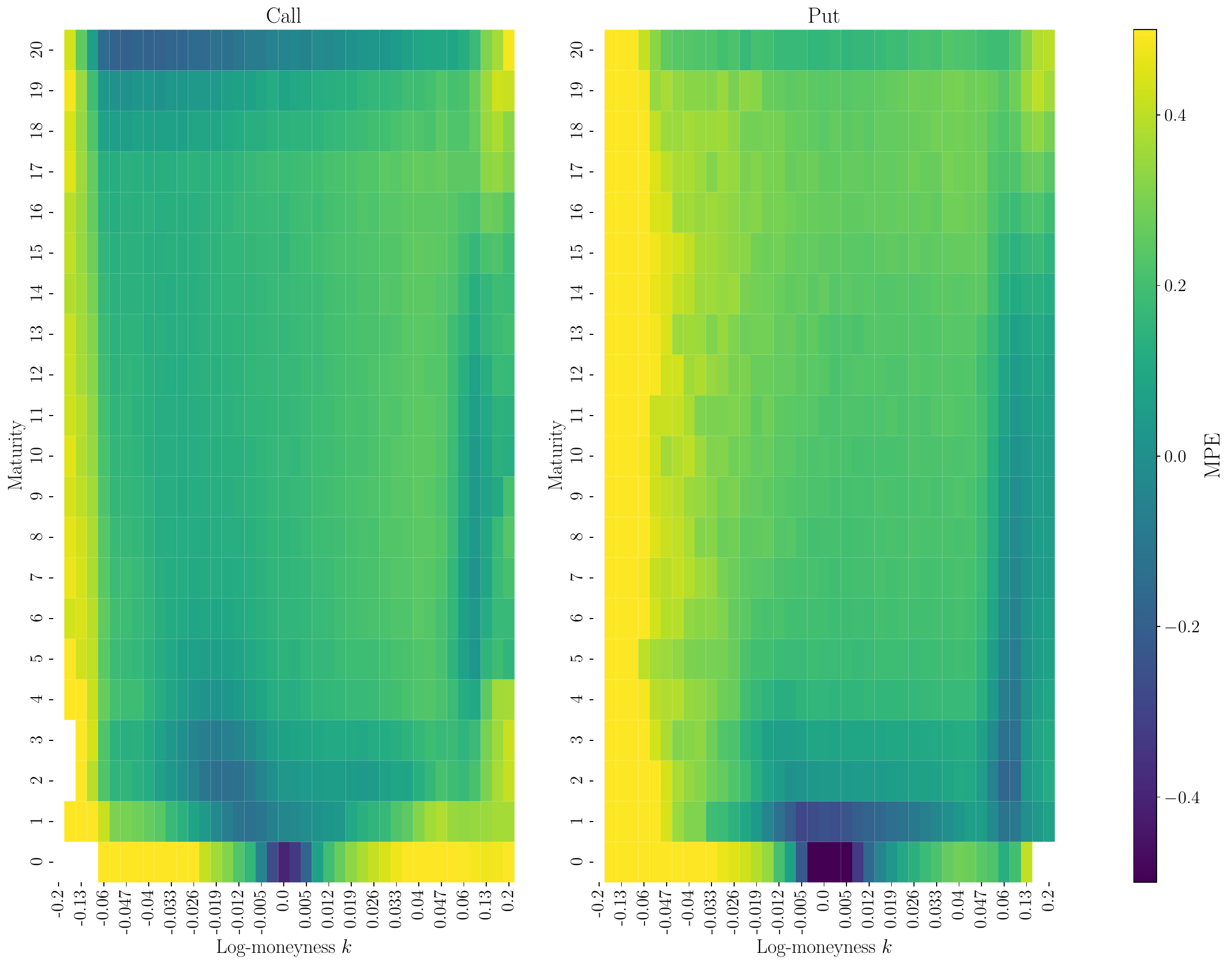}
    \caption{Out-of-sample MPE comparison between naive Random Walk predictions and convLSTM predictions for linear interpolation, prediction horizon 10.}
\end{figure}


\begin{figure}[H]
    \centering
    \includegraphics[width=\linewidth]{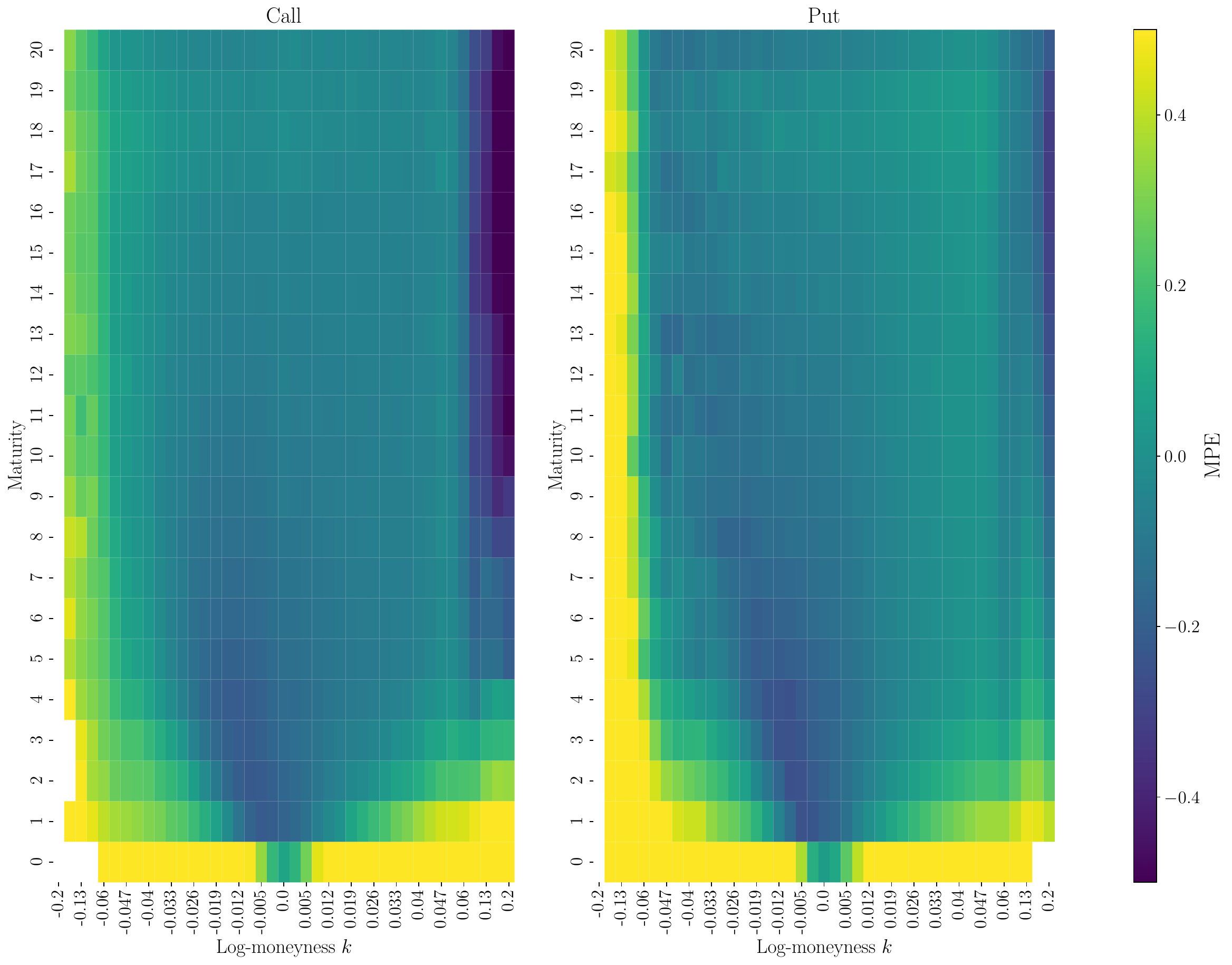}
    \caption{Out-of-sample MPE comparison between naive Random Walk predictions and convLSTM predictions for AHBS model, prediction horizon 1.}
\end{figure}

\begin{figure}[H]
    \centering
    \includegraphics[width=\linewidth]{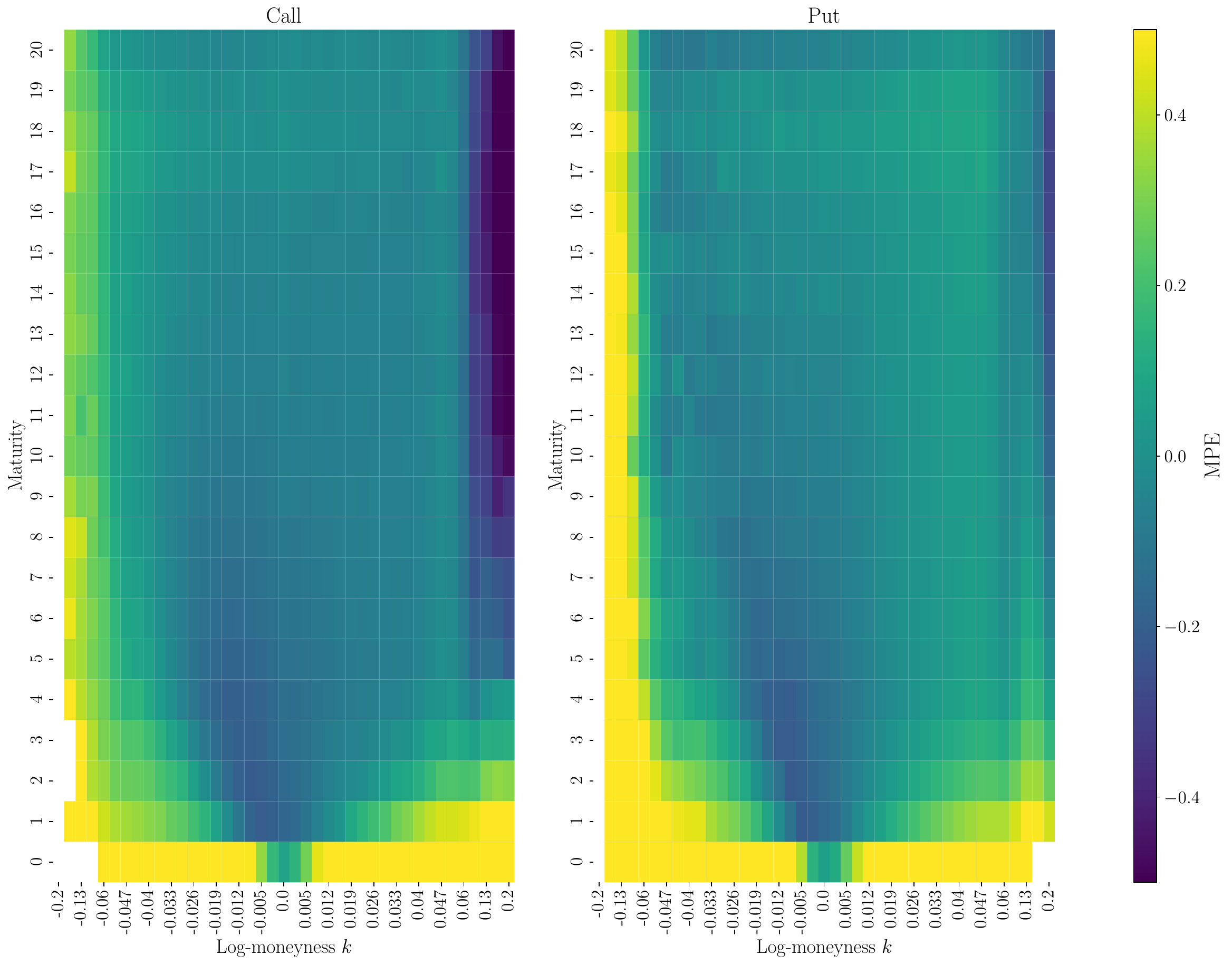}
    \caption{Out-of-sample MPE comparison between naive Random Walk predictions and convLSTM predictions for AHBS model, prediction horizon 2.}
\end{figure}

\begin{figure}[H]
    \centering
    \includegraphics[width=\linewidth]{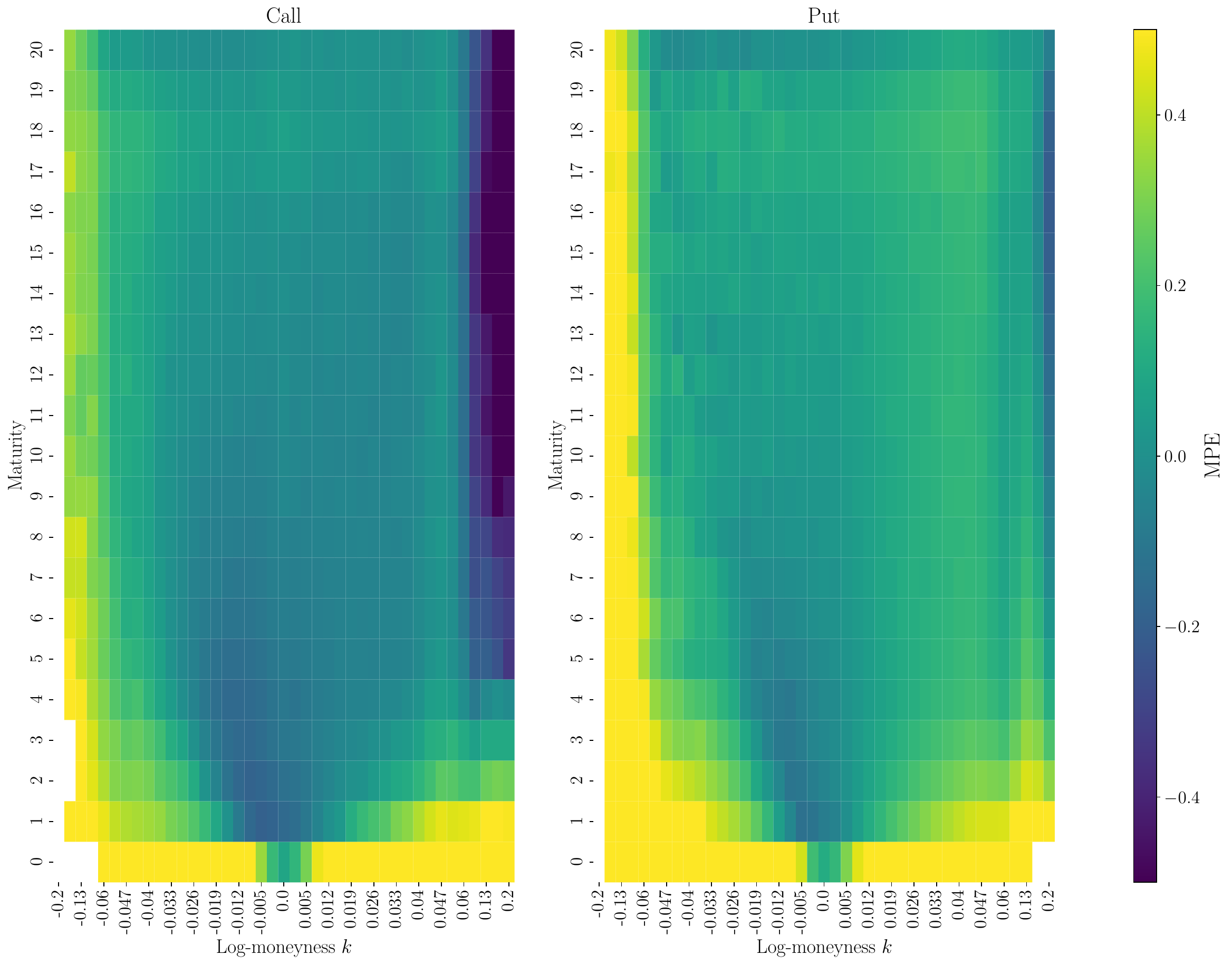}
    \caption{Out-of-sample MPE comparison between naive Random Walk predictions and convLSTM predictions for AHBS model, prediction horizon 5.}
\end{figure}

\begin{figure}[H]
    \centering
    \includegraphics[width=\linewidth]{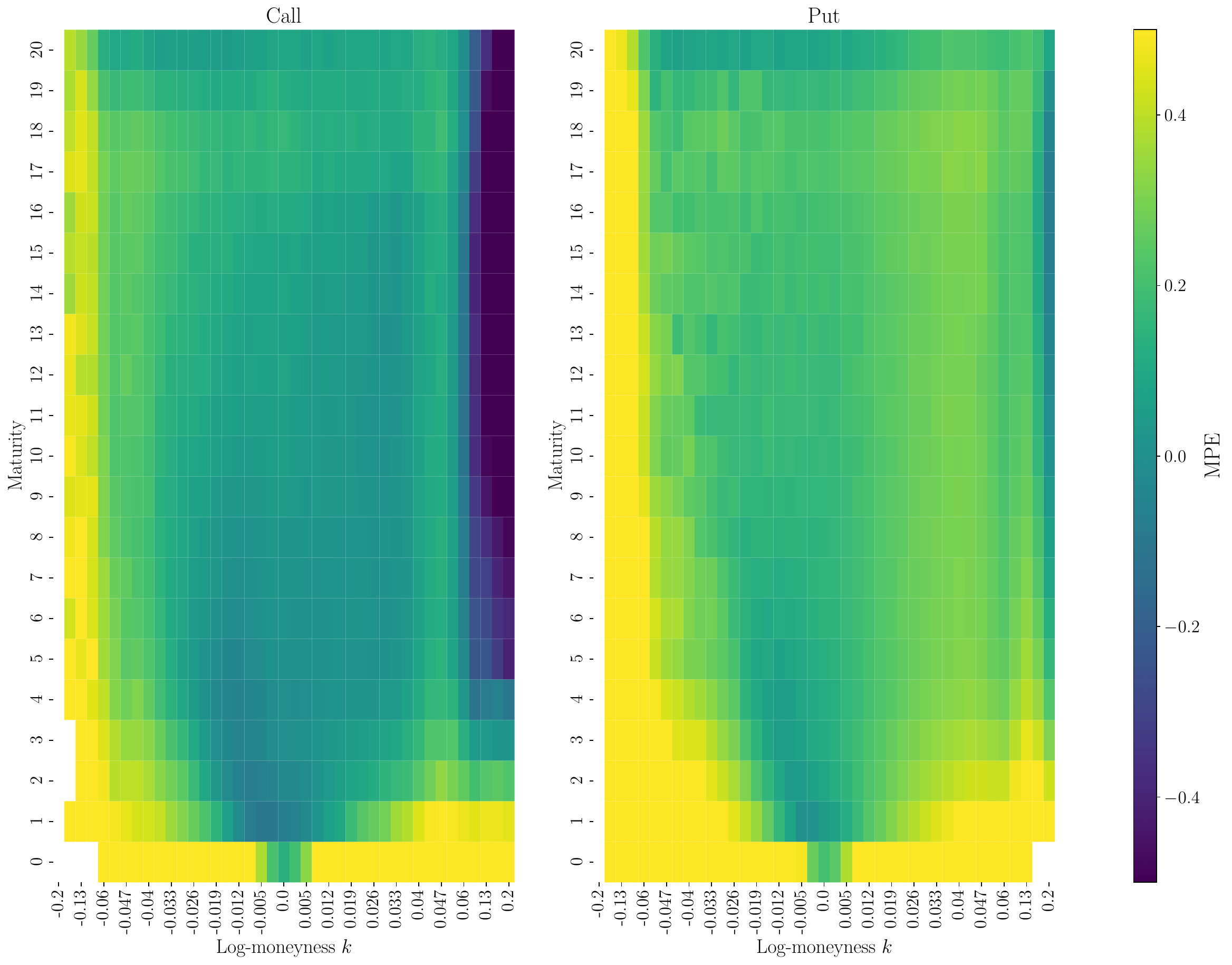}
    \caption{Out-of-sample MPE comparison between naive Random Walk predictions and convLSTM predictions for AHBS model, prediction horizon 10.}
\end{figure}


\subsection{Surface errors - DM test}\label{sec:models_diag_errors_DM}

These graphs display the fit of models without the information regarding the announcement per maturity and moneyness in terms of MPE.


\begin{figure}[H]
    \centering
    \includegraphics[width=\linewidth]{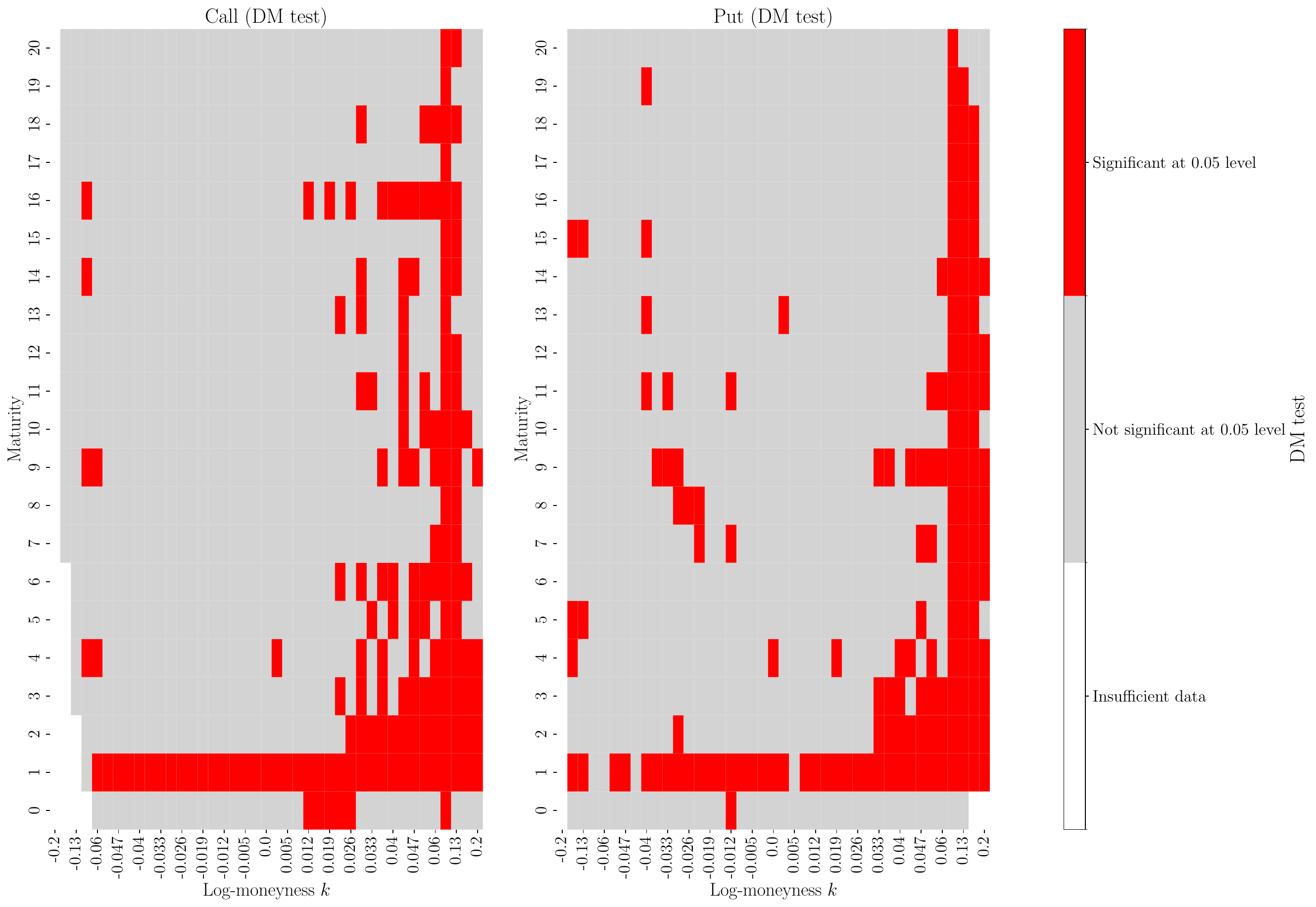}
    \caption{Out-of-sample MPE comparison between naive Random Walk predictions and convLSTM predictions for SVI model, prediction horizon 1.}
\end{figure}

\begin{figure}[H]
    \centering
    \includegraphics[width=\linewidth]{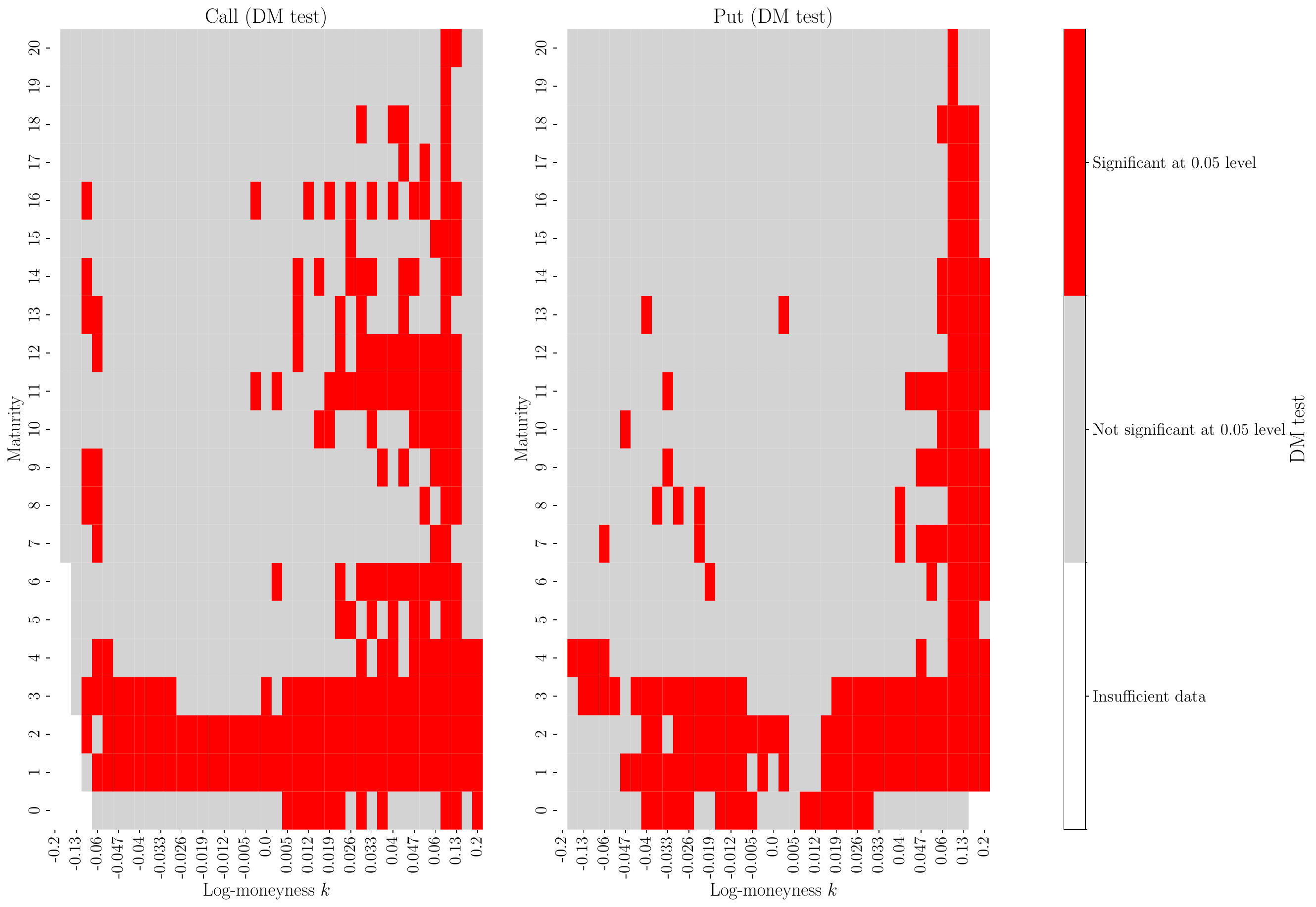}
    \caption{Out-of-sample MPE comparison between naive Random Walk predictions and convLSTM predictions for SVI model, prediction horizon 2.}
\end{figure}

\begin{figure}[H]
    \centering
    \includegraphics[width=\linewidth]{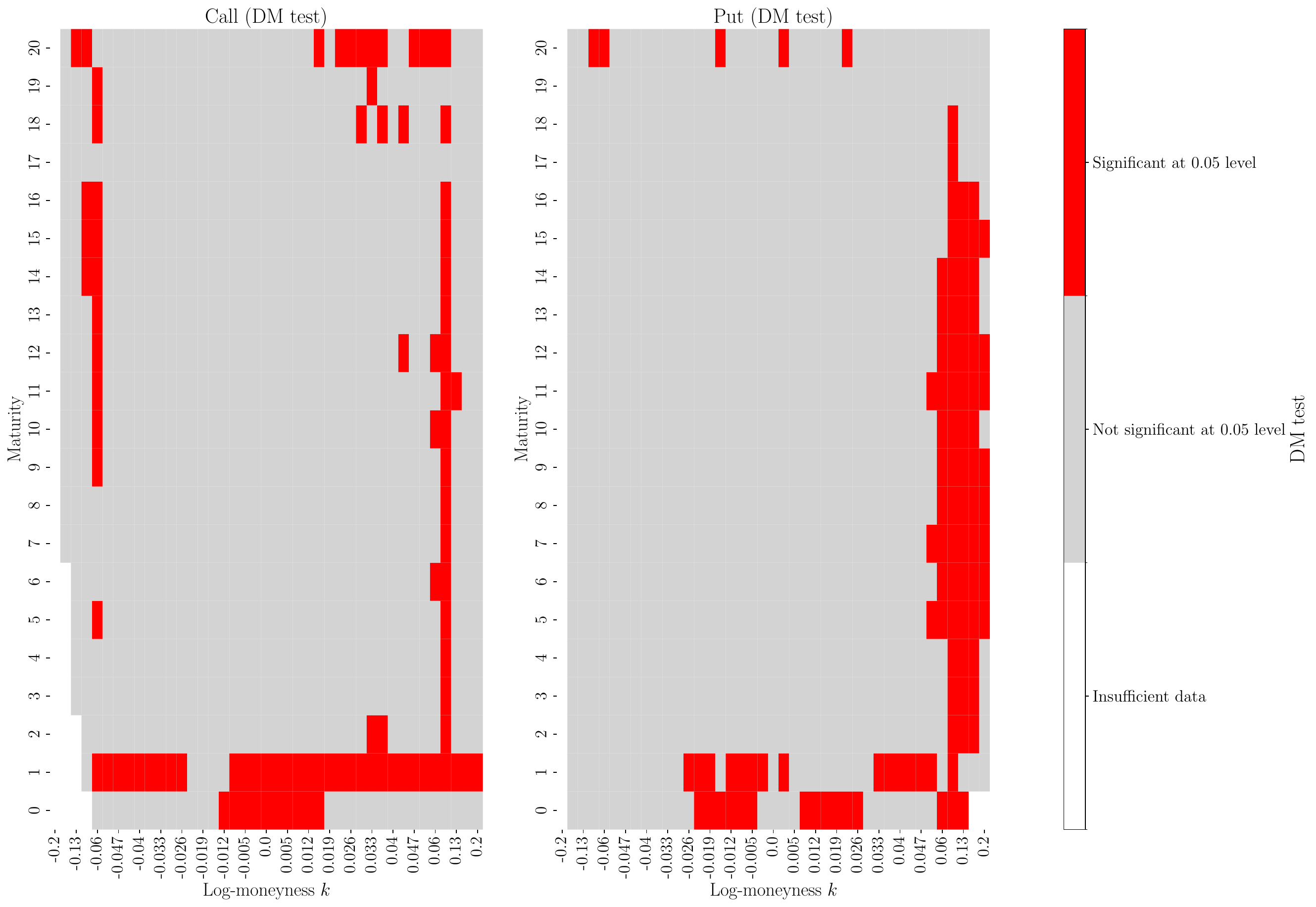}
    \caption{Diebold-Mariano test between naive Random Walk predictions and convLSTM predictions for SVI model, prediction horizon 5.}
\end{figure}

\begin{figure}[H]
    \centering
    \includegraphics[width=\linewidth]{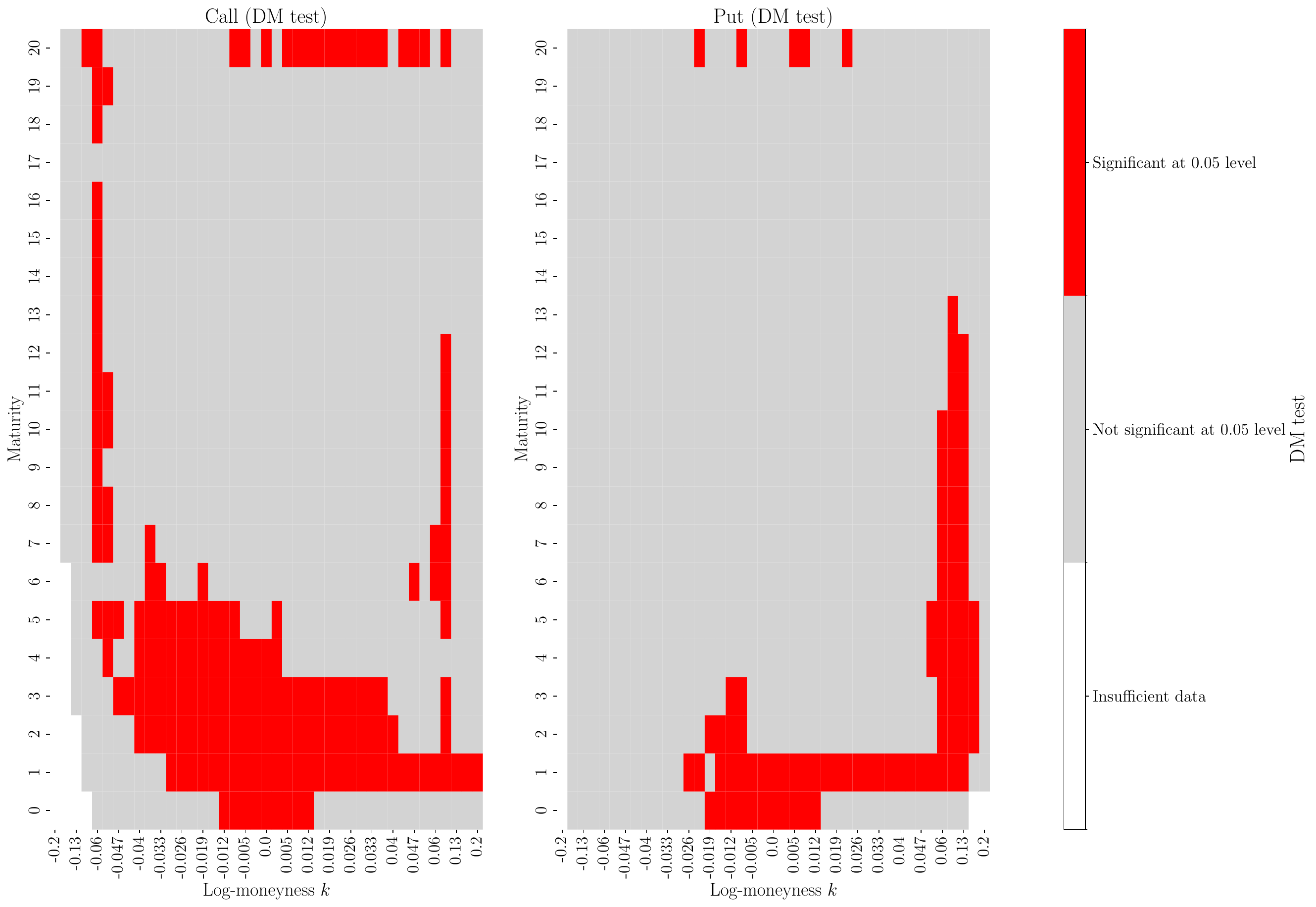}
    \caption{Diebold-Mariano test between naive Random Walk predictions and convLSTM predictions for SVI model, prediction horizon 10.}
\end{figure}


\begin{figure}[H]
    \centering
    \includegraphics[width=\linewidth]{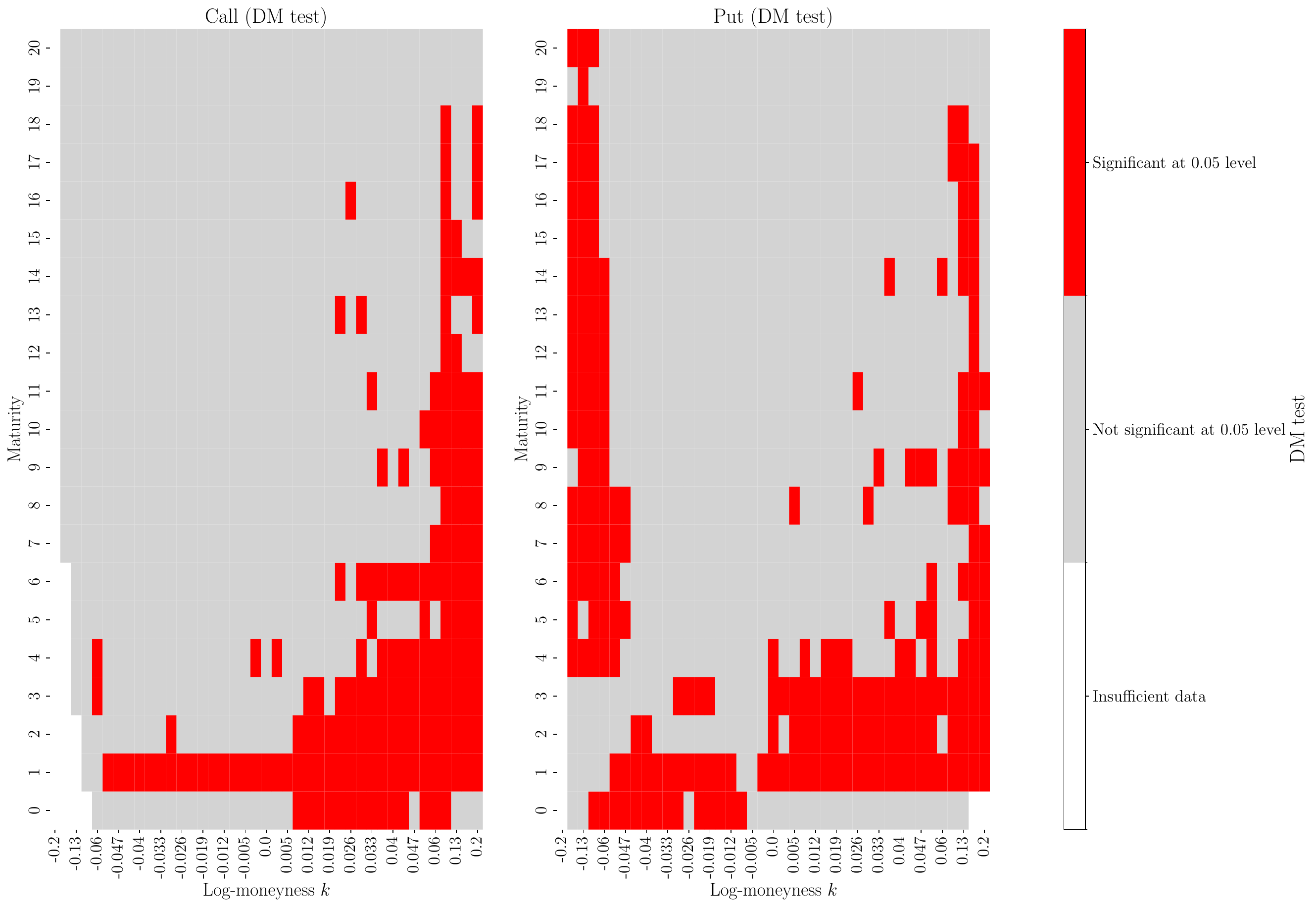}
    \caption{Diebold-Mariano test between naive Random Walk predictions and convLSTM predictions for linear interpolation, prediction horizon 1.}
\end{figure}

\begin{figure}[H]
    \centering
    \includegraphics[width=\linewidth]{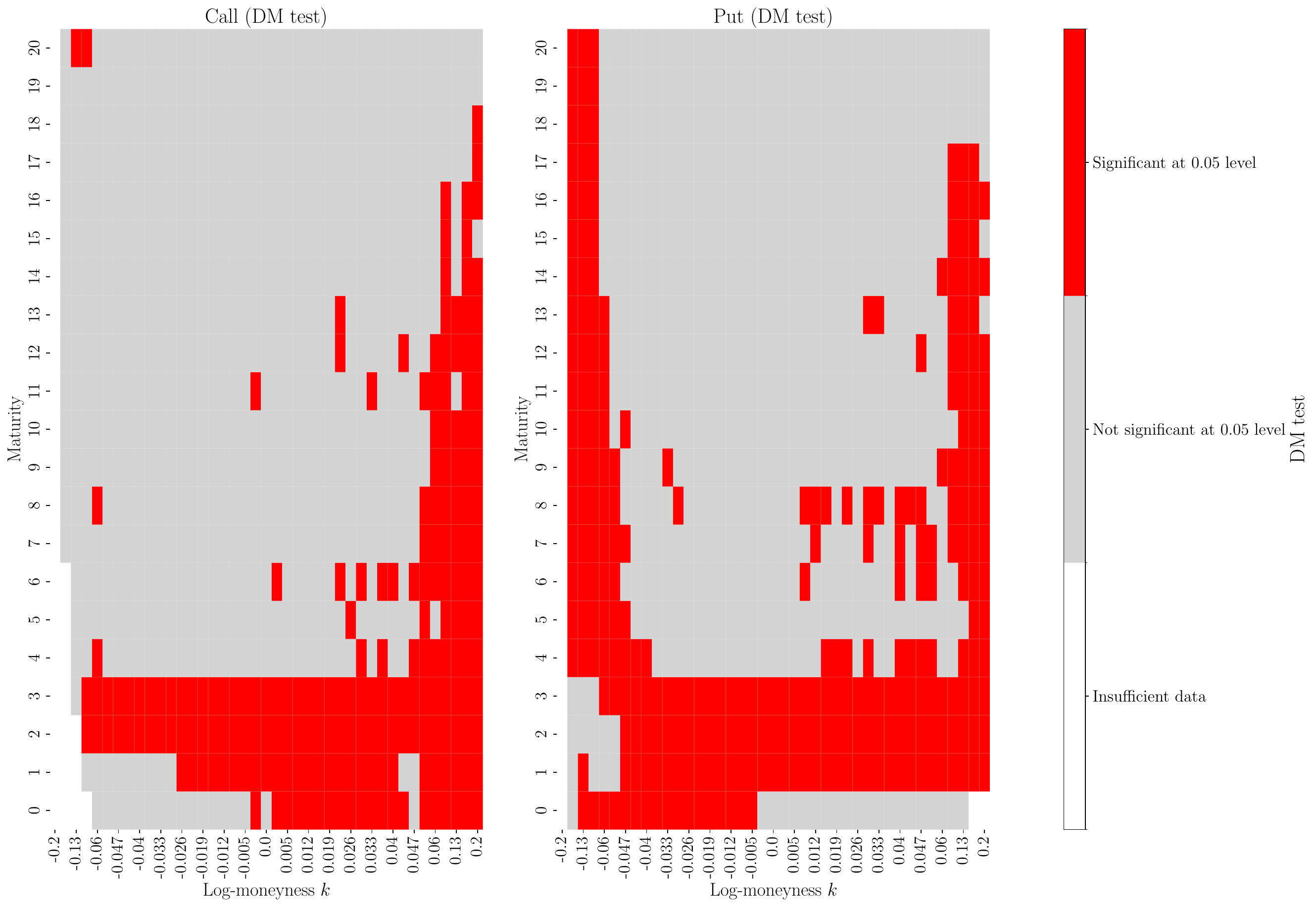}
    \caption{Diebold-Mariano test between naive Random Walk predictions and convLSTM predictions for linear interpolation, prediction horizon 2.}
\end{figure}

\begin{figure}[H]
    \centering
    \includegraphics[width=\linewidth]{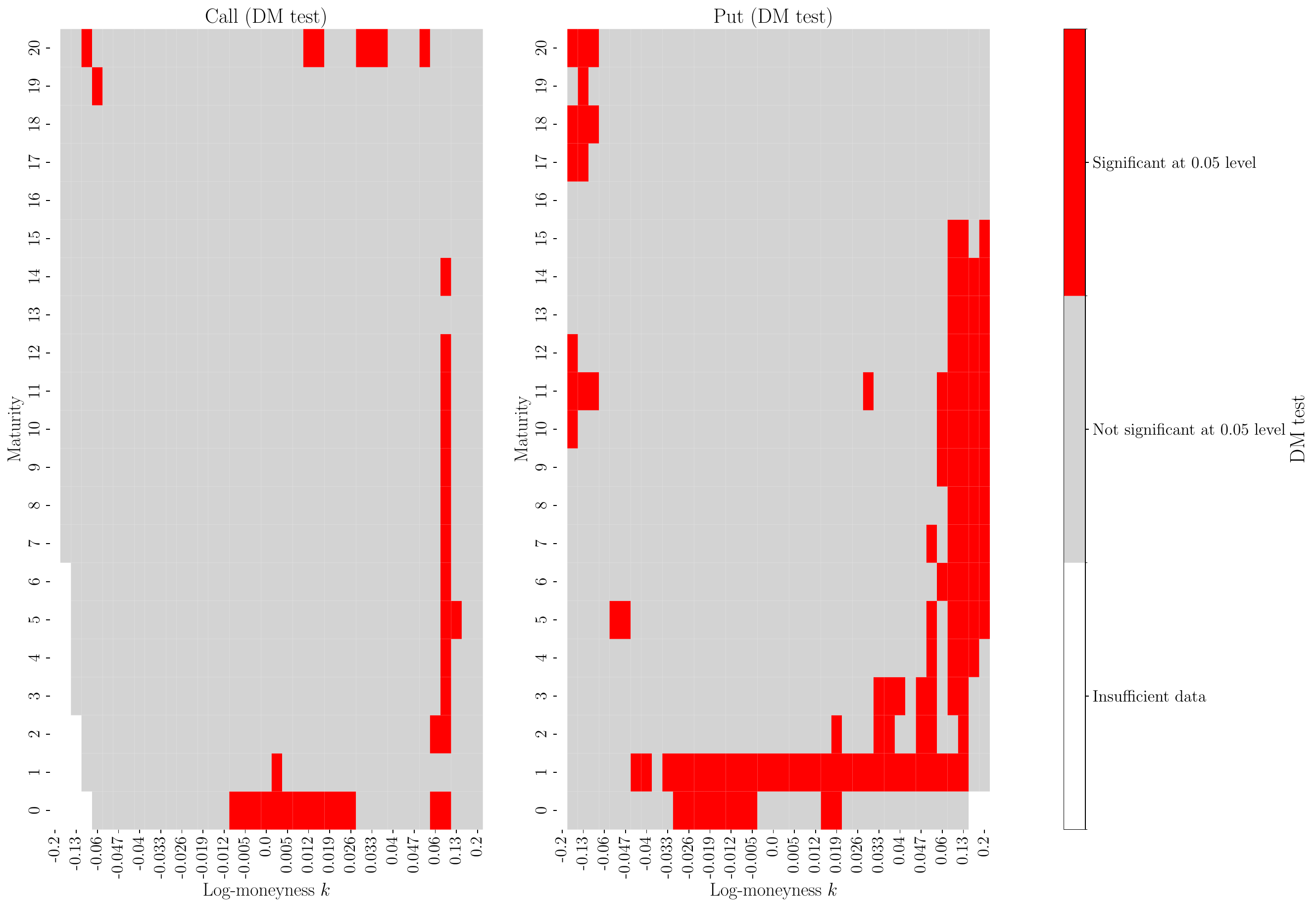}
    \caption{Diebold-Mariano test between naive Random Walk predictions and convLSTM predictions for linear interpolation, prediction horizon 5.}
\end{figure}

\begin{figure}[H]
    \centering
    \includegraphics[width=\linewidth]{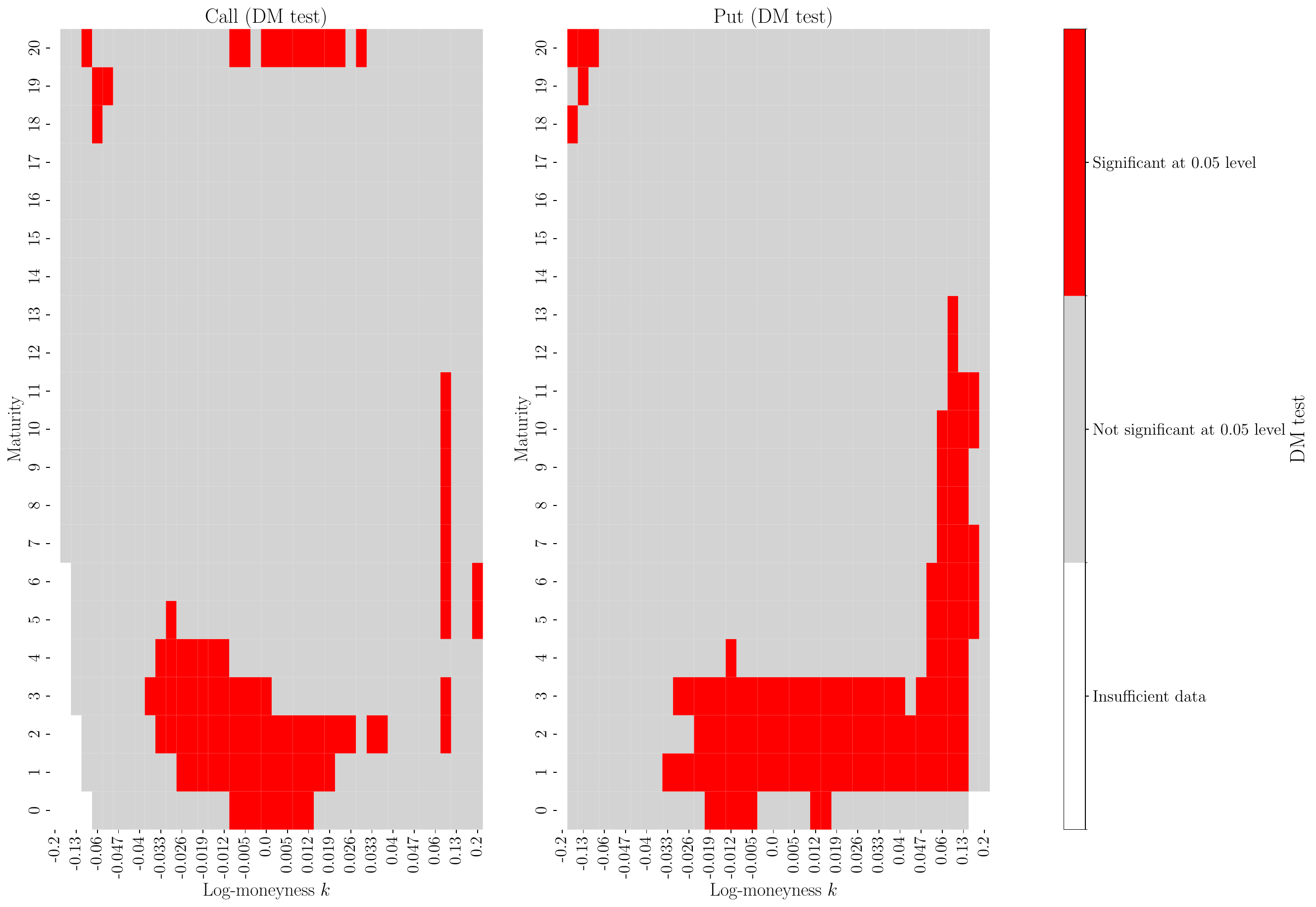}
    \caption{Diebold-Mariano test between naive Random Walk predictions and convLSTM predictions for linear interpolation, prediction horizon 10.}
\end{figure}


\begin{figure}[H]
    \centering
    \includegraphics[width=\linewidth]{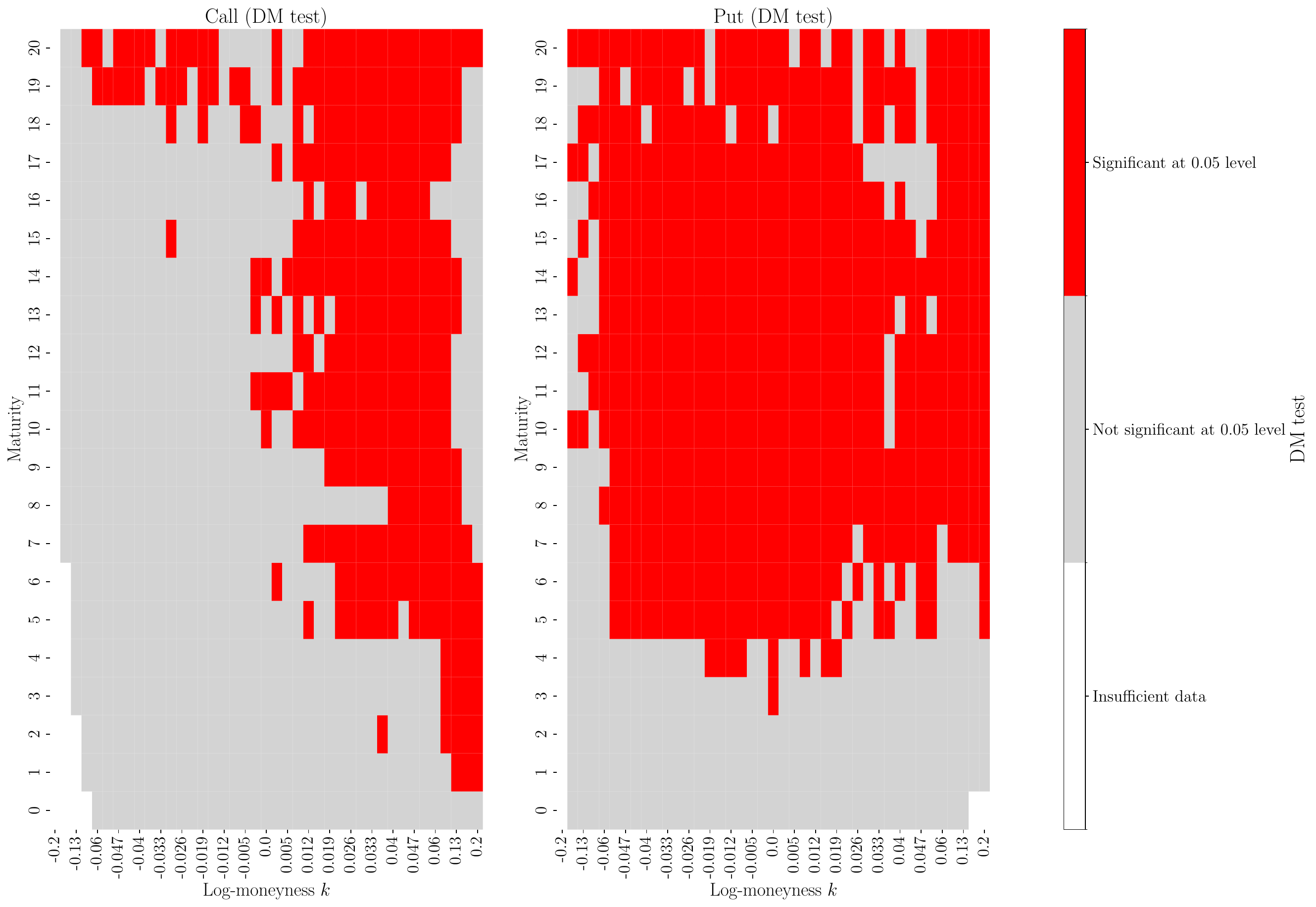}
    \caption{Diebold-Mariano test between naive Random Walk predictions and convLSTM predictions for AHBS model, prediction horizon 1.}
\end{figure}

\begin{figure}[H]
    \centering
    \includegraphics[width=\linewidth]{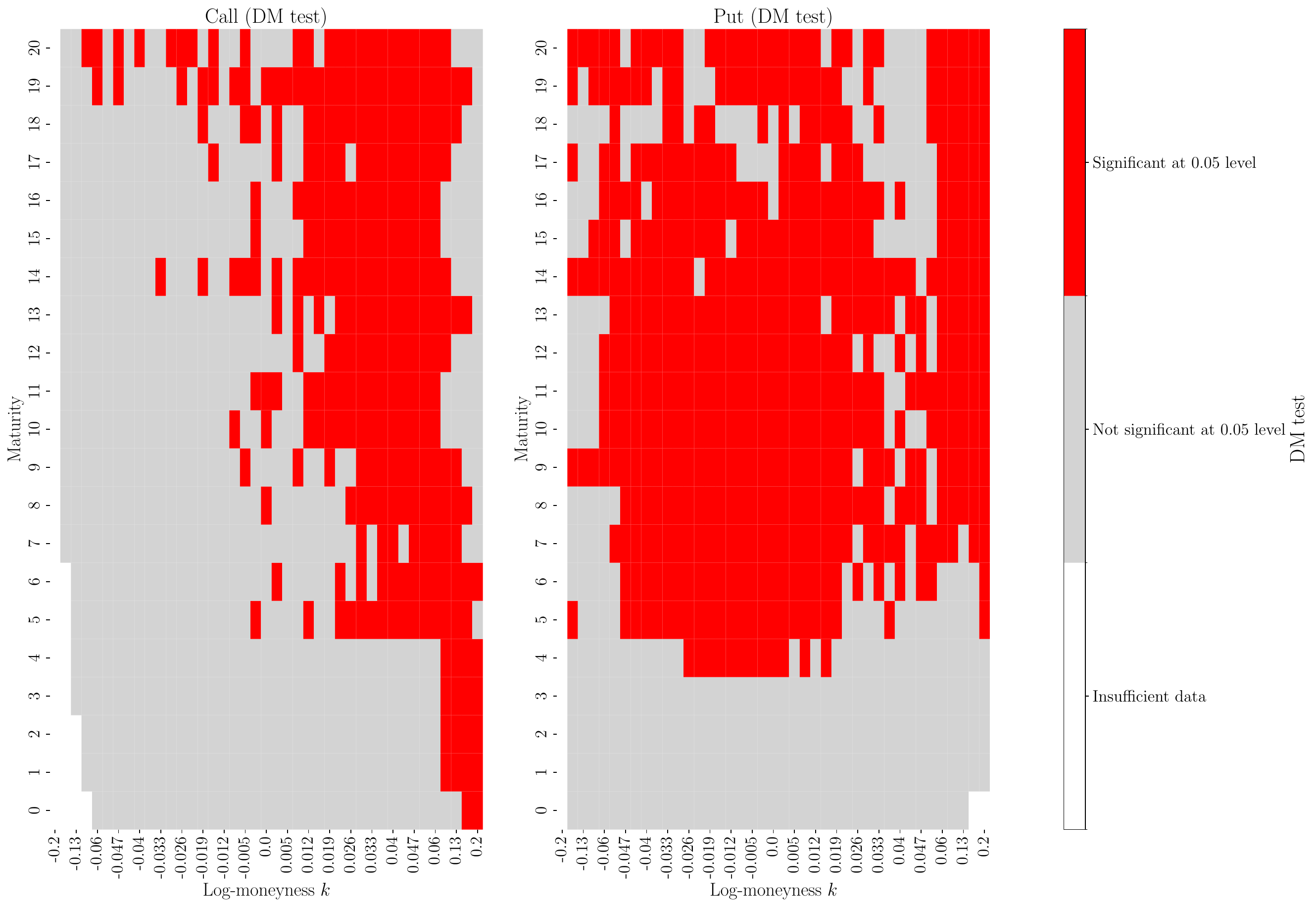}
    \caption{Diebold-Mariano test between naive Random Walk predictions and convLSTM predictions for AHBS model, prediction horizon 2.}
\end{figure}

\begin{figure}[H]
    \centering
    \includegraphics[width=\linewidth]{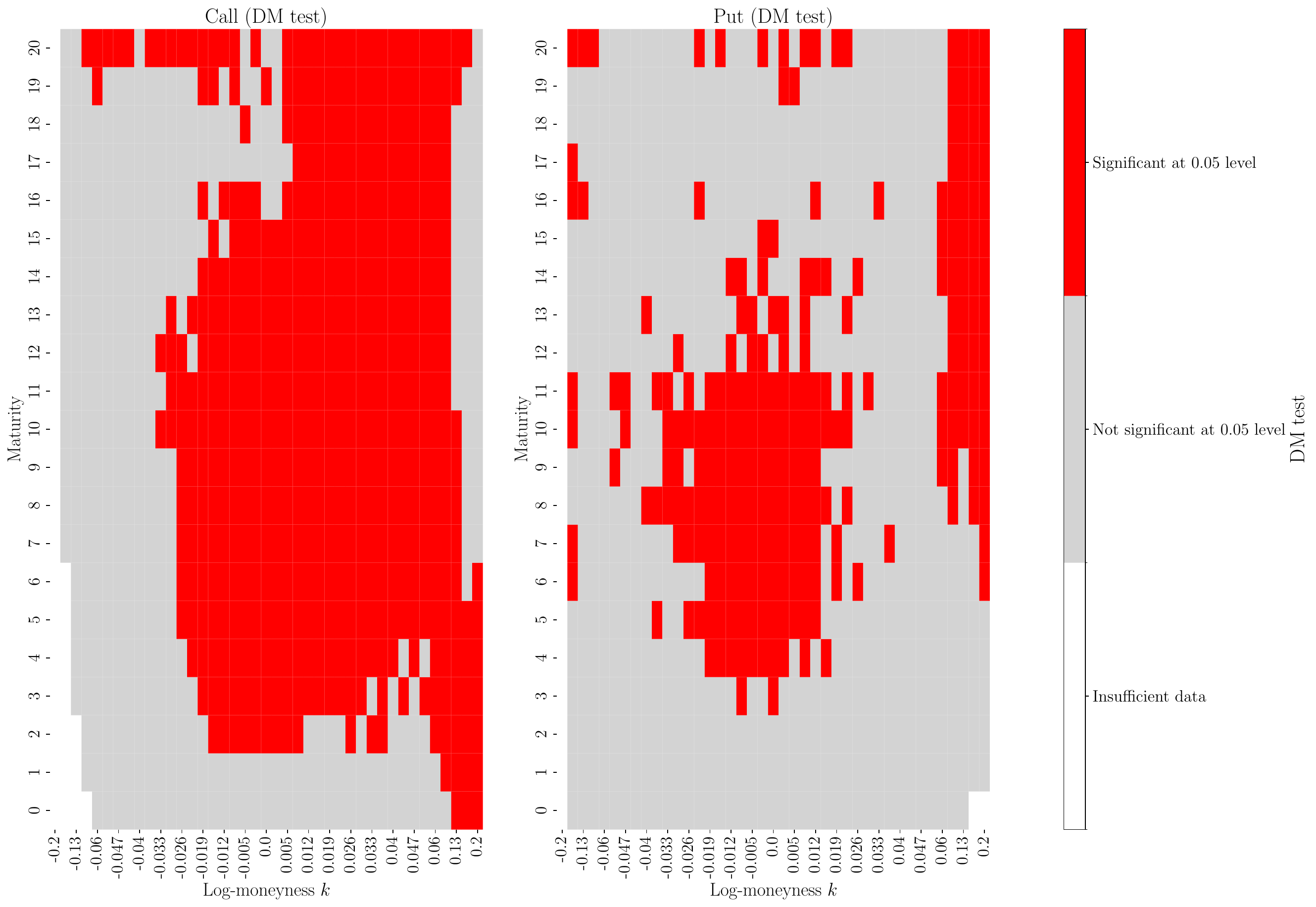}
    \caption{Diebold-Mariano test between naive Random Walk predictions and convLSTM predictions for AHBS model, prediction horizon 5.}
\end{figure}

\begin{figure}[H]
    \centering
    \includegraphics[width=\linewidth]{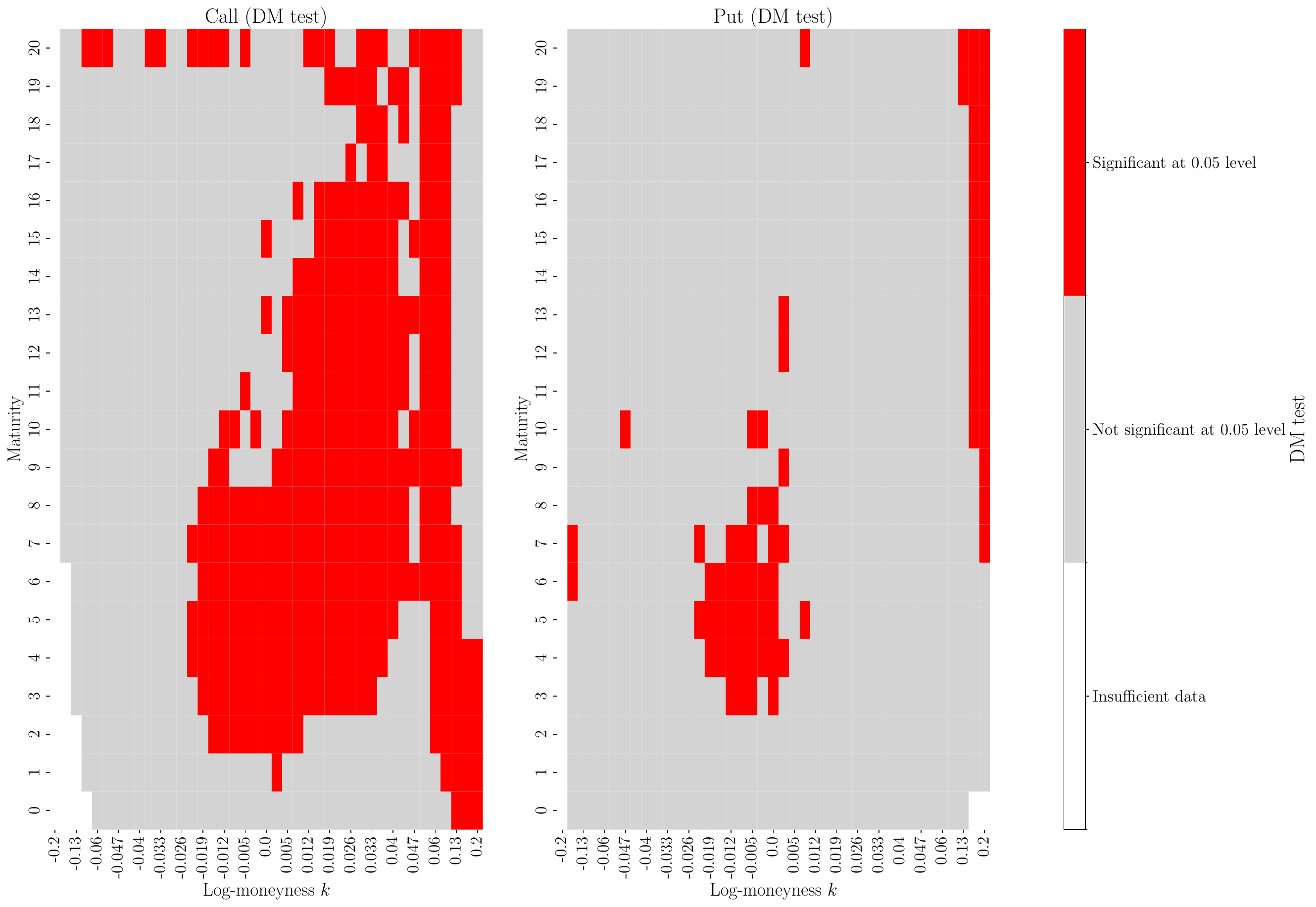}
    \caption{Diebold-Mariano test between naive Random Walk predictions and convLSTM predictions for AHBS model, prediction horizon 10.}
\end{figure}


\subsection{ML models ATM IV predictions - models without announcement dummy}\label{sec:ConvLSTM_models_call_atm_iv}

\subsubsection{Call options}

These graphs display the fit of models without the information regarding the announcement to ATM IV with announcement days highlighted in test sample for call options.

\begin{figure}[H]
    \centering
    \includegraphics[width=\linewidth]{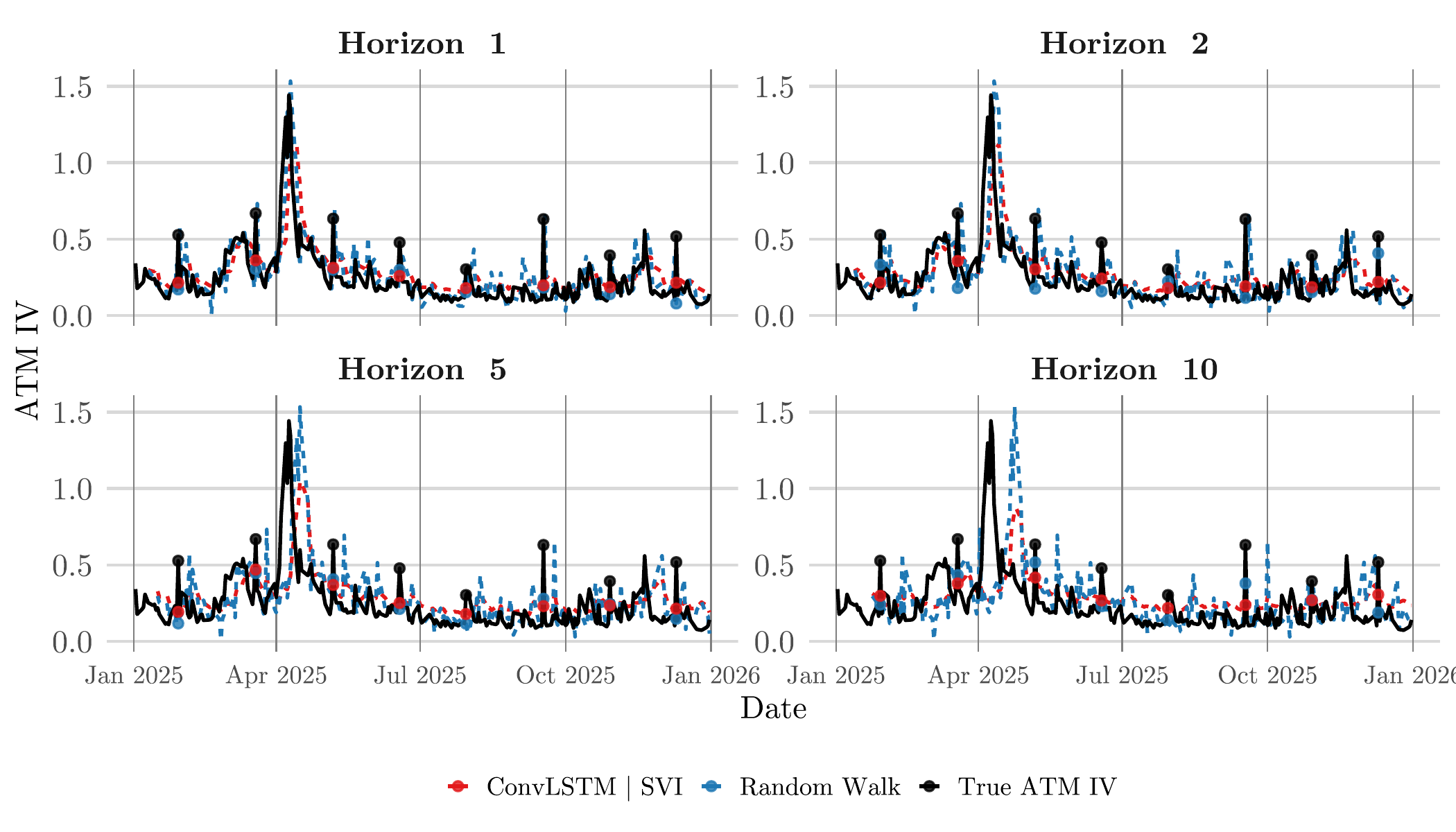}
    \caption{Out-of-sample comparison between naive Random Walk predictions and convLSTM predictions for SVI model, call options.}
\end{figure}

\begin{figure}[H]
    \centering
    \includegraphics[width=\linewidth]{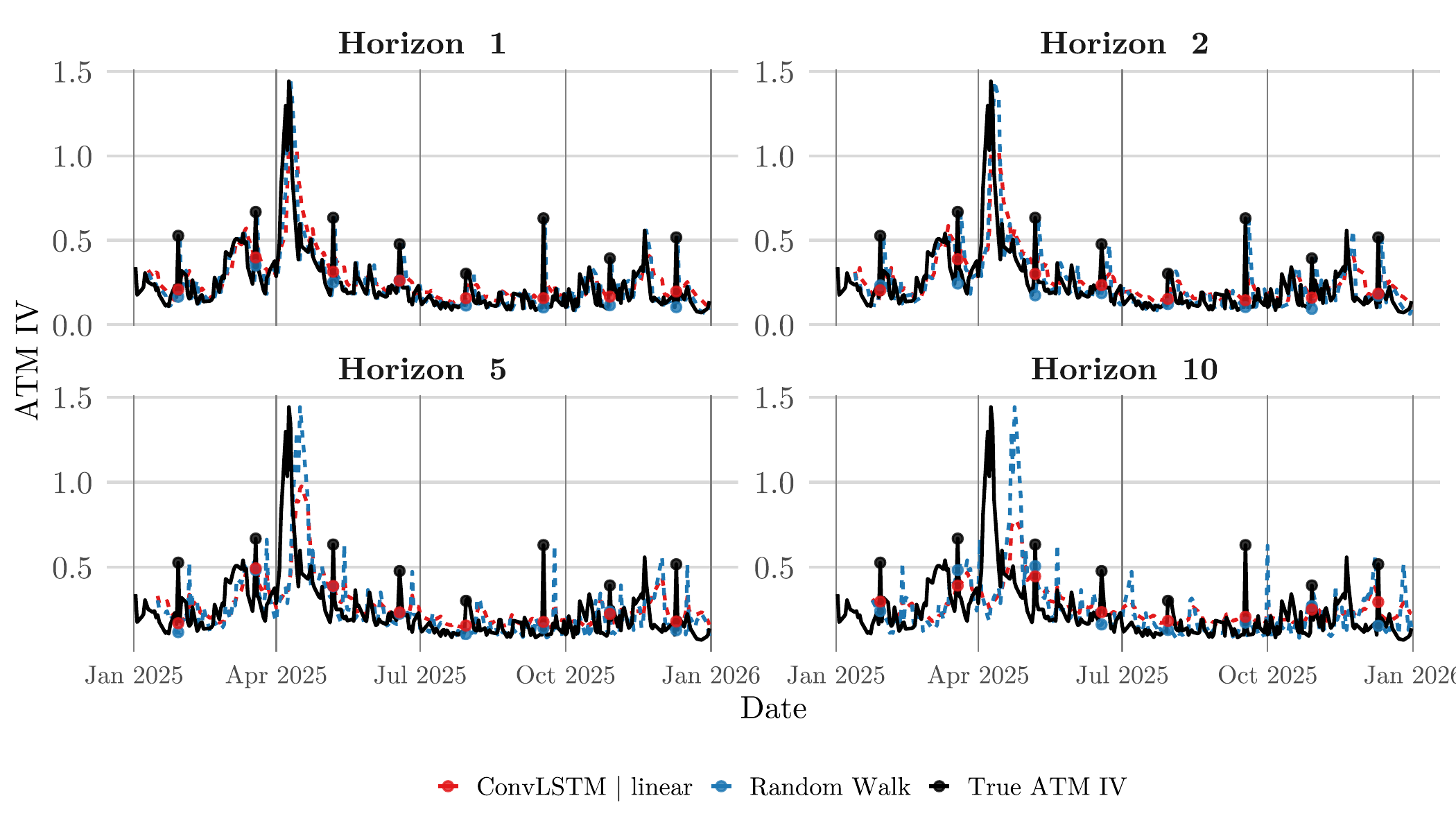}
    \caption{Out-of-sample comparison between naive Random Walk predictions and convLSTM predictions for linear interpolation, call options.}
\end{figure}

\begin{figure}[H]
    \centering
    \includegraphics[width=\linewidth]{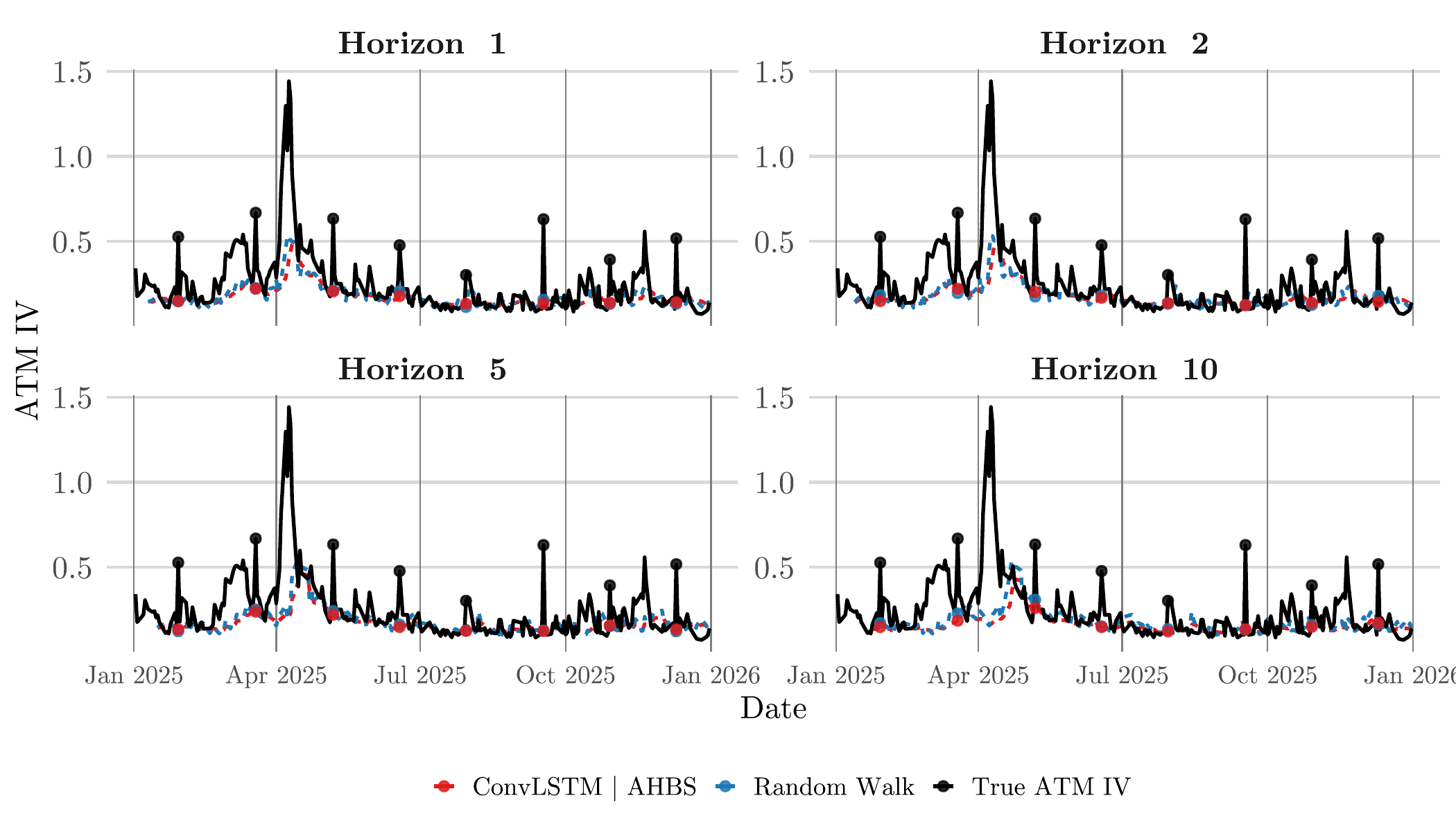}
    \caption{Out-of-sample comparison between naive Random Walk predictions and convLSTM predictions for AHBS model, call options.}
\end{figure}

\subsubsection{Put options}

These graphs display the fit of models without the information regarding the announcement to ATM IV with announcement days highlighted in test sample for put options.

\begin{figure}[H]
    \centering
    \includegraphics[width=\linewidth]{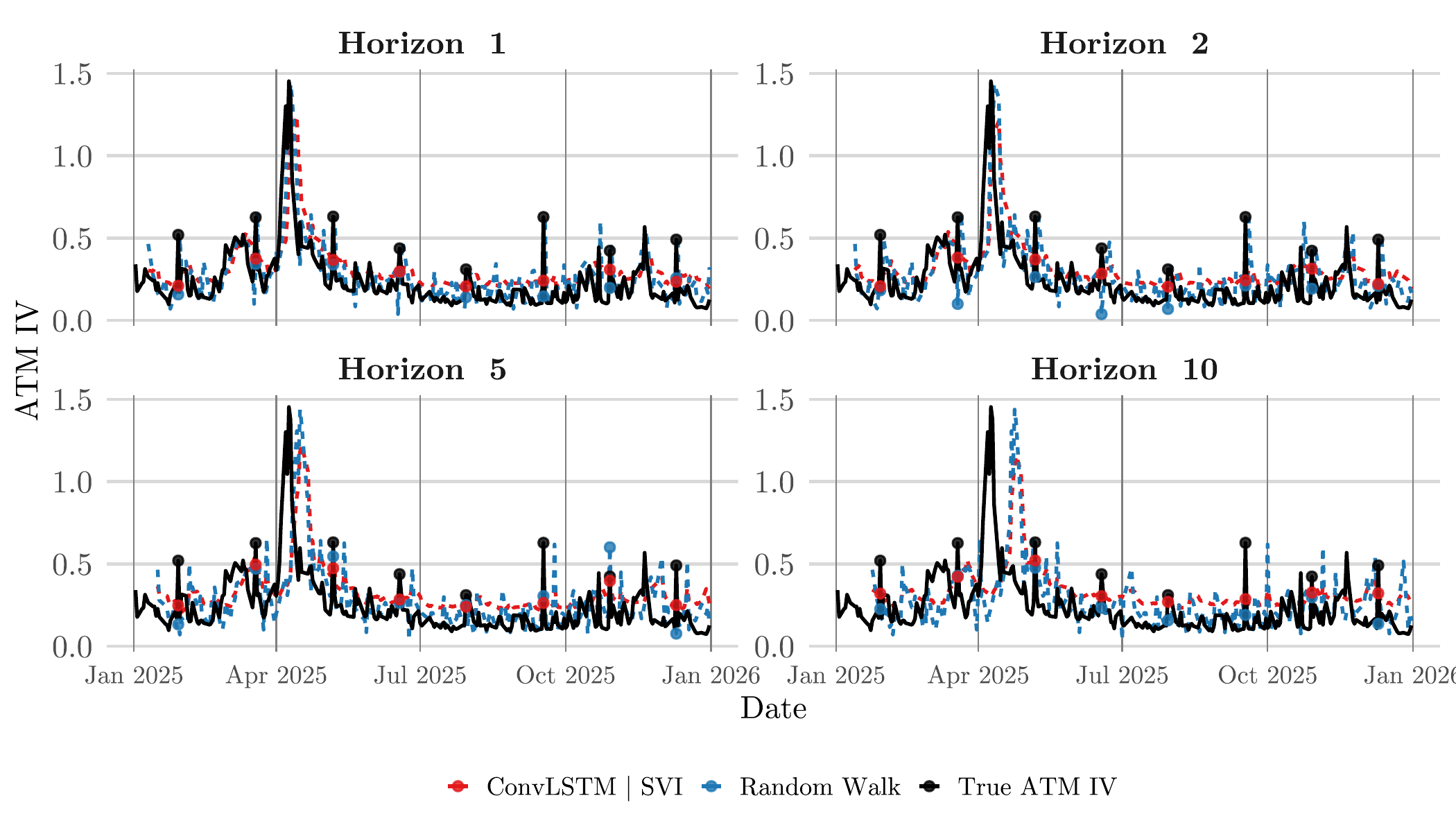}
    \caption{Out-of-sample comparison between naive Random Walk predictions and convLSTM predictions for SVI model, put options.}
\end{figure}

\begin{figure}[H]
    \centering
    \includegraphics[width=\linewidth]{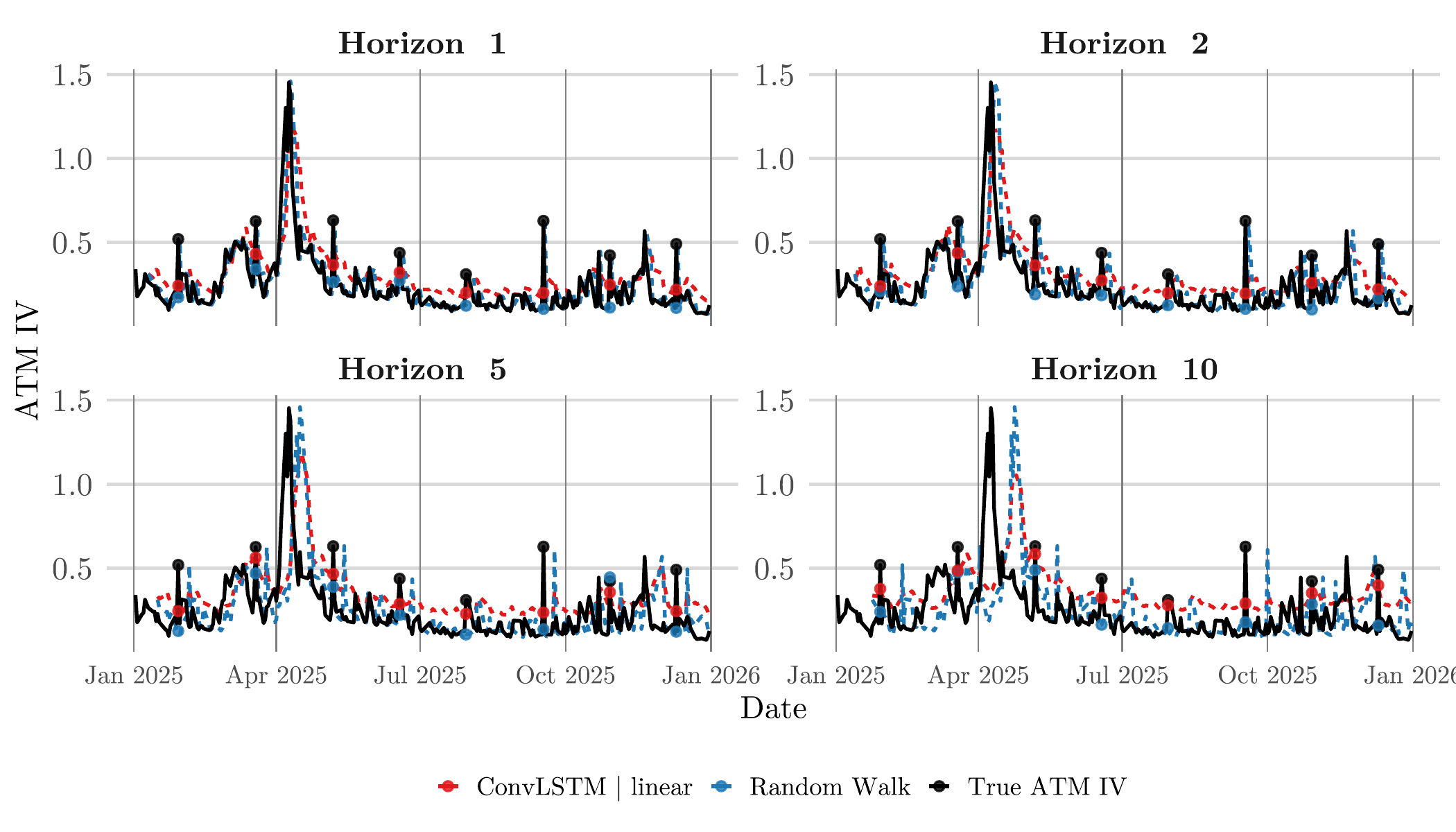}
    \caption{Out-of-sample comparison between naive Random Walk predictions and convLSTM predictions for linear interpolation, put options.}
\end{figure}

\begin{figure}[H]
    \centering
    \includegraphics[width=\linewidth]{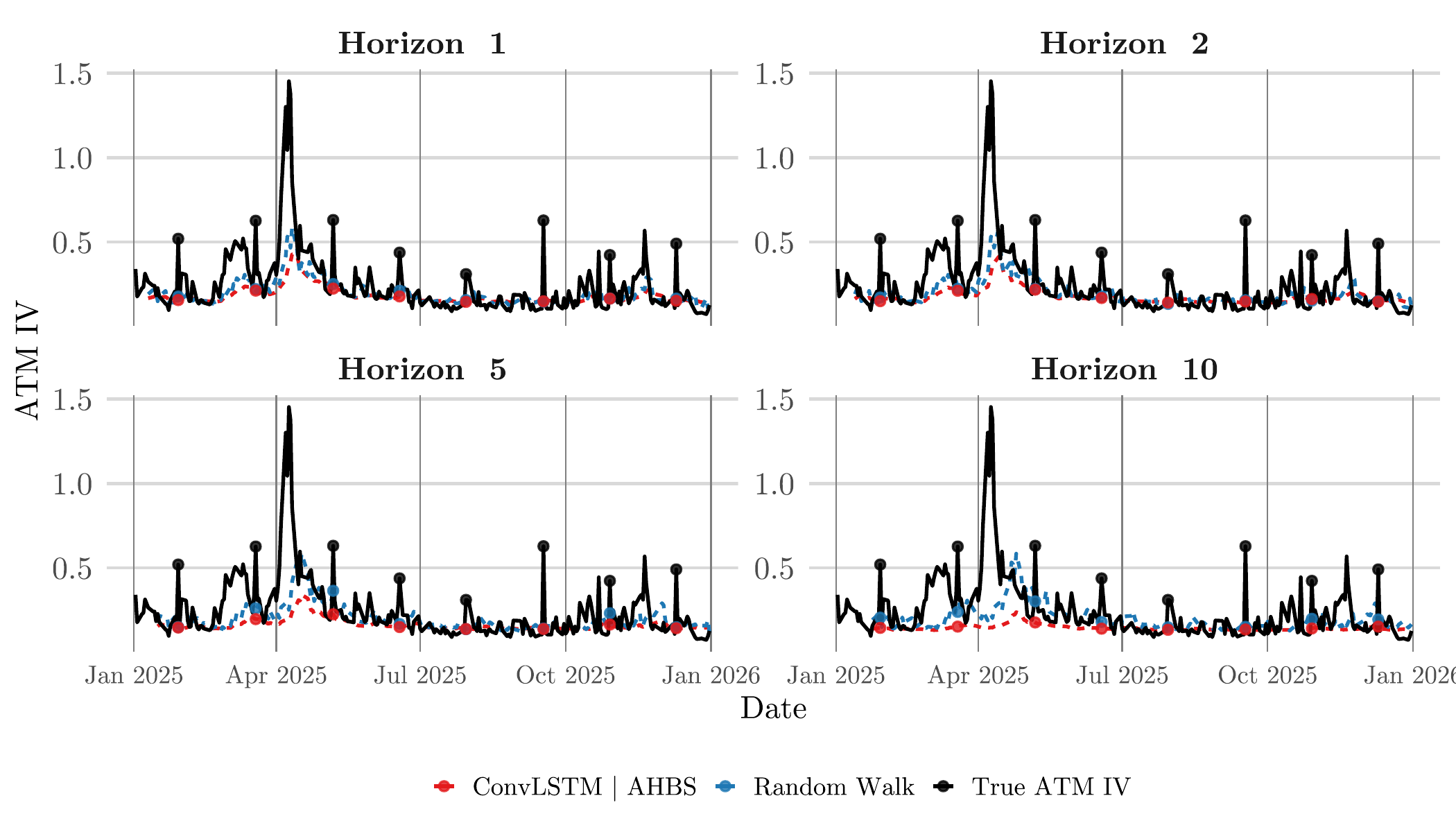}
    \caption{Out-of-sample comparison between naive Random Walk predictions and convLSTM predictions for AHBS model, put options.}
\end{figure}

\subsection{ML models ATM IV predictions - models with announcement dummy}\label{sec:ConvXLSTM_models_call_atm_iv_dummy}

\subsubsection{Call options}

These graphs display the fit of models with the announcement dummy to ATM IV with announcement days highlighted in test sample for call options.

\begin{figure}[H]
    \centering
    \includegraphics[width=\linewidth]{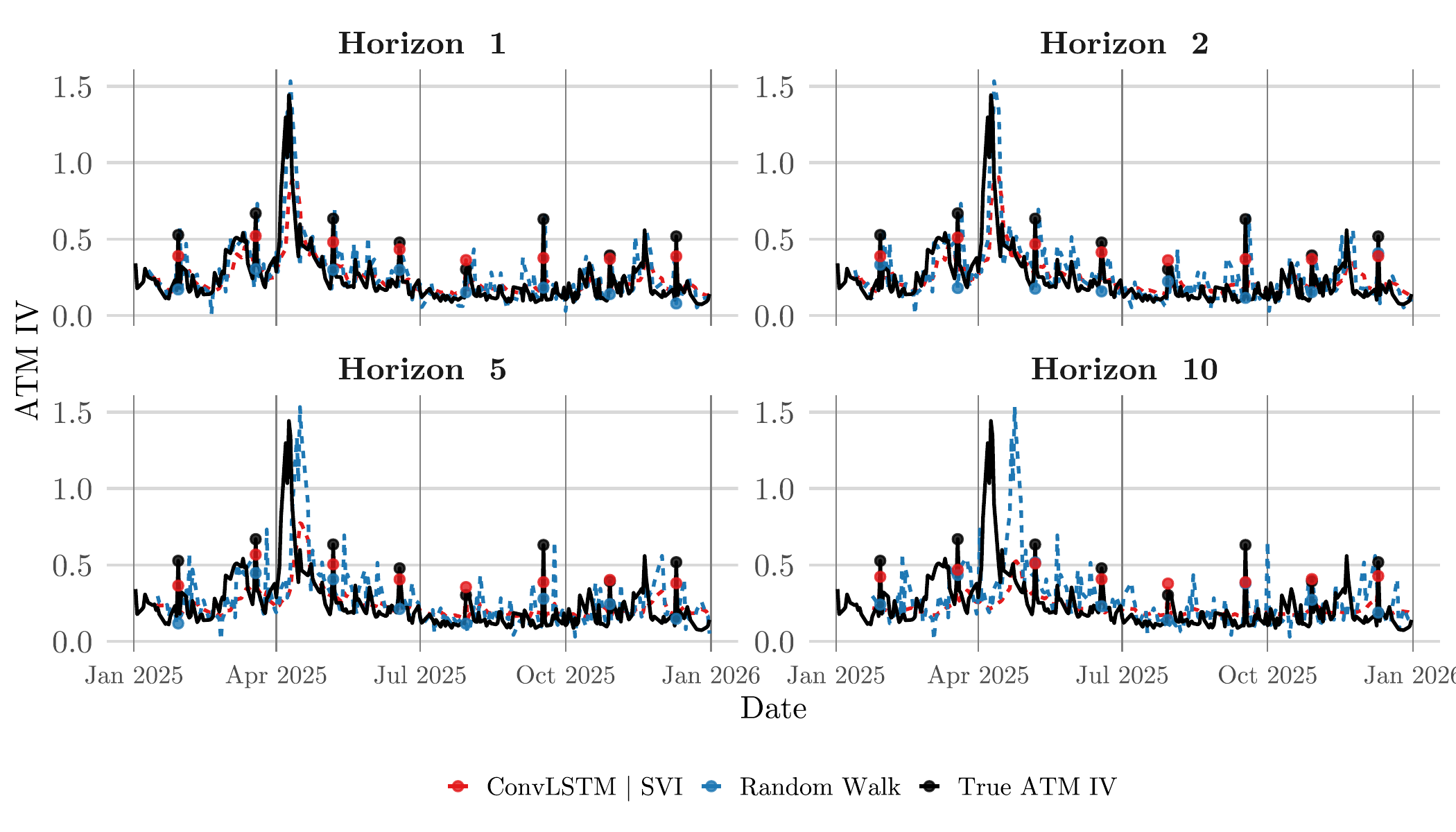}
    \caption{Out-of-sample comparison between naive Random Walk predictions and convLSTM predictions for SVI model, call options.}
\end{figure}

\begin{figure}[H]
    \centering
    \includegraphics[width=\linewidth]{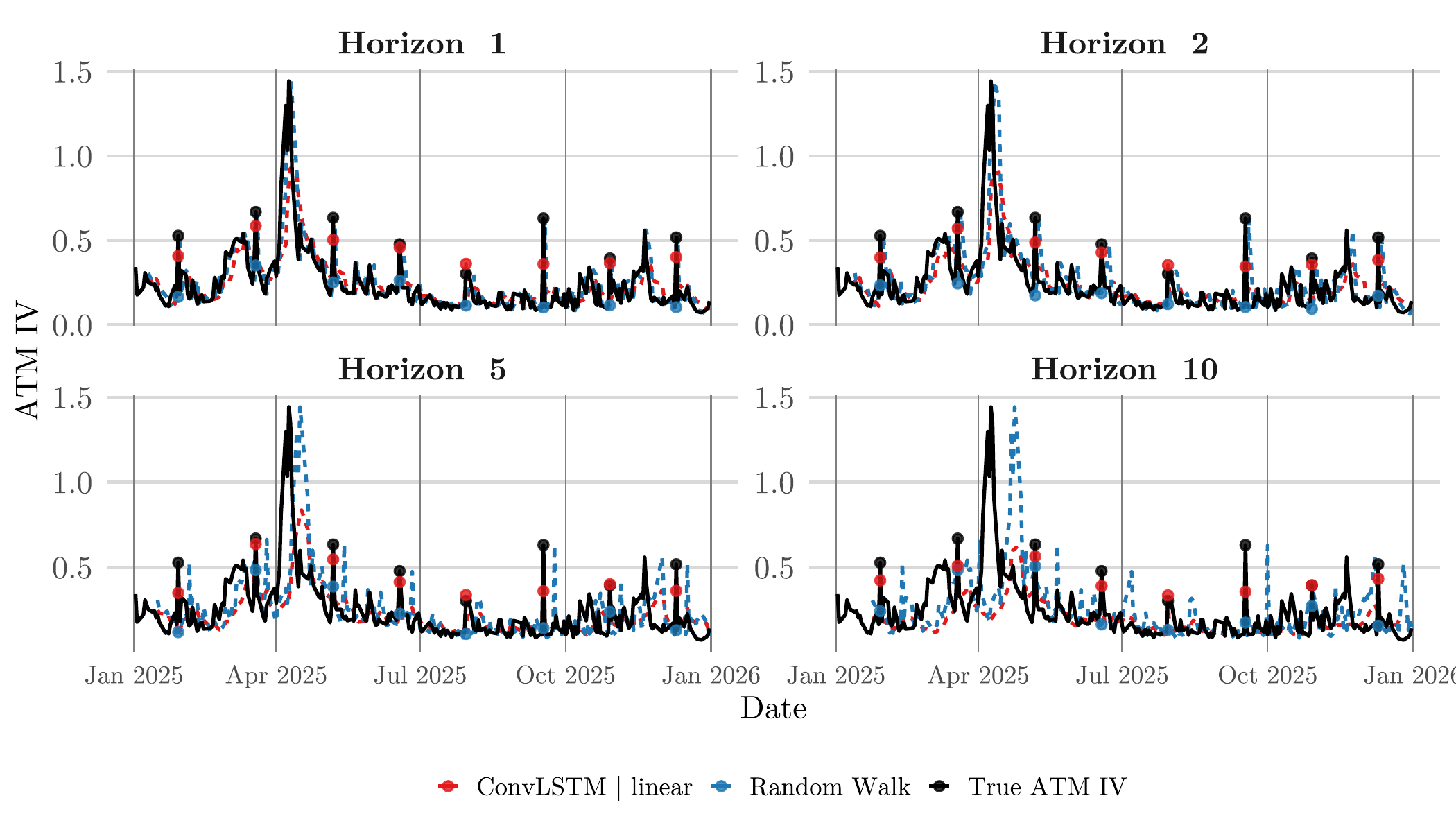}
    \caption{Out-of-sample comparison between naive Random Walk predictions and convLSTM predictions for linear interpolation, call options.}
\end{figure}

\begin{figure}[H]
    \centering
    \includegraphics[width=\linewidth]{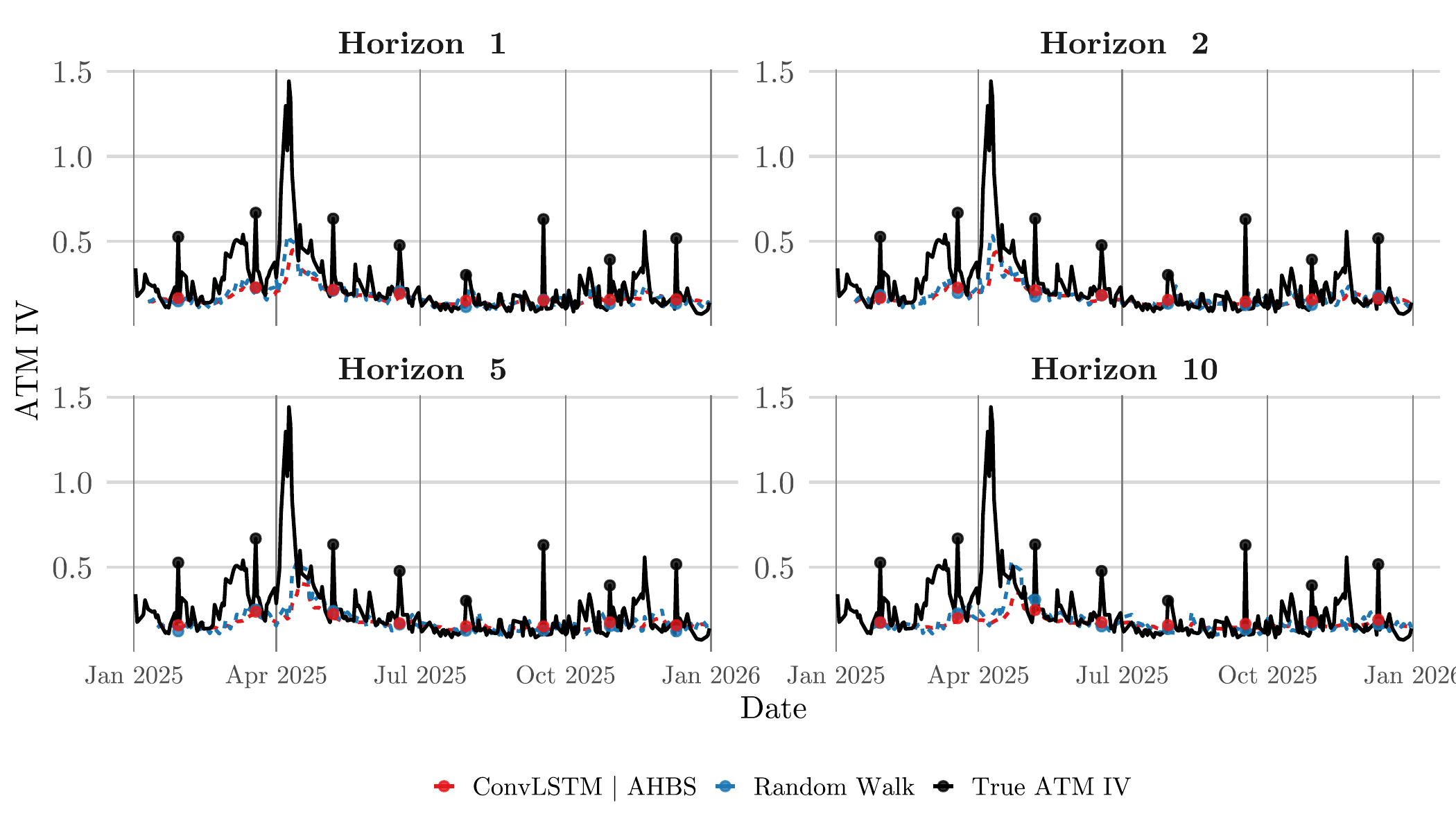}
    \caption{Out-of-sample comparison between naive Random Walk predictions and convLSTM predictions for AHBS model, call options.}
\end{figure}

\subsubsection{Put options}

These graphs display the fit of models with the announcement dummy to ATM IV with announcement days highlighted in test sample for put options.

\begin{figure}[H]
    \centering
    \includegraphics[width=\linewidth]{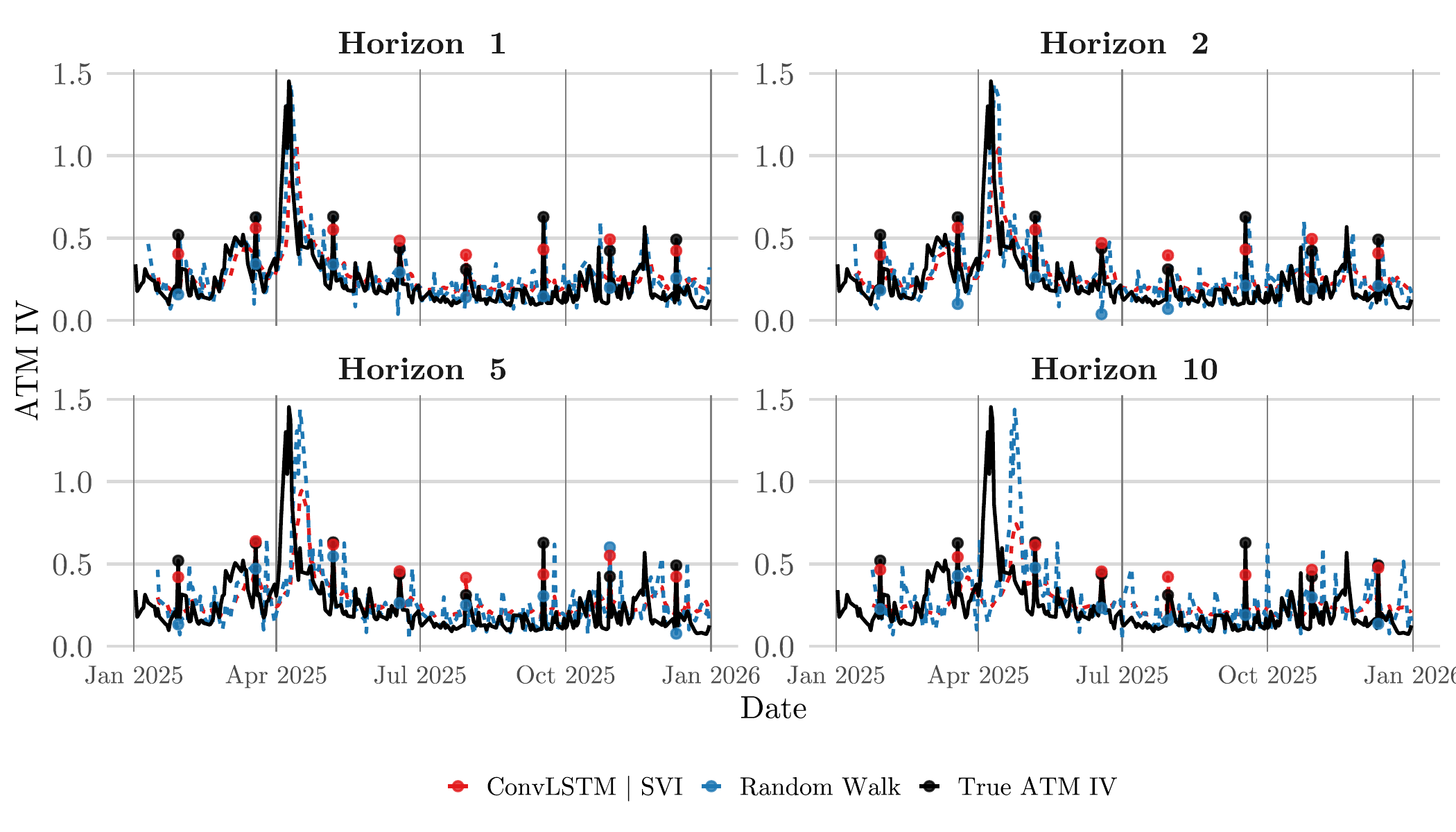}
    \caption{Out-of-sample comparison between naive Random Walk predictions and convLSTM predictions for SVI model, put options.}
\end{figure}

\begin{figure}[H]
    \centering
    \includegraphics[width=\linewidth]{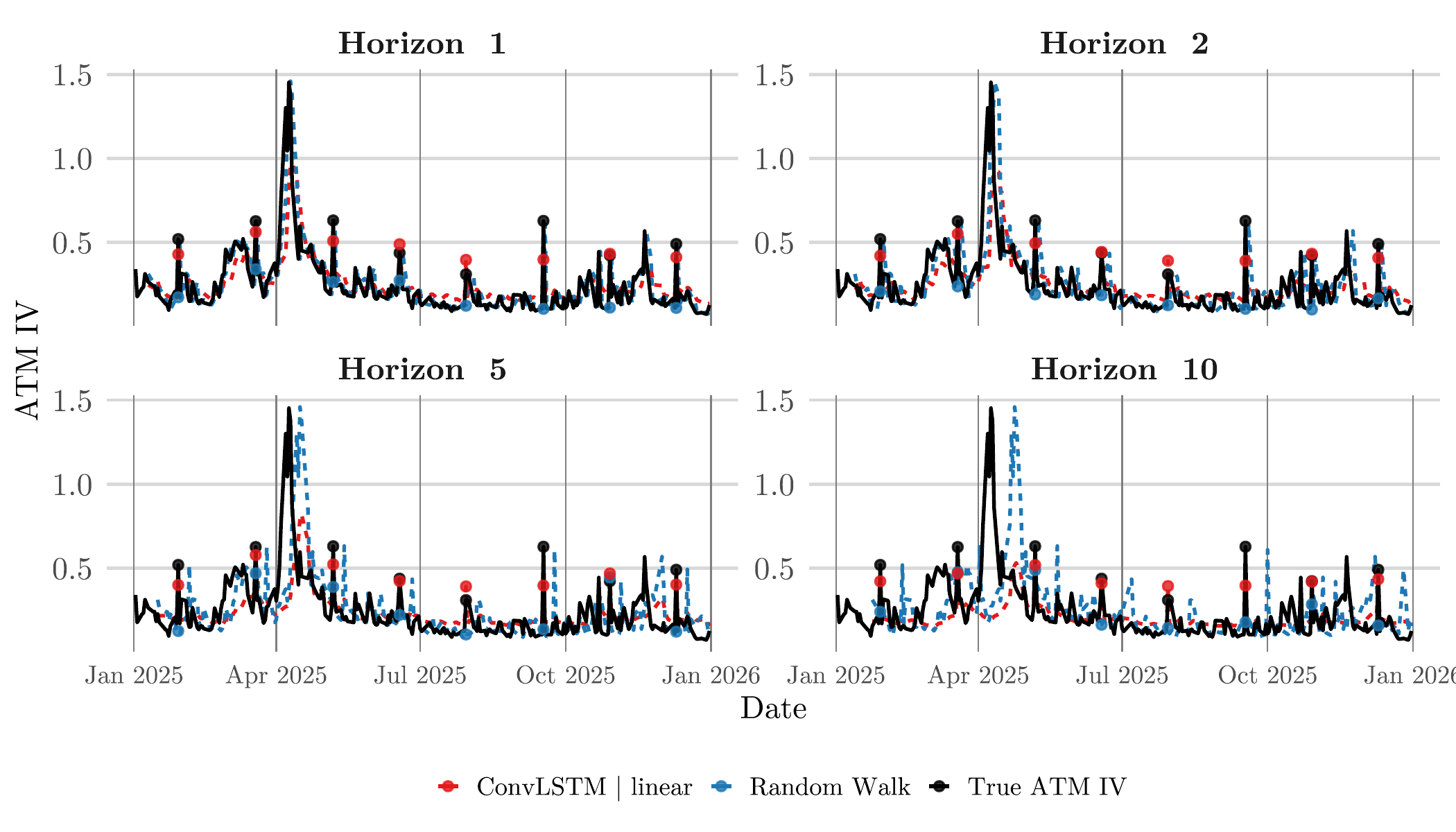}
    \caption{Out-of-sample comparison between naive Random Walk predictions and convLSTM predictions for linear interpolation, put options.}
\end{figure}

\begin{figure}[H]
    \centering
    \includegraphics[width=\linewidth]{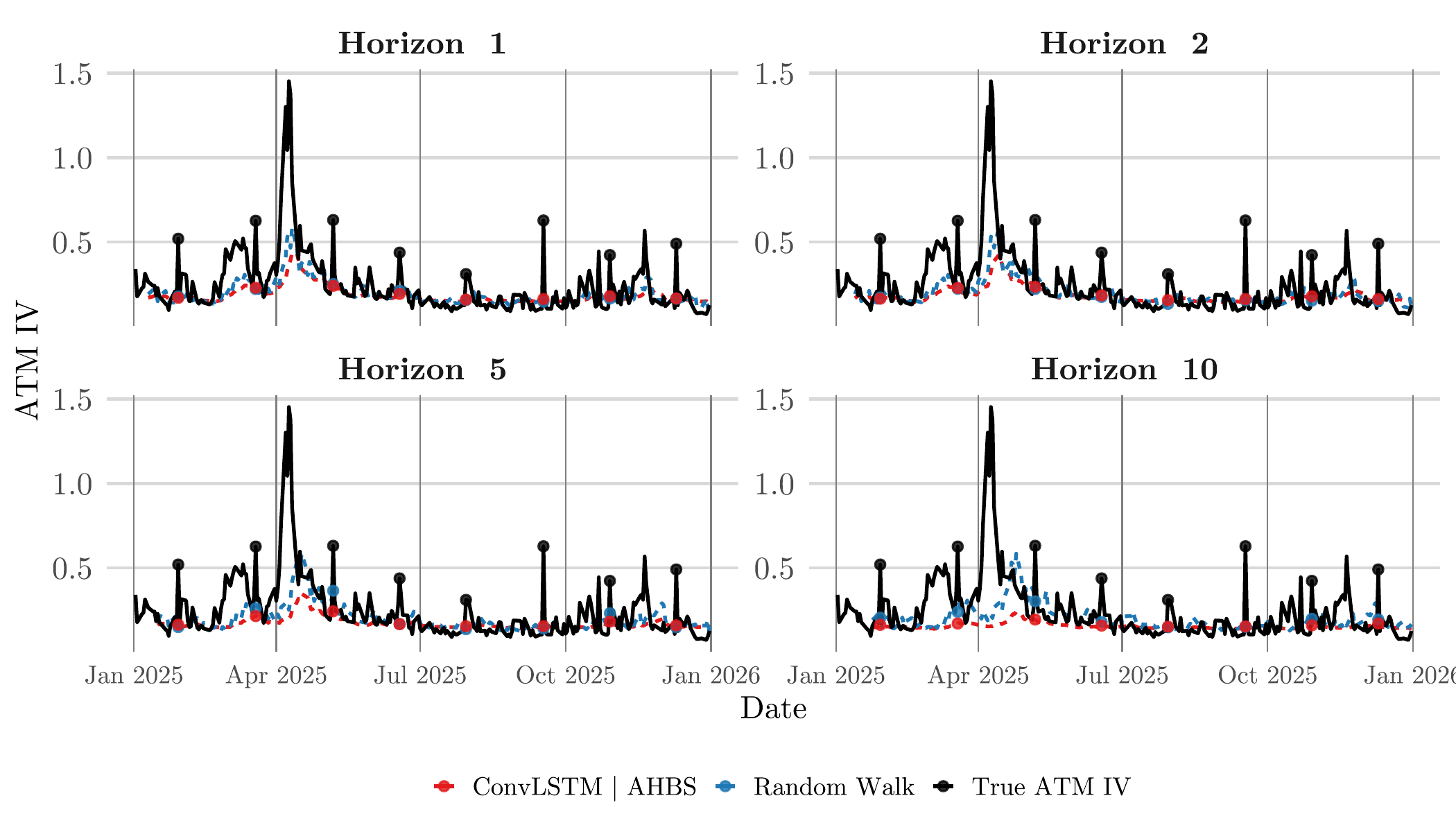}
    \caption{Out-of-sample comparison between naive Random Walk predictions and convLSTM predictions for AHBS model, put options.}
\end{figure}

\subsection{Annoucement shock fit}\label{sec:ml_models_call_shock}

\subsubsection{Call options}

These graphs show the fitted shape of the shock (defined in terms of $\bar{\Delta IV}$ where $\Delta IV = IV_\{\tau\} - IV_{\{\tau - 1\}}$ - 
therefore a mean difference between IV at an announcement and a previous day). 

\begin{figure}[H]
    \centering
    \includegraphics[width=\linewidth]{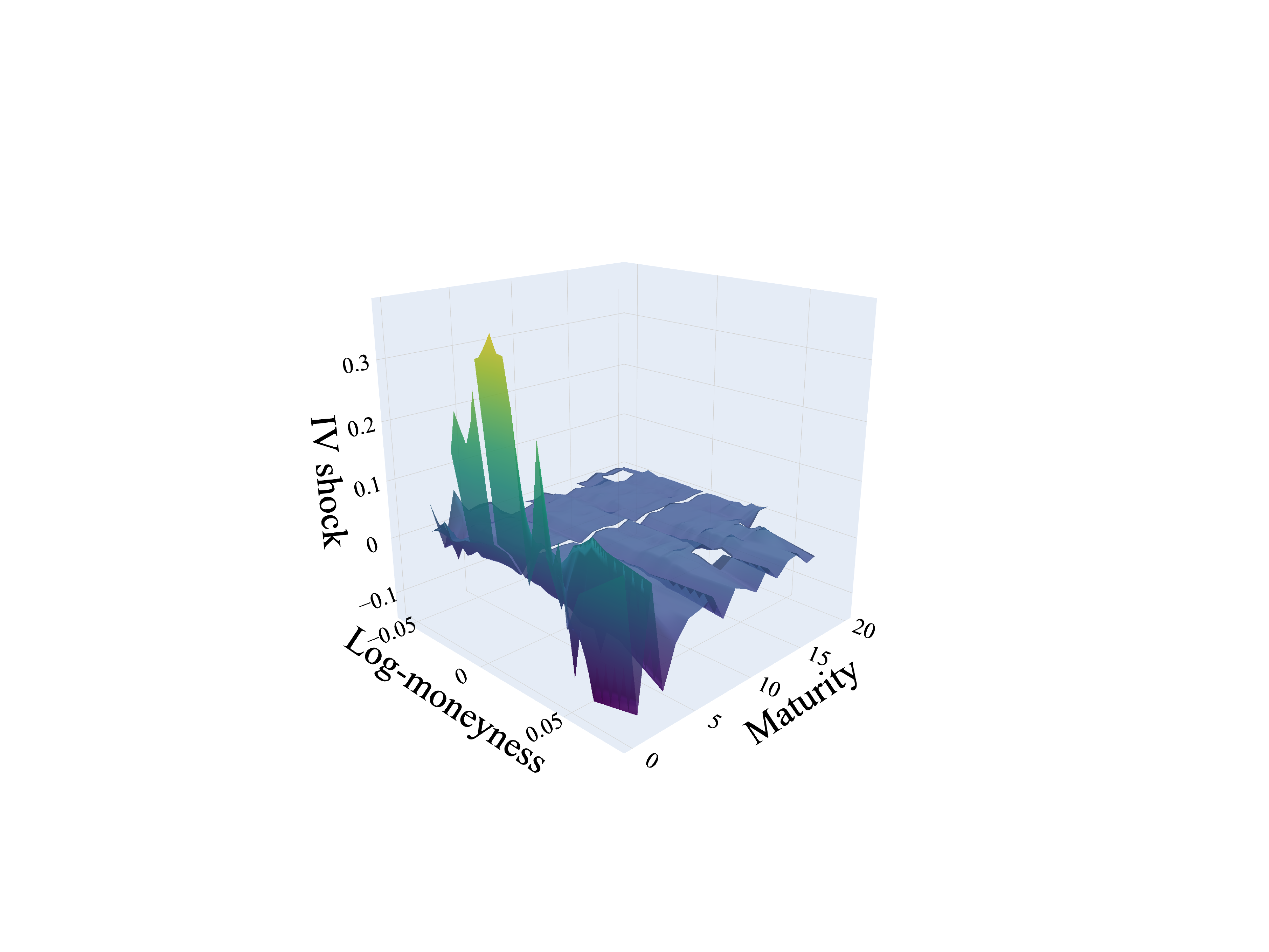}
    \caption{True announcement shock defined as a mean difference between IV at an announcement and a previous day for call options.}
\end{figure}

\begin{figure}[H]
    \centering
    \includegraphics[width=\linewidth]{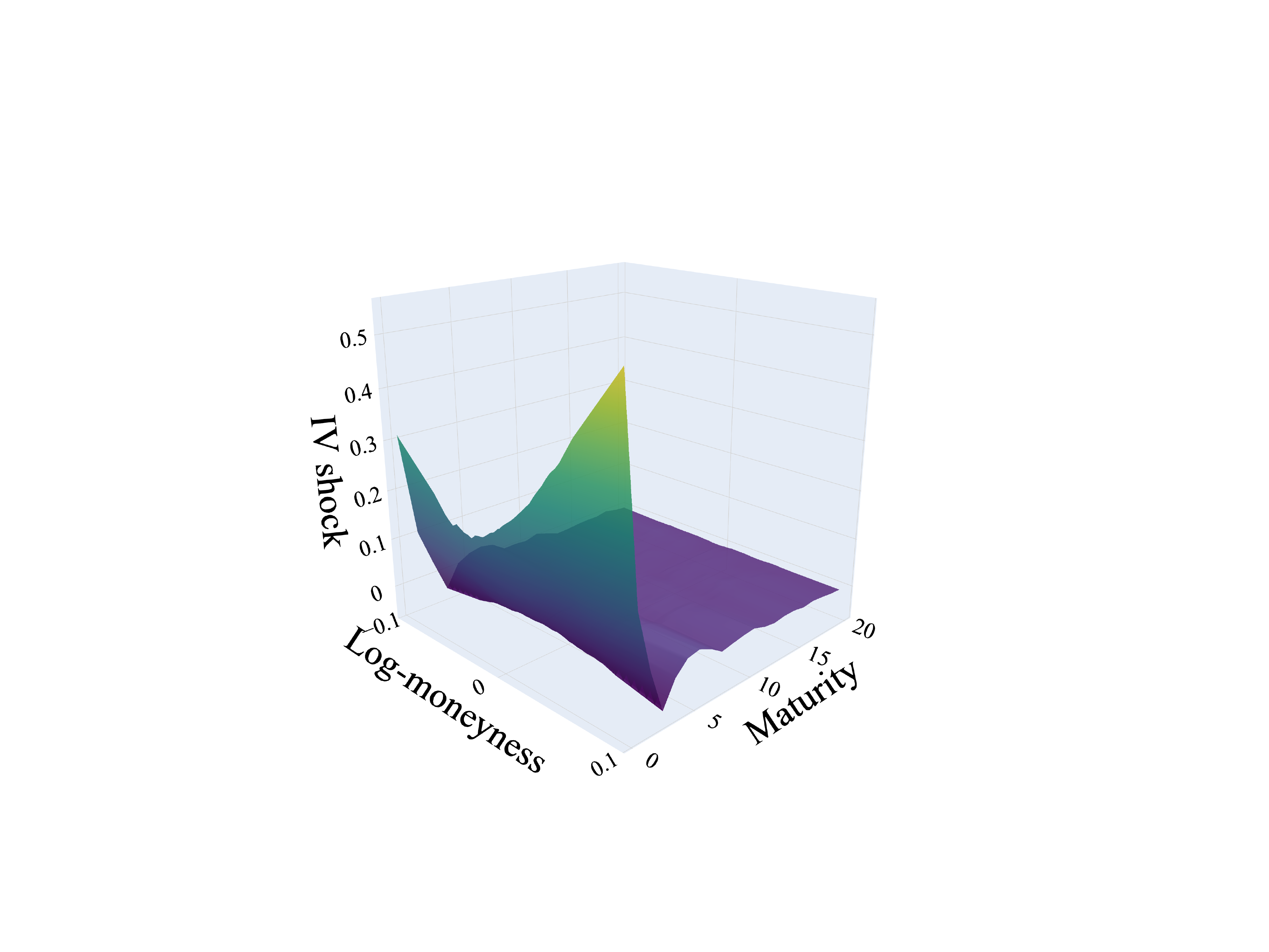}
    \caption{Out-of-sample fitted shape of the announcement shock by the model trained on SVI interpolated surface
     defined as a mean difference between IV at an announcement and a previous day for call options.}
\end{figure}

\begin{figure}[H]
    \centering
    \includegraphics[width=\linewidth]{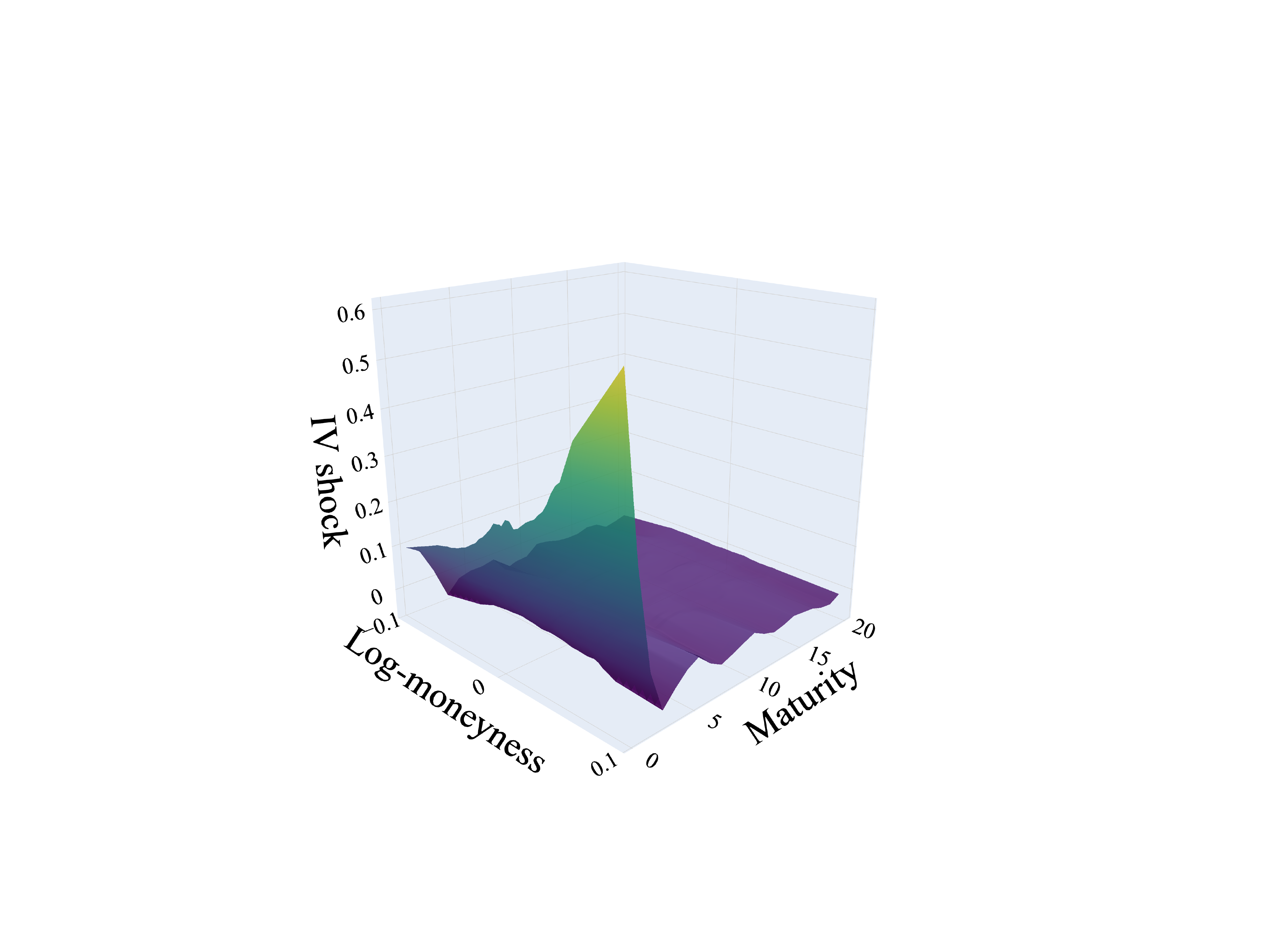}
    \caption{Out-of-sample fitted shape of the announcement shock by the model trained on linearly interpolated surface
     defined as a mean difference between IV at an announcement and a previous day for call options.}
\end{figure}

\begin{figure}[H]
    \centering
    \includegraphics[width=\linewidth]{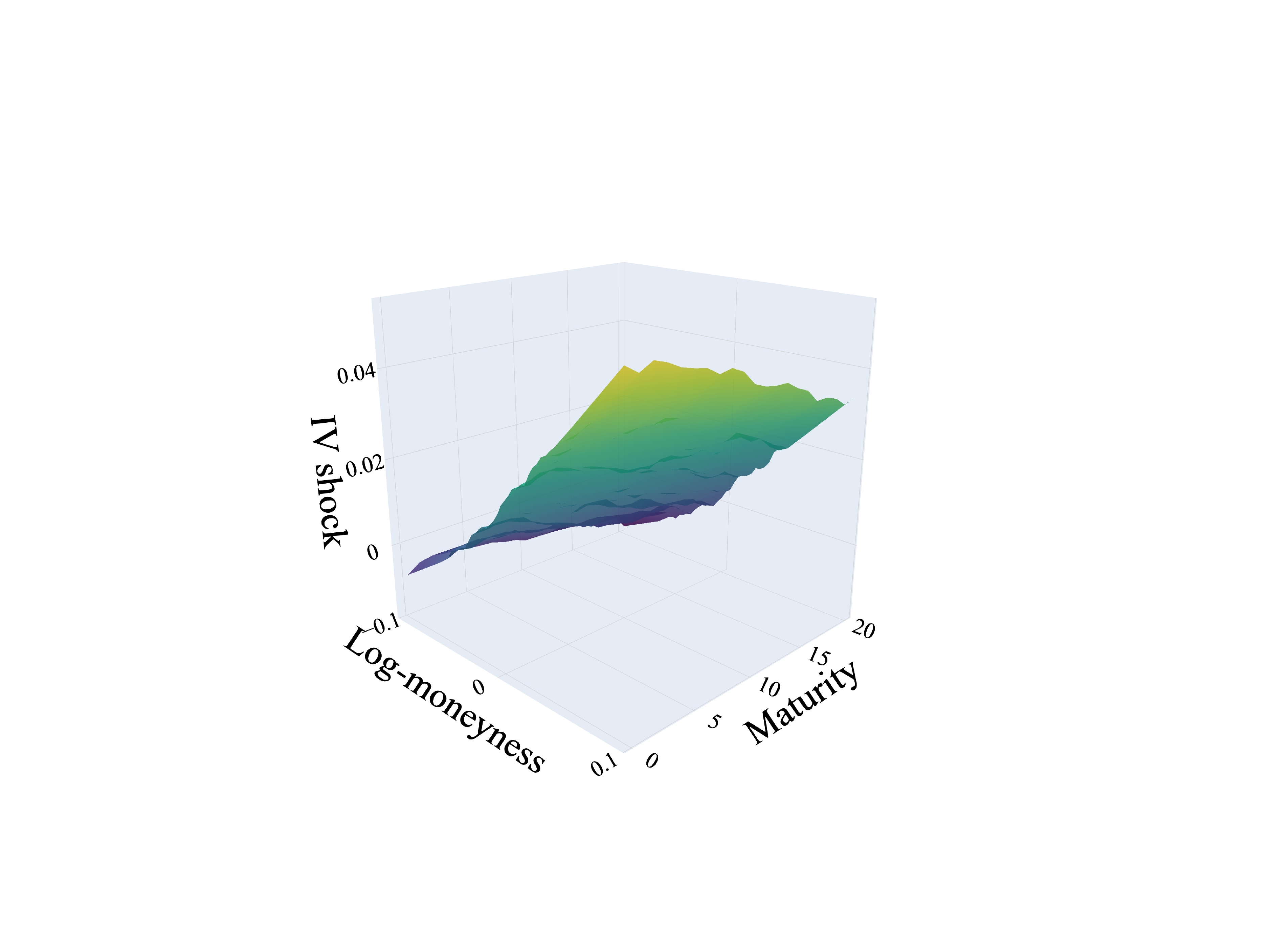}
    \caption{Out-of-sample fitted shape of the announcement shock by the model trained on AHBS interpolated surface
     defined as a mean difference between IV at an announcement and a previous day for call options.}
\end{figure}

\subsubsection{Put options}

These graphs show the fitted shape of the shock (defined in terms of $\bar{\Delta IV}$ where $\Delta IV = IV_\tau - IV_{\tau - 1}$ - 
therefore a mean difference between IV at an announcement and a previous day). 

\begin{figure}[H]
    \centering
    \includegraphics[width=\linewidth]{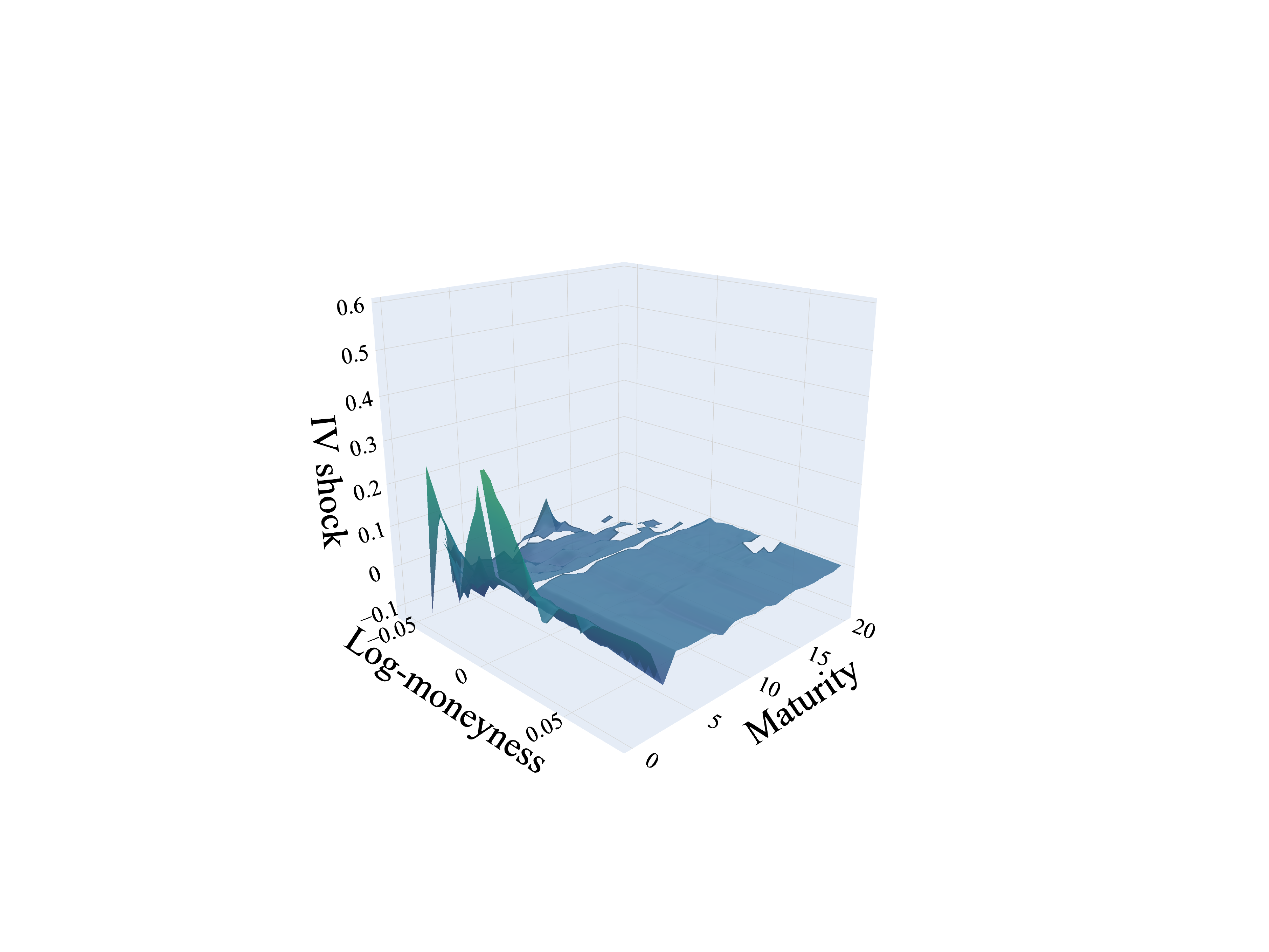}
    \caption{True announcement shock defined as a mean difference between IV at an announcement and a previous day for put options.}
\end{figure}

\begin{figure}[H]
    \centering
    \includegraphics[width=\linewidth]{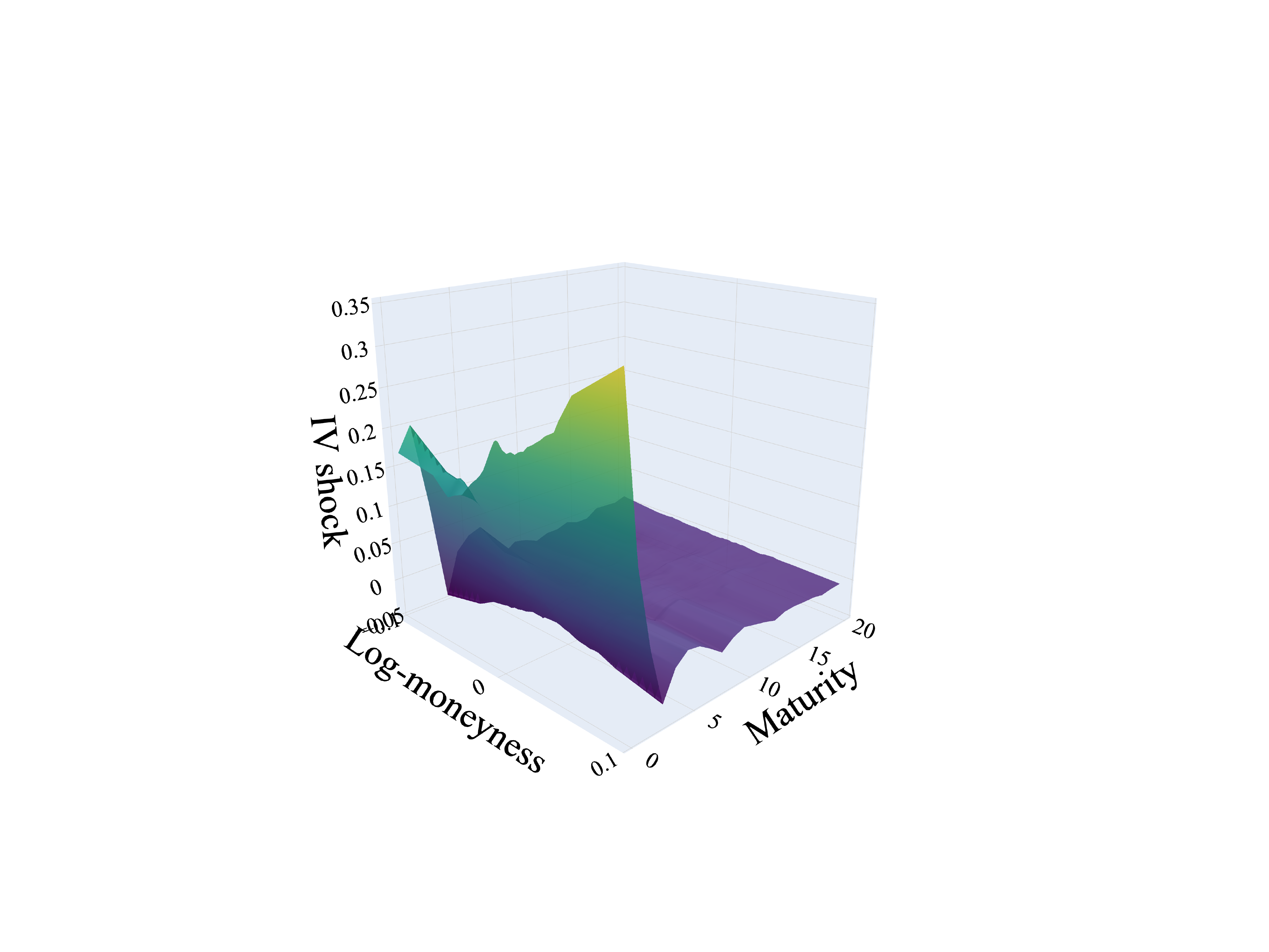}
    \caption{Out-of-sample fitted shape of the announcement shock by the model trained on SVI interpolated surface
     defined as a mean difference between IV at an announcement and a previous day for put options.}
\end{figure}

\begin{figure}[H]
    \centering
    \includegraphics[width=\linewidth]{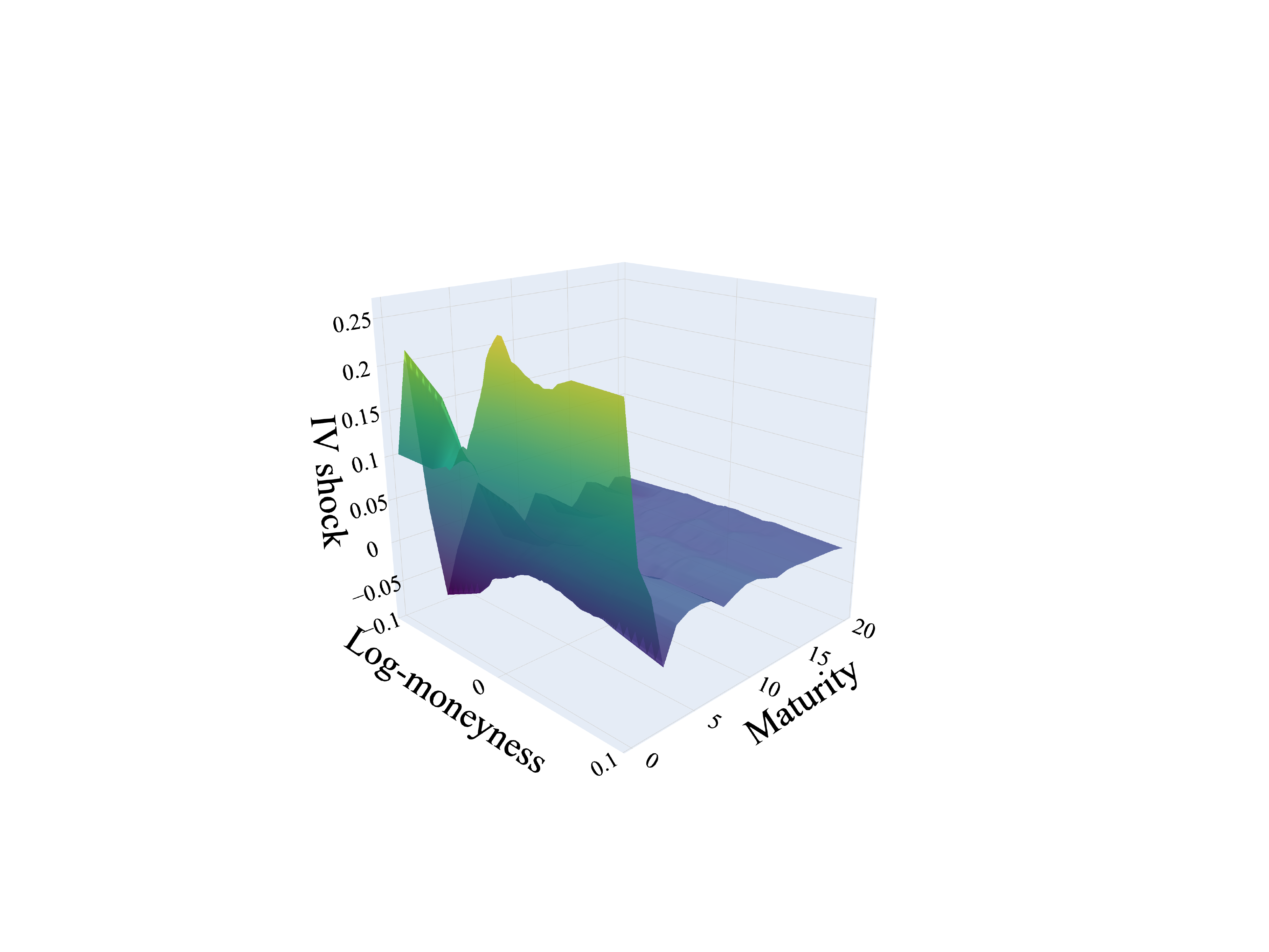}
    \caption{Out-of-sample fitted shape of the announcement shock by the model trained on linearly interpolated surface
     defined as a mean difference between IV at an announcement and a previous day for put options.}
\end{figure}

\begin{figure}[H]
    \centering
    \includegraphics[width=\linewidth]{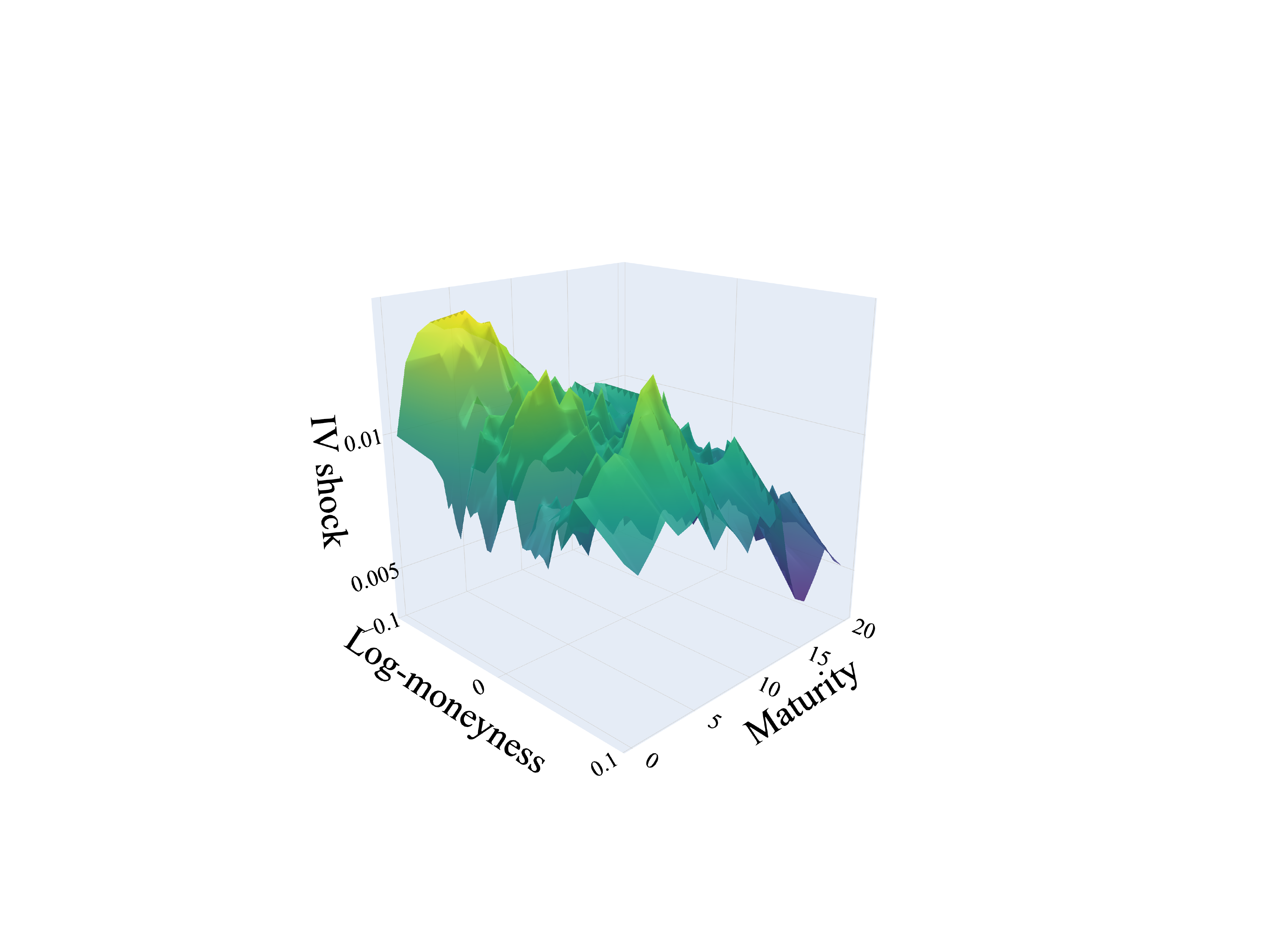}
    \caption{Out-of-sample fitted shape of the announcement shock by the model trained on AHBS interpolated surface
     defined as a mean difference between IV at an announcement and a previous day for put options.}
\end{figure}

\subsection{Impulse response functions}\label{sec:irf_call}

\subsubsection{Call options}

The average IV surface at the announcement moment is passed to the model trained on the announcement dummy, which predicts future IV based on this type of shock.
In the first case the announcement dummy is set to $1$, while in the second to $0$, which allows to take the difference between the two IRF, depicted on the graphs below.

\begin{figure}[H]
    \centering
    \includegraphics[width=\linewidth]{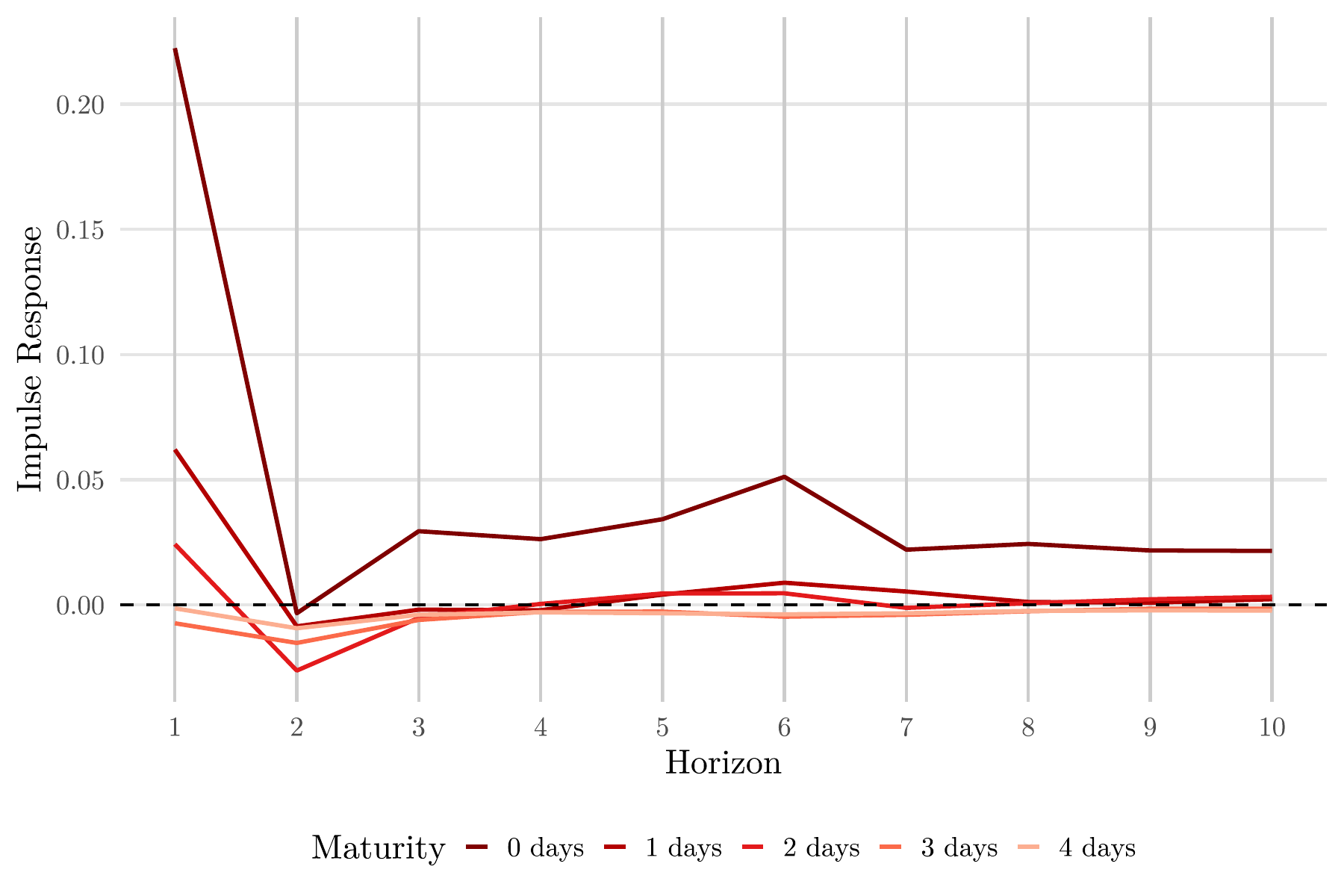}
    \caption{The difference between the announcement and non-announcement IRF for call options for the model trained on linearly interpolated surface.}
\end{figure}

\begin{figure}[H]
    \centering
    \includegraphics[width=\linewidth]{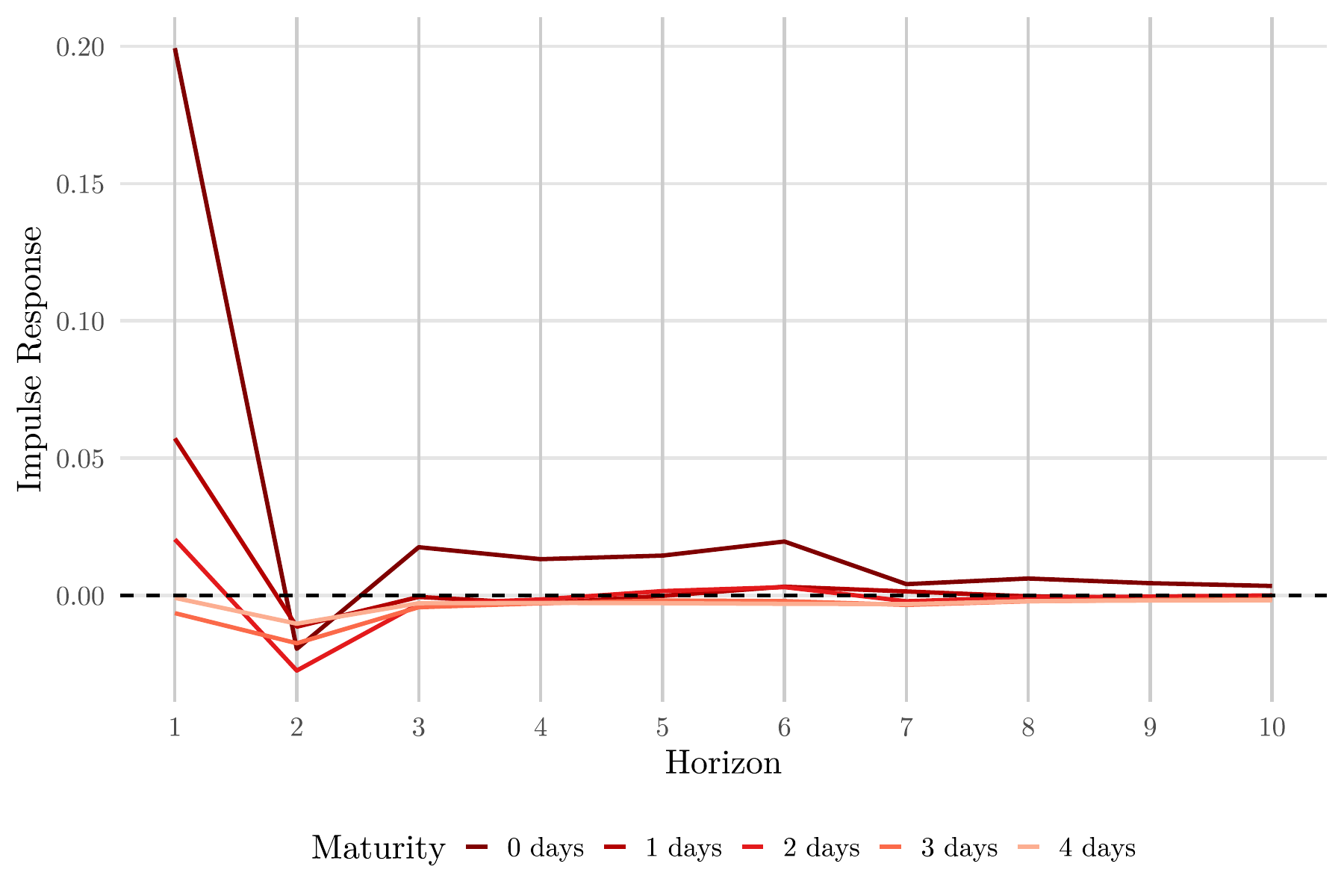}
    \caption{The difference between the announcement and non-announcement IRF for call options for the model trained on the surface interpolated by SVI.}
\end{figure}

\begin{figure}[H]
    \centering
    \includegraphics[width=\linewidth]{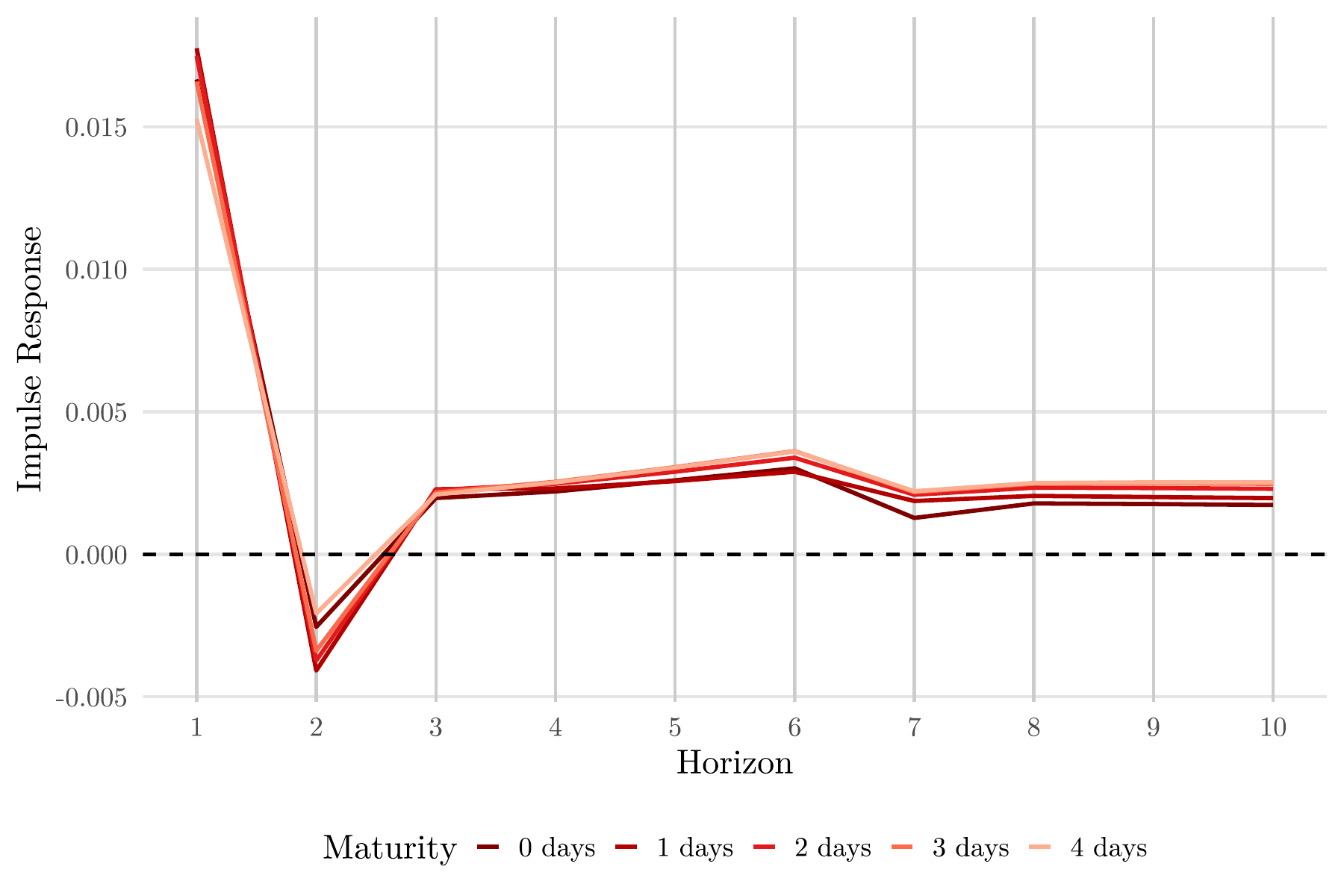}
    \caption{The difference between the announcement and non-announcement IRF for call options for the model trained on the surface interpolated by AHBS.}
\end{figure}

\subsubsection{Put options}

The average IV surface at the announcement moment is passed to the model trained on the announcement dummy, which predicts future IV based on this type of shock.
In the first case the announcement dummy is set to $1$, while in the second to $0$, which allows to take the difference between the two IRF, depicted on the graphs below.

\begin{figure}[H]
    \centering
    \includegraphics[width=\linewidth]{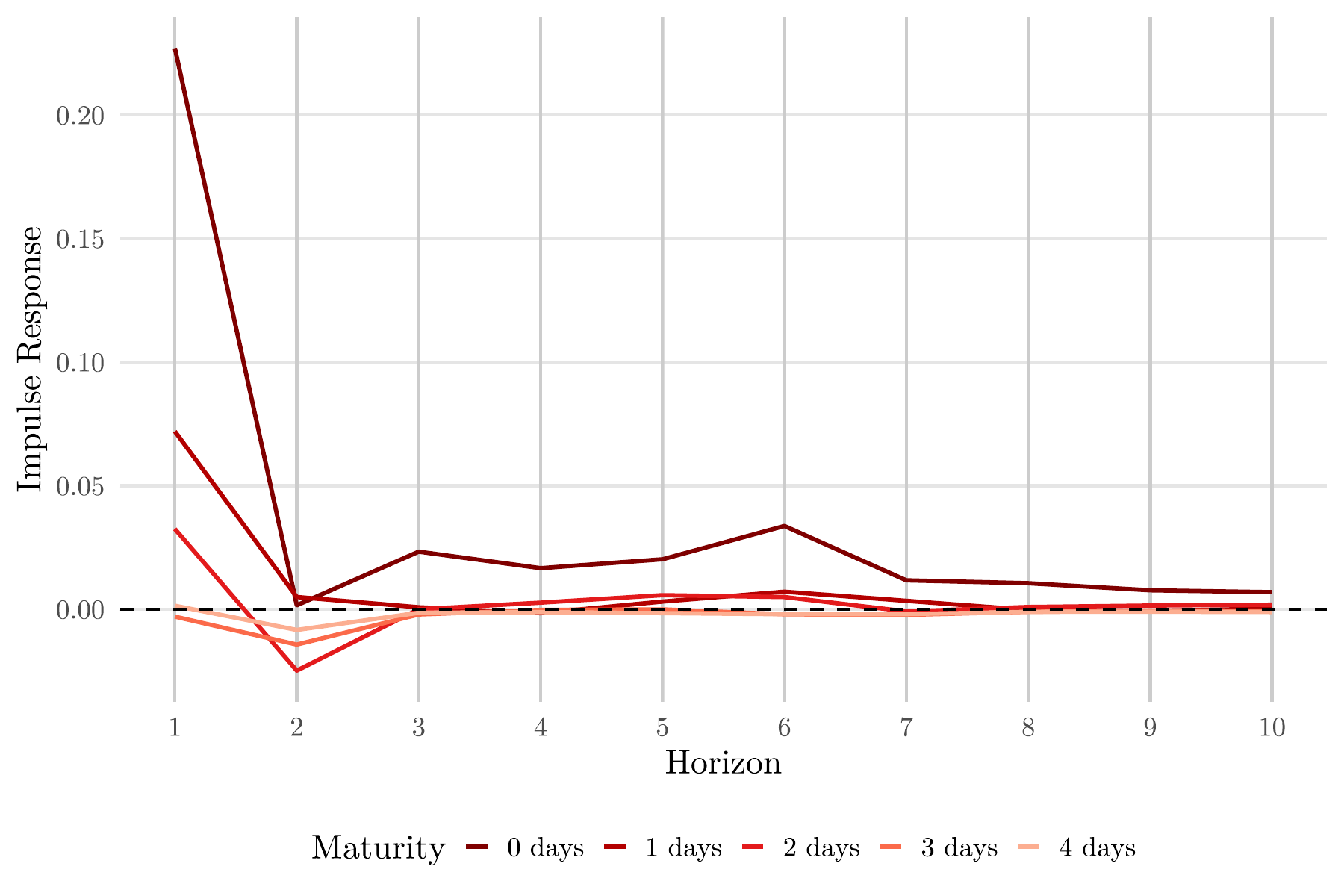}
    \caption{The difference between the announcement and non-announcement IRF for put options for the model trained on linearly interpolated surface.}
\end{figure}

\begin{figure}[H]
    \centering
    \includegraphics[width=\linewidth]{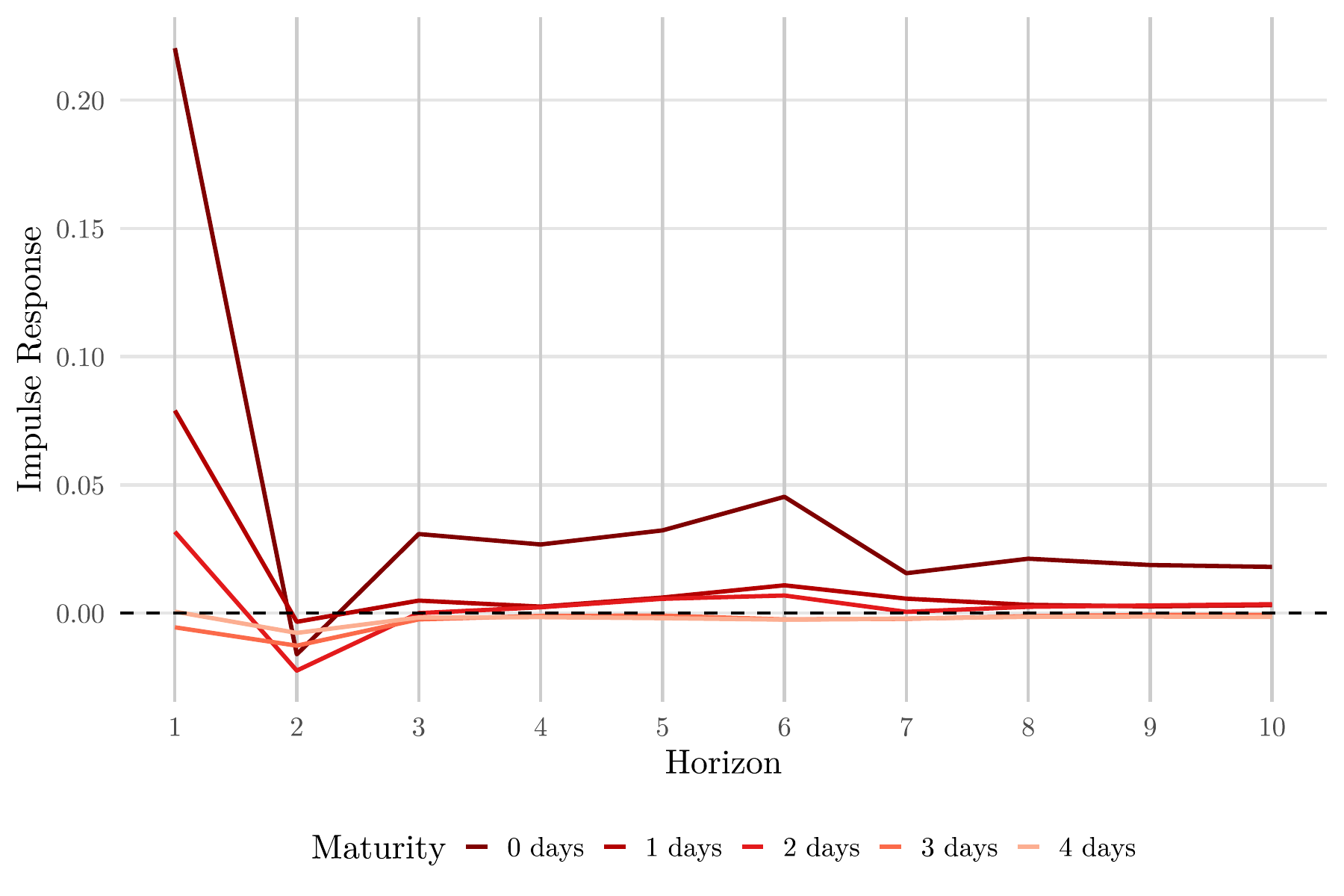}
    \caption{The difference between the announcement and non-announcement IRF for put options for the model trained on the surface interpolated by SVI.}
\end{figure}

\begin{figure}[H]
    \centering
    \includegraphics[width=\linewidth]{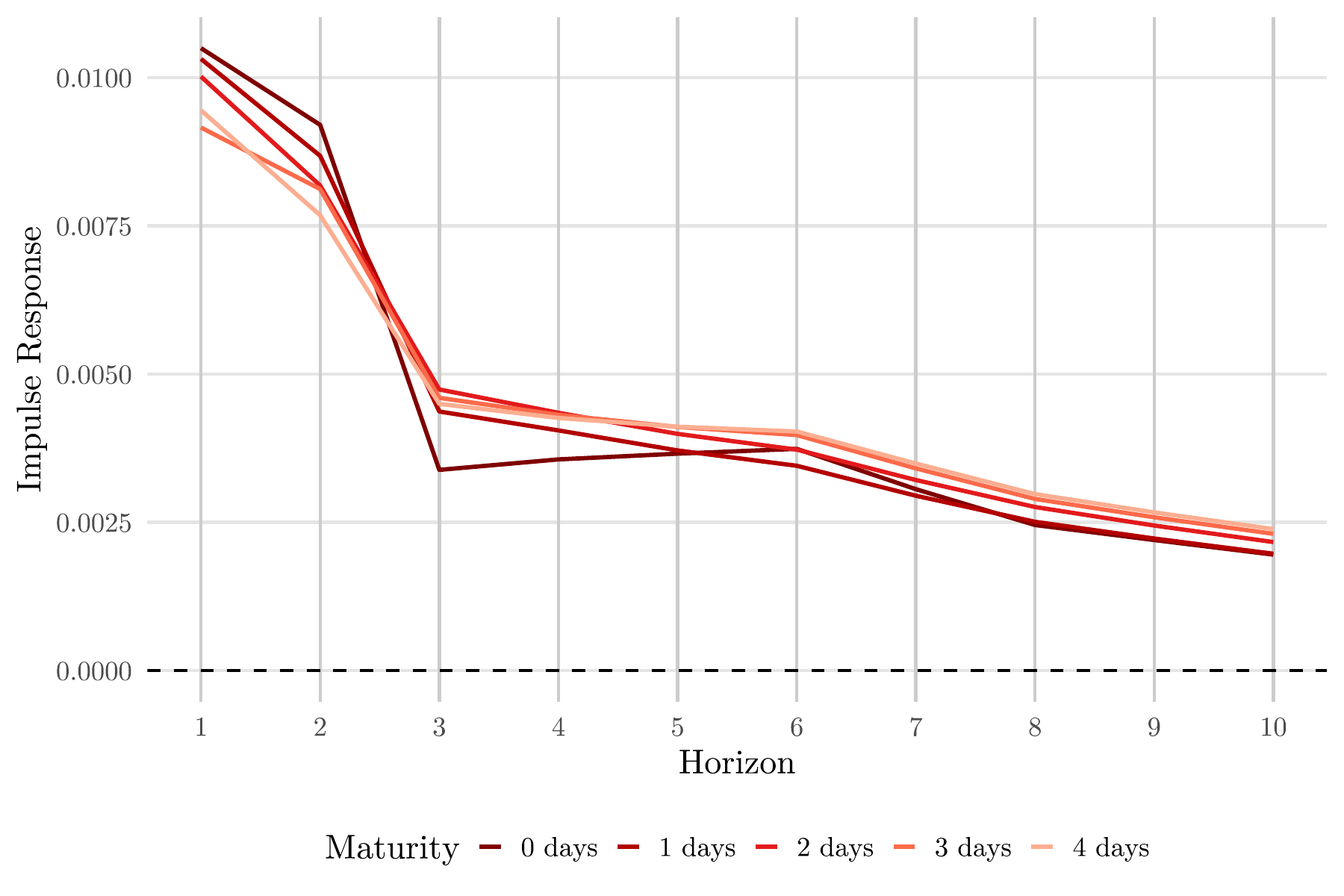}
    \caption{The difference between the announcement and non-announcement IRF for put options for the model trained on the surface interpolated by AHBS.}
\end{figure}

\newpage

\end{document}